\documentclass{article}
\usepackage{tabularx}
\usepackage{nat_comms}
\usepackage{rotating}
\usepackage{dsfont} 
\usepackage{cleveref}
\creflabelformat{equation}{#2\textup{#1}#3}
\usepackage{enumitem} 

\usepackage{pdflscape}   
\usepackage{array}       
\usepackage[utf8]{inputenc} 
\DeclareUnicodeCharacter{2009}{ } 

\usepackage[round,authoryear]{natbib}
\allowdisplaybreaks 

\title{When and where higher-resolution climate data improve impact model performance}

\author{Johanna T. Malle$^{1,2,\star}$,
Christopher P.O. Reyer$^{3}$,
Yael Amitai$^{4}$,
Andrey L. D. Augustynczik$^{5}$,
Yaron Be’eri-Shlevin$^{4}$,
Elad Ben-Zur$^{6}$,
Peter Burek$^{5}$,
Tarunsinh Chaudhari$^{7}$,
Jinfeng Chang$^{8,29}$,
Alessio Collalti$^{9}$,
Daniela Dalmonech$^{9}$,
Shouro Dasgupta$^{10,11}$,
Iulii Didovets$^{3}$,
Marc Djahangard$^{12}$,
Laura Dobor$^{13}$,
Louis Fran\c{c}ois$^{7}$,
Simon N. Gosling$^{14}$,
Fred F. Hattermann$^{3}$,
Shaoshun Huang$^{15}$,
Heike Lischke$^{2}$,
Thomas Lorimer$^{16, 26, 27}$,
Katarina Merganicova$^{13,18}$,
Francesco Minunno$^{17,19}$,
Mats Nieberg$^{3,20,21}$,
Elizabeth J. Z. Robinson$^{11}$,
Martin Schmid$^{16}$,
Mikhail Smilovic$^{5, 28}$,
Ritika Srinet$^{17,19}$,
Elia Vangi$^{9}$,
Xue Yang$^{22}$,
Rasoul Yousefpour$^{12,23}$,
Ana I. Ayala$^{25}$,
Daniel Mercado\mbox{-}Bettin$^{24}$,
D\'{a}nnell Quesada-Chac\'{o}n$^{3}$,
Dirk N. Karger$^{2}$\\
\begin{flushleft}
$^{1}$ \small{Department of Evolutionary Biology and Environmental Studies, University of Zurich, Zurich, Switzerland}\smallskip\\
$^{2}$ \small{Dynamic Macroecology, Land Change Science, Swiss Federal Institute for Forest, Snow and Landscape Research WSL, Birmensdorf, Switzerland}\smallskip\\
$^{3}$ \small{Potsdam Institute for Climate Impact Research (PIK), Member of the Leibniz Association, Potsdam, Germany}\smallskip\\
$^{4}$ \small{Kinneret Limnological Institute, Israel Oceanographic \& Limnological Research, Israel}\smallskip\\
$^{5}$ \small{Water Security Research Group, Biodiversity and Natural Resources Program, International Institute for Applied Systems Analysis (IIASA), Laxenburg, Austria}\smallskip\\
$^{6}$ \small{Institute of Earth Sciences, Hebrew University of Jerusalem, Jerusalem, Israel}\smallskip\\
$^{7}$ \small{Department of Astrophysics, Geophysics and Oceanography (AGO), SPHERES Research Unit, University of Li\`ege, Li\`ege, Belgium}\smallskip\\
$^{8}$ \small{State Key Laboratory of Soil Pollution Control and Safety, College of Environmental and Resource Sciences, Zhejiang University, Hangzhou 310058, China}\smallskip\\
$^{9}$ \small{Forest Modelling Lab., Institute for Agriculture and Forestry Systems in the Mediterranean, National Research Council of Italy (CNR-ISAFOM), Perugia 06128, Italy}\smallskip\\
$^{10}$ \small{Centro Euro-Mediterraneo sui Cambiamenti Climatici (CMCC), Venice, Italy}\smallskip\\
$^{11}$ \small{Grantham Research Institute on Climate Change and the Environment, London School of Economics and Political Science (LSE), London, UK}\smallskip\\
$^{12}$ \small{Faculty of Environment and Natural Resources, University of Freiburg, Freiburg, Germany}\smallskip\\
$^{13}$ \small{Faculty of Forestry and Wood Sciences, Czech University of Life Sciences, Prague, Czech Republic}\smallskip\\
$^{14}$ \small{School of Geography, University of Nottingham, Nottingham, UK}\smallskip\\
$^{15}$ \small{The Norwegian Water Resources and Energy Directorate (NVE), Oslo, Norway}\smallskip\\
$^{16}$ \small{Surface Waters – Research and Management, Eawag: Swiss Federal Institute of Aquatic Science and Technology, Kastanienbaum, Switzerland}\smallskip\\
$^{17}$ \small{Institute for Atmospheric and Earth System Research (INAR) \& Faculty of Agriculture and Forestry, University of Helsinki, Helsinki, Finland}\smallskip\\
$^{18}$ \small{Department of Biodiversity of Ecosystems and Landscape, Institute of Landscape Ecology, Slovak Academy of Sciences, Slovakia}\smallskip\\
$^{19}$ \small{Forest Modelling, Yucatrote}\smallskip\\
$^{20}$ \small{European Forest Institute, Bonn, Germany}\smallskip\\
$^{21}$ \small{Chair of Forest Growth and Woody Biomass Production, TU Dresden, Tharandt, Germany}\smallskip\\
$^{22}$ \small{State Key Laboratory of Water Engineering Ecology and Environment in Arid Area, Xi'an University of Technology, Xi'an, China}\smallskip\\
$^{23}$ \small{John H. Daniels Faculty of Architecture, Landscape, and Design, University of Toronto, Toronto, Canada}\smallskip\\
$^{24}$ \small{Spanish National Research Council, Centre for Advanced Studies Blanes, Blanes, Spain}\smallskip\\
$^{25}$ \small{Limnology Unit, Department of Ecology and Genetics, Uppsala University, Uppsala, Sweden}\smallskip\\
$^{26}$ \small{Laboratory of Molecular and Behavioral Neuroscience, Institute for Neuroscience, Department of Health Sciences and Technology, ETH Zurich, Switzerland}\smallskip\\
$^{27}$ \small{ETH Zurich 3R Hub, ETH Zurich, Switzerland}\smallskip\\
$^{28}$ \small{Institute of Environmental Engineering, ETH Zurich, 8093 Zürich, Switzerland}\smallskip\\
$^{29}$ \small{International Institute for Applied Systems Analysis (IIASA), Laxenburg, Austria}\smallskip\\

$^{\star}$ Corresponding author, email: \href{mailto:johanna.malle@uzh.ch}{\texttt{johanna.malle@uzh.ch}}
\end{flushleft}
}

\date{\today}

\begin{document}

\maketitle

\abstract{Climate impact assessments increasingly rely on high-resolution climate and forcing datasets, under the premise that finer detail enhances both the accuracy and the policy relevance of projections. Systematic evaluations of when and where higher resolution data improve model outcomes remain limited, and it is still unclear whether increasing spatial resolution consistently enhances climate impact model performance across application areas, regions, and forcing variables. Here we show that improvements in climate input accuracy and impact model performance are most pronounced when moving from coarse (60 km) to intermediate (10 km) resolution, while further refinement to 3 km and 1 km provides more modest and inconsistent benefits. Using the cross-sectoral model simulations from the Inter-Sectoral Impact Model Intercomparison Project (ISIMIP), we demonstrate that higher resolution substantially improves model skill in temperature-sensitive impact models and topographically complex regions, whereas precipitation-driven and low-relief systems show less consistency to increase performance with resolution. For temperature, both climate inputs and model outputs improved most strongly at the 60 km → 10 km transition, with diminishing gains at finer scales. A similar result emerged for precipitation, although some models even exhibited reduced performance when resolution increased beyond 10 km. These results highlight that optimal resolution depends on sectoral and regional context, and point to the need for improving model process representation and downscaling techniques so that added spatial detail can translate into meaningful performance gains. For data providers, this implies prioritizing investments in resolutions that maximize improvements where they matter most, while for modeling groups and users, it underscores the need for explicit benchmarking of resolution choices. More broadly, this work advances the design of consistent, efficient, and policy-relevant multi-sectoral climate impact assessments by clarifying when high-resolution data meaningfully enhance outcomes.}


\maketitle

\section{Introduction}\label{sec:introduction}

Climate change is progressing rapidly and poses serious risks to both natural ecosystems and human societies \citep{Lee2023,Chen2021}. To assess these risks, climate impact models are widely applied across multiple sectors, including agriculture \citep{Jagermeyr2021}, forestry \citep{Reyer2014, Lawrence2022}, terrestrial biodiversity \citep{Habibullah2022,Cramer2001}, hydrology \citep{Goderniaux2011,Woolway2021,Wang2025}, energy \citep{Isaac2009,Zapata2022}, water quality \citep{Jones2025}, labour \citep{Dasgupta2021} and human health \citep{Smith2024,Caminade2014}. These models are inherently complex, integrating a wide range of biophysical and socio-economic processes. As a result, they carry substantial uncertainty, which is further amplified when driven by inherently complex future climate scenarios.

To address and disentangle these uncertainties and enhance the comparability of results, international initiatives such as the Inter-Sectoral Impact Model Intercomparison Project (ISIMIP) \citep{Frieler2017} provide harmonized protocols and standardized input datasets to facilitate inter-model and cross-sectoral comparisons. This approach is inspired by practices from the climate modeling community, where coordinated model intercomparison projects like CMIP \citep{Meehl2000,Meehl2007} have long used ensembles of simulations to assess uncertainty and model performance \citep{Eyring2016}. However, truly interdisciplinary climate impact assessments that integrate models from multiple sectors under common forcing conditions are still in their infancy \citep{Frieler2024,Schipper2021,Harrison2016}, despite increasing evidence that climate impacts are interconnected and often occur simultaneously across different ecosystems \citep{Zscheischler2018,Ridder2020,Ciscar2019,Pfenning-Butterworth2024}.

An important but often overlooked source of uncertainty in climate impact modeling stems from the discrepancy between the resolution of climate input data and the spatial scale at which impacts occur. Many impact models operate at coarse spatial resolutions (e.g., 0.5° or ~50 km at the equator), which can mask fine-scale climatic variability, particularly in heterogeneous or mountainous terrain. This mismatch can significantly affect simulations of components of ecosystem dynamics such as water fluxes \citep{Wada2016}, forest carbon fluxes \citep{Harris2021}, growing seasons \citep{McMaster1997}, and species migration dynamics \citep{Engler2009,Zani2023}. For example, climate data averaged over large grid cells may lead to inaccurate estimates of local temperature and precipitation, which are critical drivers of ecological processes.

Several sector-specific studies have already demonstrated that the spatial resolution of climate forcing data can substantially influence model outcomes, including for agriculture \citep{Mearns2001}, biodiversity \citep{Randin2009,Seo2009,Koenig2021}, vegetation dynamics \citep{Hickler2012}, hydrology \citep{Hattermann2017, Kumar2022, Aerts2022}, including snow-specific processes \citep{Barnhart2024, Magnusson2019}, and climate variability \citep{Karger2020c}. However, a comprehensive, cross-sectoral analysis of spatial-scale effects using consistent climate input data has been lacking. A major barrier to such studies is the inconsistent application and, in some cases, limited availability of high-resolution climate forcing data across impact sectors and geographic regions. Models are often run only at one resolution or rely on regionally tuned downscaling approaches that vary in methodology and coverage \citep{Wood2004, Keller2022}.

To address this gap, globally standardized, high-resolution climate datasets are needed. While coarse-resolution standardized data products are already available and widely used, for example within the ISIMIP framework \citep{Lange2019,Cucchi2020}, kilometer-scale standardized datasets are still scarce due to the computational challenges of running global climate models at such fine resolution \citep{Dipankar2015,Schaer2019}. Current high-resolution simulations are limited by throughput constraints of even the most powerful supercomputers \citep{Fuhrer2018,Schulthess2018,Neumann2019, Bottazzi2024}. Available datasets and experimental setups are often restricted in spatial extent \citep{Sun2021,Bano-Medina2022, Raffa2023}, limiting comprehensive analyses of spatial-scale effects.

Substantial research has addressed the challenge of reducing computational cost and runtime for downscaling. Feasible alternatives to the most computationally expensive way of obtaining high-resolution climate information from large-scale models (dynamic downscaling) include statistical downscaling and interpolation of observational and reanalysis data \citep{Tabios1985,Daly1997,Thornton1997, Reder2025}. However, observation-based interpolation methods often suffer from spatially uneven station densities \citep{Briggs1996,Schneider2014,Kidd2017,Berndt2018}, which can distort model outputs. Downscaling reanalysis data using methods that correct for terrain and bias can offer a more consistent and scalable solution \citep{Karger2017,Karger2020b,Karger2020c}.

\begin{figure}[htbp]
\centering
\includegraphics[width=\textwidth]{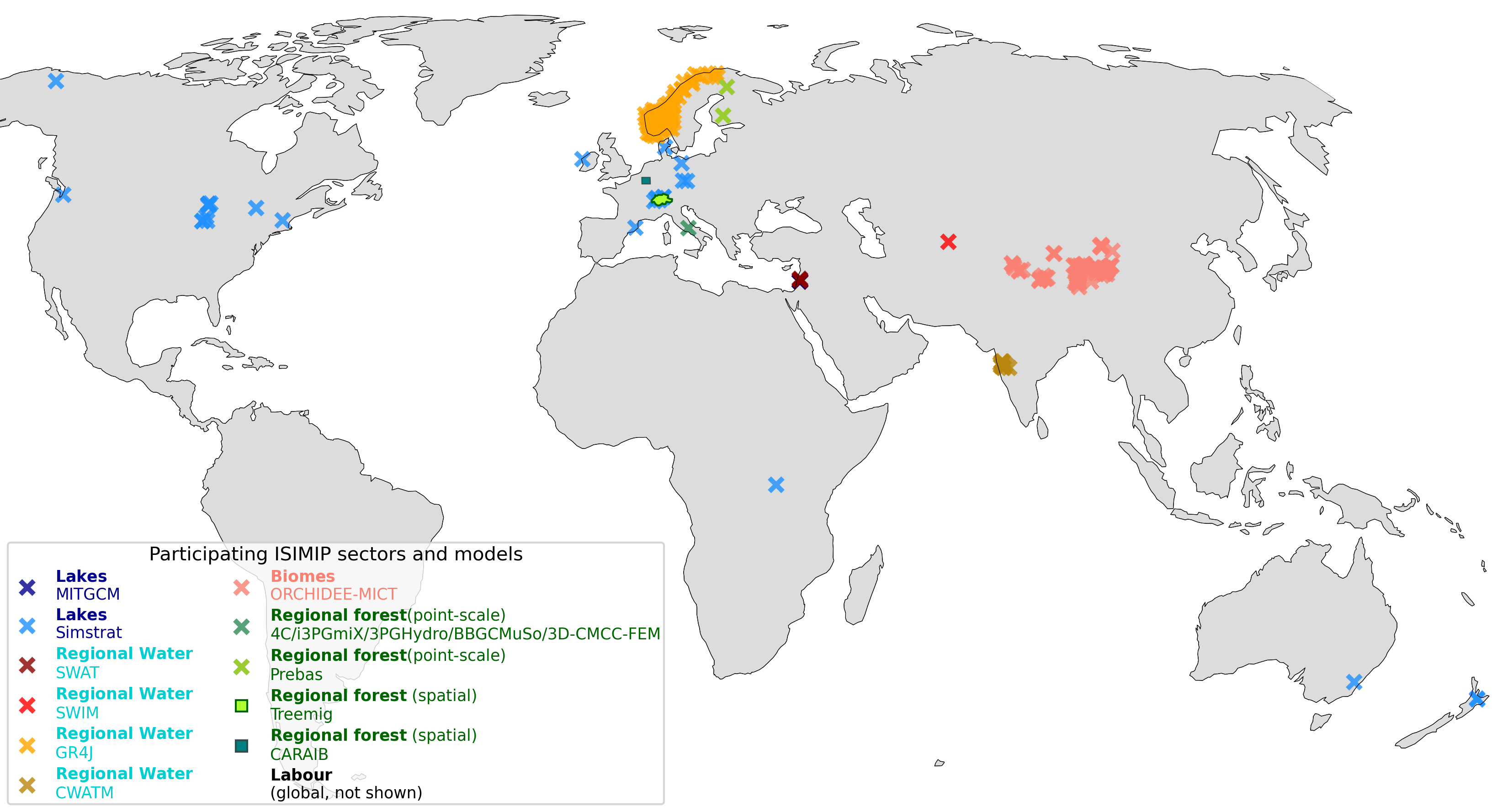} 
\caption{Overview of participating ISIMIP sectors and models, and location/geographic extent of the performed simulations.}
\label{fig:spatial_extent}
\end{figure}

In this study, we investigate how and to what extent the spatial resolution of climate forcing data affects the performance of climate impact models. Within the ISIMIP3a framework, we conducted standardized simulations using the CHELSA-W5E5 v1 climate data \citep{Karger2023} at four spatial resolutions (30$^{\prime\prime}$, 90$^{\prime\prime}$, 300$^{\prime\prime}$, and 1800$^{\prime\prime}$, corresponding to roughly 1, 3, 10, and 60 km) across 16 impact models from five different sectors (regional forest, regional water, lakes, biomes, and labour) with each model set up according to its sector-specific ISIMIP3a protocol \citep{Frieler2024}. A map illustrating the geographic extent of impact model simulations across different regions and sectors is provided in Figure~\ref{fig:spatial_extent}. For each model and sector, we assess both the accuracy of the climate forcing data (precipitation and temperature) and the agreement between model outputs and observational data across the range of resolutions, with particular focus on how terrain complexity influences the results. This work contributes to the broader goal of enhancing transparency, comparability, and robustness in climate impact assessments, and aims to provide practical guidance on the scientific value of high-resolution forcing data.

\clearpage

\section{Results}\label{sec3}
\subsection{Evaluation of CHELSA W5E5 across resolutions}

\begin{figure}[htbp]
\centering
\includegraphics[width=\textwidth]{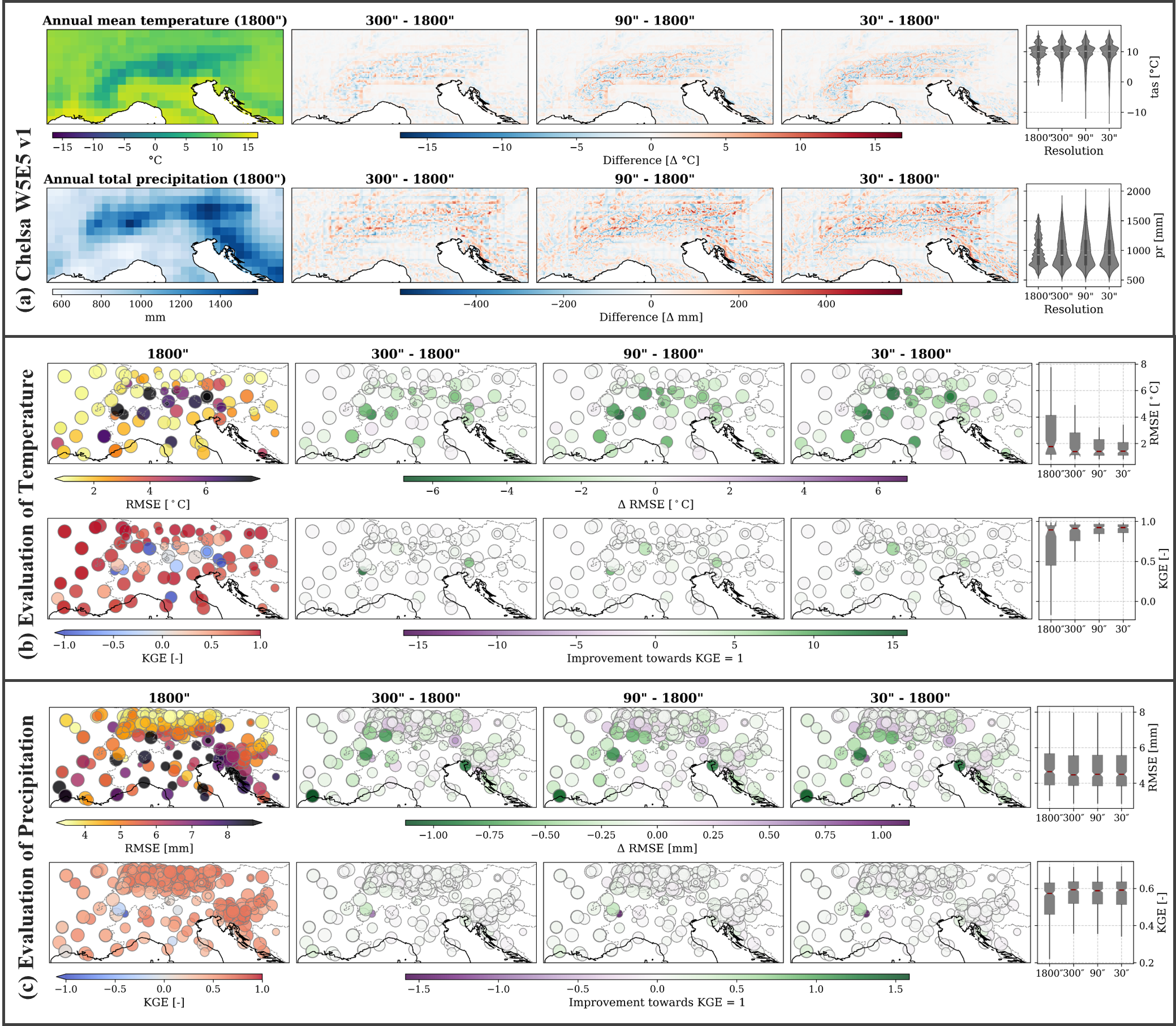} 
\caption{Evaluation of CHELSA-W5E5 v1 climate forcing data at multiple spatial resolutions over the European Alps.
Panel (a) shows CHELSA-W5E5 v1 temperature and precipitation at four spatial resolutions. The first column displays absolute values at 1800$^{\prime\prime}$  resolution; subsequent columns show differences relative to the 1800$^{\prime\prime}$ baseline. Mountainous regions, such as the European Alps, are apparent in the first column of (a) as areas with lower temperatures (bluer shading) and higher precipitation (darker blue shading). Panels (b–c) show evaluation against GHCN-D station observations, with dot size indicating the number of daily measurements available per station. The first column presents absolute error metrics for 1800$^{\prime\prime}$  resolution; the remaining columns show relative changes in error compared to this baseline. Panel (b) corresponds to temperature, and panel (c) to precipitation. For KGE in the relative-change panels of (b) and (c), the color scale shows changes in the absolute KGE value (i.e. movement toward or away from the ideal value of 1).}
\label{fig:chelsa_eval}
\end{figure}

We assessed the accuracy of the CHELSA-W5E5 v1 climate forcing dataset across four spatial resolutions by benchmarking it against station observations from the Global Historical Climatology Network Daily (GHCN-D) dataset \citep{Menne2018}. As a representative case, Fig.~\ref{fig:chelsa_eval} illustrates model performance over the European Alps, a region characterized by pronounced topographic heterogeneity.
For near-surface temperature, increasing spatial resolution leads to a systematic and spatially coherent reduction in model error, particularly in mountainous terrain. This highlights the substantial value of high-resolution climate forcing in capturing local temperature variability. In contrast, improvements in precipitation are more nuanced. While root mean square errors (RMSE) tend to decline slightly from the coarsest (1800$^{\prime\prime}$) to the finest (30$^{\prime\prime}$) resolution, mean Kling-Gupta Efficiency (KGE) scores increase marginally, and the gains are neither consistent across sites nor uniformly positive. 

\subsection{Impact model performance across resolutions}
\begin{figure}[htbp]
\centering
\includegraphics[width=\textwidth]{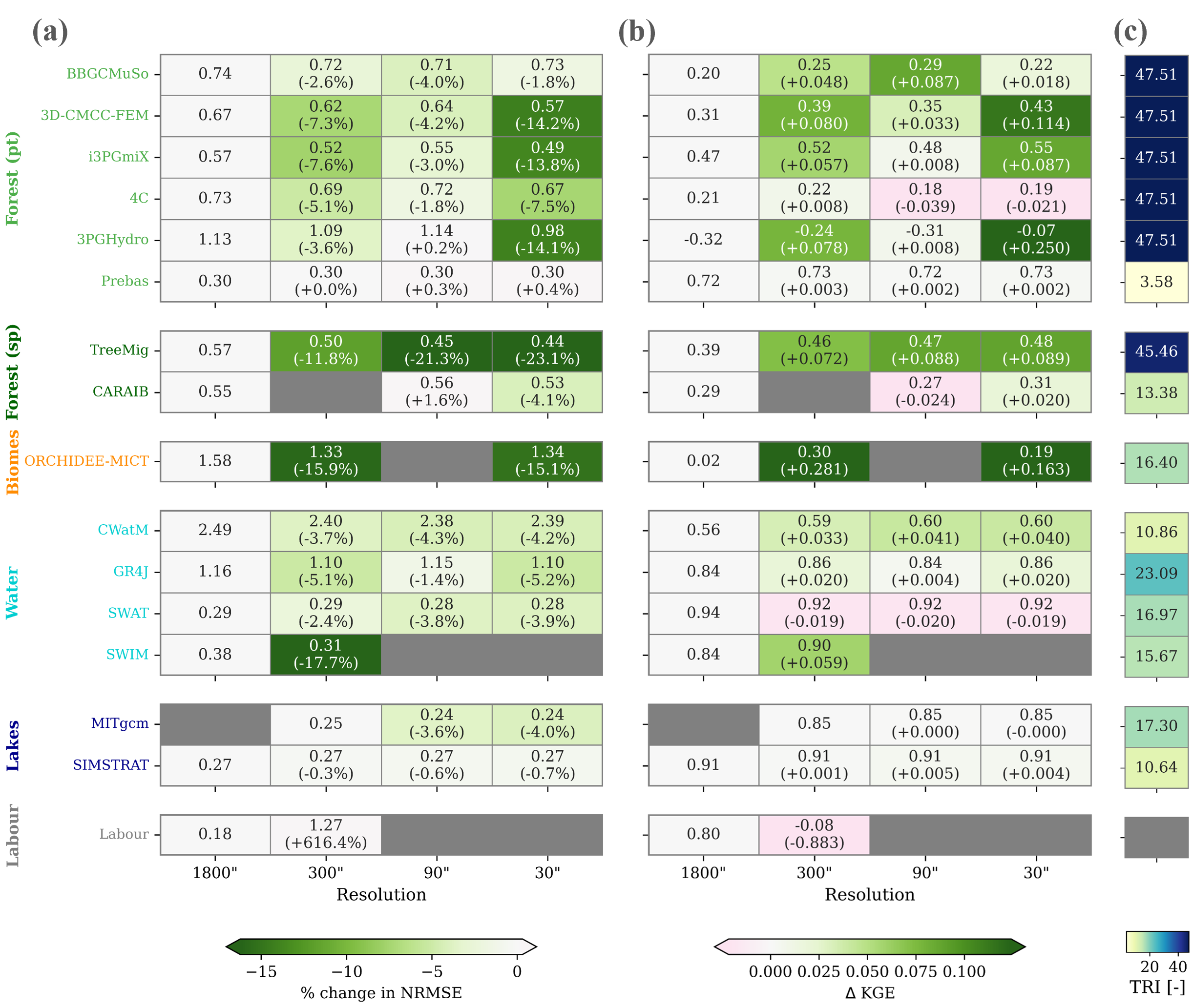} 
\caption{
Comparison of NRMSE (interquartile range–normalized) and KGE across spatial resolutions for all participating ISIMIP sectors and models using sector-specific variables and sector-specific temporal scales (daily, monthly, or annual) for the evaluation (see Table~\ref{tab:model_overview_part2}). (a) NRMSE and (b) KGE results from impact model simulations, each evaluated against observational reference data using sector-specific variables. 
For NRMSE, cell colors indicate the percentage difference in error relative to the coarse 1800$^{\prime\prime}$ ($\sim$60\, km) resolution (green indicates improvement), while absolute error values are displayed within each cell. For KGE, cell colors indicate the change in KGE relative to the coarse 1800$^{\prime\prime}$ resolution, with negative values indicating worse performance and positive values indicating improved performance. Colorbar limits are clipped to the 95th percentile; triangular shaped colorbar indicates values exceeding the scale. Panel (c) shows the mean topographic complexity at each model’s evaluation locations. We show the Terrain Ruggedness Index (TRI), which quantifies surface heterogeneity, with higher values (blue on colormap) indicating more complex terrain around the evaluation sites. As the labour sector model was evaluated globally, no TRI was computed.
}
\label{fig:model_eval}
\end{figure}

When analyzing model performance across all sectors, variables, and locations, we find that increasing the spatial resolution of climate forcing generally improves the accuracy of the impact model (Fig. \ref{fig:model_eval}). The most substantial improvements are observed in models from the forest and biome sectors, reflecting their increased sensitivity to temperature variability and their application in regions of greater topographic complexity (Fig. \ref{fig:model_eval}c). In contrast, models in the lakes and water sectors show more modest gains, likely due to their modelling domains in regions of smaller topographic complexity. Water-sector models also rely more heavily on precipitation input, whose accuracy shows less consistent improvement with increasing resolution (as noted in Fig.~\ref{fig:chelsa_eval}).

Among all models evaluated, the spatially explicit forest dynamics model TreeMig demonstrates the greatest reduction in normalized root mean square error (NRMSE) with finer resolution, underlining the value of high-resolution climate inputs in complex terrain and suggesting that landscape-scale simulations can better exploit high-resolution information than point-scale models. In contrast, the socioeconomic labour model shows reduced performance at intermediate (300$^{\prime\prime}$) resolution compared to the coarsest (1800$^{\prime\prime}$). This decline is primarily attributable to data limitations: observed labour supply data are unavailable at fine spatial scales, preventing robust validation. To enable at least partial evaluation, we used the GHCN-D station data as a proxy, spatially linking the climate records to the second administrative level of the labour supply dataset which are obtained from household and labour force surveys.

Lastly, panel (c) in Fig.~\ref{fig:model_eval} indicates a relationship between terrain complexity and model improvement, suggesting that regions with more pronounced topographic variation benefit the most from increased spatial resolution in climate forcing data.

\subsubsection*{Illustrative example: resolution effects in a forest dynamics model}

\begin{figure}[htbp]
\centering
\includegraphics[width=\textwidth]{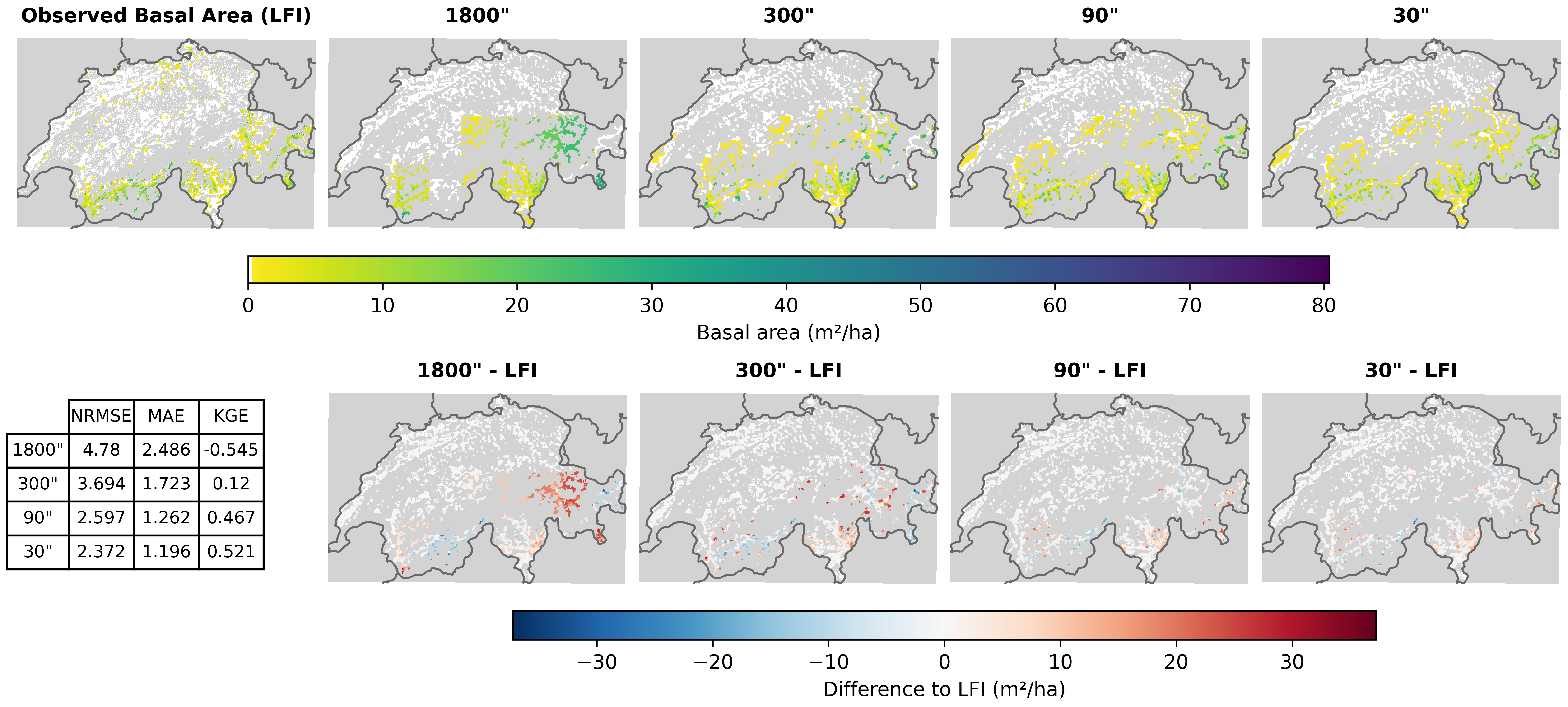} 
\caption{
Example of spatial results for the tree species Larix decidua simulated with the TreeMig model from the forest sector. Simulated basal area from TreeMig is compared to observational data from the national forest inventory (LFI) across the four spatial resolutions (1800$^{\prime\prime}$, 300$^{\prime\prime}$, 90$^{\prime\prime}$, and 30$^{\prime\prime}$). Results show spatial variation, with notable improvements in simulation accuracy at higher resolutions (see NRMSE, MAE and KGE metrics in the lower left corner) - particularly in the eastern mountainous regions. White in the first row indicates species absence; in the second row white indicates no difference between simulated and observed basal area. Grey in both rows indicates missing LFI data.  
}
\label{fig:treemig_example}
\end{figure}

To illustrate how these resolution effects manifest at the model and species level, we focus on the forest model \textit{TreeMig}, which simulates basal area per tree species (evaluated against plot-level observations from the Swiss National Forest Inventory, LFI; see Supplementary Material A and B). A spatial comparison for one predominant species (\textit{Larix decidua}) is shown in Figure~\ref{fig:treemig_example}, illustrating substantial differences in model output across spatial resolutions. This example highlights that results aggregated across multiple locations and/or variables may obscure resolution-dependent differences that are more evident at finer spatial or species-specific scales.

\subsection{Impact model performance vs climate accuracy}
\begin{figure}[htbp]
\centering
\includegraphics[width=\textwidth]{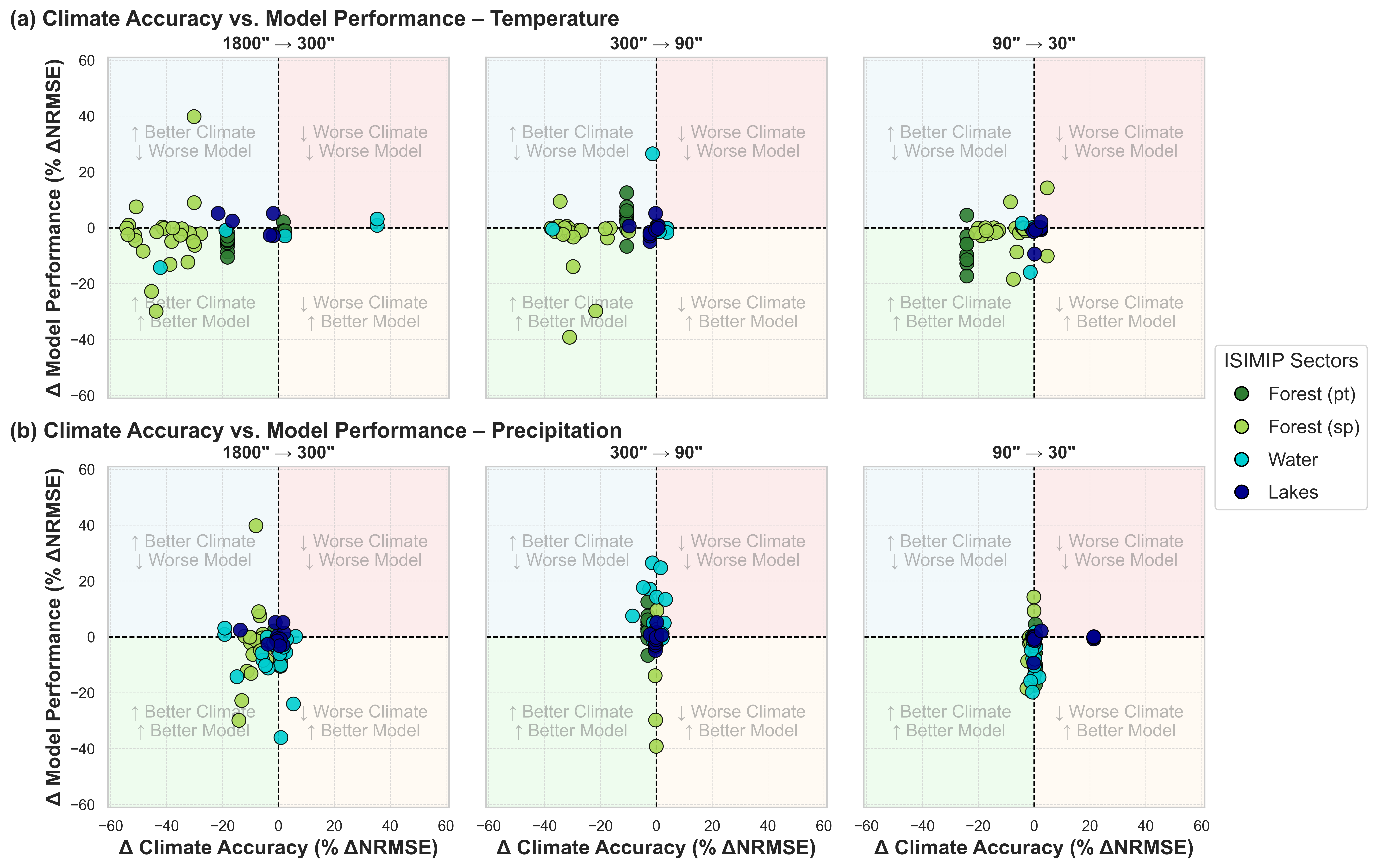} 
\caption{
Per-sector comparison of percentage changes in NRMSE for model performance and climate accuracy across spatial resolutions at all model evaluation locations where climate accuracy reference data is available. Panel (a) uses GHCNd temperature station data, and panel (b) uses GHCNd precipitation station data for climate accuracy assessment. In both panels, the subplots from left to right correspond to resolution changes from 1800$^{\prime\prime}$ (approximately 60 km) to 300$^{\prime\prime}$ (approximately 10 km), from 300$^{\prime\prime}$ to 90$^{\prime\prime}$ (approximately 3 km), and from 90$^{\prime\prime}$ to 30$^{\prime\prime}$ (approximately 1 km). Each data point represents a model-variable evaluation pair with co-located GHCNd or local station data.For spatial models, one point is shown per location with a co-located GHCNd station, which is why there can be multiple points for the same model–variable combination. The plots are divided into four quadrants: the green lower-left quadrant represents simultaneous improvement in both climate accuracy and model performance; the blue upper-left quadrant indicates improved climate accuracy but degraded model performance; the yellow lower-right quadrant indicates worse climate accuracy but better model performance; and the red upper-right quadrant indicates a decline in both climate accuracy and model performance.
}
\label{fig:all_comp}
\end{figure}

We examined how the impact model performance changes with increasing spatial resolution of climate forcing, progressing from 60 km (1800$^{\prime\prime}$) to 10 km (300$^{\prime\prime}$), then to 3 km (90$^{\prime\prime}$), and finally to 1 km (30$^{\prime\prime}$). These shifts in model accuracy were evaluated alongside corresponding changes in the accuracy of the underlying climate data (Fig.~\ref{fig:all_comp}). For each step, we computed the percentage change in normalized root mean square error (NRMSE) across all impact models with available local climate station data or GHCN-D observations within the hydrologic catchment (where a catchment/watershed outline was available), or within a +-0.125° grid around each evaluation site (see Tables 1 and 2 for what was available per model). Temperature and precipitation were assessed separately, as shown in panels (Fig.~\ref{fig:all_comp}a, b), respectively.

For temperature, the largest improvements in both climate input accuracy and model performance occurred in the transition from 60 km to 10 km resolution. Beyond this, gains from increasing resolution further to 3 km and 1 km were more modest. Most evaluation points clustered in the lower-left quadrant of the plots, indicating that improved climate accuracy was generally accompanied by improved model performance.

For precipitation, a similar pattern emerged, although changes in climate accuracy between resolutions were smaller than for temperature, consistent with the evaluation of CHELSA-W5E5 v1 (Fig.~\ref{fig:chelsa_eval}). The 60 km to 10 km transition yielded the most consistent improvements, again with most points located in the lower-left quadrant. However, in the transition from 10 km to 3 km, some points, particularly for the water sector (GR4J model; see Fig. S61 in the Supplementary Material Part C for the same plot by model rather than by sector) fell in the upper quadrants, suggesting reduced model performance despite stable climate accuracy. Intriguingly, these same locations showed marked improvement when moving from 3 km to 1 km, even though climate accuracy remained nearly unchanged. These sites lacked co-located temperature observations, pointing to temperature as a likely explanatory variable and underscoring the importance of jointly evaluating multiple climate inputs when interpreting spatial resolution effects in impact modeling.

Figure S62 in the Supplementary Material Part C presents the same analysis as \ref{fig:all_comp}, showing a direct comparison between 1800$^{\prime\prime}$ (60 km) and 30$^{\prime\prime}$ (1 km) resolution, confirming the trend of improvements in model performance aligning with better climate data.

\clearpage

\section{Discussion}\label{sec12}
We systematically evaluated how changes in the spatial resolution of climate input data affect the predictive performance of a suite of climate impact models participating in ISIMIP. A portfolio of 16 impact models was run using the same global CHELSA-W5E5 v1 climate forcing dataset, varying in spatial resolution from 1800$^{\prime\prime}$ to 30$^{\prime\prime}$. 

Our study demonstrates that increasing the spatial resolution of climate forcing data can improve the accuracy of climate impact models across a range of sectors. These improvements are particularly evident in regions with high topographic complexity, confirming long-standing concerns about the scale mismatch between climate inputs and ecological or socio-environmental processes \citep{Seo2009, Hickler2012, Thuiller2003}. While prior studies have identified resolution effects in isolated models or sectors \citep{Mearns2001, Hattermann2017, Randin2009, Kumar2022}, our cross-sectoral experimental setup provides a unique dataset to assess these effects systematically under harmonized conditions. The consistency of the improvements observed—especially for temperature-dependent variables—underscores the value of high-resolution forcing data in impact modeling.

When disaggregated by sector, the most substantial resolution-driven improvements are observed in the forest and biome models. These systems, which are often tightly coupled to temperature and elevation gradients, benefited from finer spatial representations of climate, particularly in mountainous regions. Recent studies in the forest sector have demonstrated that the impact of temperature is greater than the impact of precipitation along the environmental gradient \citep{Merganicova2024}. Models such as \textit{TreeMig}, which are spatially explicit and incorporate heterogeneity over the cells in forest composition and structure, showed notable gains in performance as resolution increased. Conversely, models in the lakes and regional water sectors exhibited more modest benefits, with some even showing reduced performance at intermediate resolutions. For the water sector, this is likely due to their stronger dependence on precipitation forcing, which showed only limited accuracy improvements with increasing resolution, consistent with previous findings on the challenges of precipitation interpolation and downscaling \citep{Daly1997, Karger2020c}. For lakes, sensitivity to spatial resolution is expected to depend primarily on wind speed and radiative fluxes rather than precipitation, because wind controls vertical mixing, entrainment and stratification stability, while shortwave radiation governs the primary heat input to the lake. Where precipitation is represented only as direct rainfall on the lake surface and lateral inflows are neglected, the impact of increasing the spatial resolution of precipitation on lake water temperature is expected to be negligible in most cases, because the associated heat flux is extremely small compared to solar radiation, sensible and latent heat fluxes, and the lake’s stored heat. Correspondingly, refinements in wind and radiation forcing may exert a stronger influence on simulated lake temperatures and stratification dynamics than further improvements in air temperature or precipitation alone; however, in our analysis the evaluation of the climate forcing was limited to temperature and precipitation, as suitable observations of shortwave radiation and wind speed at the evaluation sites were not available. Furthermore, lake model performance was evaluated at sub-hourly temporal resolution, whereas the meteorological forcing was provided as daily averages, introducing an additional temporal-scale mismatch that may further constrain detectable resolution-dependent improvements. In addition, the lakes and water sector simulations were conducted mainly in regions with relatively low topographic complexity, further limiting the potential benefits of finer spatial resolution.

The empirical labour model presents a distinct case. Its reduced performance at finer resolution can be attributed not to climate data quality but to the lack of high-resolution validation data. This example highlights that the usefulness of high-resolution forcing also depends on the availability of observational reference data at high-resolution, particularly for non-environmental systems such as labour supply and income where such data may not be spatially resolved. For the 10 km simulations, we compare this high-resolution simulation with observed labour supply data from surveys. The larger difference is due to the fact that unlike the natural sciences, there is no observed data at 10 km resolution to compare with. For the labour force, the advantage of using high-resolution data is that it enables the identification of locations where workers are particularly vulnerable, due to a combination of increasing heat, location of workplace (including indoors or outdoors) and person-specific attributes, where context specific labour force protection policies are likely to have the most impact on health and economic outcomes. Additionally, the advantage of using high-resolution data is that we are able to improve the robustness of the estimated impacts compared to using coarser resolution data. This case also raises important questions about how socioeconomic impact models can be meaningfully validated when fine-scale observational data are missing or inconsistent

From a methodological perspective, our results support the increasing emphasis in climate impact research on harmonized input data and protocol standardization, as exemplified by initiatives such as ISIMIP  \citep{Frieler2017, Frieler2024}. While previous ensemble modeling efforts have primarily focused on capturing structural and parametric uncertainty \citep{Eyring2016, Knutti2013, Collalti2019}, our findings show that spatial resolution is an equally critical factor, particularly when high-resolution data are applied consistently across sectors.

While the advantages of high-resolution climate data are evident in many contexts, our results also highlight several important limitations. First, the computational and storage demands of running impact models at high spatial resolution are substantial, especially for gridded models operating over large spatial domains. This limits practical applicability, particularly for global-scale simulations or long-term projections. Second, finer resolution does not automatically guarantee better model performance. In several cases — especially for precipitation - higher resolution introduced noise or errors, likely due to uncertainties in the downscaling process or mismatches with the spatial scale of observational validation data. These mixed results underscore the importance of local context and caution against assuming uniform benefits of high-resolution precipitation input when evaluating downstream model performance. These findings are consistent with concerns in the literature about the challenges of accurately representing small-scale precipitation variability using interpolation-based methods \citep{Karger2017, Karger2020c, Daly1997}. Targeted efforts to improve both the density and quality of precipitation observations, as well as the associated downscaling methods, are therefore crucial. Moreover, for some impact models, finer-scale climate inputs can fall below the spatial or temporal scales for which model structures and parameters were designed, causing models calibrated at coarser resolutions to respond weakly or artefactually to added small-scale variability \citep{Maraun2016, Maurer2016}. This highlights that increasing input resolution alone may be insufficient and motivates efforts to improve both model process representation and downscaling and bias-correction methods so that high-resolution climate forcings can be used effectively. Lastly, as demonstrated by the labour sector model, improvements in input data resolution are only meaningful when supported by equally resolved observational datasets for evaluation. Without appropriate validation data, the benefits of high-resolution modeling may remain speculative and risk misleading interpretation.



\clearpage

\section{Methods}\label{sec2}

\subsection{Climate Forcings and Experiments}

\subsubsection{High-resolution forcing data: CHELSA-W5E5 v1.0}

The CHELSA-W5E5 v1.0 dataset \citep{Karger2023}, as included in the ISIMIP framework, provides global land-based climate data at a spatial resolution of 30$^{\prime\prime}$ (approximately 1\,km at the equator) and daily temporal resolution, covering the period from 1979 to 2016. It includes the following variables: precipitation (pr), surface downwelling shortwave radiation (rsds), and daily mean, minimum, and maximum near-surface air temperature (tas, tasmin, tasmax). These data are derived through topographic downscaling of the W5E5 v1.0 observational dataset (original resolution 0.5$^\circ$).

CHELSA-W5E5 v1.0 is produced using the CHELSA (Climatologies at High Resolution for the Earth's Land Surface Areas) v2.0 algorithm \citep{Karger2017, Karger2021, Karger2023}. The downscaling process corrects for systematic biases associated with orographic features that are not adequately resolved in coarse-resolution data products.

The original 30$^{\prime\prime}$ data are spatially aggregated to coarser resolutions of 90$^{\prime\prime}$ ($\sim$3\,km, aggregation factor 3), 300$^{\prime\prime}$ ($\sim$10\,km, factor 10), and 1800$^{\prime\prime}$ ($\sim$60\,km, factor 60). Aggregation to 1800$^{\prime\prime}$ is necessary since the downscaled products differ from the default ISIMIP W5E5 data product at  0.5$^\circ$/1800$^{\prime\prime}$. Because all coarser grids are derived from the same underlying 30$^{\prime\prime}$ fields rather than independently downscaled, climate inputs at adjacent resolutions are not fully independent, which may reduce apparent differences in model performance between resolutions.

All climate data are available at: \url{https://data.isimip.org/search/query/chelsa/}.

Because some impact models required additional variables at higher resolution, a scripting environment was provided (\url{https://github.com/johanna-malle/w5e5_downscale}) to downscale relative humidity, surface wind, air pressure, and longwave radiation, which are not yet supported by the CHELSA downscaling approach. The method for these variables is described in detail in \cite{Frieler2024}. For the high-resolution experiments, it was crucial to maintain consistent input forcings across resolutions wherever possible.

For further information on the climate forcing data used in the high-resolution sensitivity experiments, see \cite{Karger2023} and \cite{Frieler2024}.

\subsubsection{Experimental Design}
All four resolutions included in the CHELSA-W5E5 v1.0 dataset were included in the experimental design to evaluate whether higher-resolution atmospheric climate data enhance the performance of climate impact model simulations.

These experiments aim to identify the resolution at which improvements in impact model simulations become evident when compared against observational impact indicators. Given the substantial storage and computational demands of daily 30$^{\prime\prime}$ data, this analysis is critical to determine the trade-offs between resolution and model performance.

Participating modeling groups were instructed to submit simulations for a minimum of two different spatial resolutions, along with corresponding observational data, to evaluate how the agreement between model output and observations varies with the spatial resolution of the climate forcing data and, for gridded models, with the resolution of the impact models themselves. The respective sector-specific ISIMIP3a simulation protocols were used as guidelines by all participating modeling groups. Depending on the sector and model, and constrained by the temporal resolution of the available observational data, simulated and observed variables were compared at daily, monthly, annual, or native model time-step resolution, as detailed in Supplementary Material A and B.

A more detailed description of the experimental design can also be found in \citep{Frieler2024}.

\subsection{Participating Sectors and Impact Models}

The participating impact models in this study span a wide range of complexity, from relatively simple one-dimensional (1D) point-scale representations to sophisticated three-dimensional (3D) spatially explicit systems. These models differ not only in their structural complexity but also in their spatial scale, process representation, and sensitivity to climate input data resolution - these differences occur both within and between sectors. Models simulated processes for different regions and at different resolutions (Tables~\ref{tab:model_overview_part1} and ~\ref{tab:model_overview_part2}, Figure~\ref{fig:spatial_extent}). 

Impact models from five different ISIMIP sectors were included in this analysis: Regional Forests, Regional Water, Lakes, Biomes, and Labour. Evaluation data include eddy-covariance flux tower measurements, river discharge records, reservoir level measurements, lake temperature profiles, and data on labour force outcomes from micro-surveys.

From the Regional Forests sector, five point-scale impact models participated (4C, i3PGmiX, 3PG-Hydro, Biome-BGCMuSo (BBGCMuSo), 3D-CMCC-FEM), all simulating at the same point location (Fluxnet site Collelongo, IT). One additional point-scale model (Prebas) was simulated at two locations in Finland, and two spatially distributed models (TreeMig, CARAIB) provided simulations across Switzerland and the area of Wallonia in Belgium/Netherlands, respectively. From the Regional Water sector, four models participated (CWatM, GR4J, SWAT, SWIM), each of which simulated different catchments. From the Lakes sector, two models participated (Simstrat and MITgcm), simulating 32 ISIMIP lakes and Lake Kinneret, respectively. From the Biomes sector, one model participated (ORCHIDEE), which provided point-scale simulations across the Tibetan Plateau. From the Labour sector, five empirical models provided simulations at two spatial resolutions, spanning a global extent. However, these results were combined and treated as one model result.

Tables~\ref{tab:model_overview_part1} and ~\ref{tab:model_overview_part2} provides a summary of the models, their sectors, configurations, and evaluation data. A detailed description of each model setup can be found in Supplementary Material Part A.

\begin{sidewaystable}
\caption{Model overview (Part 1 of 2): sectors, spatial/model setup, region, temporal resolution, and number of evaluation sites.}
\label{tab:model_overview_part1}
\begin{tabular*}{\textheight}{@{\extracolsep\fill}l l l l l l l}
\toprule
\parbox[t]{2.8cm}{\raggedright \textbf{Impact Model}} &
\parbox[t]{2.4cm}{\raggedright \textbf{Sector}} &
\parbox[t]{3.0cm}{\raggedright \textbf{Spatial Resolutions} \\ \textbf{(arcsec)}} &
\parbox[t]{1.8cm}{\raggedright \textbf{Model} \\ \textbf{Setup}} &
\parbox[t]{3.2cm}{\raggedright \textbf{Region}} &
\parbox[t]{3.6cm}{\raggedright \textbf{Temporal Resolution} \\ \textbf{(internal/output)}} &
\parbox[t]{2.8cm}{\raggedright \textbf{No. of} \\ \textbf{Eval. Sites}} \\
\midrule
BBGCMuSo & Reg.\ Forest & 30$''$, 90$''$, 300$''$, 1800$''$ & point & Collelongo & daily/daily & 1 \\
3D-CMCC-FEM & Reg.\ Forest & 30$''$, 90$''$, 300$''$, 1800$''$ & point & Collelongo & daily/daily & 1 \\
i3PGmiX & Reg.\ Forest & 30$''$, 90$''$, 300$''$, 1800$''$ & point & Collelongo & monthly/monthly & 1 \\
4C & Reg.\ Forest & 30$''$, 90$''$, 300$''$, 1800$''$ & point & Collelongo & daily/daily & 1 \\
3PG-Hydro & Reg.\ Forest & 30$''$, 90$''$, 300$''$, 1800$''$ & point & Collelongo & daily/daily & 1 \\
Prebas & Reg.\ Forest & 30$''$, 90$''$, 300$''$, 1800$''$ & point & Hyytiala+Sodankylae & daily/daily & 2 \\
TreeMig & Reg.\ Forest & 30$''$, 90$''$, 300$''$, 1800$''$ & spatial & Switzerland & yearly/yearly & 10080 \\
CARAIB & Reg.\ Forest & 30$''$, 90$''$, 1800$''$ & spatial & Belgium + Netherlands & 2hours / daily & 1 \\
ORCHIDEE\textendash MICT & Biomes & 30$''$, 90$''$, 300$''$, 1800$''$ & point & Tibetan Plateau & 30min / daily & 137 \\
CWatM & Regional Water & 30$''$, 90$''$, 300$''$, 1800$''$ & spatial & Bhima basin & daily/daily & 31 \\
GR4J & Regional Water & 30$''$, 90$''$, 300$''$, 1800$''$ & spatial & Norway & daily/daily & 86 \\
SWAT & Regional Water & 90$''$, 300$''$, 1800$''$ & spatial & 4 basins at Lake Kinneret & daily/monthly & 4 \\
SWIM & Regional Water & 300$''$, 1800$''$ & spatial & Zeravshan catchment & daily/daily & 1 \\
MITgcm & Lakes & 30$''$, 90$''$, 300$''$ & spatial & Lake Kinneret & 300s / daily & 5 \\
Simstrat & Lakes & 30$''$, 90$''$, 300$''$, 1800$''$ & point & 32 ISIMIP lakes & 300s / 6hours & 32 \\
Labour model & Labour & 30$''$, 90$''$, 300$''$, 1800$''$ & spatial & Global & yearly/yearly & All grid cells \\
\bottomrule
\end{tabular*}
\end{sidewaystable}

\begin{sidewaystable}
\caption{Model overview (Part 2 of 2): forcing inputs and evaluation setup.}
\label{tab:model_overview_part2}
\begin{tabular*}{\textheight}{@{\extracolsep\fill}l l l l l l}
\toprule
\parbox[t]{2.7cm}{\raggedright \textbf{Impact Model}} &
\parbox[t]{3.6cm}{\raggedright \textbf{Forcing Variables} \\ \textbf{(CHELSA-W5E5)}} &
\parbox[t]{3.6cm}{\raggedright \textbf{Add. Forcing} \\ \textbf{Variables}} &
\parbox[t]{3.0cm}{\raggedright \textbf{Evaluation} \\ \textbf{Variables}} &
\parbox[t]{3.0cm}{\raggedright \textbf{Evaluation Temporal} \\ \textbf{Resolution}} &
\parbox[t]{3.8cm}{\raggedright \textbf{Forcing Eval.} \\ \textbf{Benchmark}} \\
\midrule
BBGCMuSo & \begin{tabular}[t]{@{}l@{}}tasmax, tasmin, pr\\rsds\end{tabular} & day tas, day vpd & GPP, AET & monthly & \begin{tabular}[t]{@{}l@{}}Local climate\\data\end{tabular} \\ \addlinespace[0.4em]
3D-CMCC-FEM & \begin{tabular}[t]{@{}l@{}}tas, tasmax, tasmin\\pr, rsds\end{tabular} & rh & GPP, AET & monthly & \begin{tabular}[t]{@{}l@{}}Local climate\\data\end{tabular} \\ \addlinespace[0.4em]
i3PGmiX & \begin{tabular}[t]{@{}l@{}}tas, tasmax, tasmin\\pr, rsds\end{tabular} & - & GPP, AET & monthly & \begin{tabular}[t]{@{}l@{}}Local climate\\data\end{tabular} \\ \addlinespace[0.4em]
4C & \begin{tabular}[t]{@{}l@{}}tas, tasmax, tasmin\\pr, rsds\end{tabular} & ps, rh & GPP, AET & monthly & \begin{tabular}[t]{@{}l@{}}Local climate\\data\end{tabular} \\  \addlinespace[0.4em]
3PG-Hydro & \begin{tabular}[t]{@{}l@{}}tas, tasmax, tasmin\\pr, rsds\end{tabular} & - & GPP, AET & monthly & \begin{tabular}[t]{@{}l@{}}Local climate\\data\end{tabular} \\  \addlinespace[0.4em]
Prebas & tas, pr, rsds & rh & GPP, AET & monthly & \begin{tabular}[t]{@{}l@{}}Local climate\\data\end{tabular} \\ \addlinespace[0.4em]
TreeMig & tas, pr & - & basal area & yearly & \begin{tabular}[t]{@{}l@{}}GHCN-D stations in\\$\pm 0.125^\circ$ of eval.locations\end{tabular} \\ \addlinespace[0.4em]
CARAIB & \begin{tabular}[t]{@{}l@{}}tas, tasmax, tasmin\\pr, rsds\end{tabular} & rh, wind & GPP, AET & daily & \begin{tabular}[t]{@{}l@{}}Local climate\\data\end{tabular} \\  \addlinespace[0.4em]
ORCHIDEE\textendash MICT & \begin{tabular}[t]{@{}l@{}}tas, tasmax, tasmin\\pr, rsds\end{tabular} & qair, rlds, wind, ps & ALT, MAGT, soil temp. & daily & \begin{tabular}[t]{@{}l@{}}GHCN-D stations in\\$\pm 0.125^\circ$ of eval.locations\end{tabular} \\  \addlinespace[0.4em]
CWatM & \begin{tabular}[t]{@{}l@{}}tas, tasmax, tasmin\\pr, rsds\end{tabular} & rlds, wind, rh, ps & river discharge, res. level & daily & \begin{tabular}[t]{@{}l@{}}GHCN-D stations in\\$\pm 0.125^\circ$ of eval.locations\end{tabular} \\ \addlinespace[0.4em]
GR4J & tas, pr & - & river discharge & daily & \begin{tabular}[t]{@{}l@{}}GHCN-D stations in\\catchment of eval. location\end{tabular} \\ \addlinespace[0.4em]
SWAT & \begin{tabular}[t]{@{}l@{}}tasmax, tasmin, pr\\rsds\end{tabular} & rh, wind & river discharge & monthly & \begin{tabular}[t]{@{}l@{}}GHCN-D stations in\\$\pm 0.125^\circ$ of eval.locations\end{tabular} \\ \addlinespace[0.4em]
SWIM & \begin{tabular}[t]{@{}l@{}}tas, tasmax, tasmin\\pr, rsds\end{tabular} & - & river discharge & daily & \begin{tabular}[t]{@{}l@{}}GHCN-D stations in\\$\pm 0.125^\circ$ of eval.locations\end{tabular} \\ \addlinespace[0.4em]
MITgcm & tas, rsds, pr & rlds, wind, rh, hurs & lake temp. profile & daily & \begin{tabular}[t]{@{}l@{}}Local climate\\data\end{tabular} \\ \addlinespace[0.4em]
Simstrat & tas, rsds, pr & rlds, wind, ps, hurs & lake temp. profile & 6hours & \begin{tabular}[t]{@{}l@{}}GHCN-D stations in\\$\pm 0.125^\circ$ of eval.locations\end{tabular} \\ \addlinespace[0.4em]
Labour model & tasmin & rh & diff. to station forc. & yearly & \begin{tabular}[t]{@{}l@{}}N/A\\~\end{tabular} \\
\bottomrule
\end{tabular*}
\end{sidewaystable}

\subsection{Evaluation Framework}

\subsubsection{Evaluation Metrics}\label{eval_metrics}

To assess both the quality of the climate forcing data (CHELSA-W5E5 v1) and the performance of the impact models, we employed a set of standard evaluation metrics. These include the root mean squared error (RMSE), mean absolute error (MAE) and Kling–Gupta Efficiency (KGE).

\vspace{1em}
\noindent
The RMSE and MAE were normalized by the interquartile range (IQR), defined as the difference between the 75th (Q3) and 25th (Q1) percentiles, to reduce sensitivity to outliers compared with normalization using the full range:

\begin{align}
\text{RMSE} &= \sqrt{\frac{1}{n} \sum_{i=1}^{n} (y_i - y_{\text{obs},i})^2} \\
\text{NRMSE} &= \frac{\text{RMSE}}{Q3 - Q1} \\
\text{MAE} &= \frac{1}{n} \sum_{i=1}^{n} |y_i - y_{\text{obs},i}| \\
\text{NMAE} &= \frac{\text{MAE}}{Q3 - Q1}
\end{align}

\vspace{1em}
\noindent
To capture multiple sources of disagreement beyond magnitude alone, we computed the KGE, which integrates correlation, bias, and variability:

\begin{align}
\text{KGE} &= 1 - \sqrt{(r - 1)^2 + (\beta - 1)^2 + (\gamma - 1)^2}
\end{align}

\noindent
where:
\begin{align*}
r &= \text{Pearson correlation coefficient between } y_i \text{ and } y_{\text{obs},i} \\
\beta &= \frac{\mu_y}{\mu_{\text{obs}}} \quad \text{(bias ratio)} \\
\gamma &= \frac{CV_y}{CV_{\text{obs}}} = \frac{\sigma_y / \mu_y}{\sigma_{\text{obs}} / \mu_{\text{obs}}} \quad \text{(variability ratio)}
\end{align*}

\noindent
KGE values range from $-\infty$ to 1, with 1 indicating a perfect match between model and observation. 

Some models were evaluated based on more than one variable (e.g., forest models using both GPP and AET). In those cases, each metric was first computed separately for each variable and then averaged across variables, yielding one RMSE, one MAE, and one KGE value per model, as shown in Figure~\ref{fig:model_eval}.

\subsubsection{Evaluation of CHELSA-W5E5 Forcing Data (30$^{\prime\prime}$, 90$^{\prime\prime}$, 300$^{\prime\prime}$, 1800$^{\prime\prime}$)}

To evaluate the accuracy of the CHELSA-W5E5 climate forcing dataset at different spatial resolutions (30$^{\prime\prime}$, 90$^{\prime\prime}$, 300$^{\prime\prime}$, and 1800$^{\prime\prime}$ ), we compared the CHELSA-W5E5 climate variables against observations from the Global Historical Climatology Network Daily (GHCN-D) dataset \citep{Menne2018}. GHCN-D provides daily meteorological measurements, including 2\,m air temperature and precipitation, based on in situ weather station data, available for the same period as CHELSA-W5E5 (1979–2016).

For each evaluation site, we used local station climate data wherever possible to evaluate CHELSA-W5E5 forcing data. For the remaining models, we used GHCN-D station data for evaluation: if we had the exact coordinates of e.g. catchments, we selected all GHCN-D stations within this catchment. Otherwise, we selected all GHCN-D stations located within a +-0.125$^\circ$ grid cell around each evaluation location. From the CHELSA-W5E5 dataset, we extracted the grid cell value corresponding to the location of each local/GHCN-D station for all four spatial resolutions. Refer to Table\ref{tab:model_overview_part2} to see for which models which approach was applied.

The extracted temperature and precipitation values were then evaluated using the metrics described in Section~\ref{eval_metrics}. For each resolution and evaluation location, we computed the mean of each evaluation metric across all available GHCN-D stations within the grid cell. This yielded a single summary value per metric, per resolution, per evaluation site, representing the accuracy of the climate forcing data at that location.

\subsection{Terrain Descriptors}

To quantify topographic complexity, we used the Terrain Ruggedness Index (TRI) as described by \citet{Amatulli2018}. The TRI dataset provides near-global coverage (excluding polar regions) and is derived from the 250\,m Global Multi-resolution Terrain Elevation Data 2010 (GMTED2010) dataset \citep{Danielson2011}.

TRI is calculated as the mean of the absolute differences in elevation between a center grid cell and its eight surrounding cells. This metric captures the total elevation change within a given neighbourhood: flat terrain results in TRI values close to zero, while steep or mountainous regions yield higher values.

For each model evaluation location, TRI was averaged over a 0.25$^\circ$ grid cell centered on the evaluation point (i.e., $\pm$0.125$^\circ$ in both latitude and longitude).

\section*{Supplementary Information}
This study is accompanied by Supplementary Material, which is divided into three parts. \textbf{Part~A} provides additional details on the modelling setups, based on information collected through a standardised questionnaire completed by each participating modelling group upon submission of their simulation results. \textbf{Part~B} presents model-specific results and analyses. \textbf{Part~C} presents additional diagnostic figures that complement the main manuscript.

\section*{Data and Code availability}
The CHELSA-W5E5 v1.0 climate data used in this study are available from the ISIMIP repository at \url{https://data.isimip.org/search/query/CHELSA-W5E5%20v1.0/}, hosted and maintained by the Potsdam Institute for Climate Impact Research (PIK). The code used to evaluate the climate data and model results and to generate all figures is available on GitHub at \url{https://github.com/johanna-malle/isimip-highres-sensitivity-experiments}. The datasets required to run the code (including processed inputs used for figure reproduction) are available on Zenodo at \url{https://doi.org/10.5281/zenodo.17940720}. Simulation outputs will be made publicly available on the ISIMIP data repository upon publication of this manuscript.

\section*{Author contributions}
J.T.M., C.P.O.R., and D.N.K. conceived the study, developed the methodology, and coordinated the simulations. J.T.M. performed the analysis with input from C.P.O.R. and D.N.K. J.T.M. wrote the original draft, and D.N.K. supervised the project and acquired funding. Y.A., A.L.D.A., Y.B.-S., E.B.-Z., P.B., T.C., J.C., A.C., D.D., S.D., I.D., M.D., L.D., L.F., S.N.G., F.H., S.H., H.L., T.L., A.M., K.M., F.M., M.N., E.J.Z.R., M.S., M.S., R.S., E.V., X.Y., and R.Y. provided simulation results and observational data used in this study and are listed in alphabetical order. A.I.A. and D.M.-B. coordinated the sectoral modelling efforts and are listed in alphabetical order. D.Q.-C. supported ISIMIP’s provision and development of climate data. All authors contributed to manuscript review and editing.

\section*{Acknowledgments}
This article benefited from COST Action CA19139 PROCLIAS (PROcess-based models for CLimate Impact Attribution across Sectors), supported by COST (European Cooperation in Science and Technology; https://www.cost.eu, last access: 15 October 2025). Johanna Malle and Dirk Nikolaus Karger received funding from the funding organization of the Swiss National Science Foundation (SNF; project Adohris, 205530). Xue Yang would like to acknowledge grant support from the National Natural Science Foundation of China (Grant No. 52209035). Elad Ben-Zur was supported by a grant from the Israeli Ministry of Science and Technology (\#4755). Iulii Didovets and Fred F. Hattermann would like to acknowledge funding from the Green Central Asia initiative – Enhancing environment, climate and water resilience. Laura Dobor was supported by the Czech Science Foundation (GAČR), grant number 26-20790N. Mergani\v{c}ov\'a, Katar\'ina was funded by the EU NextGenerationEU through the Recovery and Resilience Plan for Slovakia under the project No. 09I03-03-V04- 00130.

Observational data for the TreeMig model was made available by the Schweizerisches Landesforstinventar LFI (Daten der Erhebung 1983-85 (LFI1), EstherThürig-DL1619. Eidg. Forschungsanstalt WSL, Birmensdorf). We thank Katja Frieler, Matthias Mengel, Jacob Schewe and Stefan Lange for insightful discussions during this project, and Lisa Novak for her assistance with data harmonisation.

\clearpage

\bibliography{isimip_high_res_exp}

@article{Amatulli2018,
  title = {A suite of global, cross-scale topographic variables for environmental and biodiversity modeling},
  author = {Amatulli, Giuseppe and Domisch, Sami and Tuanmu, Mao-Ning and Parmentier, Benoit and Ranipeta, Ajay and Malczyk, Jeremy and Jetz, Walter},
  year = {2018},
  journal = {Scientific Data},
  volume = {5},
  number = {1},
  pages = {180040},
  publisher = {Nature Publishing Group},
  issn = {2052-4463},
  doi = {10.1038/sdata.2018.40},
  copyright = {2018 The Author(s)}
}

@article{Augustynczik2017,
  title = {Productivity of \textit{Fagus Sylvatica} under Climate Change -- A Bayesian Analysis of Risk and Uncertainty Using the Model 3-PG},
  author = {Augustynczik, Andrey L. D. and Hartig, Florian and Minunno, Francesco and Kahle, Hans-Peter and Diaconu, Daniela and Hanewinkel, Marc and Yousefpour, Rasoul},
  year = {2017},
  month = oct,
  journal = {Forest Ecology and Management},
  volume = {401},
  pages = {192--206},
  issn = {0378-1127},
  doi = {10.1016/j.foreco.2017.06.061}
}

@article{Augustynczik2025,
  title = {Adapting forest management to climate change impacts and policy targets in the EU: Insights from the coupled GLOBIOM/G4M-i3PGmiX model},
  author = {Augustynczik, Andrey Lessa Derci and Gusti, Mykola and Deppermann, Andre and Nakhavali, Mahdi (Andr{\'e}) and Jia, Fanqi and {di Fulvio}, Fulvio and Havl{\'i}k, Petr},
  year = {2025},
  month = jun,
  journal = {One Earth},
  volume = {8},
  number = {6},
  pages = {101313},
  issn = {2590-3322},
  doi = {10.1016/j.oneear.2025.101313}
}

@article{Collalti2020,
  title = {Forest Production Efficiency Increases with Growth Temperature},
  author = {Collalti, A. and Ibrom, A. and Stockmarr, A. and Cescatti, A. and Alkama, R. and {Fern{\'a}ndez-Mart{\'i}nez}, M. and Matteucci, G. and Sitch, S. and Friedlingstein, P. and Ciais, P. and Goll, D. S. and Nabel, J. E. M. S. and Pongratz, J. and Arneth, A. and Haverd, V. and Prentice, I. C.},
  year = {2020},
  month = oct,
  journal = {Nature Communications},
  volume = {11},
  number = {1},
  pages = {5322},
  publisher = {Nature Publishing Group},
  issn = {2041-1723},
  doi = {10.1038/s41467-020-19187-w},
  copyright = {2020 The Author(s)},
  langid = {english}
}

@article{Dalmonech2022,
  title = {Feasibility of Enhancing Carbon Sequestration and Stock Capacity in Temperate and Boreal European Forests via Changes to Management Regimes},
  author = {Dalmonech, D. and Marano, G. and Amthor, J. S. and Cescatti, A. and Lindner, M. and Trotta, C. and Collalti, A.},
  year = {2022},
  month = dec,
  journal = {Agricultural and Forest Meteorology},
  volume = {327},
  pages = {109203},
  issn = {0168-1923},
  doi = {10.1016/j.agrformet.2022.109203}
}

@article{Eyring2016,
  title = {Overview of the Coupled Model Intercomparison Project Phase 6 (CMIP6) Experimental Design and Organization},
  author = {Eyring, Veronika and Bony, Sandrine and Meehl, Gerald A. and Senior, Catherine A. and Stevens, Bjorn and Stouffer, Ronald J. and Taylor, Karl E.},
  year = {2016},
  month = may,
  journal = {Geoscientific Model Development},
  volume = {9},
  number = {5},
  pages = {1937--1958},
  publisher = {Copernicus GmbH},
  issn = {1991-959X},
  doi = {10.5194/gmd-9-1937-2016},
  langid = {english}
}

@article{Frieler2017,
  title = {Assessing the Impacts of 1.5 {$^\circ$}C Global Warming -- Simulation Protocol of the Inter-Sectoral Impact Model Intercomparison Project (ISIMIP2b)},
  author = {Frieler, Katja and Lange, Stefan and Piontek, Franziska and Reyer, Christopher P. O. and Schewe, Jacob and Warszawski, Lila and Zhao, Fang and Chini, Louise and Denvil, Sebastien and Emanuel, Kerry and Geiger, Tobias and Halladay, Kate and Hurtt, George and Mengel, Matthias and Murakami, Daisuke and Ostberg, Sebastian and Popp, Alexander and Riva, Riccardo and Stevanovic, Miodrag and Suzuki, Tatsuo and Volkholz, Jan and Burke, Eleanor and Ciais, Philippe and Ebi, Kristie and Eddy, Tyler D. and Elliott, Joshua and Galbraith, Eric and Gosling, Simon N. and Hattermann, Fred and Hickler, Thomas and Hinkel, Jochen and Hof, Christian and Huber, Veronika and J{\"a}germeyr, Jonas and Krysanova, Valentina and Marc{\'e}, Rafael and M{\"u}ller Schmied, Hannes and Mouratiadou, Ioanna and Pierson, Don and Tittensor, Derek P. and Vautard, Robert and {van Vliet}, Michelle and Biber, Matthias F. and Betts, Richard A. and Bodirsky, Benjamin Leon and Deryng, Delphine and Frolking, Steve and Jones, Chris D. and Lotze, Heike K. and {Lotze-Campen}, Hermann and Sahajpal, Ritvik and Thonicke, Kirsten and Tian, Hanqin and Yamagata, Yoshiki},
  year = {2017},
  month = nov,
  journal = {Geoscientific Model Development},
  volume = {10},
  number = {12},
  pages = {4321--4345},
  publisher = {Copernicus GmbH},
  issn = {1991-959X},
  doi = {10.5194/gmd-10-4321-2017},
  langid = {english}
}

@article{Frieler2024,
  title = {Scenario Setup and Forcing Data for Impact Model Evaluation and Impact Attribution within the Third Round of the Inter-Sectoral Impact Model Intercomparison Project (ISIMIP3a)},
  author = {Frieler, Katja and Volkholz, Jan and Lange, Stefan and Schewe, Jacob and Mengel, Matthias and {del Roc{\'i}o Rivas L{\'o}pez}, Mar{\'i}a and Otto, Christian and Reyer, Christopher P. O. and Karger, Dirk Nikolaus and Malle, Johanna T. and Treu, Simon and Menz, Christoph and Blanchard, Julia L. and Harrison, Cheryl S. and Petrik, Colleen M. and Eddy, Tyler D. and {Ortega-Cisneros}, Kelly and Novaglio, Camilla and Rousseau, Yannick and Watson, Reg A. and Stock, Charles and Liu, Xiao and Heneghan, Ryan and Tittensor, Derek and Maury, Olivier and B{\"u}chner, Matthias and Vogt, Thomas and Wang, Tingting and Sun, Fubao and Sauer, Inga J. and Koch, Johannes and Vanderkelen, Inne and J{\"a}germeyr, Jonas and M{\"u}ller, Christoph and Rabin, Sam and Klar, Jochen and {Vega del Valle}, Iliusi D. and Lasslop, Gitta and Chadburn, Sarah and Burke, Eleanor and {Gallego-Sala}, Angela and Smith, Noah and Chang, Jinfeng and Hantson, Stijn and Burton, Chantelle and G{\"a}deke, Anne and Li, Fang and Gosling, Simon N. and M{\"u}ller Schmied, Hannes and Hattermann, Fred and Wang, Jida and Yao, Fangfang and Hickler, Thomas and Marc{\'e}, Rafael and Pierson, Don and Thiery, Wim and {Mercado-Bett{\'i}n}, Daniel and Ladwig, Robert and {Ayala-Zamora}, Ana Isabel and Forrest, Matthew and Bechtold, Michel},
  year = {2024},
  month = jan,
  journal = {Geoscientific Model Development},
  volume = {17},
  number = {1},
  pages = {1--51},
  publisher = {Copernicus GmbH},
  issn = {1991-959X},
  doi = {10.5194/gmd-17-1-2024},
  langid = {english}
}

@article{Hidy2012,
  title = {Development of the Biome-BGC Model for Simulation of Managed Herbaceous Ecosystems},
  author = {Hidy, D. and Barcza, Z. and Haszpra, L. and Churkina, G. and Pint{\'e}r, K. and Nagy, Z.},
  year = {2012},
  month = feb,
  journal = {Ecological Modelling},
  volume = {226},
  pages = {99--119},
  issn = {0304-3800},
  doi = {10.1016/j.ecolmodel.2011.11.008}
}

@article{Hidy2016,
  title = {Terrestrial Ecosystem Process Model Biome-BGCMuSo v4.0: Summary of Improvements and New Modeling Possibilities},
  author = {Hidy, D{\'o}ra and Barcza, Zolt{\'a}n and Marjanovi{\'c}, Hrvoje and Ostrogovi{\'c} Sever, Ma{\v s}a Zorana and Dobor, Laura and Gelyb{\'o}, Gy{\"o}rgyi and Fodor, N{\'a}ndor and Pint{\'e}r, Krisztina and Churkina, Galina and Running, Steven and Thornton, Peter and Bellocchi, Gianni and Haszpra, L{\'a}szl{\'o} and Horv{\'a}th, Ferenc and Suyker, Andrew and Nagy, Zolt{\'a}n},
  year = {2016},
  month = dec,
  journal = {Geoscientific Model Development},
  volume = {9},
  number = {12},
  pages = {4405--4437},
  publisher = {Copernicus GmbH},
  issn = {1991-959X},
  doi = {10.5194/gmd-9-4405-2016},
  langid = {english}
}

@article{Hidy2022,
  title = {Soil-Related Developments of the Biome-BGCMuSo v6.2 Terrestrial Ecosystem Model},
  author = {Hidy, D{\'o}ra and Barcza, Zolt{\'a}n and Holl{\'o}s, Roland and Dobor, Laura and {\'A}cs, Tam{\'a}s and Zach{\'a}ry, D{\'o}ra and Filep, Tibor and P{\'a}sztor, L{\'a}szl{\'o} and Incze, D{\'o}ra and Dencs{\H o}, M{\'a}rton and T{\'o}th, Eszter and Mergani{\v c}ov{\'a}, Katar{\'i}na and Thornton, Peter and Running, Steven and Fodor, N{\'a}ndor},
  year = {2022},
  month = mar,
  journal = {Geoscientific Model Development},
  volume = {15},
  number = {5},
  pages = {2157--2181},
  publisher = {Copernicus GmbH},
  issn = {1991-959X},
  doi = {10.5194/gmd-15-2157-2022},
  langid = {english}
}

@article{Hungerford1989,
  title = {MTCLIM: A Mountain Microclimate Simulation Model},
  author = {Hungerford, Roger D. and Nemani, Ramakrishna R. and Running, Steven W. and Coughlan, Joseph C.},
  year = {1989},
  journal = {Res. Pap. INT-RP-414. Ogden, UT: U.S. Department of Agriculture, Forest Service, Intermountain Research Station. 52 p.},
  volume = {414},
  doi = {10.2737/INT-RP-414}
}

@article{Karger2017,
  title = {Climatologies at High Resolution for the Earth's Land Surface Areas},
  author = {Karger, Dirk Nikolaus and Conrad, Olaf and B{\"o}hner, J{\"u}rgen and Kawohl, Tobias and Kreft, Holger and {Soria-Auza}, Rodrigo Wilber and Zimmermann, Niklaus E. and Linder, H. Peter and Kessler, Michael},
  year = {2017},
  month = sep,
  journal = {Scientific Data},
  volume = {4},
  number = {1},
  pages = {170122},
  publisher = {Nature Publishing Group},
  issn = {2052-4463},
  doi = {10.1038/sdata.2017.122},
  copyright = {2017 The Author(s)},
  langid = {english}
}

@article{Karger2021,
  title = {Global Daily 1 Km Land Surface Precipitation Based on Cloud Cover-Informed Downscaling},
  author = {Karger, Dirk Nikolaus and Wilson, Adam M. and Mahony, Colin and Zimmermann, Niklaus E. and Jetz, Walter},
  year = {2021},
  month = nov,
  journal = {Scientific Data},
  volume = {8},
  number = {1},
  pages = {307},
  publisher = {Nature Publishing Group},
  issn = {2052-4463},
  doi = {10.1038/s41597-021-01084-6},
  copyright = {2021 The Author(s)},
  langid = {english}
}

@article{Karger2023,
  title = {CHELSA-W5E5: daily 1 km meteorological forcing data for climate impact studies},
  author = {Karger, Dirk Nikolaus and Lange, Stefan and Hari, Chantal and Reyer, Christopher P. O. and Conrad, Olaf and Zimmermann, Niklaus E. and Frieler, Katja},
  year = {2023},
  month = jun,
  journal = {Earth System Science Data},
  volume = {15},
  number = {6},
  pages = {2445--2464},
  publisher = {Copernicus GmbH},
  issn = {1866-3508},
  doi = {10.5194/essd-15-2445-2023},
  langid = {english}
}

@article{Landsberg1997,
  title = {A Generalised Model of Forest Productivity Using Simplified Concepts of Radiation-Use Efficiency, Carbon Balance and Partitioning},
  author = {Landsberg, J. J. and Waring, R. H.},
  year = {1997},
  month = aug,
  journal = {Forest Ecology and Management},
  volume = {95},
  number = {3},
  pages = {209--228},
  issn = {0378-1127},
  doi = {10.1016/S0378-1127(97)00026-1}
}

@article{Lasch-Born2020,
  title = {Description and Evaluation of the Process-Based Forest Model 4C v2.2 at Four European Forest Sites},
  author = {Lasch-Born, Petra and Suckow, Felicitas and Reyer, Christopher P. O. and Gutsch, Martin and Kollas, Chris and Badeck, Franz-Werner and Bugmann, Harald K. M. and Grote, R{\"u}diger and F{\"u}rstenau, Cornelia and Lindner, Marcus and Schaber, J{\"o}rg},
  year = {2020},
  month = nov,
  journal = {Geoscientific Model Development},
  volume = {13},
  number = {11},
  pages = {5311--5343},
  publisher = {Copernicus GmbH},
  issn = {1991-959X},
  doi = {10.5194/gmd-13-5311-2020},
  langid = {english}
}

@article{Lischke2006a,
  title = {TreeMig: A Forest-Landscape Model for Simulating Spatio-Temporal Patterns from Stand to Landscape Scale},
  author = {Lischke, Heike and Zimmermann, Niklaus E. and Bolliger, Janine and Rickebusch, Sophie and L{\"o}ffler, Thomas J.},
  year = {2006},
  month = dec,
  journal = {Ecological Modelling},
  volume = {199},
  number = {4},
  pages = {409--420},
  issn = {0304-3800},
  doi = {10.1016/j.ecolmodel.2005.11.046},
}

@article{Liu2022,
  title = {Mapping High Resolution National Soil Information Grids of China},
  author = {Liu, Feng and Wu, Huayong and Zhao, Yuguo and Li, Decheng and Yang, Jin-Ling and Song, Xiaodong and Shi, Zhou and Zhu, A-Xing and Zhang, Gan-Lin},
  year = {2022},
  month = feb,
  journal = {Science bulletin},
  volume = {67},
  number = {3},
  pages = {328--340},
  issn = {2095-9281},
  doi = {10.1016/j.scib.2021.10.013},
  copyright = {cc by-nc-nd},
  langid = {english},
  pmid = {36546081}
}

@article{Minunno2019,
  title = {Bayesian Calibration of a Carbon Balance Model PREBAS Using Data from Permanent Growth Experiments and National Forest Inventory},
  author = {Minunno, Francesco and Peltoniemi, Mikko and H{\"a}rk{\"o}nen, Sanna and Kalliokoski, Tuomo and Makinen, Harri and M{\"a}kel{\"a}, Annikki},
  year = {2019},
  month = may,
  journal = {Forest Ecology and Management},
  volume = {440},
  pages = {208--257},
  issn = {0378-1127},
  doi = {10.1016/j.foreco.2019.02.041}
}

@article{Reyer2020,
  title = {The PROFOUND Database for evaluating vegetation models and simulating climate impacts on European forests},
  author = {Reyer, Christopher P. O. and Silveyra Gonzalez, Ramiro and Dolos, Klara and Hartig, Florian and Hauf, Ylva and Noack, Matthias and {Lasch-Born}, Petra and R{\"o}tzer, Thomas and Pretzsch, Hans and Meesenburg, Henning and Fleck, Stefan and Wagner, Markus and Bolte, Andreas and Sanders, Tanja G. M. and Kolari, Pasi and M{\"a}kel{\"a}, Annikki and Vesala, Timo and Mammarella, Ivan and Pumpanen, Jukka and Collalti, Alessio and Trotta, Carlo and Matteucci, Giorgio and D'Andrea, Ettore and Folt{\'y}nov{\'a}, Lenka and Krejza, Jan and Ibrom, Andreas and Pilegaard, Kim and Loustau, Denis and Bonnefond, Jean-Marc and Berbigier, Paul and Picart, Delphine and Lafont, S{\'e}bastien and Dietze, Michael and Cameron, David and Vieno, Massimo and Tian, Hanqin and {Palacios-Orueta}, Alicia and Cicuendez, Victor and Recuero, Laura and Wiese, Klaus and B{\"u}chner, Matthias and Lange, Stefan and Volkholz, Jan and Kim, Hyungjun and Horemans, Joanna A. and Bohn, Friedrich and Steinkamp, J{\"o}rg and Chikalanov, Alexander and Weedon, Graham P. and Sheffield, Justin and Babst, Flurin and {Vega del Valle}, Iliusi and Suckow, Felicitas and Martel, Simon and Mahnken, Mats and Gutsch, Martin and Frieler, Katja},
  year = {2020},
  journal = {Earth System Science Data},
  volume = {12},
  number = {2},
  pages = {1295--1320},
  publisher = {Copernicus GmbH},
  issn = {1866-3508},
  doi = {10.5194/essd-12-1295-2020},
}

@article{Trotsiuk2020,
  title = {Assessing the Response of Forest Productivity to Climate Extremes in Switzerland Using Model--Data Fusion},
  author = {Trotsiuk, Volodymyr and Hartig, Florian and Cailleret, Maxime and Babst, Flurin and Forrester, David I. and Baltensweiler, Andri and Buchmann, Nina and Bugmann, Harald and Gessler, Arthur and Gharun, Mana and Minunno, Francesco and Rigling, Andreas and Rohner, Brigitte and Stillhard, Jonas and Th{\"u}rig, Esther and Waldner, Peter and Ferretti, Marco and Eugster, Werner and Schaub, Marcus},
  year = {2020},
  journal = {Global Change Biology},
  volume = {26},
  number = {4},
  pages = {2463--2476},
  issn = {1365-2486},
  doi = {10.1111/gcb.15011}
}

@article{Yousefpour2021,
  title = {Simulating the Effects of Thinning Events on Forest Growth and Water Services Asks for Daily Analysis of Underlying Processes},
  author = {Yousefpour, Rasoul and Djahangard, Marc},
  year = {2021},
  month = dec,
  journal = {Forests},
  volume = {12},
  number = {12},
  pages = {1729},
  publisher = {Multidisciplinary Digital Publishing Institute},
  issn = {1999-4907},
  doi = {10.3390/f12121729},
  copyright = {http://creativecommons.org/licenses/by/3.0/},
  langid = {english}
}

@article{Lee2023,
  author    = {Lee, Hoesung and Calvin, Katherine and Dasgupta, Dipak and Krinner, Gerhard and Mukherji, Aditi and Thorne, Peter and Trisos, Christopher and Romero, Jos{\'e} and Aldunce, Paulina and Barret, Ko and Blanco, Gabriel and Cheung, William W. L. and Connors, Sarah L. and Denton, Fatima and Diongue-Niang, A{\"i}da and Dodman, David and Garschagen, Matthias and Geden, Oliver and Hayward, Bronwyn and Jones, Christopher and Jotzo, Frank and Krug, Thelma and Lasco, Rodel and Lee, Yune-Yi and Masson-Delmotte, Val{\'e}rie and Meinshausen, Malte and Mintenbeck, Katja and Mokssit, Abdalah and Otto, Friederike E. L. and Pathak, Minal and Pirani, Anna and Poloczanska, Elvira and P{\"o}rtner, Hans-Otto and Revi, Aromar and Roberts, Debra C. and Roy, Joyashree and Ruane, Alex C. and Skea, Jim and Shukla, Priyadarshi R. and Slade, Raphael and Slangen, Aim{\'e}e and Sokona, Youba and S{\"o}rensson, Anna A. and Tignor, Melinda and van Vuuren, Detlef and Wei, Yi-Ming and Winkler, Harald and Zhai, Panmao and Zommers, Zinta and Hourcade, Jean-Charles and Johnson, Francis X. and Pachauri, Shonali and Simpson, Nicholas P. and Singh, Chandni and Thomas, Adelle and Totin, Edmond and Arias, Paola and Bustamante, Mercedes and Elgizouli, Ismail and Flato, Gregory and Howden, Mark and M{\'e}ndez-Vallejo, Carlos and Pereira, Joy Jacqueline and Pichs-Madruga, Ram{\'o}n and Rose, Steven K. and Saheb, Yamina and S{\'a}nchez Rodr{\'i}guez, Roberto and {\"U}rge-Vorsatz, Diana and Xiao, Cunde and Yassaa, Noureddine and Alegr{\'i}a, Andr{\'e}s and Armour, Kyle and Bednar-Friedl, Birgit and Blok, Kornelis and Ciss{\'e}, Gu{\'e}ladio and Dentener, Frank and Eriksen, Siri and Fischer, Erich and Garner, Gregory and Guivarch, C{\'e}line and Haasnoot, Marjolijn and Hansen, Gerrit and Hauser, Mathias and Hawkins, Ed and Hermans, Tim and Kopp, Robert and Leprince-Ringuet, No{\"e}mie and Lewis, Jared and Ley, Debora and Ludden, Chlo{\'e} and Niamir, Leila and Nicholls, Zebedee and Some, Shreya and Szopa, Sophie and Trewin, Blair and van der Wijst, Kaj-Ivar and Winter, Gundula and Witting, Maximilian and Birt, Arlene and Ha, Meeyoung and Romero, Jos{\'e} and Kim, Jinmi and Haites, Erik F. and Jung, Yonghun and Stavins, Robert and Birt, Arlene and Ha, Meeyoung and Orendain, Dan Jezreel A. and Ignon, Lance and Park, Semin and Park, Youngin},
  title     = {Climate change 2014: synthesis report. Contribution of Working Groups I, II and III to the fifth assessment report of the Intergovernmental Panel on Climate Change},
  journal   = {IPCC},
  year      = {2014}
}

@article{Reyer2014,
  author    = {Reyer, Christopher P. O. and Lasch-Born, Petra and Suckow, Fritz and Gutsch, Matthias and Murawski, Andreas and Pilz, Tobias},
  title     = {Projections of regional changes in forest net primary productivity for different tree species in Europe driven by climate change and carbon dioxide},
  journal   = {Annals of Forest Science},
  volume    = {71},
  number    = {2},
  pages     = {211--225},
  year      = {2014}
}

@article{Cramer2001,
  author    = {Cramer, Wolfgang and Bondeau, Alberte and Woodward, Fiona I. and Prentice, I. Colin and Betts, Richard A. and Brovkin, Victor and Cox, Peter M. and Fisher, Vanessa and Foley, Jonathan A. and Friend, Andrew D. and Kucharik, Christopher and Lomas, Mark R. and Ramankutty, Navin and Sitch, Stephen and Smith, Benjamin and White, Andy and Young‐Molling, Chris},
  title     = {Global response of terrestrial ecosystem structure and function to CO2 and climate change: results from six dynamic global vegetation models},
  journal   = {Global Change Biology},
  volume    = {7},
  number    = {4},
  pages     = {357--373},
  year      = {2001}
}

@article{Goderniaux2011,
  author    = {Goderniaux, Pierre and Brouy\`ere, Serge and Blenkinsop, Stephen and Burton, Ann and Fowler, Hayley J. and Orban, Philippe and Dassargues, Alain},
  title     = {Modeling climate change impacts on groundwater resources using transient stochastic climatic scenarios},
  journal   = {Water Resources Research},
  volume    = {47},
  year      = {2011}
}

@article{Isaac2009,
  author    = {Isaac, M. and van Vuuren, D. P.},
  title     = {Modeling global residential sector energy demand for heating and air conditioning in the context of climate change},
  journal   = {Energy Policy},
  volume    = {37},
  pages     = {507--521},
  year      = {2009}
}

@article{Caminade2014,
  author    = {Caminade, Cyril and Kovats, Sari and Rocklov, Joacim and Tompkins, Adrian M. and Morse, Andy P. and Col\'on-Gonz\'alez, Felipe J. and Stenlund, Hans and Martens, Pim and Lloyd, Simon J.},
  title     = {Impact of climate change on global malaria distribution},
  journal   = {Proceedings of the National Academy of Sciences},
  volume    = {111},
  number    = {9},
  pages     = {3286--3291},
  year      = {2014}
}

@article{Harrison2016,
  author    = {Harrison, Paula A. and Dunford, Robert W. and Holman, Ian P. and Rounsevell, Mark D. A.},
  title     = {Climate change impact modelling needs to include cross-sectoral interactions},
  journal   = {Nature Climate Change},
  volume    = {6},
  number    = {9},
  pages     = {885--890},
  year      = {2016}
}

@article{Zscheischler2018,
  author    = {Zscheischler, Jakob and Westra, Seth and Van Den Hurk, Bart J. J. M. and Seneviratne, Sonia I. and Ward, Philip J. and Pitman, Andy and AghaKouchak, Amir and Bresch, David N. and Leonard, Michael and Wahl, Thomas},
  title     = {Future climate risk from compound events},
  journal   = {Nature Climate Change},
  volume    = {8},
  pages     = {469--477},
  year      = {2018}
}

@article{Ridder2020,
  author    = {Ridder, N. N. and Pitman, A. J. and Westra, S. and Ukkola, A. and Do, H. X. and Bador, M. and Hirsch, A. L. and Evans, J. P. and Di Luca, A. and Zscheischler, J.},
  title     = {Global hotspots for the occurrence of compound events},
  journal   = {Nature Communications},
  volume    = {11},
  pages     = {5956},
  year      = {2020}
}

@article{Ciscar2019,
  author    = {Ciscar, Juan-Carlos and Rising, James and Kopp, Robert E. and Feyen, Luc},
  title     = {Assessing future climate change impacts in the EU and the USA: insights and lessons from two continental-scale projects},
  journal   = {Environmental Research Letters},
  volume    = {14},
  number    = {8},
  pages     = {084010},
  year      = {2019}
}

@article{Hattermann2017,
  author    = {Hattermann, Fred F. and Krysanova, Valentina and Gosling, Simon N. and Dankers, Rutger and Daggupati, Prabhakar and Donnelly, Chantal and Fl{\"o}rke, Martina and Huang, Shaochun and Motovilov, Yuri and Buda, Suhail and others},
  title     = {Cross‐scale intercomparison of climate change impacts simulated by regional and global hydrological models in eleven large river basins},
  journal   = {Climatic Change},
  volume    = {141},
  number    = {3},
  pages     = {561--576},
  year      = {2017}
}

@article{Meehl2000,
  author    = {Meehl, Gerald A. and Boer, Gerrit J. and Covey, Curt and Latif, Mojib and Stouffer, Ronald J.},
  title     = {The Coupled Model Intercomparison Project (CMIP)},
  journal   = {Bulletin of the American Meteorological Society},
  volume    = {81},
  number    = {2},
  pages     = {313--318},
  year      = {2000}
}

@article{Meehl2007,
  author    = {Meehl, Gerald A. and Covey, Curt and Delworth, Thomas and Latif, Mojib and McAvaney, Bob and Mitchell, John F. B. and Stouffer, Ronald J. and Taylor, Karl E.},
  title     = {THE WCRP CMIP3 Multimodel Dataset: A New Era in Climate Change Research},
  journal   = {Bulletin of the American Meteorological Society},
  volume    = {88},
  pages     = {1383--1394},
  year      = {2007}
}

@article{Knutti2013,
  author    = {Knutti, Reto and Sedláček, Jan},
  title     = {Robustness and uncertainties in the new CMIP5 climate model projections},
  journal   = {Nature Climate Change},
  volume    = {3},
  pages     = {369--373},
  year      = {2013}
}

@techreport{Krysanova2000,
  type = {Report},
  title = {SWIM (Soil and Water Integrated Model)},
  author = {Krysanova, V. and Wechsung, F. and Arnold, J. and Srinivasan, R. and Williams, J.},
  date = {2000-12-01T05:00:00Z},
  year = {2000},
  number = {69},
  institution = {Potsdam},
  url = {https://www.osti.gov/etdeweb/biblio/20170564},
  urldate = {2025-09-03}
}

@article{McMaster1997,
  author    = {McMaster, G.S. and Wilhelm, W.},
  title     = {Growing degree-days: one equation, two interpretations},
  journal   = {Agricultural and Forest Meteorology},
  volume    = {87},
  pages     = {291--300},
  year      = {1997}
}

@article{Mearns2001,
  author    = {Mearns, Linda O. and Easterling, William and Hays, Cynthia and Marx, Donald},
  title     = {Comparison of Agricultural Impacts of Climate Change Calculated from High and Low Resolution Climate Change Scenarios: Part I. The Uncertainty Due to Spatial Scale},
  journal   = {Climatic Change},
  volume    = {51},
  pages     = {131--172},
  year      = {2001}
}

@article{Menne2018,
  author    = {Menne, Matthew J. and Bryant, Jared A. and Korzeniewski, Scott M. and Kristy, Tamara and Xungang, Yin and Anthony, Shannon and Ray, Rob and Vose, Russell S. and Gleason, Byron E. and Houston, Thomas G.},
  title     = {Global Historical Climatology Network - Daily (GHCN-Daily), Version 3},
  journal   = {NOAA National Climatic Data Center},
  year      = {2018},
  note      = {doi:10.7289/V5D21VHZ}
}

@article{Neumann2019,
  author    = {Neumann, Philipp and Düben, Peter and Adamidis, Panagiotis and Bauer, Peter and Brück, Matthias and Kornblueh, Lars and Klocke, Daniel and Stevens, Bjorn and Wedi, Nils and Biercamp, Jan},
  title     = {Assessing the scales in numerical weather and climate predictions: will exascale be the rescue?},
  journal   = {Philosophical Transactions of the Royal Society A: Mathematical, Physical and Engineering Sciences},
  volume    = {377},
  number    = {2142},
  pages     = {20180148},
  year      = {2019}
}

@article{Randin2009,
  author    = {Randin, Christophe F. and Engler, Robin and Normand, Signe and Zappa, Massimiliano and Zimmermann, Niklaus E. and Pearman, Peter B. and Vittoz, Pascal and Thuiller, Wilfried and Guisan, Antoine},
  title     = {Climate change and plant distribution: local models predict high-elevation persistence},
  journal   = {Global Change Biology},
  volume    = {15},
  number    = {6},
  pages     = {1557--1569},
  year      = {2009}
}

@article{Schulthess2018,
  author    = {Schulthess, Thomas C. and Bauer, Peter and Wedi, Nils and Fuhrer, Oliver and Hoefler, Torsten and Schär, Christoph},
  title     = {Reflecting on the goal and baseline for exascale computing: a roadmap based on weather and climate simulations},
  journal   = {Computing in Science and Engineering},
  volume    = {21},
  number    = {1},
  pages     = {30--41},
  year      = {2018}
}

@article{Seo2009,
  author    = {Seo, Cheol and Thorne, James H. and Hannah, Lee and Thuiller, Wilfried},
  title     = {Scale effects in species distribution models: implications for conservation planning under climate change},
  journal   = {Biology Letters},
  volume    = {5},
  number    = {1},
  pages     = {39--43},
  year      = {2009}
}

@article{Tabios1985,
  author    = {Tabios, G. Q. and Salas, J. D.},
  title     = {A Comparative Analysis of Techniques for Spatial Interpolation of Precipitation},
  journal   = {JAWRA Journal of the American Water Resources Association},
  volume    = {21},
  number    = {3},
  pages     = {365--380},
  year      = {1985}
}

@article{Thornton1997,
  author    = {Thornton, Peter E. and Running, Steven W. and White, Michael A.},
  title     = {Generating surfaces of daily meteorological variables over large regions of complex terrain},
  journal   = {Journal of Hydrology},
  volume    = {190},
  pages     = {214--251},
  year      = {1997}
}

@article{Thuiller2003,
  author    = {Thuiller, Wilfried},
  title     = {BIOMOD – optimizing predictions of species distributions and projecting potential future shifts under global change},
  journal   = {Global Change Biology},
  volume    = {9},
  number    = {8},
  pages     = {1353--1362},
  year      = {2003}
}

@article{Wood2004,
  author    = {Wood, Andrew W. and Leung, Lai-Yung R. and Sridhar, Venkataraman and Lettenmaier, Dennis P.},
  title     = {Hydrologic Implications of Dynamical and Statistical Approaches to Downscaling Climate Model Outputs},
  journal   = {Climatic Change},
  volume    = {62},
  number    = {1-3},
  pages     = {189--216},
  year      = {2004}
}

@article{Hickler2012,
  author    = {Hickler, Thomas and Vohland, Katrin and Feehan, John and Miller, Paul A. and Smith, Benjamin and Costa, Lu{\'\i}s and Giesecke, Thomas and Fronzek, Stefan and Carter, Timothy R. and Cramer, Wolfgang and others},
  title     = {Projecting the future distribution of European potential natural vegetation zones with a generalized, tree species‐based dynamic vegetation model},
  journal   = {Global Ecology and Biogeography},
  volume    = {21},
  number    = {1},
  pages     = {50--63},
  year      = {2012}
}

@article{Koenig2021,
  author    = {K{\"o}nig, Christian and Karger, Dirk N. and Cord, Anna F. and Sauermann, Julian and Kreft, Holger and Zimmermann, Niklaus E. and Thuiller, Wilfried and Zurell, Damaris},
  title     = {Scale dependency of joint species distribution models challenges interpretation of biotic interactions},
  journal   = {Journal of Biogeography},
  volume    = {48},
  number    = {4},
  pages     = {839--850},
  year      = {2021}
}

@article{Karger2020c,
  author    = {Karger, Dirk N. and Conrad, Olaf and B\"ohner, J\"urgen and Kawohl, Tobias and Kreft, Holger and Soria‐Auza, Rodrigo W. and Zimmermann, Niklaus E. and Linder, Hans Peter and Kessler, Michael and Guisan, Antoine and others},
  title     = {Climatologies at high resolution for the earth’s land surface areas},
  journal   = {Scientific Data},
  volume    = {4},
  pages     = {170122},
  year      = {2020}
}

@article{Lange2019,
  author    = {Lange, Stefan},
  title     = {WFDE5 over land merged with ERA5 over the ocean (W5E5)},
  journal   = {ISIMIP},
  year      = {2019},
  note      = {\url{https://doi.org/10.48364/ISIMIP.342217}}
}

@article{Cucchi2020,
  author    = {Cucchi, Martina and Weedon, Graham P. and Amici, Alessandra and Bellouin, Nicolas and Lange, Stefan and Müller Schmied, Hannes and Hersbach, Hans and Buontempo, Carlo},
  title     = {WFDE5: bias-adjusted ERA5 reanalysis data for impact studies},
  journal   = {Earth System Science Data},
  volume    = {12},
  pages     = {2097--2120},
  year      = {2020}
}

@article{Dipankar2015,
  author    = {Dipankar, Anurag and Stevens, Bjorn and Heinze, Ralf and Moseley, Christopher and Zängl, Günther and Giorgetta, Marco and Brdar, Slavko},
  title     = {Large eddy simulation using the general circulation model ICON},
  journal   = {Journal of Advances in Modeling Earth Systems},
  volume    = {7},
  number    = {3},
  pages     = {963--986},
  year      = {2015}
}

@article{Schaer2019,
  author    = {Schär, Christoph and Fuhrer, Oliver and Arteaga, Andres and Ban, Nikolina and Charpilloz, Christophe and Di Girolamo, Sara and Hentgen, Lukas and Hoefler, Torsten and Lapillonne, Xavier and Leutwyler, David and others},
  title     = {Kilometer-scale climate models: Prospects and challenges},
  journal   = {Bulletin of the American Meteorological Society},
  volume    = {101},
  number    = {5},
  pages     = {E567--E587},
  year      = {2019}
}

@article{Fuhrer2018,
  author    = {Fuhrer, Oliver and Osuna, Carlos and Lapillonne, Xavier and Gysi, Thomas and Cumming, Bruce and Bianco, Mario and Arteaga, Andres and Schär, Christoph and Wedi, Nils},
  title     = {Near-global climate simulation at 1 km resolution: Establishing a performance baseline on 4888 GPUs with COSMO 5.0},
  journal   = {Geoscientific Model Development},
  volume    = {11},
  number    = {4},
  pages     = {1665--1681},
  year      = {2018}
}

@article{Daly1997,
  author    = {Daly, Christopher and Taylor, George H. and Gibson, Wayne P. and Parzybok, Thomas W. and Johnson, G. H. and Pasteris, Paul A.},
  title     = {High-quality spatial climate data sets for the United States and beyond},
  journal   = {Transactions of the ASAE},
  volume    = {43},
  number    = {6},
  pages     = {1957--1962},
  year      = {1997}
}

@article{Briggs1996,
  author    = {Briggs, William M. and Cogley, J. Graham},
  title     = {Topographic bias in climatological observations of precipitation due to coarseness of the grid},
  journal   = {Journal of Geophysical Research: Atmospheres},
  volume    = {101},
  number    = {D3},
  pages     = {2611--2619},
  year      = {1996}
}

@article{Schneider2014,
  author    = {Schneider, Udo and Becker, Andreas and Finger, Peter and Meyer-Christoffer, Anja and Ziese, Markus and Rudolf, Bruno},
  title     = {GPCC’s new land surface precipitation climatology based on quality-controlled in situ data and its role in quantifying the global water cycle},
  journal   = {Theoretical and Applied Climatology},
  volume    = {115},
  pages     = {15--40},
  year      = {2014}
}

@article{Kidd2017,
  author    = {Kidd, Chris and Huffman, George and Wilheit, Thomas and Ferraro, Ralph and Joyce, Robert and Hsu, Kuolin and Braithwaite, Dan},
  title     = {GPM satellite constellation overview and applications to precipitation monitoring},
  journal   = {Remote Sensing},
  volume    = {9},
  number    = {4},
  pages     = {382},
  year      = {2017}
}

@article{Berndt2018,
  author    = {Berndt, Christian and Diedrich, Henrik and Früh, Barbara and Hofmann, Martin and Kruschke, Tim and Leuprecht, Andrea and Zimmer, Julia},
  title     = {Climate services for Germany: first experiences of national provision of climate information},
  journal   = {Climate Services},
  volume    = {9},
  pages     = {1--9},
  year      = {2018}
}

@article{Karger2020b,
  author    = {Karger, Dirk N. and Conrad, Olaf and B\"ohner, J\"urgen and Kawohl, Tobias and Kreft, Holger and Soria‐Auza, Rodrigo W. and Zimmermann, Niklaus E. and Linder, Hans Peter and Kessler, Michael and Guisan, Antoine and others},
  title     = {Data descriptor: Climatologies at high resolution for the earth’s land surface areas},
  journal   = {Scientific Data},
  volume    = {4},
  pages     = {170122},
  year      = {2020}
}

@article{Wada2016,
  title = {High-Resolution Modeling of Human and Climate Impacts on Global Water Resources},
  author = {Wada, Yoshihide and Graaf, Inge E. M. and Beek, Ludovicus P. H.},
  year = {2016},
  journal = {Journal of Advances in Modeling Earth Systems},
  volume = {8},
  number = {2},
  pages = {735--763},
  issn = {1942-2466},
  doi = {10.1002/2015MS000618},
  langid = {english}
}

@article{Harris2021,
  title = {Global Maps of Twenty-First Century Forest Carbon Fluxes},
  author = {Harris, Nancy L. and Gibbs, David A. and Baccini, Alessandro and Birdsey, Richard A. and Bruin, Sytze and Farina, Mary and Fatoyinbo, Lola and Hansen, Matthew C. and Herold, Martin and Houghton, Richard A. and Potapov, Peter V. and Suarez, Daniela Requena and Roman-Cuesta, Rosa M. and Saatchi, Sassan S. and Slay, Christy M. and Turubanova, Svetlana A. and Tyukavina, Alexandra},
  date = {2021},
  year = {2021},
  journal = {Nature Climate Change},
  shortjournal = {Nat. Clim. Chang.},
  volume = {11},
  number = {3},
  pages = {234--240},
  publisher = {Nature Publishing Group},
  issn = {1758-6798},
  doi = {10.1038/s41558-020-00976-6},
  langid = {english}
}

@book{Foken2024,
  title = {Micrometeorology},
  author = {Foken, Thomas and Mauder, Matthias},
  year = {2024},
  publisher = {Springer International Publishing},
  location = {Cham},
  doi = {10.1007/978-3-031-47526-9},
  isbn = {978-3-031-47525-2 978-3-031-47526-9}
}

@article{Mahnken2022,
  title = {Accuracy, Realism and General Applicability of European Forest Models},
  author = {Mahnken, Mats and Cailleret, Maxime and Collalti, Alessio and Trotta, Carlo and Biondo, Corrado and D'Andrea, Ettore and Dalmonech, Daniela and Marano, Gina and Mäkelä, Annikki and Minunno, Francesco and Peltoniemi, Mikko and Trotsiuk, Volodymyr and Nadal-Sala, Daniel and Sabaté, Santiago and Vallet, Patrick and Aussenac, Raphaël and Cameron, David R. and Bohn, Friedrich J. and Grote, Rüdiger and Augustynczik, Andrey L. D. and Yousefpour, Rasoul and Huber, Nica and Bugmann, Harald and Merganičová, Katarina and Merganic, Jan and Valent, Peter and Lasch-Born, Petra and Hartig, Florian and Vega del Valle, Iliusi D. and Volkholz, Jan and Gutsch, Martin and Matteucci, Giorgio and Krejza, Jan and Ibrom, Andreas and Meesenburg, Henning and Rötzer, Thomas and van der Maaten-Theunissen, Marieke and van der Maaten, Ernst and Reyer, Christopher P. O.},
  year = {2022},
  journal = {Global Change Biology},
  volume = {28},
  number = {23},
  pages = {6921--6943},
  issn = {1365-2486},
  doi = {10.1111/gcb.16384},
}

@article{Pastorello2020,
  title = {The FLUXNET2015 Dataset and the ONEFlux Processing Pipeline for Eddy Covariance Data},
  author = {Pastorello, Gilberto and Trotta, Carlo and Canfora, Eleonora and Chu, Housen and Christianson, Danielle and Cheah, You-Wei and Poindexter, Cristina and Chen, Jiquan and Elbashandy, Abdelrahman and Humphrey, Marty and Isaac, Peter and Polidori, Diego and Reichstein, Markus and Ribeca, Alessio and Vuichard, Nicolas and Zhang, Leiming and Amiro, Brian and Ammann, Christof and Arain, M. Altaf and Ardö, Jonas and Arkebauer, Timothy and Arndt, Stefan K. and Arriga, Nicola and Aubinet, Marc and Aurela, Mika and Baldocchi, Dennis and Barr, Alan and Beamesderfer, Eric and Marchesini, Luca Belelli and Bergeron, Onil and Beringer, Jason and Bernhofer, Christian and Berveiller, Daniel and Billesbach, Dave and Black, Thomas Andrew and Blanken, Peter D. and Bohrer, Gil and Boike, Julia and Bolstad, Paul V. and Bonal, Damien and Bonnefond, Jean-Marc and Bowling, David R. and Bracho, Rosvel and Brodeur, Jason and Brümmer, Christian and Buchmann, Nina and Burban, Benoit and Burns, Sean P. and Buysse, Pauline and Cale, Peter and Cavagna, Mauro and Cellier, Pierre and Chen, Shiping and Chini, Isaac and Christensen, Torben R. and Cleverly, James and Collalti, Alessio and Consalvo, Claudia and Cook, Bruce D. and Cook, David and Coursolle, Carole and Cremonese, Edoardo and Curtis, Peter S. and D’Andrea, Ettore and Dai, Xiaoqin and Davis, Kenneth J. and Cinti, Bruno De and Ligne, Anne De and De Oliveira, Raimundo C. and Delpierre, Nicolas and Desai, Ankur R. and Di Bella, Carlos Marcelo and Dolman, Han and Domingo, Francisco and Dong, Gang and Dore, Sabina and Duce, Pierpaolo and Dufrêne, Eric and Dunn, Allison and Dušek, Jiří and Eamus, Derek and Eichelmann, Uwe and ElKhidir, Hatim Abdalla M. and Eugster, Werner and Ewenz, Cacilia M. and Ewers, Brent and Famulari, Daniela and Fares, Silvano and Feigenwinter, Iris and Feitz, Andrew and Fensholt, Rasmus and Filippa, Gianluca and Fischer, Marc and Frank, John and Galvagno, Marta and Gharun, Mana and Gianelle, Damiano and Gielen, Bert and Gioli, Beniamino and Gitelson, Anatoly and Goded, Ignacio and Goeckede, Mathias and Goldstein, Allen H. and Gough, Christopher M. and Goulden, Michael L. and Graf, Alexander and Griebel, Anne and Gruening, Carsten and Grünwald, Thomas and Hammerle, Albin and Han, Shijie and Han, Xingguo and Hansen, Birger Ulf and Hanson, Chad and Hatakka, Juha and He, Yongtao and Hehn, Markus and Heinesch, Bernard and Hinko-Najera, Nina and Hörtnagl, Lukas and Hutley, Lindsay and Ibrom, Andreas and Ikawa, Hiroki and Jackowicz-Korczynski, Marcin and Janouš, Dalibor and Jans, Wilma and Jassal, Rachhpal and Jiang, Shicheng and Kato, Tomomichi and Khomik, Myroslava and Klatt, Janina and Knohl, Alexander and Knox, Sara and Kobayashi, Hideki and Koerber, Georgia and Kolle, Olaf and Kosugi, Yoshiko and Kotani, Ayumi and Kowalski, Andrew and Kruijt, Bart and Kurbatova, Julia and Kutsch, Werner L. and Kwon, Hyojung and Launiainen, Samuli and Laurila, Tuomas and Law, Bev and Leuning, Ray and Li, Yingnian and Liddell, Michael and Limousin, Jean-Marc and Lion, Marryanna and Liska, Adam J. and Lohila, Annalea and López-Ballesteros, Ana and López-Blanco, Efrén and Loubet, Benjamin and Loustau, Denis and Lucas-Moffat, Antje and Lüers, Johannes and Ma, Siyan and Macfarlane, Craig and Magliulo, Vincenzo and Maier, Regine and Mammarella, Ivan and Manca, Giovanni and Marcolla, Barbara and Margolis, Hank A. and Marras, Serena and Massman, William and Mastepanov, Mikhail and Matamala, Roser and Matthes, Jaclyn Hatala and Mazzenga, Francesco and McCaughey, Harry and McHugh, Ian and McMillan, Andrew M. S. and Merbold, Lutz and Meyer, Wayne and Meyers, Tilden and Miller, Scott D. and Minerbi, Stefano and Moderow, Uta and Monson, Russell K. and Montagnani, Leonardo and Moore, Caitlin E. and Moors, Eddy and Moreaux, Virginie and Moureaux, Christine and Munger, J. William and Nakai, Taro and Neirynck, Johan and Nesic, Zoran and Nicolini, Giacomo and Noormets, Asko and Northwood, Matthew and Nosetto, Marcelo and Nouvellon, Yann and Novick, Kimberly and Oechel, Walter and Olesen, Jørgen Eivind and Ourcival, Jean-Marc and Papuga, Shirley A. and Parmentier, Frans-Jan and Paul-Limoges, Eugenie and Pavelka, Marian and Peichl, Matthias and Pendall, Elise and Phillips, Richard P. and Pilegaard, Kim and Pirk, Norbert and Posse, Gabriela and Powell, Thomas and Prasse, Heiko and Prober, Suzanne M. and Rambal, Serge and Rannik, {\"U}llar and Raz-Yaseef, Naama and Rebmann, Corinna and Reed, David and Restrepo-Coupe, Natalia and Reverter, Borja R. and Roland, Marilyn and Sabbatini, Simone and Sachs, Torsten and Saleska, Scott R. and Sánchez-Cañete, Enrique P. and Sanchez-Mejia, Zulia M. and Schmid, Hans Peter and Schmidt, Marius and Schneider, Karl and Schrader, Frederik and Schroder, Ivan and Scott, Russell L. and Sedlák, Pavel and Serrano-Ortíz, Penélope and Shao, Changliang and Shi, Peili and Shironya, Ivan and Siebicke, Lukas and Šigut, Ladislav and Silberstein, Richard and Sirca, Costantino and Spano, Donatella and Steinbrecher, Rainer and Stevens, Robert M. and Sturtevant, Cove and Suyker, Andy and Tagesson, Torbern and Takanashi, Satoru and Tang, Yanhong and Tapper, Nigel and Thom, Jonathan and Tomassucci, Michele and Tuovinen, Juha-Pekka and Urbanski, Shawn and Valentini, Riccardo and Varlagin, Andrej and Verfaillie, Joseph and Vesala, Timo and Vincke, Caroline and Vitale, Domenico and Vygodskaya, Natalia and Walker, Jeffrey P. and Walter-Shea, Elizabeth and Wang, Huimin and Weber, Robin and Westermann, Sebastian and Wille, Christian and Wofsy, Steven and Wohlfahrt, Georg and Wolf, Sebastian and Woodgate, William and Li, Yuelin and Zampedri, Roberto and Zhang, Junhui and Zhou, Guoyi and Zona, Donatella and Agarwal, Deb and Biraud, Sebastien and Torn, Margaret and Papale, Dario},
  date = {2020},
  year = {2020},
  journal = {Scientific Data},
  shortjournal = {Sci Data},
  volume = {7},
  number = {1},
  pages = {225},
  publisher = {Nature Publishing Group},
  issn = {2052-4463},
  doi = {10.1038/s41597-020-0534-3},
  langid = {english}
}

@article{Lasslop2010,
  title = {Separation of Net Ecosystem Exchange into Assimilation and Respiration Using a Light Response Curve Approach: Critical Issues and Global Evaluation},
  shorttitle = {Separation of Net Ecosystem Exchange into Assimilation and Respiration Using a Light Response Curve Approach},
  author = {Lasslop, Gitta and Reichstein, Markus and Papale, Dario and Richardson, Andrew D. and Arneth, Almut and Barr, Alan and Stoy, Paul and Wohlfahrt, Georg},
  year = {2010},
  journal = {Global Change Biology},
  volume = {16},
  number = {1},
  pages = {187--208},
  issn = {1365-2486},
  doi = {10.1111/j.1365-2486.2009.02041.x},
  langid = {english}
}

@article{Goudsmit2002,
  title = {Application of K-$\varepsilon$ Turbulence Models to Enclosed Basins: The Role of Internal Seiches},
  shorttitle = {Application of K-ϵ Turbulence Models to Enclosed Basins},
  author = {Goudsmit, G.-H. and Burchard, H. and Peeters, F. and Wüest, A.},
  year = {2002},
  journal = {Journal of Geophysical Research: Oceans},
  volume = {107},
  number = {C12},
  pages = {3230},
  issn = {2156-2202},
  doi = {10.1029/2001JC000954},
  langid = {english}
}

@article{Gaudard2019,
  title = {Toward an Open Access to High-Frequency Lake Modeling and Statistics Data for Scientists and Practitioners – the Case of Swiss Lakes Using Simstrat v2.1},
  author = {Gaudard, Adrien and Råman Vinnå, Love and Bärenbold, Fabian and Schmid, Martin and Bouffard, Damien},
  year = {2019},
  journal = {Geoscientific Model Development},
  volume = {12},
  number = {9},
  pages = {3955--3974},
  publisher = {Copernicus GmbH},
  issn = {1991-959X},
  doi = {10.5194/gmd-12-3955-2019},
  langid = {english}
}

@article{Golub2022,
  title = {A Framework for Ensemble Modelling of Climate Change Impacts on Lakes Worldwide: The ISIMIP Lake Sector},
  shorttitle = {A Framework for Ensemble Modelling of Climate Change Impacts on Lakes Worldwide},
  author = {Golub, Malgorzata and Thiery, Wim and Marcé, Rafael and Pierson, Don and Vanderkelen, Inne and Mercado-Bettin, Daniel and Woolway, R. Iestyn and Grant, Luke and Jennings, Eleanor and Kraemer, Benjamin M. and Schewe, Jacob and Zhao, Fang and Frieler, Katja and Mengel, Matthias and Bogomolov, Vasiliy Y. and Bouffard, Damien and Côté, Marianne and Couture, Raoul-Marie and Debolskiy, Andrey V. and Droppers, Bram and Gal, Gideon and Guo, Mingyang and Janssen, Annette B. G. and Kirillin, Georgiy and Ladwig, Robert and Magee, Madeline and Moore, Tadhg and Perroud, Marjorie and Piccolroaz, Sebastiano and Raaman Vinnaa, Love and Schmid, Martin and Shatwell, Tom and Stepanenko, Victor M. and Tan, Zeli and Woodward, Bronwyn and Yao, Huaxia and Adrian, Rita and Allan, Mathew and Anneville, Orlane and Arvola, Lauri and Atkins, Karen and Boegman, Leon and Carey, Cayelan and Christianson, Kyle and Eyto, Elvira and DeGasperi, Curtis and Grechushnikova, Maria and Hejzlar, Josef and Joehnk, Klaus and Jones, Ian D. and Laas, Alo and Mackay, Eleanor B. and Mammarella, Ivan and Markensten, Hampus and McBride, Chris and Özkundakci, Deniz and Potes, Miguel and Rinke, Karsten and Robertson, Dale and Rusak, James A. and Salgado, Rui and Linden, Leon and Verburg, Piet and Wain, Danielle and Ward, Nicole K. and Wollrab, Sabine and Zdorovennova, Galina},
  year = {2022},
  journal = {Geoscientific Model Development},
  volume = {15},
  number = {11},
  pages = {4597--4623},
  publisher = {Copernicus GmbH},
  issn = {1991-959X},
  doi = {10.5194/gmd-15-4597-2022},
  langid = {english}
}

@article{Marshall1997,
  title = {A Finite-Volume, Incompressible Navier Stokes Model for Studies of the Ocean on Parallel Computers},
  author = {Marshall, John and Adcroft, Alistair and Hill, Chris and Perelman, Lev and Heisey, Curt},
  year = {1997},
  journal = {Journal of Geophysical Research: Oceans},
  volume = {102},
  number = {C3},
  pages = {5753--5766},
  issn = {2156-2202},
  doi = {10.1029/96JC02775},
  langid = {english}
}

@article{Marshall1997a,
  title = {Hydrostatic, Quasi-Hydrostatic, and Nonhydrostatic Ocean Modeling},
  author = {Marshall, John and Hill, Chris and Perelman, Lev and Adcroft, Alistair},
  year = {1997},
  journal = {Journal of Geophysical Research: Oceans},
  volume = {102},
  number = {C3},
  pages = {5733--5752},
  issn = {2156-2202},
  doi = {10.1029/96JC02776},
  langid = {english}
}

@article{Arnold1998,
  title = {Large Area Hydrologic Modeling and Assessment Part I: Model Development},
  author = {Arnold, J. G. and Srinivasan, R. and Muttiah, R. S. and Williams, J. R.},
  year = {1998},
  journal = {JAWRA Journal of the American Water Resources Association},
  volume = {34},
  number = {1},
  pages = {73--89},
  issn = {1752-1688},
  doi = {10.1111/j.1752-1688.1998.tb05961.x},
}

@article{Bieger2017,
  title = {Introduction to SWAT+, A Completely Restructured Version of the Soil and Water Assessment Tool},
  author = {Bieger, Katrin and Arnold, Jeffrey G. and Rathjens, Hendrik and White, Michael J. and Bosch, David D. and Allen, Peter M. and Volk, Martin and Srinivasan, Raghavan},
  year = {2017},
  journal = {JAWRA Journal of the American Water Resources Association},
  volume = {53},
  number = {1},
  pages = {115--130},
  issn = {1752-1688},
  doi = {10.1111/1752-1688.12482},
  langid = {english}
}

@book{Doherty2018,
  title = {PEST User Manual},
  author = {Doherty, J},
  year = {2018},
  edition = {7},
  publisher = {Watermark Numerical Computing},
  location = {Brisbane, Queensland, Australia}
}

@article{Perrin2003,
  title = {Improvement of a Parsimonious Model for Streamflow Simulation},
  author = {Perrin, Charles and Michel, Claude and Andréassian, Vazken},
  date = {2003},
  year = {2003},
  journal = {Journal of Hydrology},
  shortjournal = {Journal of Hydrology},
  volume = {279},
  number = {1},
  pages = {275--289},
  issn = {0022-1694},
  doi = {10.1016/S0022-1694(03)00225-7}
}

@article{Kapoor2023,
  title = {DeepGR4J: A deep learning hybridization approach for conceptual rainfall-runoff modelling},
  author = {Kapoor, Arpit and Pathiraja, Sahani and Marshall, Lucy and Chandra, Rohitash},
  year = {2023},
  journal = {Environmental Modelling \& Software},
  shortjournal = {Environmental Modelling \& Software},
  volume = {169},
  pages = {105831},
  issn = {1364-8152},
  doi = {10.1016/j.envsoft.2023.105831}
}

@article{Burek2020,
  title = {Development of the Community Water Model (CWatM v1.04) – a high-resolution hydrological model for global and regional assessment of integrated water resources management},
  author = {Burek, Peter and Satoh, Yusuke and Kahil, Taher and Tang, Ting and Greve, Peter and Smilovic, Mikhail and Guillaumot, Luca and Zhao, Fang and Wada, Yoshihide},
  year = {2020},
  journal = {Geoscientific Model Development},
  volume = {13},
  number = {7},
  pages = {3267--3298},
  publisher = {Copernicus GmbH},
  issn = {1991-959X},
  doi = {10.5194/gmd-13-3267-2020}
}

@article{Guillaumot2022,
  title = {Coupling a Large-Scale Hydrological Model (CWatM v1.1) with a High-Resolution Groundwater Flow Model (MODFLOW 6) to Assess the Impact of Irrigation at Regional Scale},
  author = {Guillaumot, Luca and Smilovic, Mikhail and Burek, Peter and Bruijn, Jens and Greve, Peter and Kahil, Taher and Wada, Yoshihide},
  year = {2022},
  journal = {Geoscientific Model Development},
  volume = {15},
  number = {18},
  pages = {7099--7120},
  publisher = {Copernicus GmbH},
  issn = {1991-959X},
  doi = {10.5194/gmd-15-7099-2022},
  langid = {english}
}

@article{Lee2022,
  title = {Mapping Sugarcane in Central India with Smartphone Crowdsourcing},
  author = {Lee, Ju Young and Wang, Sherrie and Figueroa, Anjuli Jain and Strey, Rob and Lobell, David B. and Naylor, Rosamond L. and Gorelick, Steven M.},
  year = {2022},
  journal = {Remote Sensing},
  volume = {14},
  number = {3},
  pages = {703},
  publisher = {Multidisciplinary Digital Publishing Institute},
  issn = {2072-4292},
  doi = {10.3390/rs14030703},
  langid = {english}
}

@article{Meier2018a,
  title = {A Global Approach to Estimate Irrigated Areas – a Comparison between Different Data and Statistics},
  author = {Meier, Jonas and Zabel, Florian and Mauser, Wolfram},
  year = {2018},
  journal = {Hydrology and Earth System Sciences},
  volume = {22},
  number = {2},
  pages = {1119--1133},
  publisher = {Copernicus GmbH},
  issn = {1027-5606},
  doi = {10.5194/hess-22-1119-2018},
  langid = {english}
}

@article{Chang2013,
  title = {Incorporating Grassland Management in ORCHIDEE: Model Description and Evaluation at 11 Eddy-Covariance Sites in Europe},
  author = {Chang, J. F. and Viovy, N. and Vuichard, N. and Ciais, P. and Wang, T. and Cozic, A. and Lardy, R. and Graux, A.-I. and Klumpp, K. and Martin, R. and Soussana, J.-F.},
  year = {2013},
  journal = {Geoscientific Model Development},
  volume = {6},
  number = {6},
  pages = {2165--2181},
  publisher = {Copernicus GmbH},
  issn = {1991-959X},
  doi = {10.5194/gmd-6-2165-2013},
}

@article{Chang2021,
  title = {Climate Warming from Managed Grasslands Cancels the Cooling Effect of Carbon Sinks in Sparsely Grazed and Natural Grasslands},
  author = {Chang, Jinfeng and Ciais, Philippe and Gasser, Thomas and Smith, Pete and Herrero, Mario and Havlík, Petr and Obersteiner, Michael and Guenet, Bertrand and Goll, Daniel S. and Li, Wei and Naipal, Victoria and Peng, Shushi and Qiu, Chunjing and Tian, Hanqin and Viovy, Nicolas and Yue, Chao and Zhu, Dan},
  year = {2021},
  journal = {Nature Communications},
  shortjournal = {Nat Commun},
  volume = {12},
  number = {1},
  pages = {118},
  publisher = {Nature Publishing Group},
  issn = {2041-1723},
  doi = {10.1038/s41467-020-20406-7},
  langid = {english}
}

@article{Ciais2005,
  title = {Europe-Wide Reduction in Primary Productivity Caused by the Heat and Drought in 2003},
  author = {Ciais, Ph and Reichstein, M. and Viovy, N. and Granier, A. and Ogée, J. and Allard, V. and Aubinet, M. and Buchmann, N. and Bernhofer, Chr and Carrara, A. and Chevallier, F. and De Noblet, N. and Friend, A. D. and Friedlingstein, P. and Grünwald, T. and Heinesch, B. and Keronen, P. and Knohl, A. and Krinner, G. and Loustau, D. and Manca, G. and Matteucci, G. and Miglietta, F. and Ourcival, J. M. and Papale, D. and Pilegaard, K. and Rambal, S. and Seufert, G. and Soussana, J. F. and Sanz, M. J. and Schulze, E. D. and Vesala, T. and Valentini, R.},
  year = {2005},
  journal = {Nature},
  volume = {437},
  number = {7058},
  pages = {529--533},
  publisher = {Nature Publishing Group},
  issn = {1476-4687},
  doi = {10.1038/nature03972},
  langid = {english}
}

@article{Guimberteau2018,
  title = {ORCHIDEE-MICT (v8.4.1), a Land Surface Model for the High Latitudes: Model Description and Validation},
  author = {Guimberteau, Matthieu and Zhu, Dan and Maignan, Fabienne and Huang, Ye and Yue, Chao and Dantec-Nédélec, Sarah and Ottlé, Catherine and Jornet-Puig, Albert and Bastos, Ana and Laurent, Pierre and Goll, Daniel and Bowring, Simon and Chang, Jinfeng and Guenet, Bertrand and Tifafi, Marwa and Peng, Shushi and Krinner, Gerhard and Ducharne, Agnès and Wang, Fuxing and Wang, Tao and Wang, Xuhui and Wang, Yilong and Yin, Zun and Lauerwald, Ronny and Joetzjer, Emilie and Qiu, Chunjing and Kim, Hyungjun and Ciais, Philippe},
  year = {2018},
  journal = {Geoscientific Model Development},
  volume = {11},
  number = {1},
  pages = {121--163},
  publisher = {Copernicus GmbH},
  issn = {1991-959X},
  doi = {10.5194/gmd-11-121-2018},
  langid = {english}
}

@article{Krinner2005,
  title = {A Dynamic Global Vegetation Model for Studies of the Coupled Atmosphere-Biosphere System},
  author = {Krinner, G. and Viovy, Nicolas and Noblet-Ducoudré, Nathalie and Ogée, Jérôme and Polcher, Jan and Friedlingstein, Pierre and Ciais, Philippe and Sitch, Stephen and Prentice, I. Colin},
  year = {2005},
  journal = {Global Biogeochemical Cycles},
  volume = {19},
  number = {1},
  issn = {1944-9224},
  doi = {10.1029/2003GB002199},
  langid = {english}
}

@article{Piao2007,
  title = {Growing Season Extension and Its Impact on Terrestrial Carbon Cycle in the Northern Hemisphere over the Past 2 Decades},
  author = {Piao, Shilong and Friedlingstein, Pierre and Ciais, Philippe and Viovy, Nicolas and Demarty, Jérôme},
  year = {2007},
  journal = {Global Biogeochemical Cycles},
  volume = {21},
  number = {3},
  issn = {1944-9224},
  doi = {10.1029/2006GB002888},
}

@article{Qiu2018,
  title = {ORCHIDEE-PEAT (Revision 4596), a Model for Northern Peatland CO \textsubscript{2}, Water, and Energy Fluxes on Daily to Annual Scales},
  author = {Qiu, Chunjing and Zhu, Dan and Ciais, Philippe and Guenet, Bertrand and Krinner, Gerhard and Peng, Shushi and Aurela, Mika and Bernhofer, Christian and Brümmer, Christian and Bret-Harte, Syndonia and Chu, Housen and Chen, Jiquan and Desai, Ankur R. and Dušek, Jiří and Euskirchen, Eugénie S. and Fortuniak, Krzysztof and Flanagan, Lawrence B. and Friborg, Thomas and Grygoruk, Mateusz and Gogo, Sébastien and Grünwald, Thomas and Hansen, Birger U. and Holl, David and Humphreys, Elyn and Hurkuck, Miriam and Kiely, Gerard and Klatt, Janina and Kutzbach, Lars and Largeron, Chloé and Laggoun-Défarge, Fatima and Lund, Magnus and Lafleur, Peter M. and Li, Xuefei and Mammarella, Ivan and Merbold, Lutz and Nilsson, Mats B. and Olejnik, Janusz and Ottosson-Löfvenius, Mikaell and Oechel, Walter and Parmentier, Frans-Jan W. and Peichl, Matthias and Pirk, Norbert and Peltola, Olli and Pawlak, Włodzimierz and Rasse, Daniel and Rinne, Janne and Shaver, Gaius and Schmid, Hans Peter and Sottocornola, Matteo and Steinbrecher, Rainer and Sachs, Torsten and Urbaniak, Marek and Zona, Donatella and Ziemblinska, Klaudia},
  year = {2018},
  journal = {Geoscientific Model Development},
  volume = {11},
  number = {2},
  pages = {497--519},
  publisher = {Copernicus GmbH},
  issn = {1991-959X},
  doi = {10.5194/gmd-11-497-2018}
}

@article{Qiu2022,
  title = {A Strong Mitigation Scenario Maintains Climate Neutrality of Northern Peatlands},
  author = {Qiu, Chunjing and Ciais, Philippe and Zhu, Dan and Guenet, Bertrand and Chang, Jinfeng and Chaudhary, Nitin and Kleinen, Thomas and Li, XinYu and Müller, Jurek and Xi, Yi and Zhang, Wenxin and Ballantyne, Ashley and Brewer, Simon C. and Brovkin, Victor and Charman, Dan J. and Gustafson, Adrian and Gallego-Sala, Angela V. and Gasser, Thomas and Holden, Joseph and Joos, Fortunat and Kwon, Min Jung and Lauerwald, Ronny and Miller, Paul A. and Peng, Shushi and Page, Susan and Smith, Benjamin and Stocker, Benjamin D. and Sannel, A. Britta K. and Salmon, Elodie and Schurgers, Guy and Shurpali, Narasinha J. and Wårlind, David and Westermann, Sebastian},
  date = {2022},
  year = {2022},
  journal = {One Earth},
  shortjournal = {One Earth},
  volume = {5},
  number = {1},
  pages = {86--97},
  issn = {2590-3322},
  doi = {10.1016/j.oneear.2021.12.008}
}

@misc{ESB2004,
  author       = {European Commission, Joint Research Centre, European Soil Bureau Network},
  title        = {The European Soil Database Distribution Version 2.0},
  year         = {2004},
  note         = {EUR 19945 EN}
}

@article{Panagos2006,
  title = {The European Soil Database},
  author = {Panagos, Panos},
  date = {2006},
  year = {2006},
  journal = {GEO: connexion},
  volume = {5},
  number = {7},
  pages = {32--33}
}

@article{Dasgupta2021,
  title = {Effects of Climate Change on Combined Labour Productivity and Supply: An Empirical, Multi-Model Study},
  shorttitle = {Effects of Climate Change on Combined Labour Productivity and Supply},
  author = {Dasgupta, Shouro and Maanen, Nicole and Gosling, Simon N and Piontek, Franziska and Otto, Christian and Schleussner, Carl-Friedrich},
  date = {2021-07-01},
  year = {2021},
  journal = {The Lancet Planetary Health},
  shortjournal = {The Lancet Planetary Health},
  volume = {5},
  number = {7},
  pages = {e455-e465},
  issn = {2542-5196},
  doi = {10.1016/S2542-5196(21)00170-4}
}

@article{Dasgupta2024,
  title = {Heat Stress and the Labour Force},
  author = {Dasgupta, Shouro and Robinson, Elizabeth J. Z. and Shayegh, Soheil and Bosello, Francesco and Park, R. Jisung and Gosling, Simon N.},
  date = {2024-12},
  year = {2024},
  journal = {Nature Reviews Earth \& Environment},
  shortjournal = {Nat Rev Earth Environ},
  volume = {5},
  number = {12},
  pages = {859--872},
  publisher = {Nature Publishing Group},
  issn = {2662-138X},
  doi = {10.1038/s43017-024-00606-1},
  langid = {english}
}

@article{Stocker2020,
  title = {P-Model v1.0: An Optimality-Based Light Use Efficiency Model for Simulating Ecosystem Gross Primary Production},
  shorttitle = {P-Model v1.0},
  author = {Stocker, Benjamin D. and Wang, Han and Smith, Nicholas G. and Harrison, Sandy P. and Keenan, Trevor F. and Sandoval, David and Davis, Tyler and Prentice, I. Colin},
  date = {2020-03-26},
  year = {2020},
  journal = {Geoscientific Model Development},
  volume = {13},
  number = {3},
  pages = {1545--1581},
  publisher = {Copernicus GmbH},
  issn = {1991-959X},
  doi = {10.5194/gmd-13-1545-2020},
  langid = {english}
}

@article{Collalti2014,
  title = {A Process-Based Model to Simulate Growth in Forests with Complex Structure: Evaluation and Use of 3D-CMCC Forest Ecosystem Model in a Deciduous Forest in Central Italy},
  shorttitle = {A Process-Based Model to Simulate Growth in Forests with Complex Structure},
  author = {Collalti, Alessio and Perugini, Lucia and Santini, Monia and Chiti, Tommaso and Nolè, Angelo and Matteucci, Giorgio and Valentini, Riccardo},
  year = {2014},
  journal = {Ecological Modelling},
  shortjournal = {Ecological Modelling},
  volume = {272},
  pages = {362--378},
  issn = {0304-3800},
  doi = {10.1016/j.ecolmodel.2013.09.016}
}

@article{Collalti2016,
  title = {Validation of 3D-CMCC Forest Ecosystem Model (v.5.1) against Eddy Covariance Data for 10 European Forest Sites},
  author = {Collalti, A. and Marconi, S. and Ibrom, A. and Trotta, C. and Anav, A. and D'Andrea, E. and Matteucci, G. and Montagnani, L. and Gielen, B. and Mammarella, I. and Grünwald, T. and Knohl, A. and Berninger, F. and Zhao, Y. and Valentini, R. and Santini, M.},
  year = {2016},
  journal = {Geoscientific Model Development},
  volume = {9},
  number = {2},
  pages = {479--504},
  publisher = {Copernicus GmbH},
  issn = {1991-959X},
  doi = {10.5194/gmd-9-479-2016}
}

@article{Collalti2018,
  title = {Thinning Can Reduce Losses in Carbon Use Efficiency and Carbon Stocks in Managed Forests Under Warmer Climate},
  author = {Collalti, Alessio and Trotta, Carlo and Keenan, Trevor F. and Ibrom, Andreas and Bond-Lamberty, Ben and Grote, Ruediger and Vicca, Sara and Reyer, Christopher P. O. and Migliavacca, Mirco and Veroustraete, Frank and Anav, Alessandro and Campioli, Matteo and Scoccimarro, Enrico and Šigut, Ladislav and Grieco, Elisa and Cescatti, Alessandro and Matteucci, Giorgio},
  year = {2018},
  journal = {Journal of Advances in Modeling Earth Systems},
  volume = {10},
  number = {10},
  pages = {2427--2452},
  issn = {1942-2466},
  doi = {10.1029/2018MS001275}
}

@article{Collalti2019,
  title = {The Sensitivity of the Forest Carbon Budget Shifts across Processes along with Stand Development and Climate Change},
  author = {Collalti, Alessio and Thornton, Peter E. and Cescatti, Alessandro and Rita, Angelo and Borghetti, Marco and Nolè, Angelo and Trotta, Carlo and Ciais, Philippe and Matteucci, Giorgio},
  year = {2019},
  journal = {Ecological Applications},
  volume = {29},
  number = {2},
  pages = {e01837},
  issn = {1939-5582},
  doi = {10.1002/eap.1837},
  langid = {english}
}

@article{Luo2012,
  author    = {D. Luo and Hui-jun Jin and Lin Lin and Rui-xia He and Si-zhong Yang and Xiao-li Chang},
  title     = {New Progress on Permafrost Temperature and Thickness in the Source Area of the Huanghe River},
  journal   = {Scientia Geographica Sinica},
  year      = {2012},
  volume    = {32},
  number    = {7},
  pages     = {898--904},
  doi       = {10.13249/j.cnki.sgs.2012.07.898}
}

@article{Wu2012,
  author    = {Q. Wu and T. Zhang and Y. Liu},
  title     = {Thermal state of the active layer and permafrost along the Qinghai-Xizang (Tibet) Railway from 2006 to 2010},
  journal   = {The Cryosphere},
  year      = {2012},
  volume    = {6},
  pages     = {607--612},
  doi       = {10.5194/tc-6-607-2012}
}

@article{Chen2015,
  author    = {H. Chen and Z. Nan and L. Zhao and Y. Ding and J. Chen and Q. Pang},
  title     = {Noah modelling of the permafrost distribution and characteristics in the West Kunlun area, Qinghai-Tibet Plateau, China},
  journal   = {Permafrost and Periglacial Processes},
  year      = {2015},
  volume    = {26},
  pages     = {160--174},
  doi       = {10.1002/ppp.1848}
}

@article{Chen2016,
  author    = {J. Chen and L. Zhao and Y. Sheng and J. Li and X.-d. Wu and E.-j. Du and G.-y. Liu and Q.-q. Pang},
  title     = {Some characteristics of permafrost and its distribution in the Gaize area on the Qinghai—Tibet Plateau, China},
  journal   = {Arctic, Antarctic, and Alpine Research},
  year      = {2016},
  volume    = {48},
  pages     = {395--409},
  doi       = {10.1657/AAAR0015-071}
}

@article{Qin2017,
  author    = {Y. Qin and T. Wu and L. Zhao and X. Wu and R. Li and C. Xie and Q. Pang and G. Hu and Y. Qiao and G. Zhao and G. Liu and X. Zhu and J. Hao},
  title     = {Numerical modeling of the active layer thickness and permafrost thermal state across Qinghai-Tibetan Plateau},
  journal   = {Journal of Geophysical Research: Atmospheres},
  year      = {2017},
  volume    = {122},
  pages     = {11604--11620},
  doi       = {10.1002/2017JD027424}
}

@article{Cao2018,
  author    = {B. Cao and T. Zhang and X. Peng and C. Mu and Q. Wang and L. Zheng and K. Wang and X. Y. Zhong},
  title     = {Thermal characteristics and recent changes of permafrost in the upper reaches of the Heihe River basin, Western China},
  journal   = {Journal of Geophysical Research: Atmospheres},
  year      = {2018},
  volume    = {123},
  pages     = {7935--7949},
  doi       = {10.1029/2018JD028442}
}

@article{Wu2015CALM,
  author    = {Q. Wu and Y. Hou and H. Yun and Y. Liu},
  title     = {Changes in active-layer thickness and near-surface permafrost between 2002 and 2012 in alpine ecosystems, Qinghai–Xizang (Tibet) plateau, China},
  journal   = {Cold Regions Science and Technology},
  year      = {2015},
  volume    = {116},
  pages     = {76--85},
  doi       = {10.1016/j.coldregions.2015.04.014}
}

@article{Wang2013,
  author    = {Q. F. Wang and T. J. Zhang and J. C. Wu and X. Q. Peng and X. Y. Zhong and C. Mou and G. D. Cheng},
  title     = {Investigation on permafrost distribution over the upper reaches of the Heihe River in the Qilian Mountains (in Chinese with English abstract)},
  journal   = {Journal of Glaciology and Geocryology},
  year      = {2013},
  volume    = {35},
  pages     = {19--25}
}

@article{Wu2008,
  author    = {Q. Wu and T. Zhang},
  title     = {Recent permafrost warming on the Qinghai-Tibetan Plateau},
  journal   = {Journal of Geophysical Research: Atmospheres},
  year      = {2008},
  volume    = {113},
  pages     = {D13108},
  doi       = {10.1029/2007JD009539}
}

@article{Wu2016,
  author    = {Q. Wu and Z. Zhang and S. Gao and W. Ma},
  title     = {Thermal impacts of engineering activities and vegetation layer on permafrost in different alpine ecosystems of the Qinghai–Tibet Plateau, China},
  journal   = {The Cryosphere},
  year      = {2016},
  volume    = {10},
  pages     = {1695--1706},
  doi       = {10.5194/tc-10-1695-2016}
}

@article{Wu2010,
  author    = {Q. Wu and T. Zhang},
  title     = {Changes in active layer thickness over the Qinghai-Tibetan Plateau from 1995 to 2007},
  journal   = {Journal of Geophysical Research: Atmospheres},
  year      = {2010},
  volume    = {115},
  pages     = {D09107},
  doi       = {10.1029/2009JD012974}
}

@article{Yu2008,
  author    = {H. Yu and Q. B. Wu and Y. Z. Liu},
  title     = {The long-term monitoring system on permafrost regions along the Qinghai-Tibet Railway (in Chinese with English abstract)},
  journal   = {Journal of Glaciology and Geocryology},
  year      = {2008},
  volume    = {30},
  pages     = {475--481}
}

@article{Li2016,
  author    = {J. Li and Y. Sheng and J. C. Wu and Z. L. Feng and Z. J. Ning and X. Y. Hu and X. M. Zhang},
  title     = {Mapping frozen soil distribution and modeling permafrost stability in the source area of the Yellow River (in Chinese with English abstract)},
  journal   = {Scientia Geographica Sinica},
  year      = {2016},
  volume    = {36},
  pages     = {588--596}
}

@article{Luo2018,
  author    = {Dongliang Luo and Huijun Jin and Xiaoying Jin and Ruixia He and Xiaoying Li and Reginald R. Muskett and Sergey S. Marchenko and Vladimir E. Romanovsky},
  title     = {Elevation‐dependent thermal regime and dynamics of frozen ground in the Bayan Har Mountains, northeastern Qinghai‐Tibet Plateau, southwest China},
  journal   = {Permafrost and Periglacial Processes},
  year      = {2018},
  volume    = {29},
  number    = {4},
  pages     = {257--270},
  doi       = {10.1002/ppp.1988}
}

@article{Wu2010Sci,
  author    = {J. Wu and Y. Sheng and Q. Wu and others},
  title     = {Processes and modes of permafrost degradation on the Qinghai-Tibet Plateau},
  journal   = {Science in China Series D: Earth Sciences},
  year      = {2010},
  volume    = {53},
  number    = {1},
  pages     = {150--158},
  doi       = {10.1007/s11430-009-0198-5}
}

@article{Xie2012,
  author    = {Changwei Xie and Lin Zhao and Tonghua Wu and Xicheng Dong},
  title     = {Changes in the thermal and hydraulic regime within the active layer in the Qinghai-Tibet Plateau},
  journal   = {Journal of Mountain Science},
  year      = {2012},
  volume    = {9},
  number    = {4},
  pages     = {483--491},
  doi       = {10.1007/s11629-012-2352-3}
}

@article{Jin2009,
  author    = {Huijun Jin and Ruixia He and Guodong Cheng and Qingbai Wu and Shaoling Wang and Lanzhi Lü and Xiaoli Chang},
  title     = {Changes in frozen ground in the Source Area of the Yellow River on the Qinghai–Tibet Plateau, China, and their eco-environmental impacts},
  journal   = {Environmental Research Letters},
  year      = {2009},
  volume    = {4},
  number    = {4},
  pages     = {045020},
  doi       = {10.1088/1748-9326/4/4/04520}
}

@article{Sheng2010,
  author    = {Y. Sheng and J. Li and J. C. Wu and B. S. Ye and J. Wang},
  title     = {Distribution patterns of permafrost in the upper area of Shule River with the application of GIS technique},
  journal   = {Journal of China University of Mining and Technology},
  year      = {2010},
  volume    = {39},
  pages     = {32--39}
}

@article{Li2011,
  author    = {J. Li and Y. Sheng and J. Cheng and B. Zhang and J. C. Wu and X. M. Zhang},
  title     = {Characteristics of ground temperatures and influencing factors of permafrost development and distribution in the source region of Datong River (in Chinese with English abstract)},
  journal   = {Progress in Geography},
  year      = {2011},
  volume    = {30},
  pages     = {827--836}
}

@report{Siebert2014,
  type = {Project report},
  title = {Mapping of Rainfed and Irrigated Agriculture in India – Data Inventory and Documentation},
  author = {Siebert, Stefan and Zhao, Gang},
  year = {2014},
  number = {8000054884-AG},
  institution = {University of Bonn},
  location = {Germany}
}

@misc{HWSD2009,
  title        = {Harmonized World Soil Database Version 1.1},
  author       = {Freddy Nachtergaele and Harrij van Velthuizen and Luc Verelst 
                  and Niels Batjes and Koos Dijkshoorn and Vincent van Engelen 
                  and Guenther Fischer and Arwyn Jones and Luca Montanarella 
                  and Monica Petri and Sylvia Prieler and Edmar Teixeira 
                  and David Wiberg and Xuezheng Shi},
  year         = {2009},
  month        = {March},
  note         = {Food and Agriculture Organization of the United Nations (FAO), 
                  International Institute for Applied Systems Analysis (IIASA), 
                  ISRIC-World Soil Information, Institute of Soil Science -- Chinese Academy of Sciences (ISSCAS), 
                  Joint Research Centre of the European Commission (JRC)},
  url          = {http://www.fao.org/soils-portal/data-hub/soil-maps-and-databases/hwsd/en/}
}

@techreport{Danielson2011,
  author       = {Danielson, Jeffrey J. and Gesch, Dean B.},
  title        = {Global multi-resolution terrain elevation data 2010 (GMTED2010)},
  year         = {2011},
  institution  = {U.S. Geological Survey},
  type         = {Open-File Report},
  number       = {2011-1073},
  doi          = {10.3133/ofr20111073},
  note         = {Earth Resources Observation and Science (EROS) Center}
}

@article{Engler2009,
  title = {MigClim: Predicting Plant Distribution and Dispersal in a Changing Climate},
  author = {Engler, Robin and Guisan, Antoine},
  year = {2009},
  journal = {Diversity and Distributions},
  volume = {15},
  number = {4},
  pages = {590--601},
  issn = {1472-4642},
  doi = {10.1111/j.1472-4642.2009.00566.x},
}

@book{Collalti2024,
  title = {Monitoring and Predicting Forest Growth and Dynamics},
  author = {Collalti, Alessio and Dalmonech, D. and Vangi, E and Marano, G. and Puchi, P. and Morichetti, M. and Saponaro, V. and Orrico, M.R. and Grieco, E.},
  year = {2024},
  edition = {CNR Edizioni},
  location = {Roma},
  url = {https://doi.org/10.32018/FORMODLAB- BOOK-2024},
  isbn = {978-88-8080-655-4}
}

@article{Dury2011,
  title = {Responses of European Forest Ecosystems to 21st Century Climate: Assessing Changes in Interannual Variability and Fire Intensity},
  author = {Dury, M. and Hambuckers, A. and Warnant, P. and Henrot, A. and Favre, E. and Ouberdous, M. and François, L.},
  year = {2011},
  journal = {iForest - Biogeosciences and Forestry},
  volume = {4},
  number = {2},
  eprint = {ifor0572-004},
  eprinttype = {pmid},
  pages = {82},
  publisher = {SISEF - Italian Society of Silviculture and Forest Ecology},
  issn = {1971-7458},
  doi = {10.3832/ifor0572-004},
}

@article{Hubert1998,
  title = {Stochastic Generation of Meteorological Variables and Effects on Global Models of Water and Carbon Cycles in Vegetation and Soils},
  author = {Hubert, B. and Francois, L. and Warnant, P. and Strivay, D.},
  year = {1998},
  journal = {Journal of Hydrology},
  shortjournal = {Journal of Hydrology},
  volume = {212--213},
  pages = {318--334},
  issn = {0022-1694},
  doi = {10.1016/S0022-1694(98)00214-5}
}

@article{Nemry1996,
  title = {The Seasonality of the CO2 Exchange between the Atmosphere and the Land Biosphere: A Study with a Global Mechanistic Vegetation Model},
  author = {Nemry, B. and François, L. and Warnant, P. and Robinet, F. and Gérard, J.-C.},
  year = {1996},
  journal = {Journal of Geophysical Research: Atmospheres},
  volume = {101},
  number = {D3},
  pages = {7111--7125},
  issn = {2156-2202},
  doi = {10.1029/95JD03656}
}

@article{Otto2002,
  title = {Biospheric Carbon Stocks Reconstructed at the Last Glacial Maximum: Comparison between General Circulation Models Using Prescribed and Computed Sea Surface Temperatures},
  author = {Otto, Dominique and Rasse, Daniel and Kaplan, Jed and Warnant, Pierre and François, Louis},
  year = {2002},
  journal = {Global and Planetary Change},
  shortjournal = {Global and Planetary Change},
  series = {The Global Carbon Cycle and Its Changes over Glacial-Interglacial Cycles},
  volume = {33},
  number = {1},
  pages = {117--138},
  issn = {0921-8181},
  doi = {10.1016/S0921-8181(02)00066-8}
}

@article{Warnant1994,
  title = {CARAIB: A Global Model of Terrestrial Biological Productivity},
  author = {Warnant, P. and François, L. and Strivay, D. and Gérard, J.-C.},
  year = {1994},
  journal = {Global Biogeochemical Cycles},
  volume = {8},
  number = {3},
  pages = {255--270},
  issn = {1944-9224},
  doi = {10.1029/94GB00850}
}

@article{Saxton1986,
  title = {Estimating Generalized Soil-water Characteristics from Texture},
  author = {Saxton, K. E. and Rawls, W. J. and Romberger, J. S. and Papendick, R. I.},
  year = {1986},
  journal = {Soil Science Society of America Journal},
  volume = {50},
  number = {4},
  pages = {1031--1036},
  issn = {1435-0661},
  doi = {10.2136/sssaj1986.03615995005000040039x}
}

@incollection{Chen2021,
  author = {Chen, Deliang and Rojas, Maisa and Samset, Bjørn H. and Cobb, Kim and Diongue Niang, Aida and Edwards, Paul and Emori, Seita and Faria, Sergio Henrique and Hawkins, Ed and Hope, Pandora and Huybrechts, Philippe and Meinshausen, Malte and Mustafa, Sawsan K. and Plattner, Gian-Kasper and Tréguier, Anne-Marie and others},
  title = {Framing, Context, and Methods},
  booktitle = {Climate Change 2021: The Physical Science Basis. Contribution of Working Group I to the Sixth Assessment Report of the Intergovernmental Panel on Climate Change},
  editor = {Masson-Delmotte, V. and Zhai, P. and Pirani, A. and Connors, S.L. and Péan, C. and Berger, S. and Caud, N. and Chen, Y. and Goldfarb, L. and Gomis, M.I. and Huang, M. and Leitzell, K. and Lonnoy, E. and Matthews, J.B.R. and Maycock, T.K. and Waterfield, T. and Yelek{\c{c}}i, {\"O}. and Yu, R. and Zhou, B. and others},
  publisher = {Cambridge University Press},
  address = {Cambridge, United Kingdom and New York, NY, USA},
  pages = {147--286},
  year = {2021},
  doi = {10.1017/9781009157896.003}
}

@article{Woolway2021,
  title={Lake heatwaves under climate change},
  author={Woolway, R Iestyn and Jennings, Eleanor and Shatwell, Tom and Golub, Malgorzata and Pierson, Don C and Maberly, Stephen C},
  journal={Nature},
  volume={589},
  number={7842},
  pages={402--407},
  year={2021},
  publisher={Nature Publishing Group UK London},
  doi = {10.1038/s41586-020-03119-1}
}

@article{Wang2025,
  title = {China’s Nationwide Streamflow Decline Driven by Landscape Changes and Human Interventions},
  author = {Wang, Kaiwen and Liu, Xiaomang and Cui, Peng and Zhang, Yangjian and Xie, Jiaxin and Liu, Changming and Gosling, Simon N.},
  year = {2025},
  journal = {Science Advances},
  volume = {11},
  number = {32},
  pages = {eadu8032},
  publisher = {American Association for the Advancement of Science},
  doi = {10.1126/sciadv.adu8032}
}

@article{Jones2025,
  title = {A Multi-Model Assessment of Global Freshwater Temperature and Thermoelectric Power Supply under Climate Change},
  author = {Jones, Edward R and Beek, Rens and Cárdenas Belleza, Gabriel and Burek, Peter and Dugdale, Stephen J and Flörke, Martina and Fridman, Dor and Gosling, Simon N and Kumar, Rohini and Mercado-Bettin, Daniel and Müller Schmied, Hannes and Tan, Zeli and Thiery, Wim and Tilahun, Ammanuel B and Wanders, Niko and Vliet, Michelle T H},
  year = {2025},
  journal = {Environmental Research: Water},
  shortjournal = {Environ. Res.: Water},
  volume = {1},
  number = {2},
  pages = {025002},
  publisher = {IOP Publishing},
  issn = {3033-4942},
  doi = {10.1088/3033-4942/addffa},
  langid = {english}
}

@article{Smith2024,
  title = {Future Malaria Environmental Suitability in Africa Is Sensitive to Hydrology},
  author = {Smith, Mark W. and Willis, Thomas and Mroz, Elizabeth and James, William H. M. and Klaar, Megan J. and Gosling, Simon N. and Thomas, Christopher J.},
  year = {2024},
  journal = {Science},
  volume = {384},
  number = {6696},
  pages = {697--703},
  publisher = {American Association for the Advancement of Science},
  doi = {10.1126/science.adk8755}
}

@article{Zani2023,
  title = {Climate and Dispersal Limitation Drive Tree Species Range Shifts in Post-Glacial Europe: Results from Dynamic Simulations},
  author = {Zani, Deborah and Lischke, Heike and Lehsten, Veiko},
  year = {2023},
  journal = {Frontiers in Ecology and Evolution},
  shortjournal = {Front. Ecol. Evol.},
  volume = {11},
  publisher = {Frontiers},
  issn = {2296-701X},
  doi = {10.3389/fevo.2023.1321104}
}

@article{Aerts2022,
  title = {Large-Sample Assessment of Varying Spatial Resolution on the Streamflow Estimates of the Wflow\_sbm Hydrological Model},
  author = {Aerts, Jerom P. M. and Hut, Rolf W. and Giesen, Nick C. and Drost, Niels and Verseveld, Willem J. and Weerts, Albrecht H. and Hazenberg, Pieter},
  year = {2022},
  journal = {Hydrology and Earth System Sciences},
  volume = {26},
  number = {16},
  pages = {4407--4430},
  publisher = {Copernicus GmbH},
  issn = {1027-5606},
  doi = {10.5194/hess-26-4407-2022},
  langid = {english}
}

@article{Barnhart2024,
  title = {Evaluating Distributed Snow Model Resolution and Meteorology Parameterizations Against Streamflow Observations: Finer Is Not Always Better},
  author = {Barnhart, Theodore B. and Putman, Annie L. and Heldmyer, Aaron J. and Rey, David M. and Hammond, John C. and Driscoll, Jessica M. and Sexstone, Graham A.},
  year = {2024},
  journal = {Water Resources Research},
  volume = {60},
  number = {7},
  pages = {e2023WR035982},
  issn = {1944-7973},
  doi = {10.1029/2023WR035982}
}

@article{Kumar2022,
  title = {Multi-Model Evaluation of Catchment- and Global-Scale Hydrological Model Simulations of Drought Characteristics across Eight Large River Catchments},
  author = {Kumar, Amit and Gosling, Simon N. and Johnson, Matthew F. and Jones, Matthew D. and Zaherpour, Jamal and Kumar, Rohini and Leng, Guoyong and Schmied, Hannes Müller and Kupzig, Jenny and Breuer, Lutz and Hanasaki, Naota and Tang, Qiuhong and Ostberg, Sebastian and Stacke, Tobias and Pokhrel, Yadu and Wada, Yoshihide and Masaki, Yoshimitsu},
  year = {2022},
  journal = {Advances in Water Resources},
  shortjournal = {Advances in Water Resources},
  volume = {165},
  pages = {104212},
  issn = {0309-1708},
  doi = {10.1016/j.advwatres.2022.104212}
}

@article{Magnusson2019,
  title = {Influence of Spatial Resolution on Snow Cover Dynamic for a Coastal and Mountainous Region at High Latitudes (Norway)},
  author = {Magnusson, Jan and Eisner, Stephanie and Huang, Shaochun and Lussana, Cristian and Mazzotti, Giulia and Essery, Richard and Saloranta, Tuomo and Beldring, Stein},
  year = {2019},
  journal = {Water Resources Research},
  volume = {55},
  number = {7},
  pages = {5612--5630},
  issn = {1944-7973},
  doi = {10.1029/2019WR024925}
}

@article{Jagermeyr2021,
  title = {Climate Impacts on Global Agriculture Emerge Earlier in New Generation of Climate and Crop Models},
  author = {Jägermeyr, Jonas and Müller, Christoph and Ruane, Alex C. and Elliott, Joshua and Balkovic, Juraj and Castillo, Oscar and Faye, Babacar and Foster, Ian and Folberth, Christian and Franke, James A. and Fuchs, Kathrin and Guarin, Jose R. and Heinke, Jens and Hoogenboom, Gerrit and Iizumi, Toshichika and Jain, Atul K. and Kelly, David and Khabarov, Nikolay and Lange, Stefan and Lin, Tzu-Shun and Liu, Wenfeng and Mialyk, Oleksandr and Minoli, Sara and Moyer, Elisabeth J. and Okada, Masashi and Phillips, Meridel and Porter, Cheryl and Rabin, Sam S. and Scheer, Clemens and Schneider, Julia M. and Schyns, Joep F. and Skalsky, Rastislav and Smerald, Andrew and Stella, Tommaso and Stephens, Haynes and Webber, Heidi and Zabel, Florian and Rosenzweig, Cynthia},
  year = {2021},
  journal = {Nature Food},
  shortjournal = {Nat Food},
  volume = {2},
  number = {11},
  pages = {873--885},
  publisher = {Nature Publishing Group},
  issn = {2662-1355},
  doi = {10.1038/s43016-021-00400-y},
  langid = {english}
}

@article{Lawrence2022,
  title = {The Unseen Effects of Deforestation: Biophysical Effects on Climate},
  author = {Lawrence, Deborah and Coe, Michael and Walker, Wayne and Verchot, Louis and Vandecar, Karen},
  year = {2022},
  journal = {Frontiers in Forests and Global Change},
  shortjournal = {Front. For. Glob. Change},
  volume = {5},
  publisher = {Frontiers},
  issn = {2624-893X},
  doi = {10.3389/ffgc.2022.756115}
}

@article{Habibullah2022,
  title = {Impact of Climate Change on Biodiversity Loss: Global Evidence},
  shorttitle = {Impact of Climate Change on Biodiversity Loss},
  author = {Habibullah, Muzafar Shah and Din, Badariah Haji and Tan, Siow-Hooi and Zahid, Hasan},
  year = {2022},
  journal = {Environmental Science and Pollution Research},
  shortjournal = {Environ Sci Pollut Res},
  volume = {29},
  number = {1},
  pages = {1073--1086},
  issn = {1614-7499},
  doi = {10.1007/s11356-021-15702-8},
  langid = {english}
}

@article{Zapata2022,
  title = {Climate Change Impacts on the Energy System: A Model Comparison},
  shorttitle = {Climate Change Impacts on the Energy System},
  author = {Zapata, Victhalia and Gernaat, David E H J and Yalew, Seleshi G and Santos da Silva, Silvia R and Iyer, Gokul and Hejazi, Mohamad and Vuuren, Detlef P},
  year = {2022},
  journal = {Environmental Research Letters},
  shortjournal = {Environ. Res. Lett.},
  volume = {17},
  number = {3},
  pages = {034036},
  publisher = {IOP Publishing},
  issn = {1748-9326},
  doi = {10.1088/1748-9326/ac5141},
  langid = {english}
}

@article{Schipper2021,
  title = {Climate Change Research and the Search for Solutions: Rethinking Interdisciplinarity},
  shorttitle = {Climate Change Research and the Search for Solutions},
  author = {Schipper, E. Lisa F. and Dubash, Navroz K. and Mulugetta, Yacob},
  year = {2021},
  journal = {Climatic Change},
  shortjournal = {Climatic Change},
  volume = {168},
  number = {3},
  pages = {18},
  issn = {1573-1480},
  doi = {10.1007/s10584-021-03237-3},
  langid = {english}
}

@article{Pfenning-Butterworth2024,
  title = {Interconnecting Global Threats: Climate Change, Biodiversity Loss, and Infectious Diseases},
  shorttitle = {Interconnecting Global Threats},
  author = {Pfenning-Butterworth, Alaina and Buckley, Lauren B. and Drake, John M. and Farner, Johannah E. and Farrell, Maxwell J. and Gehman, Alyssa-Lois M. and Mordecai, Erin A. and Stephens, Patrick R. and Gittleman, John L. and Davies, T. Jonathan},
  year = {2024},
  journal = {The Lancet Planetary Health},
  shortjournal = {The Lancet Planetary Health},
  volume = {8},
  number = {4},
  eprint = {38580428},
  eprinttype = {pmid},
  pages = {e270-e283},
  publisher = {Elsevier},
  issn = {2542-5196},
  doi = {10.1016/S2542-5196(24)00021-4}
}

@article{Keller2022,
  title = {Downscaling Approaches of Climate Change Projections for Watershed Modeling: Review of Theoretical and Practical Considerations},
  author = {Keller, Arturo A. and Garner, Kendra L. and Rao, Nalini and Knipping, Eladio and Thomas, Jeffrey},
  year = {2022},
  journal = {PLOS Water},
  shortjournal = {PLOS Water},
  volume = {1},
  number = {9},
  pages = {e0000046},
  publisher = {Public Library of Science},
  issn = {2767-3219},
  doi = {10.1371/journal.pwat.0000046},
  langid = {english}
}

@article{Merganicova2024,
  title = {The Biogeochemical Model Biome-BGCMuSo v6.2 Provides Plausible and Accurate Simulations of the Carbon Cycle in Central European Beech Forests},
  author = {Merganičová, Katarína and Merganič, Ján and Dobor, Laura and Hollós, Roland and Barcza, Zoltán and Hidy, Dóra and Sitková, Zuzana and Pavlenda, Pavel and Marjanovic, Hrvoje and Kurjak, Daniel and Bošel'a, Michal and Bitunjac, Doroteja and Ostrogović Sever, Maša Zorana and Novák, Jiří and Fleischer, Peter and Hlásny, Tomáš},
  year = {2024},
  journal = {Geoscientific Model Development},
  volume = {17},
  number = {20},
  pages = {7317--7346},
  publisher = {Copernicus GmbH},
  issn = {1991-959X},
  doi = {10.5194/gmd-17-7317-2024},
  langid = {english}
}

@article{Bano-Medina2022,
  title = {Downscaling Multi-Model Climate Projection Ensembles with Deep Learning (DeepESD): Contribution to CORDEX EUR-44},
  author = {Baño-Medina, Jorge and Manzanas, Rodrigo and Cimadevilla, Ezequiel and Fernández, Jesús and González-Abad, Jose and Cofiño, Antonio S. and Gutiérrez, José Manuel},
  year = {2022-09-06},
  journal = {Geoscientific Model Development},
  volume = {15},
  number = {17},
  pages = {6747--6758},
  publisher = {Copernicus GmbH},
  issn = {1991-959X},
  doi = {10.5194/gmd-15-6747-2022},
  langid = {english}
}

@article{Bottazzi2024,
  title = {High Performance Computing to Support Land, Climate, and User-Oriented Services: The HIGHLANDER Data Portal},
  author = {Bottazzi, Michele and Rodríguez-Muñoz, Lucía and Chiavarini, Beatrice and Caroli, Cinzia and Trotta, Giuseppe and Dellacasa, Chiara and Marras, Gian Franco and Montanari, Margherita and Santini, Monia and Mancini, Marco and D'Anca, Alessandro and Mercogliano, Paola and Raffa, Mario and Villani, Giulia and Tomei, Fausto and Loglisci, Nicola and Gascón, Estíbaliz and Hewson, Timothy and Chillemi, Giovanni and Valentini, Riccardo and Gianelle, Damiano and Massarenti, Elena and Forconi, Martina and Mazzoni, Lucia and Scipione, Gabriella},
  year = {2024},
  journal = {Meteorological Applications},
  volume = {31},
  number = {2},
  pages = {e2166},
  issn = {1469-8080},
  doi = {10.1002/met.2166}
}

@article{Raffa2023,
  title = {Very High Resolution Projections over Italy under Different CMIP5 IPCC Scenarios},
  author = {Raffa, Mario and Adinolfi, Marianna and Reder, Alfredo and Marras, Gian Franco and Mancini, Marco and Scipione, Gabriella and Santini, Monia and Mercogliano, Paola},
  year = {2023},
  journal = {Scientific Data},
  shortjournal = {Sci Data},
  volume = {10},
  number = {1},
  pages = {238},
  publisher = {Nature Publishing Group},
  issn = {2052-4463},
  doi = {10.1038/s41597-023-02144-9}
}

@article{Reder2025,
  title = {Estimating Pros and Cons of Statistical Downscaling Based on EQM Bias Adjustment as a Complementary Method to Dynamical Downscaling},
  author = {Reder, Alfredo and Fedele, Giusy and Manco, Ilenia and Mercogliano, Paola},
  year = {2025},
  journal = {Scientific Reports},
  shortjournal = {Sci Rep},
  volume = {15},
  number = {1},
  pages = {621},
  publisher = {Nature Publishing Group},
  issn = {2045-2322},
  doi = {10.1038/s41598-024-84527-5},
}

@article{Sun2021,
  title = {Statistical downscaling of daily temperature and precipitation over China using deep learning neural models: Localization and comparison with other methods},
  author = {Sun, Lei and Lan, Yufeng},
  year = {2021},
  journal = {International Journal of Climatology},
  volume = {41},
  number = {2},
  pages = {1128--1147},
  issn = {1097-0088},
  doi = {10.1002/joc.6769}
}

@article{Coron2014,
  title = {On the Lack of Robustness of Hydrologic Models Regarding Water Balance Simulation: A Diagnostic Approach Applied to Three Models of Increasing Complexity on 20 Mountainous Catchments},
  shorttitle = {On the Lack of Robustness of Hydrologic Models Regarding Water Balance Simulation},
  author = {Coron, L. and Andréassian, V. and Perrin, C. and Bourqui, M. and Hendrickx, F.},
  year = {2014},
  journal = {Hydrology and Earth System Sciences},
  volume = {18},
  number = {2},
  pages = {727--746},
  publisher = {Copernicus GmbH},
  issn = {1027-5606},
  doi = {10.5194/hess-18-727-2014}
}

@article{Hublart2016,
  title = {Reliability of Lumped Hydrological Modeling in a Semi-Arid Mountainous Catchment Facing Water-Use Changes},
  author = {Hublart, Paul and Ruelland, Denis and García de Cortázar-Atauri, Inaki and Gascoin, Simon and Lhermitte, Stef and Ibacache, Antonio},
  year = {2016},
  journal = {Hydrology and Earth System Sciences},
  volume = {20},
  number = {9},
  pages = {3691--3717},
  publisher = {Copernicus GmbH},
  issn = {1027-5606},
  doi = {10.5194/hess-20-3691-2016}
}

@article{Didovets2021,
  title = {Central Asian Rivers under Climate Change: Impacts Assessment in Eight Representative Catchments},
  author = {Didovets, Iulii and Lobanova, Anastasia and Krysanova, Valentina and Menz, Christoph and Babagalieva, Zhanna and Nurbatsina, Aliya and Gavrilenko, Nadejda and Khamidov, Vohid and Umirbekov, Atabek and Qodirov, Sobir and Muhyyew, Dowletgeldi and Hattermann, Fred Fokko},
  year = {2021},
  journal = {Journal of Hydrology: Regional Studies},
  shortjournal = {Journal of Hydrology: Regional Studies},
  volume = {34},
  pages = {100779},
  issn = {2214-5818},
  doi = {10.1016/j.ejrh.2021.100779}
}

@online{Roy2016,
  title = {Vegetation Collection Decadal Land Use and Land Cover Classifications across India, 1985, 1995, 2005},
  author = {Roy, P.S. and Meiyappan, P. and Joshi, P.K. and Kale, M.P. and Srivastav, V.K. and Srivasatava, S.K. and Behera, M.D. and Roy, A. and Sharma, Y. and Ramachandran, R.M. and Bhavani, P. and Jain, A.K. and Krishnamurthy, Y.V.N.},
  namea = {ORNL DAAC},
  nameatype = {collaborator},
  year = {2016},
  eprinttype = {ORNL Distributed Active Archive Center},
  pages = {128.949 MB},
  doi = {10.3334/ORNLDAAC/1336},
  url = {https://www.earthdata.nasa.gov/data/catalog/ornl-cloud-decadal-lulc-india-1336-1},
  urldate = {2025-12-09},
  pubstate = {prepublished},
  version = {1}
}

@online{Smilovic_in_prep,
  title = {Simulating Water-Limited Crop-Specific Water Use and Relative Yield Spatiotemporally in the Hydrological Model CWatM},
  author = {Smilovic, Mikhail},
  year = {2025},
  note = {In preparation},
  pubstate = {prepublished}
}

@online{Smilovicinprep1,
  title = {Seasonal Rhythms in Collective Decision-Making Revealed through Global Water Infrastructure},
  author = {Smilovic, Mikhail},
  year = {2025},
  note = {In preparation},
  pubstate = {prepublished}
}

@article{Maraun2016,
  title = {Bias {{Correcting Climate Change Simulations}} - a {{Critical Review}}},
  author = {Maraun, Douglas},
  year = 2016,
  month = dec,
  journal = {Current Climate Change Reports},
  volume = {2},
  number = {4},
  pages = {211--220},
  issn = {2198-6061},
  doi = {10.1007/s40641-016-0050-x},
  urldate = {2023-08-25},
  langid = {english}
}

@article{Maurer2016,
  title = {Technical {{Note}}: {{The}} impact of spatial scale in bias correction of climate model output for hydrologic impact studies},
  shorttitle = {Technical {{Note}}},
  author = {Maurer, E. P. and Ficklin, D. L. and Wang, W.},
  year = 2016,
  month = feb,
  journal = {Hydrology and Earth System Sciences},
  volume = {20},
  number = {2},
  pages = {685--696},
  publisher = {Copernicus GmbH},
  issn = {1027-5606},
  doi = {10.5194/hess-20-685-2016},
  urldate = {2025-12-11},
  langid = {english}
}

\end{document}


\maketitle

\section{Part A}
\setcounter{figure}{0}
\renewcommand\thefigure{S\arabic{figure}}

\subsection*{Introduction}

This supplementary material provides additional details on the modeling setups and observational datasets used in this study. The information presented here was collected through a standardized questionnaire, which each participating modeling group completed when submitting their simulation results. The questionnaire comprised six components, outlined below:

\begin{enumerate}
    \item \textbf{Model information:} Full model name and acronym, version number, main reference(s), and other identifying information.
    \item \textbf{Model description:} A brief description of the model, including the ISIMIP sector it belongs to.
    \item \textbf{Model setup:} Description of the spatial and temporal setup, including:
    \begin{itemize}
        \item Whether the model was run in point-mode or spatially distributed;
        \item The simulation location(s), including coordinates or spatial extent;
        \item Whether the spatial resolution of the forcing data matched the model's internal resolution;
        \item Temporal resolution of simulations and outputs (e.g., 6-hourly, daily, monthly);
        \item Whether a spin-up was performed, and if so, a brief description;
        \item Whether calibration was applied, and if so, how.
    \end{itemize}
    \item \textbf{Input data:} Description of input datasets, including:
    \begin{itemize}
        \item Which official ISIMIP CHELSA-HighRes climate variables were used (tas, tasmax, tasmin, pr, rsds);
        \item Any additional variables used, and how they were calculated;
        \item Any non-ISIMIP climate data used, and rationale for their inclusion;
        \item Land use and soil datasets, and how higher-resolution data were obtained (if applicable).
    \end{itemize}
    \item \textbf{Evaluation data:} Description of model evaluation, including:
    \begin{itemize}
        \item Output variables evaluated (with units);
        \item Datasets used for comparison;
        \item Temporal extent and coverage of the evaluation data.
    \end{itemize}
    \item \textbf{Additional information:} Any further model-specific context, including:
    \begin{itemize}
        \item Why spatial resolution of input data matters for the model;
        \item A self-assessment of model complexity (low, medium, or high) relative to others in the same sector.
    \end{itemize}
\end{enumerate}

\subsection{Regional Forests - coordinated effort}

A coordinated effort was undertaken for five of the seven forest models to simulate at the same location (Collelongo, Italy, 41.83°N, 13.58°E) in order to capture a broader range of model performance with many models simulating at the same locations. All models were evaluated against flux tower observations of a carbon-related variable (Gross Primary Productivity, GPP) and a water-related variable (Actual Evapotranspiration, AET) from the Profound Database \citep{Reyer2020}, available for 1996–2014. Because the models differ in their internal time resolution (see Table 1 in the main manuscript), evaluations were conducted at monthly time steps. Additional required input variables (relative humidity, air pressure, longwave radiation, wind speed) were prepared in a coordinated effort using the methods described by \citet{Frieler2024}, available at \url{https://github.com/johanna-malle/w5e5_downscale}. As all models were evaluated in the same manner, we do not repeat the evaluation-data description for each individual model. No forest model performed site-specific calibration.

The ISIMIP regional forest simulation protocol is available at \url{https://protocol.isimip.org/#/ISIMIP3a/forestry}.

\subsubsection*{4C}

4C (FORESt Ecosystems in a changing Environment, FORESEE) is a process-based forest model. The version used in this study is 4C v2.2, described in detail by \cite{Lasch-Born2020}.

4C is part of the Regional Forests ISIMIP sector and simulates tree species composition, growth, mortality and regeneration, forest structure, and the carbon, water, and nitrogen balances of forest stands. It is a distance-independent model that groups identical trees into cohorts evenly distributed in space. These cohorts compete for light, water, and nutrients. The model operates across multiple timescales: daily (e.g., soil water dynamics, phenology), weekly (e.g., photosynthesis using weekly averaged climate data), and annually (e.g., carbon allocation, growth, mortality). GPP and AET are simulated at a daily timescale.

The internal time step varies by process, while outputs are generated daily (carbon and water variables) and yearly (structural variables). No spin-up or calibration was performed; simulations were initialized directly using observed stand structure data from the Profound Database \citep{Reyer2020}.

The model was forced with \texttt{tas}, \texttt{tasmax}, \texttt{tasmin}, \texttt{pr}, and \texttt{rsds} from the official ISIMIP CHELSA-HighRes climate dataset. Additional required variables included air pressure and relative humidity, which were obtained from the coordinated forcing data preparation, as described above. Soil data was derived from the European Soil Database \citep{Panagos2006, ESB2004}.


The accuracy of climate input data is critical for 4C, as many physiological processes such as photosynthesis, phenology, and soil dynamics are highly sensitive to climate variables. In areas with complex topography, higher resolution input data may improve agreement with observations. However, model structural and parameter uncertainties can outweigh benefits from refined inputs. Among models in the Regional Forests ISIMIP sector, 4C is considered high in complexity.

\subsubsection*{i3PGmiX}
The IIASA-3PGmix-X model (i3PGmiX) is a reduced complexity process-based forest growth model. The full description of the model is available in \cite{Augustynczik2025}. i3PGmiX (formerly 3PGN-BW) is part of the Regional Forests ISIMIP sector that describes tree species composition, tree growth, mortality and regeneration, forest structure, and the carbon, water, and nitrogen balance of a forest stand. The model is a further development of the 3PGmix model, with representation of stand or cohorts, running at monthly to yearly time steps, depending on the process. i3PGmiX employs a big leaf approach to compute photosynthesis, based on the P-model \citep{Stocker2020}, and couples alternative soil (ICBM/2N and Yasso20) and disturbance models (ForestGALES, LandClim bark beetle damage). The model was set up in point-mode, representing a one-hectare forest stand, at a location in Italy (41.83°N, 13.58°E), as mentioned above. The temporal resolution of the model output is monthly for carbon and water variables and yearly for mortality. The model was set up with initial tree data from observations aggregated at the plot level, with no spin-up performed for vegetation or soil initial state. Furthermore, no calibration was done. 

The model was forced with \texttt{tas}, \texttt{tasmax}, \texttt{tasmin}, \texttt{pr}, and \texttt{rsds} from the official ISIMIP CHELSA-HighRes database. No additional climate variables were used. Soil parameters for the sites, including maximum available soil water and texture, were retrieved from the PROFOUND database \citep{Reyer2020}. 

Forest productivity in i3PGmiX is controlled by forcing climate and management, including photosynthesis and respiration rates. Simulations with higher-resolution meteorological input data should yield model output that is more closely aligned with observations, particularly in locations with high topographical variability. i3PGmiX falls in the low (to mid) complexity category.

\subsubsection*{3PG-Hydro}
3PG-Hydro is an updated version of the process-based forest growth model 3PG \citep{Landsberg1997}, with the full description available in \citet{Yousefpour2021}. 3PG-Hydro is part of the Regional Forests ISIMIP sector that describes tree species composition, tree growth, mortality and regeneration, forest structure, and the carbon, water and nitrogen balance of a forest stand. 3PG simulates forest growth at the plot level with two main processes: biomass growth influenced by environmental factors and allocation to the different compartments. 3PG-Hydro includes an improved soil-water model, plant phenology, and operates at daily time steps to simulate processes on a higher temporal resolution. 
The model was set up in point-mode, representing a one-hectare forest stands, at a location in Italy (41.83°N, 13.58°E), as mentioned above.

Temporal resolution of the model output was daily for all variables. 3PG-Hydro was initialized with plot-level data (diameter, age, stem number), and species-specific parameters were obtained from \citet{Augustynczik2017} and \citet{Trotsiuk2020} without further calibration. The model was forced by the variables \texttt{tas}, \texttt{tasmax}, \texttt{tasmin}, \texttt{pr}, and \texttt{rsds} from the official ISIMIP CHELSA-HighRes database. Soil data (soil type, soil depth) was obtained from the PROFOUND database \citep{Reyer2020}. 

In 3PG-Hydro, forest growth is influenced by environmental factors that are derived from climate data, for example, critical temperature range and soil available water content. Thus, climate input data with higher resolution should improve the accuracy of simulations compared to observations, especially in locations with high topographical variability. The version of 3PG-Hydro applied here is a mid-complexity model.

\subsubsection*{Biome-BGCMuSo}
Biome-BGCMuSo is a big-leaf, process-based model. Here we used version 7.0. The full model description of the model structure is available in \citet{Hidy2012, Hidy2016, Hidy2022}, and the source code is available at \url{https://nimbus.elte.hu/bbgc/}. Biome-BGCMuSo is part of the Regional Forests ISIMIP sector. The model is built to simulate carbon, nitrogen, and water cycles in forest ecosystems under scenarios of climate change and disturbances (e.g., forest management), and is parameterized at the species level \citep{Merganicova2024}. All physiological processes are simulated at a daily time step. The model simulates a homogeneous representative area of a stand and provides model outputs per square meter. 

The model simulations were conducted in three phases: (1) spin-up, (2) transient, and (3) normal runs. The spin-up phase began with bare ground, minimal carbon in vegetation, and soil water at full capacity. Carbon stock was accumulated until a steady state was reached in the soil. This phase used constant pre-industrial CO\textsubscript{2} (277.15 ppm) and nitrogen deposition (0.002 kgN m\textsuperscript{-2} yr\textsuperscript{-1}). In the transient phase, starting in 1850, CO\textsubscript{2} and nitrogen levels increased annually to current values. Both spin-up and transient phases had no management, fire-induced mortality, or variable natural mortality. The normal run began at stand establishment (age 0) and continued to the present. In the first year, a clearcut removed 90–95\% of aboveground woody biomass, leaving non-woody biomass on site. This phase included varying CO\textsubscript{2} and nitrogen deposition, and management based on forestry records and generic rules to simulate earlier stand history with no on-site inventory data. Natural mortality rates decreased exponentially for the first 30 years, after which they were kept constant at 0.5\% annually. No calibration was performed for the presented study. The model was forced by the variables \texttt{tasmax}, \texttt{tasmin}, \texttt{pr}, and \texttt{rsds} from the official ISIMIP CHELSA-HighRes database. Additional required climate variables, namely daylight temperature and daylight vapor pressure deficit, were derived using MT-CLIM \citep{Hungerford1989} based on \texttt{tasmax}, \texttt{tasmin}, and \texttt{pr}. Soil parameters for the sites, including soil depth and soil texture, were retrieved from the PROFOUND database \citep{Reyer2020}. 

Meteorological input data influence Biome-BGCMuSo output since many physiological processes in the model are sensitive to changes in air temperature, precipitation, radiation, and vapor pressure deficit. These processes include photosynthesis, ecosystem respiration (both autotrophic and heterotrophic), water balance, decomposition, and the beginning and end of the growing season. 

The Biome-BGCMuSo model ranks high in terms of the complexity of simulated eco-physiological processes within the Regional Forests ISIMIP sector; however, from the point of forest stand representation, it is a low-complexity model as it does not account for species composition, or horizontal or vertical structure of a stand.

\subsubsection*{3D-CMCC-FEM}

3D-CMCC-FEM (Three Dimensional - Coupled Model Carbon Cycle - Forest Ecosystem Module) is a biogeochemical, biophysical, process-based model. The model version used in this study is v5.6, with a full model description provided in \citet{Collalti2014, Collalti2016, Collalti2018, Collalti2019, Collalti2020, Collalti2024} and \cite{Dalmonech2022}.

3D-CMCC-FEM is part of the Regional Forests ISIMIP sector and simulates tree species composition, growth, mortality, and regeneration, as well as forest structure. It also represents the carbon, water, and nitrogen balance of forest stands, which can be uneven-aged, multilayered, and mixed. The 3D-CMCC-FEM model simulates carbon, nitrogen, and water cycles in forest ecosystems, including forest dynamics under climate change and disturbances (e.g., forest management), and is parameterized at the species level. It simulates the average tree within the stand, while also accounting for vertical and horizontal competition between individuals.

The temporal resolution of the model output is daily for carbon, nitrogen, and water variables, while structural variables are updated monthly or annually. The model was initialized using observed tree data (i.e., DBH, tree height, age, and tree density). No spin-up was performed for vegetation or soil, and no site-level calibration was applied.

The model was forced with \texttt{tas}, \texttt{tasmax}, \texttt{tasmin}, \texttt{pr}, \texttt{rh}, and \texttt{rsds} from the official ISIMIP CHELSA-HighRes climate dataset.

Meteorological input data significantly influence 3D-CMCC-FEM output, as many physiological processes in the model are sensitive to changes in air temperature, precipitation, radiation, and relative humidity. These processes include the photosynthetic rate, ecosystem respiration (both autotrophic and heterotrophic), the timing of the growing season, and the allocation of carbon and nitrogen linked to phenological stage and soil dynamics. Higher-resolution meteorological data are expected to yield more accurate outputs, particularly in topographically complex regions such as Italy. Nevertheless, model structural and parameter uncertainties (see \cite{Collalti2019}) can sometimes outweigh the benefits of increased input resolution. The 3D-CMCC-FEM model is classified as a high-complexity model within the ISIMIP Regional Forests sector, featuring a predominantly mechanistic representation of ecological processes.

\subsection{Regional Forests - other models}\label{reg_for_2}
The following models are also part of the regional forest sector but were set up, run, and evaluated individually at specific locations.  

\subsubsection*{PREBAS}

PREBAS is a process-based forest model consisting of two modules: PRELES, a canopy photosynthesis model estimating gross primary productivity (GPP) and a simplified water balance; and CROBAS, a tree growth model. A full description of PREBAS used in this study is available in \citet{Minunno2019}.

PREBAS is part of the Regional Forests ISIMIP sector. It simulates tree species composition, growth, mortality, regeneration, forest structure, and the carbon and water balance of forest stands. It operates on multiple timescales, with daily time steps for photosynthesis and evapotranspiration, and annual time steps for tree carbon allocation, growth, and mortality.

The model was set up in point mode, with independent simulations at two forest stands in Finland: Hyytiälä (61.84741°N, 24.29477°E) and Sodankylä (67.36239°N, 26.63859°E). The model outputs carbon and water fluxes at daily resolution and structural variables at annual resolution. Simulations were initialized using observed tree data; no spin-up was performed for vegetation or soil state.

PREBAS was forced using \texttt{tas}, \texttt{pr}, and \texttt{rsds} from the official ISIMIP CHELSA-HighRes dataset. Relative humidity (rH), required to compute vapor pressure deficit, was derived following the method described in \cite{Frieler2024} using the downscaling approach available at \url{https://github.com/johanna-malle/w5e5_downscale}. No additional climate or land/soil datasets were used in this study.

Model evaluation considered the same output variables as were used for the coordinated forest effort: gross primary productivity (kgC m\textsuperscript{-2} y\textsuperscript{-1}, gC m\textsuperscript{-2} day\textsuperscript{-1}), and evapotranspiration (mm y\textsuperscript{-1}, mm day\textsuperscript{-1}). Observations came from eddy-covariance data from flux towers. Temporal coverage was from 2008–2016 at Hyytiälä and 2015–2016 at Sodankylä.

Meteorological input resolution matters for PREBAS because many physiological processes—such as photosynthesis and water balance—are sensitive to temperature, precipitation, radiation, and vapor pressure deficit. Higher resolution inputs can improve simulation fidelity, especially in heterogeneous landscapes. However, in Finland, with relatively low topographic variability, input resolution may have a more limited effect. PREBAS is considered an intermediate-complexity model within the Regional Forests ISIMIP sector.

\subsubsection*{TreeMig}
TreeMig v1.0 \citep{Lischke2006a} is the core of the TreeMig framework (\url{https://treemig.wsl.ch}), which also includes an R-wrapper for preprocessing and plotting, a graphical user interface (GUI), and a starter kit (\url{https://doi.org/10.16904/envidat.533}).

TreeMig belongs to the Regional Forests ISIMIP sector. It is a forest landscape model simulating forest dynamics on a grid of square cells ranging from 25~m to 10~km in width. The main state variables are population densities per species, height class, and grid cell. Dynamics are driven by demographic processes—regeneration (including seed dispersal), growth, mortality, and competition for light—which depend on bioclimate and light availability. Within-cell heterogeneity is represented by assuming a random tree distribution, leading to a frequency distribution of light conditions, which feeds back into demographic processes.

We applied TreeMig in spatially distributed mode across Switzerland on a 1~km grid, with a yearly time step. The simulation started from bare ground in the year 1700, with seed input to all grid cells for two years, followed by standard simulation with seed dispersal during a spin-up period from 1700–1978. The focus period was 1979–2016. All input data of different resolutions were disaggregated to 1~km to isolate climate data resolution effects from those of other inputs and of TreeMig's intrinsic processes, particularly seed dispersal.

The model was forced using monthly temperature means and monthly precipitation sums, aggregated from daily \texttt{tas} and \texttt{pr} from the official ISIMIP CHELSA-HighRes dataset. Climate data were prepared at 1~km, 3~km, 10~km, and 60~km resolutions but projected onto a 1~km grid in WGS84, and subsequently interpolated (bilinear) to Swiss national coordinates (CH1903 LV03). These were then converted into annual bioclimate variables: Day Degree Sum (DD), minimum winter temperature (MinWiT), and drought stress (DrStr), using additional data on soil water holding capacity (Swiss Soil Suitability Map), latitude, and exposition (DEM DHM25/200). For the spin-up period (1700–1978), bioclimate data were derived by sampling from cold and wet years (1979, 1980, 1981, 1984). Land use data from the Swiss areal statistics NOAS04 were used to restrict simulations to current forest areas (categories 50–60); no afforestation or expansion into meadows/pastures was considered.

Model output includes yearly biomass (t/ha), leaf area index (LAI, m\textsuperscript{2}/m\textsuperscript{2}), and basal area (m\textsuperscript{2}/ha) for each grid cell. Simulation results at the various resolutions were evaluated using species-specific basal area data from the Swiss National Forest Inventory (LFI/NFI) of 1983. This evaluation process is expanded upon in the TreeMig section of the supplementary material part 2. 

In topographically heterogeneous regions like Switzerland, spatial aggregation of bioclimate variables can result in strong deviations from fine-scale conditions. Because of the model’s nonlinear dynamics, these differences likely persist even after up- or downscaling outputs. TreeMig is considered a medium-complexity model. It is relatively low in complexity regarding demographic processes (which are semi-empirical and non-ecophysiological), but higher in complexity with respect to structure, owing to its spatially explicit implementation and inclusion of within-stand horizontal and vertical heterogeneity.

\subsubsection*{CARAIB}
CARAIB (CARbon Assimilation In the Biosphere) is a mechanistic dynamic vegetation model (DVM) written in FORTRAN and developed at the University of Liège. This study uses CARAIB User’s Guide v2.0 (Nov 2021) with key references including \citep{Warnant1994, Nemry1996, Hubert1998, Otto2002, Dury2011}. CARAIB simulates vegetation responses to climate and soils across local to global scales and from diurnal to centennial timescales. It couples a hydrological module (Improved Bucket Model) with a carbon module for fluxes and pools, and represents vegetation dynamics (establishment, competition, mortality, fire) across land-use types (natural vegetation, croplands, pastures). 

For this study, CARAIB was run in spatially distributed mode for Wallonia, Belgium, using CHELSA climate forcing at 30, 90, and 1800 arcsec resolutions. The spatial resolution matched the forcing grid for each experiment (i.e. one independent column per grid cell). Climatic inputs were provided at daily resolution. Internally, biomass and litter reservoirs (leaf, structural, litter, soil organic carbon, snow, soil water) were updated daily, while vegetation carbon fluxes (photosynthesis and autotrophic respiration) were evaluated at a 2-hourly time step and summed daily to account for diurnal solar cycles. Outputs were produced at monthly resolution for this study. Spin-up was performed by cycling 1901–1920 climate (GSWP3-W5E5 interpolated to the high-resolution CARAIB grid with elevation correction) over 1790–1900, followed by historical climate (GSWP3-W5E5) for 1901–1978, both with historical CO\textsubscript{2}. The run then continued for 1979–2016 with CHELSA climate and historical CO\textsubscript{2}, the period used in the analysis. No site-specific calibration was performed; the standard CARAIB parameter set was applied, with only protocol-driven adjustments to forcing and output handling.

Forcings included \texttt{tas}, \texttt{tasmax}, \texttt{tasmin}, \texttt{pr}, and \texttt{rsds} from the official ISIMIP CHELSA HighRes dataset at the respective resolution. In addition, relative humidity and wind speed at the respective resolution were derived following the method described in \cite{Frieler2024} using the downscaling approach available at \url{https://github.com/johanna-malle/w5e5_downscale}.

Land use and soils were represented using ESA-CCI Land Cover (300 m) to derive land-use/land-cover fractions per grid cell, and GSWP3–HWSD for soil texture and properties. ESA-CCI was aggregated to the CHELSA grids (30–1800 arcsec) as fractional LULC, and HWSD soils were regridded to the same grids with area-weighted mean texture fractions. Hydraulic parameters (conductivity, wilting point, field capacity, saturation) were then computed via CARAIB’s standard pedotransfer functions \citep{Saxton1986}.

Evaluation focused on gross primary productivity (GPP; g C m$^{-2}$ mo$^{-1}$, aggregated to g C m$^{-2}$ yr$^{-1}$) and evapotranspiration (mm mo$^{-1}$, aggregated to mm yr$^{-1}$). Simulations were compared with eddy-covariance measurements from the BE-Vie ICOS/FLUXNET site (1996--2019), using overlapping periods with monthly aggregates derived from half-hourly observations. Input data resolution matters for CARAIB because fluxes depend strongly on climate, especially temperature, which varies with elevation. In comparison to other models in the ISIMIP forests sector, CARAIB can be considered a mid-complexity model.

\subsection{Biomes}
The ISIMIP protocol for the biomes sector is available at \url{https://protocol.isimip.org/#/ISIMIP3a/biomes}.
\subsubsection*{ORCHIDEE-MICT}

ORCHIDEE is a process-based land surface model developed to simulate carbon, water, and energy fluxes in ecosystems from site-level to the global scale \citep{Krinner2005,Ciais2005,Piao2007}. ORCHIDEE-MICT is a variant with improved coupling of soil carbon, temperature, and hydrology processes, as well as an integrated fire module \citep{Guimberteau2018}. The version used in this study further includes advanced representations of grassland management \citep{Chang2013,Chang2021} and northern peatlands \citep{Qiu2018,Qiu2022}.

In this study, the model was run in point-mode at grid cells where site-level observations were available, although it is typically used in spatially distributed mode. The internal spatial resolution matches that of the climate forcings (30 arcsec, 300 arcsec, and 1800 arcsec). Simulations operate at a half-hourly internal timestep. Spin-up involved: (i) repeating a 10-year climate cycle with pre-industrial CO\textsubscript{2} concentrations for ~15,000 years to reach quasi-equilibrium; (ii) a pre-transient phase with CO\textsubscript{2} ramped from 1860 to 1978; (iii) a transient run from 1979–2016. No model calibration was applied for this study.

The model was driven by \texttt{tas}, \texttt{tasmax}, \texttt{tasmin}, \texttt{pr}, and \texttt{rsds} from the ISIMIP CHELSA-HighRes database, supplemented with \texttt{qair}, \texttt{rlds}, \texttt{wind}, and \texttt{ps} following CHELSA downscaling ISIMIP guidelines. Soil texture information was taken from \citet{Liu2022}, and land cover for the Tibetan Plateau grid cells was specified as C3 grasses.

Evaluation targeted active layer thickness (ALT), mean annual ground temperature (MAGT), and soil temperatures at various depths across sites on the Qinghai–Xizang (Tibetan) Plateau. Observational data were extracted from literature sources, including digitization of figures using GetData. Evaluation was based on matching the simulation to the observation date for each variable. 

The following sources were used for evaluation of ALT: \cite{Luo2012, Wu2012, Chen2015, Chen2016, Qin2017, Cao2018, Wu2015CALM, Wang2013, Wu2008, Wu2016}. 

The following sources were used for MAGT: \cite{Luo2012, Wu2012, Chen2015, Chen2016, Qin2017, Cao2018, Wu2010, Wu2015CALM, Yu2008, Li2016, Wang2013, Wu2008, Wu2016}. 

Finally, the following sources were used for soil temperature: \cite{Jin2009, Xie2012, Sheng2010, Li2011, Luo2018, Wu2010Sci, Wang2013}.

ORCHIDEE-MICT is a high-complexity land surface model, capable of simulating the coupled carbon, water, and energy cycles. Its accuracy in permafrost regions benefits from high-resolution climate forcings due to strong climate-topography interactions in the Tibetan Plateau.

\subsection{Regional Water}
The ISIMIP protocol for the regional water sector is available at \url{https://protocol.isimip.org/#/ISIMIP3a/water_regional}.

\subsubsection*{CWatM}
CWatM (Community Water Model) is an open-source process-based hydrological and water resources model, available at \url{https://github.com/iiasa/CWatM}. CWatM version 1.5 as used in this study is based on standard CWatM \citep{Burek2020}, with additional developments related to groundwater \citep{Guillaumot2022}, reservoirs \citep{Smilovicinprep1}, and crop-specific water use \citep{Smilovic_in_prep}.

CWatM simulates surface and subsurface hydrological processes, including water use, management, and infrastructure. 

The model was applied over the Bhima basin, India (approx. 40,000 km$^2$), spatially distributed at 30 arc seconds (~1 km) for surface and 500 m for groundwater processes. A decade-long spin-up using available climate data was conducted. Calibration using an evolutionary algorithm improved the agreement with observed discharge, groundwater levels, and reservoir storage.

Model forcing included all Chelsa HighRes variables (tas, tasmax, tasmin, pr, rsds). Additional variables (longwave radiation, wind, humidity, pressure) were prepared using methods described in \citep{Frieler2024}, available at \url{https://github.com/johanna-malle/w5e5_downscale}. Land use data were derived from a 30 m LULC map \citep{Roy2016}. Irrigated areas were based on \citet{Meier2018a, Lee2022}, crop areas from \citet{Siebert2014}. Soil was derived from the Harmonised World Database v1.2 \citep{HWSD2009}.

Model evaluation was based on discharge (m$^3$/s), groundwater head (m), reservoir storage (m$^3$), and water withdrawals/consumption. Observational data were provided by local authorities and spanned 1960--2015 across ~20 discharge sites, 15 reservoirs, and ~350 groundwater wells.

CWatM is a mid- to high-complexity model. It is sensitive to the spatial resolution of precipitation, especially in mountainous terrain where precipitation misallocation impacts reservoir performance.

\subsubsection*{GR4J }
GR4J (Génie Rural à 4 paramètres Journalier) is a daily lumped rainfall-runoff model with four free parameters, based on unit hydrograph principles \citep{Perrin2003}. 

GR4J is a conceptual catchment water balance model that relates catchment runoff to rainfall and evapotranspiration using daily data. The model consists of a production storage and a routing storage/reservoir and belongs to the family of soil moisture accounting models \citep{kapoor2023}. It belongs to the Regional Water ISIMIP sector. In order to simulate snow accumulation and melting processes, GR4J is coupled with  a degree-day snow module called CemaNeige (Valéry, 2010). This coupling has been tested in other studies (e.g., \cite{Hublart2016, Coron2014}).

The coupled model was applied for 86 catchments across Norway (57\textdegree58'N--71\textdegree11'N; 4\textdegree56'E--31\textdegree01'E). It operates on a daily timestep. Calibration was performed over two periods (1980--1990 and 2005--2016), with a spin-up year in 1979.

Input data included tas and pr from ISIMIP Chelsa HighRes. No additional variables or land/soil data were needed, except for a 30 m resolution DEM.

Model output included daily discharge (m$^3$/s), and performance of simulations at different resolutions was validated using the hydrometric observation network of the Norwegian Water Resources and Energy Directorate (NVE). 

The input data for each catchment in this study are derived from ten distinct elevation bands, each representing an equal fraction of the catchment area. The spatial resolution of these elevation bands is critical for the GR4J model, as it directly influences the accuracy and detail of simulated climatic and hydrological processes within the catchments. This is especially important in mountainous regions, where precise representation of snowmelt dynamics is essential. Higher resolution data also improve the model’s capability to capture the complex internal distribution within the elevation bands. GR4J is a simple-complexity model within its sector.

\subsubsection*{SWAT}
SWAT+ (version 60.5.4) is a revised version of the original SWAT model \citep{Arnold1998, Bieger2017}. It is a semi-distributed, continuous watershed model that simulates water balance, streamflow, and land use impacts. It is part of the Regional Water ISIMIP sector.

The model was applied to a 1800 km$^2$ watershed in Israel (35.3--35.9\textdegree E; 32.75--33.35\textdegree N), comprising 546 Hydrologic Response Units (HRUs). It operates on a daily timestep, with daily and monthly outputs. A one-year spin-up (1995) preceded calibration using manual tuning followed by  automatic calibration using the Parameter ESTimation (PEST) software \citep{Doherty2018}.

Forcings included precipitation (\texttt{pr}), solar radiation (\texttt{rsds}) and minimum and maximum temperature (\texttt{tasmin}, \texttt{tasmax}) from the official ISIMIP data portal. Additionally relative humidity (\texttt{hurs}) and wind speed (\texttt{sfcWind}) were obtained using methods described in \cite{Frieler2024}, available at \url{https://github.com/johanna-malle/w5e5_downscale}. No additional Land Use or Soil Datasets were used.

Output included stream discharge (10$^6$ m$^3$) at the watershed's four streams: Hermon, Meshushim, Dishon, and Jordan. Observational data were provided by the Israeli Water Authority, spanning 1996-2016.

SWAT+ is a mid-complexity model. Its performance benefits from high-resolution input due to the spatial heterogeneity of geology, topography, and land use in the watershed.

\subsubsection*{SWIM}
The Soil and Water Integrated Model (SWIM) is a process-based eco-hydrological model developed for mesoscale watershed analysis. The version used in this study is based on \cite{Krysanova2000}. The model integrates components such as hydrology, vegetation growth, sediment transport, and nutrient cycling.  It is used in the ISIMIP regional water sector.

SWIM is spatially distributed and operates at the watershed scale. It runs on a daily timestep and employs a three-level spatial disaggregation scheme: basin – subbasins – hydrotopes. The model simulates hydrological dynamics based on the water balance equation, taking into account precipitation (including snow), evapotranspiration, percolation, surface runoff, and subsurface runoff for the soil column subdivided into several layers.

SWIM was applied to the Zeravshan River catchment in Central Asia (Tajikistan and Uzbekistan), focused around the Dupuly gauge (Lon: 67.474141, Lat: 39.525736). The bounding box of the model domain spans from Lon: 67.2915 to 71.2018 and Lat: 38.7915 to 40.2018.  

The model was forced using climate data from the ISIMIP Chelsa HighRes dataset, including air temperature (\texttt{tas}), minimum and maximum temperature (\texttt{tasmin}, \texttt{tasmax}), precipitation (\texttt{pr}), and solar radiation (\texttt{rsds}). No additional climate variables were required. Land use information was sourced from the Copernicus Land Cover dataset at 30-meter resolution. Soil map together with soil parametrization data was obtained from the Harmonized World Soil Database (HWSD) at 1 km resolution. A digital elevation model (DEM) at 90 m resolution was used to define terrain characteristics.

The model calibration and validation was performed using observed daily river discharge data at multiple hydrological stations to ensure accurate reproduction of the discharge dynamics. The main evaluation variable was river discharge (m\textsuperscript{3}/s). Observational data were sourced from the National Hydrometeorological Services and the GRDC (Global Runoff Data Centre). Due to limited availability of continuous time series, both mean daily and monthly discharge data were used, covering various periods between the 1960s and 2013, depending on station.

High-resolution input data is essential for SWIM, especially in mountainous catchments like the Zeravshan basin, to accurately represent spatial variability in precipitation, temperature, and snowmelt processes.These aspects are very important for mountainous catchments like Zeravshan, where the elevation ranges from 800 to over 5,000 metres, and the catchment has complex topography \cite{Didovets2021}. It also improves the simulation of runoff routing and internal basin structures. Considering its spatial structure, process complexity, and inclusion of vegetation dynamics, SWIM can be classified as a mid- to high-complexity hydrological model.

\subsection*{Local Lakes Sector}
The ISIMIP protocol for the local lakes sector is available at \url{https://protocol.isimip.org/#/ISIMIP3a/lakes_local}.

\subsubsection*{Simstrat}
Simstrat (Version 3.02) is a one-dimensional, vertically resolved physical lake model \citep{Goudsmit2002, Gaudard2019}.

Simstrat simulates vertical temperature profiles, mixing, stratification, and ice cover in lakes. It has been used within the ISIMIP Lakes Sector to simulate a set of lakes at local scale with individual calibration, and globally across all grid cells containing lakes at 0.5° resolution \citep{Golub2022}.

In this project, Simstrat was used for 32 lakes from the ISIMIP regional lakes dataset, each associated with its specific geographic coordinates. The model was run in point mode with an internal time step of 300 seconds and produced output at 6-hour intervals. 
Calibration was conducted individually for each lake and repeated for each resolution of climate forcing data. Three model parameters were used for calibration: a parameter scaling the incoming longwave radiation; a parameter scaling the wind speed; and a parameter determining the fraction of wind energy that is transferred to internal seiche energy \citep{Golub2022}. Parameter values were estimated using the software package PEST (https://pesthomepage.org/), minimizing the RMSE between simulated and observed lake water temperatures. Data used for calibration was not used for model evaluation purposes. 

Meteorological forcing included: air temperature (\texttt{tas}), shortwave radiation (\texttt{rsds}), longwave radiation (\texttt{rlds}), specific humidity (\texttt{huss}), wind speed (\texttt{sfcWind}), precipitation (\texttt{pr}), and surface pressure (\texttt{pa}). All variables were provided by ISIMIP. \texttt{tas}, \texttt{rsds}, and \texttt{pr} were adjusted to higher resolution using the ISIMIP Chelsa HighRes dataset. The remaining variables were downscaled to the repective resolutions using the methods described by \citet{Frieler2024}, available at \url{https://github.com/johanna-malle/w5e5_downscale}. Lake bathymetry data (area vs. depth) was taken from the ISIMIP protocol.

Simulated lake temperatures were evaluated against observed temperature profiles from various lakes. The number of observations ranged from 181 to 700195, and the number of sampled depths ranged from 8 to 252. The temporal extent of the observational data varied between lakes.

The effect of input resolution is expected to be minimal due to the dominance of large-scale drivers such as air temperature and incoming radiation, which vary little at the sub-grid scale. Furthermore, model calibration offsets much of the forcing bias. Simstrat is a low- to mid-complexity model due to its one-dimensional framework and limited number of calibration parameters.

\subsubsection*{MITgcm}

MITgcm (Massachusetts Institute of Technology general circulation model, \citep{Marshall1997,Marshall1997a}) is a 3D hydrodynamic model with both hydrostatic and non-hydrostatic modes. It supports a wide range of parameterizations and boundary-layer schemes, and is capable of simulating physical processes across spatial scales, from ocean basins to small lakes and reservoirs. Within ISIMIP, MITgcm was applied to Lake Kinneret.

Simulations were carried out for Lake Kinneret, located between 32.7°–32.9°N and 35.51°–35.65°E. The horizontal resolution was 400~m~$\times$~400~m, with 47 vertical layers that are finely resolved (0.25~m) in the top 2~m, becoming coarser (1~m) below. The numerical time step was 300~s. Input and output were processed at a daily time resolution. Initial conditions were taken from a previous long-term simulation starting in January 1980.

The model was forced using air temperature (\texttt{tas}), shortwave radiation (\texttt{rsds}), and precipitation (\texttt{pr}) from ISIMIP. Additional variables (\texttt{sfcWind}, \texttt{rlds}, \texttt{hurs}) were derived using the scripts provided in the W5E5 downscaling repository (\url{https://github.com/johanna-malle/w5e5_downscale}) as described in \cite{Frieler2024}. The model also included climatological discharge, temperature, and salinity data for the Jordan River inflow, which were not available from ISIMIP and were held constant throughout the simulation.

Simulated temperature (°C) was evaluated against weekly temperature profiles at Station A (Lat: 32°48'50.76"N, Lon: 35°36'15.84"E), Station K (Lat: 32°45'0.72"N, Lon: 35°34'43.0"E), Station G (Lat: 32°52'5.52"N, Lon: 35°36'17.0"E), Station D (Lat: 32°43'3.36"N, Lon: 35°35'55.32"E), and Station H (Lat: 32°51'53.28"N, Lon: 35°32'37.32"E). Observations, binned at 1-meter depth resolution, span from 1969 onward. Instruments used include the Whitney-Montedoro thermometer (1969–1986) and the STD-12 (since 1987).

High-resolution input data is crucial for accurately capturing spatial temperature gradients and simulating realistic lake circulation patterns. MITgcm is considered a high-complexity model due to its fully 3D structure, fine-scale resolution, and comprehensive physical process representation.

\subsection{Labour}
The ISIMIP protocol for the local lakes sector is available at \url{https://protocol.isimip.org/#/ISIMIP3a/labour}.

For labour simulations empirically estimated labour supply response functions combined with an augmented mean response function of existing labour productivity impact models were used to compute a single compound metric for effective labour effects \citep{Dasgupta2021, Dasgupta2024}.

The model applies econometric analysis to more than 300 micro-survey datasets to estimate global and regional temperature and Wet-Bulb Globe Temperature exposure–response functions (ERFs) for labour supply. Labour productivity ERFs from the literature are synthesized into an augmented mean function. These two components are combined into a single compound metric—effective labour effects—capturing both changes in working hours and productivity losses under future climate change. The model is part of the ISIMIP labour sector.

The model is spatially distributed and run globally at its internal spatial resolution. Simulations are annual, and because the approach is fully empirical, no spin-up phase or calibration is required. The model is forced with mean near-surface air temperature (\texttt{tas}) from ISIMIP’s CHELSA HighRes dataset, as well as relative humidity, which was downscaled using the methods described by \citet{Frieler2024}, available at \url{https://github.com/johanna-malle/w5e5_downscale}. No other land-use datasets, soil datasets, or non-ISIMIP forcing data are used.

Direct validation with labour observations is infeasible. Instead, temperature data from the Global Historical Climatology Network daily (GHCNd) are spatially merged to the labour force dataset at the second administrative level for indirect evaluation. High-resolution input data are important for representing spatial variations in heat exposure that affect the labour force. 

\section{Part B}
\subsection{Introduction}
This supplementary material provides information on the modeling setups and observational datasets used in this study, as well as model-specific results. For each model, observations are compared against model outputs across different spatial resolutions for each evaluation site, with the overall results, which are used in the main manuscript being highlighted.

\subsection{Regional Forests - Coordinated Effort}

\subsubsection{Observed Data}\label{reg_for_1:obs}

We utilized FLUXNET data compiled in the Profound database \citep{Reyer2020}. The variables of interest were monthly gross primary productivity (GPP) and actual evapotranspiration (AET).

\paragraph{Gross Primary Productivity (GPP)}  
Multiple FLUXNET GPP products exist, reflecting different estimation methods. We used data derived with a constant friction velocity threshold (USTAR), selecting the reference product based on model efficiency following \cite{Pastorello2020}. For partitioning GPP from net ecosystem exchange (NEE), we applied the daytime (DT) method as described by \cite{Lasslop2010}.

Half-hourly measurements of the variable \texttt{gppDtCutRef} were initially reported in \(\mu\mathrm{mol}\,\mathrm{CO}_2\,\mathrm{m}^{-2}\,\mathrm{s}^{-1}\), and converted to \(\mathrm{kg\,C\,m}^{-2}\,\mathrm{s}^{-1}\) to ensure comparability with model outputs. Observations were filtered to retain only data with FLUXNET quality flags 0 (measured) and 1 (good quality gap-filled).

When aggregating from half-hourly to daily resolution, only days with at least 80\% of valid data points were used. Similarly, monthly means were calculated only for months with at least 80\% valid daily observations.

\paragraph{Actual Evapotranspiration (AET)}  
AET is not directly measured at FLUXNET sites. Instead, it was estimated using latent heat flux and air temperature as follows:

\begin{equation}
\mathrm{AET} = \frac{LHF_{\mathrm{daily}}}{\lambda \times 10^{6}} \times 86400
\end{equation}

where \(LHF_{\mathrm{daily}}\) (W\,m\(^{-2}\)) is the daily mean latent heat flux, and \(\lambda\) (J\,g\(^{-1}\)) is the latent heat of vaporization approximated by

\[
\lambda = 2.501 - 0.00237 \times T_{\mathrm{air}}
\]

with \(T_{\mathrm{air}}\) being the daily mean air temperature in °C. The factor \(10^{6}\) converts Joules to Megajoules, and multiplication by 86400 converts seconds to days, yielding AET in mm day\(^{-1}\). This approach follows recommendations by \cite{Foken2024} and is applied by \cite{Mahnken2022}.

Daily latent heat flux and air temperature data were filtered similarly to GPP to include only measured and good quality gap-filled observations. Daily AET values were linearly interpolated to fill missing days; however, monthly AET sums were calculated only for months with more than 80\% valid daily data coverage.

\subsubsection{Per-model comparison to observed data}
\paragraph{Time-series comparison}

\begin{figure}[ht!]
\centering
\includegraphics[width=\textwidth,height=0.95\textheight,keepaspectratio]{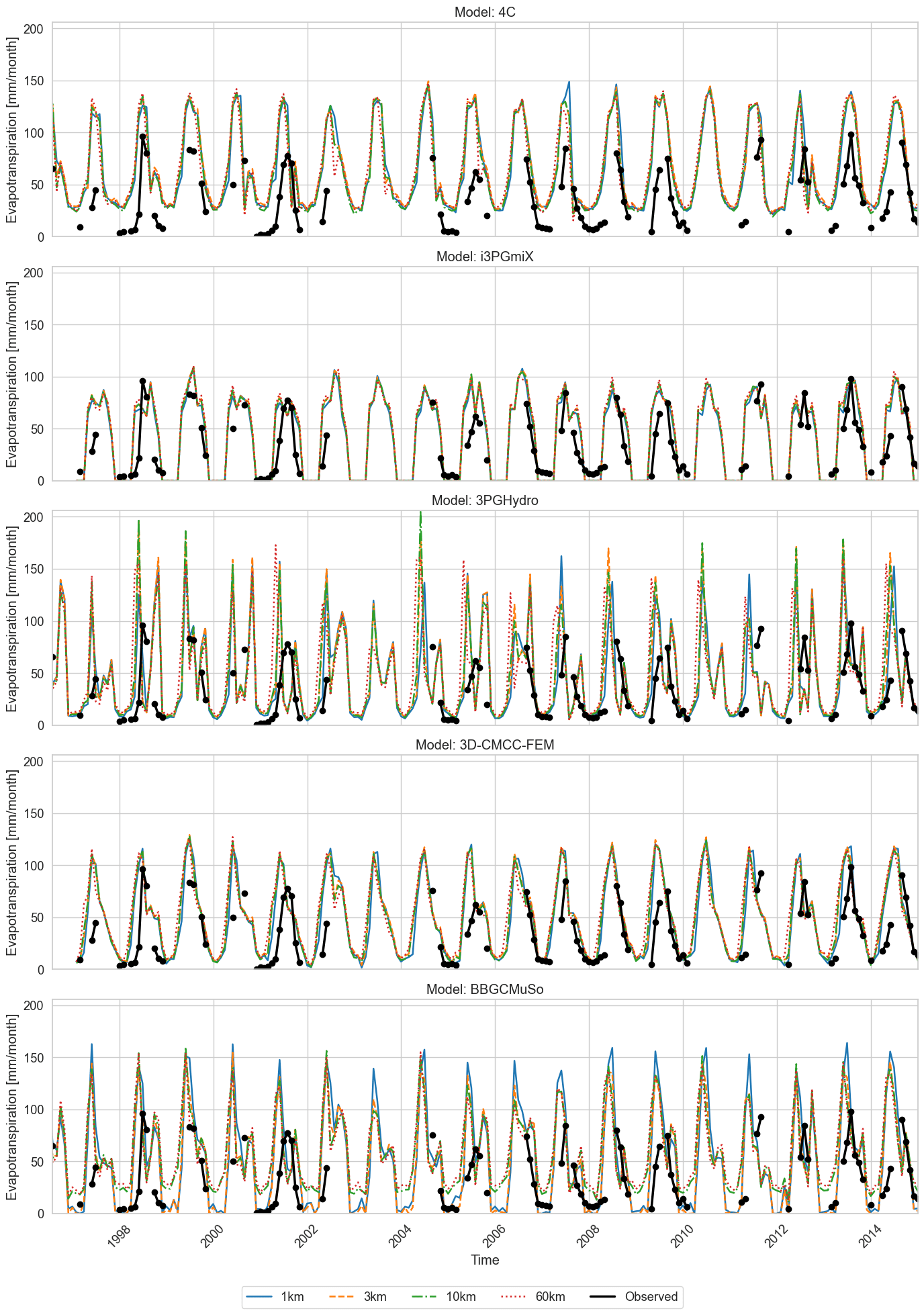} 
\caption{Per-model comparison of monthly summed AET at multiple spatial resolutions against observations from the Collelongo flux tower site.}
\label{fig:supp:for_evap}
\end{figure}

\begin{figure}[ht!]
\centering
\includegraphics[width=\textwidth,height=0.95\textheight,keepaspectratio]{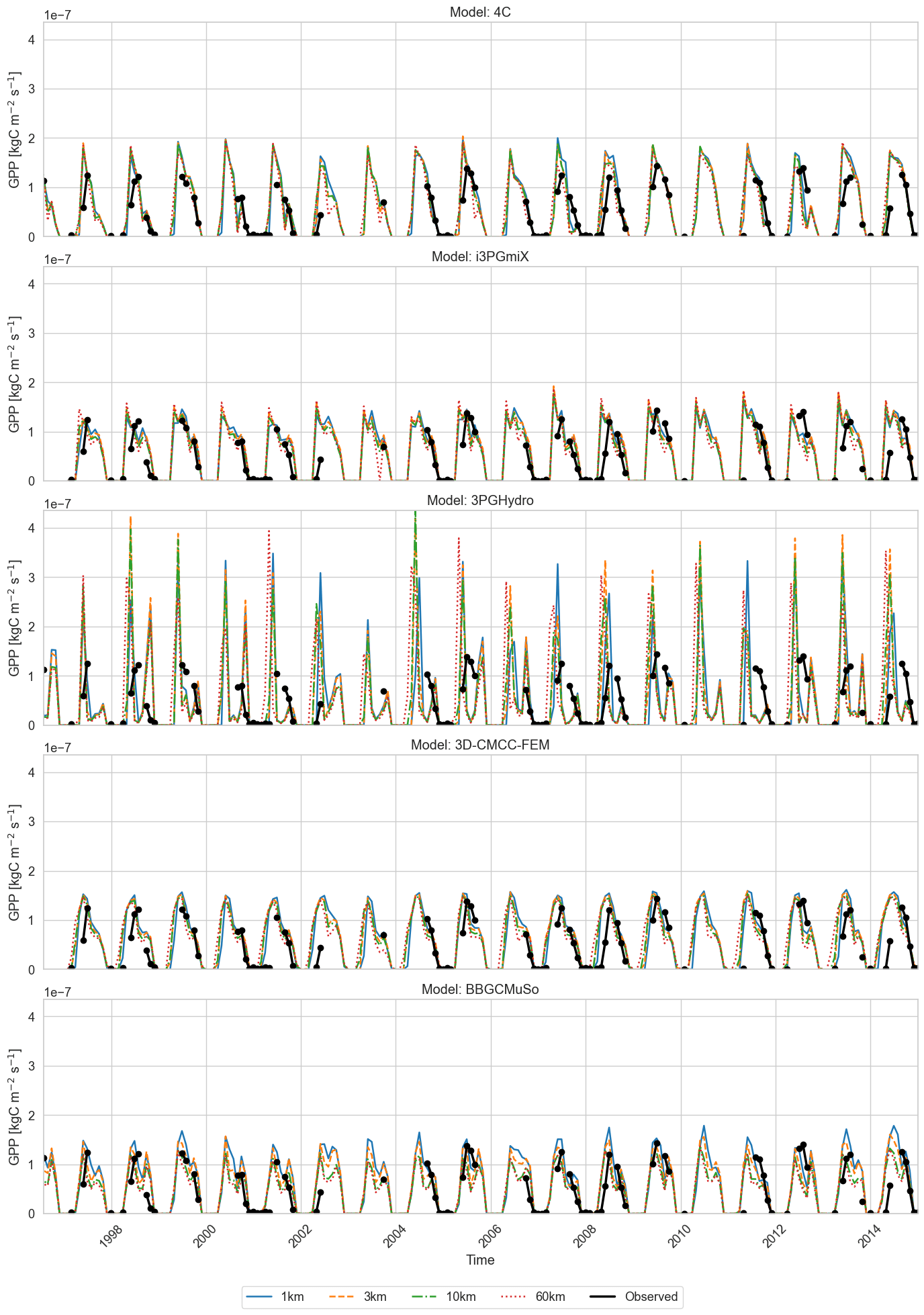} 
\caption{Per-model comparison of monthly averaged GPP at multiple spatial resolutions against observations from the Collelongo flux tower site.}
\label{fig:supp:for_gpp}
\end{figure}

Figures~\ref{fig:supp:for_evap} and~\ref{fig:supp:for_gpp} present time-series comparisons between model simulations and flux tower observations from the Collelongo site. 
For each model, simulations at all spatial resolutions are shown alongside the observations.  
Figure~\ref{fig:supp:for_evap} shows monthly summed actual evapotranspiration (AET), while Figure~\ref{fig:supp:for_gpp} presents monthly averaged gross primary productivity (GPP). In general, simulated GPP aligned more closely with observations than AET. The 3PGHydro model consistently and substantially overestimated GPP magnitudes. Visual differences between spatial resolutions were subtle, as simulation outputs appeared broadly similar; however, quantitative evaluation using statistical metrics revealed notable resolution-dependent differences (see next section).

\begin{figure}[ht!]
\centering
\includegraphics[width=\textwidth]{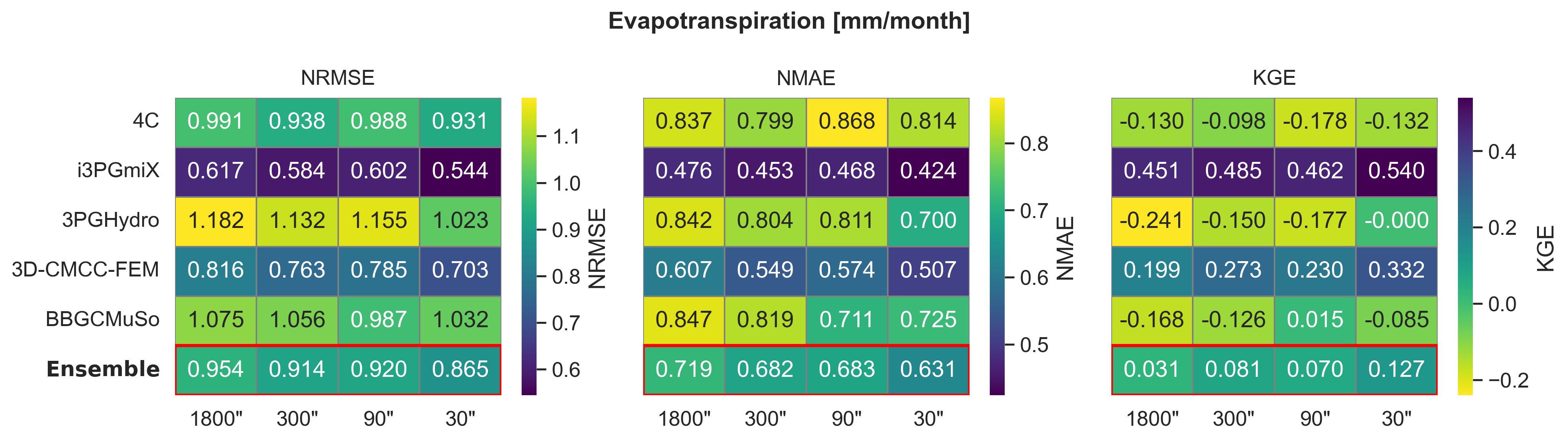}
\caption{Heatmap of error metrics (NRMSE, NMAE, KGE) for the simulated resolutions of the five forest models in the coordinated effort, compared to observed monthly summed ET.}
\label{fig:supp:for_evap_hm}
\end{figure}

\begin{figure}[ht!]
\centering
\includegraphics[width=\textwidth]{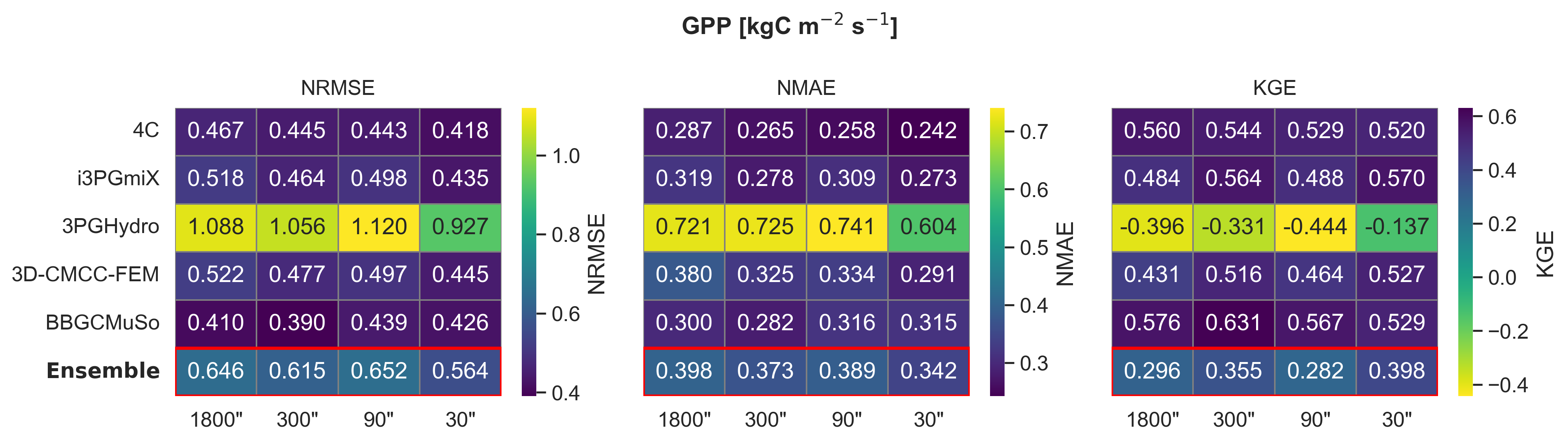}
\caption{Heatmap of error metrics (NRMSE, NMAE, KGE) for the simulated resolutions of the five forest models in the coordinated effort, compared to observed monthly averaged GPP.}
\label{fig:supp:for_gpp_hm}
\end{figure}

\begin{figure}[ht!]
\centering
\includegraphics[width=\textwidth]{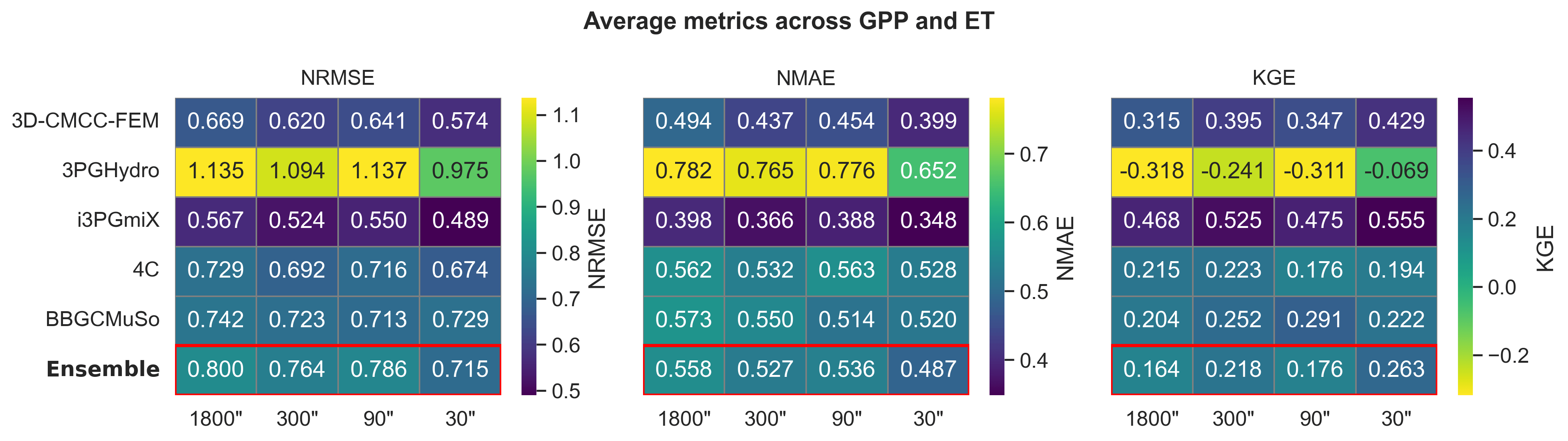}
\caption{Heatmap of error metrics (NRMSE, NMAE, KGE) for the simulated resolutions of the five forest models in the coordinated effort, averaged between GPP and ET.}
\label{fig:supp:for_hm}
\end{figure}

\paragraph{Statistical Comparison}
Figures~\ref{fig:supp:for_evap_hm} and~\ref{fig:supp:for_gpp_hm} summarize error metrics - NRMSE, NMAE, and KGE - for each model and simulated resolution, as well as the overall ensemble across all models. These heatmaps allow a direct comparison of model performance against observed monthly summed AET and monthly averaged GPP, respectively. KGE values for AET were below 0 for the 4C, 3PGHydro, and BBGCMuSo models, whereas KGEs for GPP were below 0 only for 3PGHydro, indicating generally better model performance for GPP. NRMSE and NMAE were substantially higher for AET than for GPP for all models, except 3PGHydro, highlighting that AET is more challenging to simulate accurately. For the overall analysis presented in the main paper, GPP and AET metrics were averaged, resulting in the error metrics shown in Figure~\ref{fig:supp:for_hm}.

\subsubsection{Local climate data}
To evaluate the accuracy of climate inputs across resolutions, we compared CHELSA temperature and precipitation against local measurements from the Collelongo FLUXNET site. Only high-quality observations (measured data and reliable gap fills) were used.

\begin{figure}[ht!]
\centering
\includegraphics[width=\textwidth]{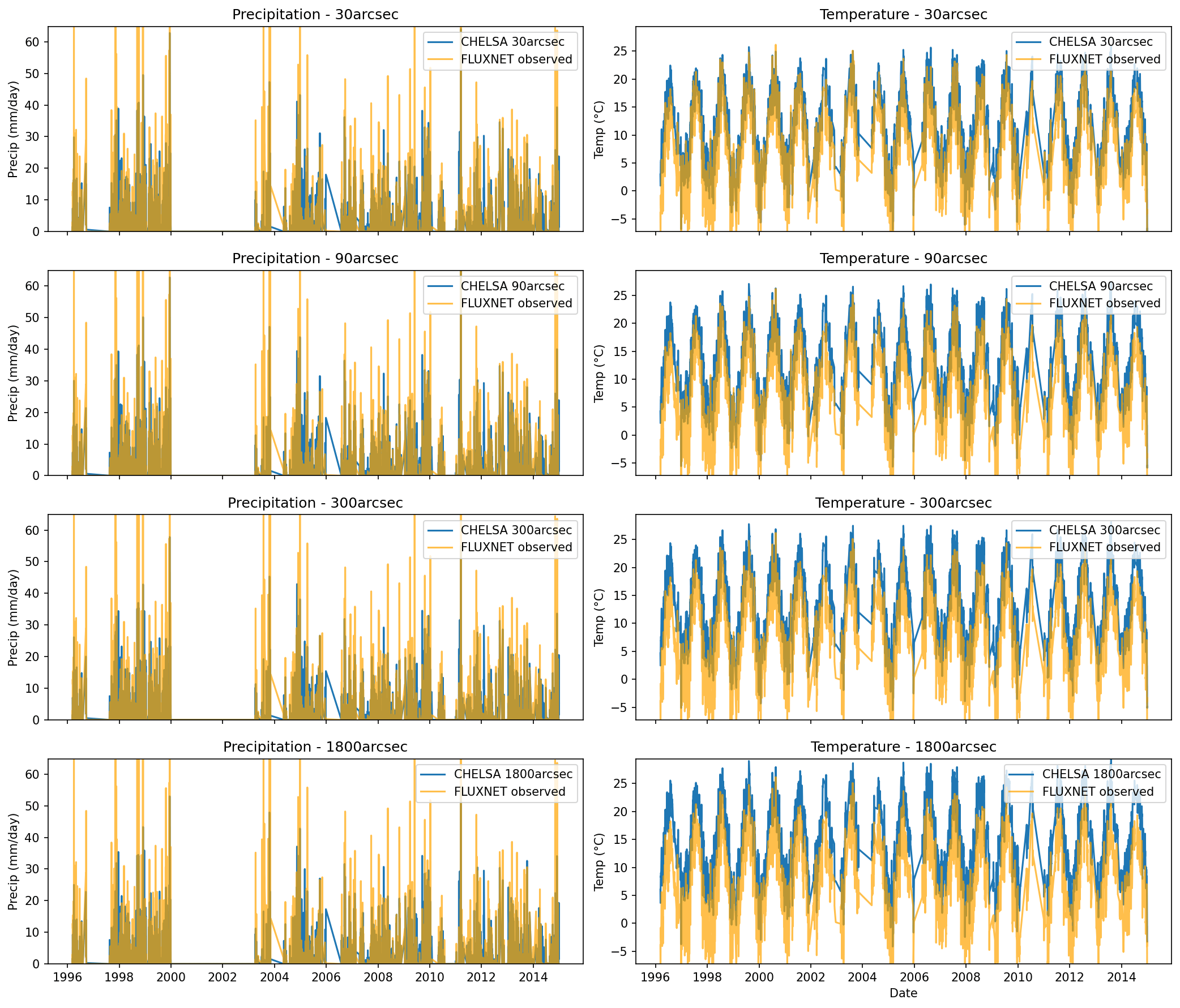}
\caption{Time series of CHELSA daily precipitation and mean daily temperature at four spatial resolutions, compared with FLUXNET observations at Collelongo.}
\label{fig:supp:for_clim_tm}
\end{figure}

\begin{figure}[ht!]
\centering
\includegraphics[width=\textwidth]{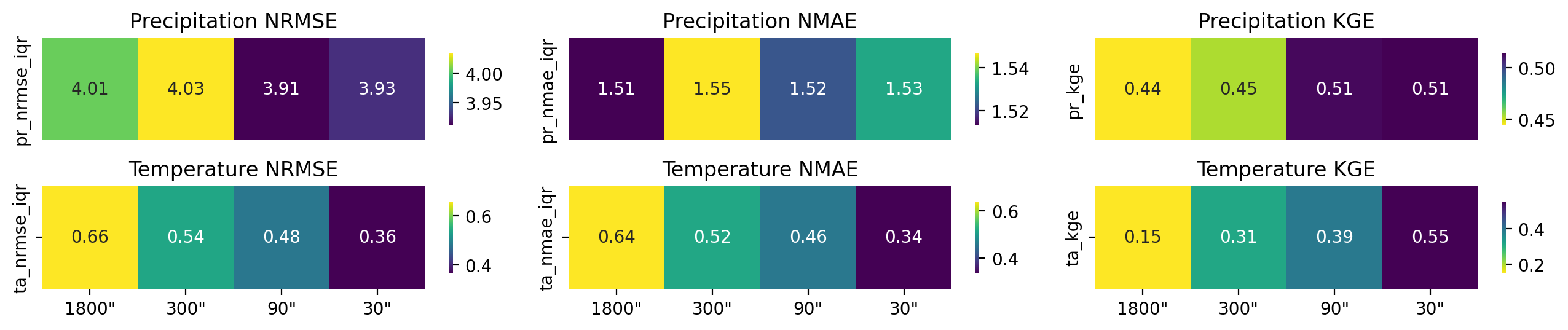}
\caption{Error metrics for precipitation (top row) and temperature (bottom row): first column shows NRMSE, second column MAE, and third column KGE.}
\label{fig:supp:for_clim_hm}
\end{figure}

Figure~\ref{fig:supp:for_clim_tm} shows time-series comparisons across all four resolutions, while Figure~\ref{fig:supp:for_clim_hm} summarizes the corresponding error metrics. Precipitation errors are substantially larger than temperature errors and do not improve consistently with resolution (e.g., performance degrades when moving from 1800" to 300", then improves again). In contrast, temperature errors decrease steadily with increasing resolution for all metrics.

\clearpage

\subsection{Regional Forests – other models}
\subsubsection{PREBAS}
PREBAS simulations were conducted at two FLUXNET sites in Finland: Hyytiälä and Sodankylä.

\paragraph{Observed data}
Observational data for both sites were obtained directly from the two stations. Steps taken to calculate these values follow the standard FLUXNET/ICOS protocol. Quality flags were used to filter out low-quality data. Because PREBAS operates at a daily time step - unlike several coordinated models that simulated only monthly values - we used daily observational products for analysis.The daily GPP values are in gC per m2 per day, whereas the ET values are in mm per day.

\begin{figure}[ht!]
\centering
\includegraphics[width=\textwidth]{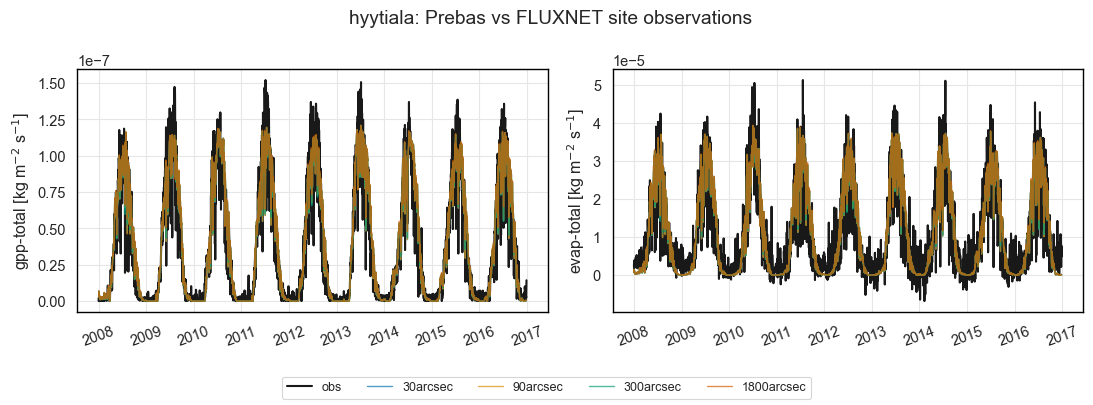} 
\caption{Comparison of daily averaged GPP and ET at the four spatial resolutions against observations from the Hyytiala flux tower site.}
\label{fig:supp:for_tm_hyy}
\end{figure}

\begin{figure}[ht!]
\centering
\includegraphics[width=\textwidth]{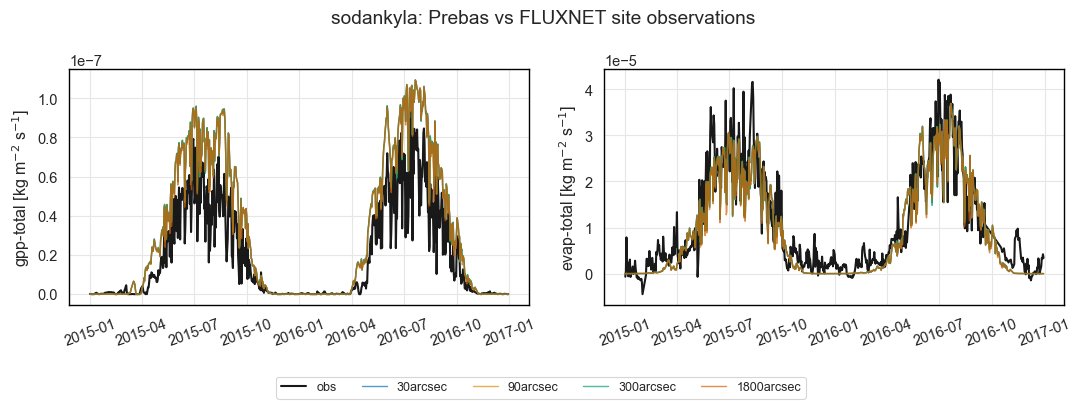} 
\caption{Comparison of daily averaged GPP and ET at the four spatial resolutions against observations from the Sodankylae flux tower site.}
\label{fig:supp:for_tm_sod}
\end{figure}

\paragraph{Per-location comparison to observed data}
\paragraph*{Time-series comparison}
Figures~\ref{fig:supp:for_tm_hyy} and~\ref{fig:supp:for_tm_sod} show that PREBAS results varied only slightly between spatial resolutions. At Hyytiälä, simulated GPP and ET magnitudes agreed reasonably well with observations, although the model did not fully capture the observed peak values. At Sodankylä, GPP was generally overestimated, especially during the summer months, and the simulations tended to underestimate the highest observed peaks of ET.

\begin{figure}[ht!]
\centering
\includegraphics[width=\textwidth]{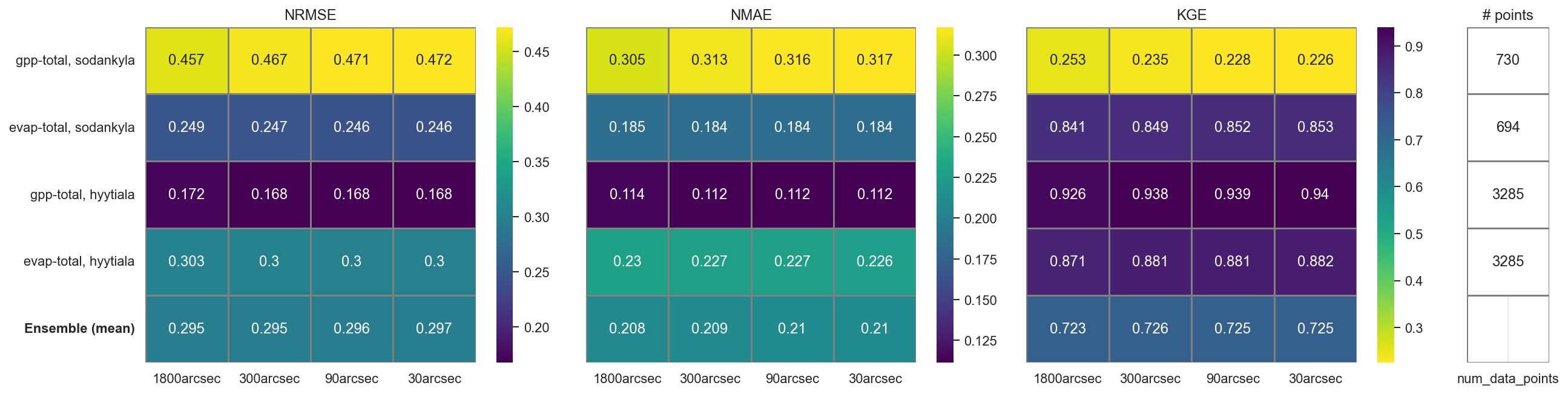} 
\caption{Heatmap of error metrics (NRMSE, NMAE, KGE) for the PREBAS simulation at four resolutions (1800", 300", 90", 30") for GPP and ET at the Sodankyl\"a and Hyyti\"al\"a FLUXNET sites, compared to observations. The last column shows the number of data points available for each variable and site.}
\label{fig:supp:for_hm_prebas}
\end{figure}

\paragraph{Statistical comparison}
Figure~\ref{fig:supp:for_hm_prebas} summarizes error metrics - NRMSE, NMAE, and KGE - across the four resolutions for both sites and variables (GPP and ET). Overall, the PREBAS simulations show minimal sensitivity to spatial resolution, with performance remaining largely consistent across resolutions. The last row ('Ensemble (mean)') represents the mean across all sites and is the data used in the main paper's overview figure.

\begin{figure}[ht!]
\centering
\includegraphics[width=\textwidth]{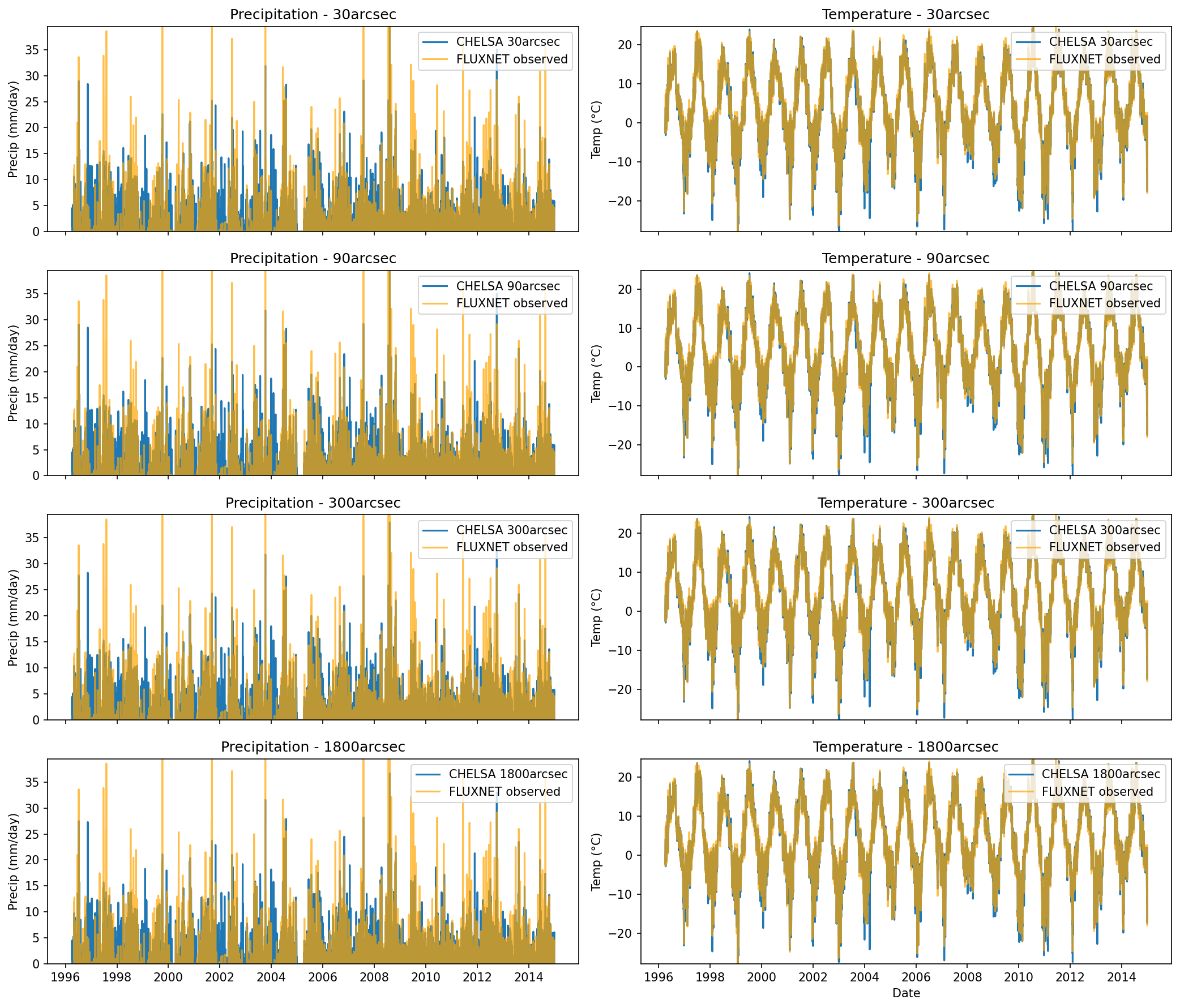}
\caption{Time series of CHELSA daily precipitation and mean daily temperature at four spatial resolutions, compared with FLUXNET observations at Hyytiälä.}
\label{fig:supp:for_clim_tm_hyy}
\end{figure}

\begin{figure}[ht!]
\centering
\includegraphics[width=\textwidth]{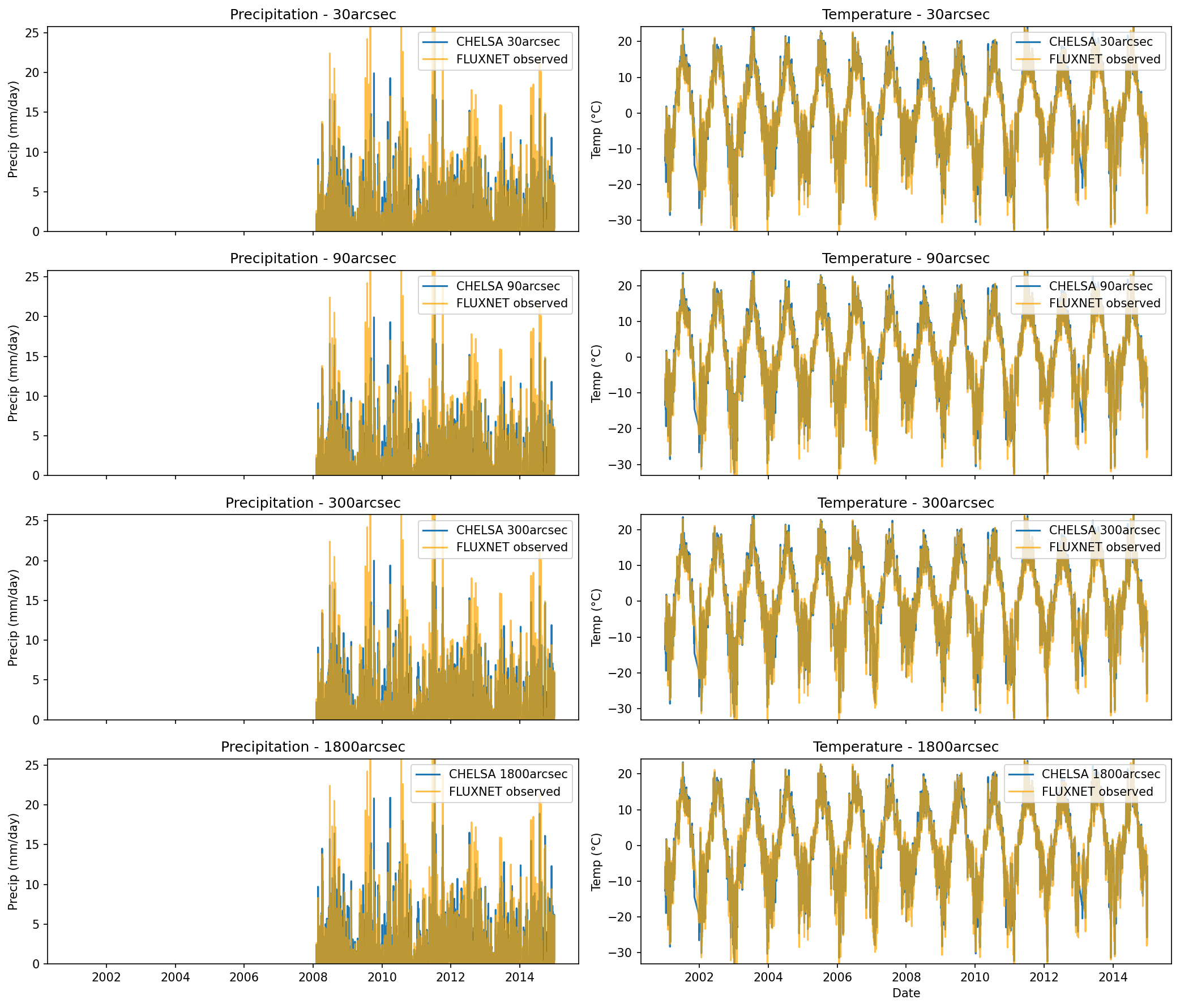}
\caption{Time series of CHELSA daily precipitation and mean daily temperature at four spatial resolutions, compared with FLUXNET observations at Sodankylae.}
\label{fig:supp:for_clim_tm_sod}
\end{figure}

\begin{figure}[ht!]
\centering
\includegraphics[width=\textwidth]{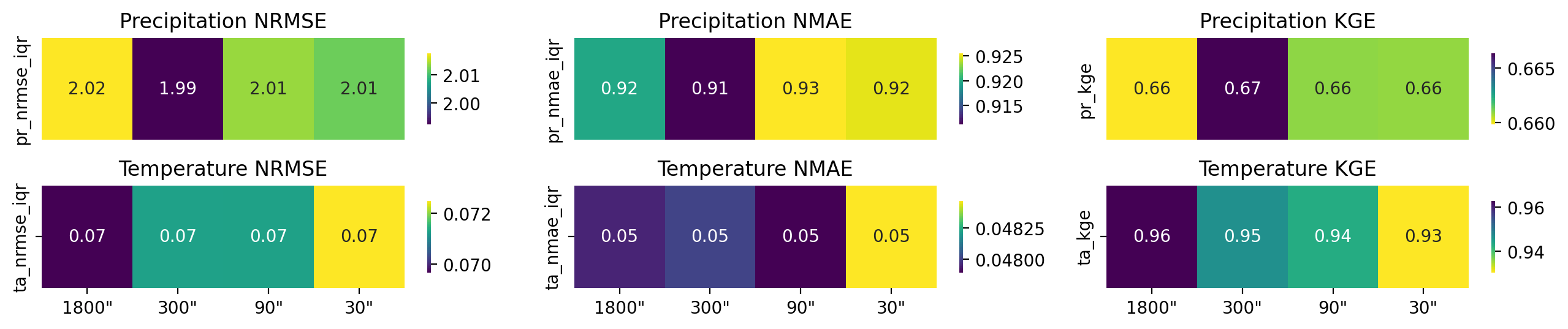}
\caption{Error metrics for precipitation (top row) and temperature (bottom row): first column shows NRMSE, second column MAE, and third column KGE. Location: Hyytiala.}
\label{fig:supp:for_clim_hm_hyy}
\end{figure}

\begin{figure}[ht!]
\centering
\includegraphics[width=\textwidth]{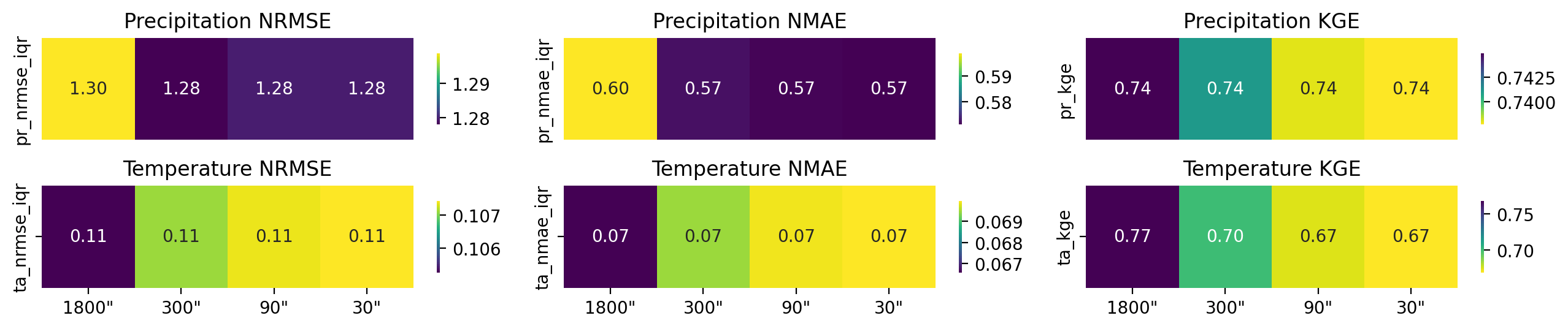}
\caption{Error metrics for precipitation (top row) and temperature (bottom row): first column shows NRMSE, second column MAE, and third column KGE. Location: Sodankylae.}
\label{fig:supp:for_clim_hm_sod}
\end{figure}

\paragraph{Local climate data}
Following the same approach used for the coordinated forest models, we evaluated the accuracy of CHELSA temperature and precipitation across different resolutions at the two FLUXNET sites of interest: Hyytiälä and Sodankyl\"a. Figures~\ref{fig:supp:for_clim_tm_hyy} and~\ref{fig:supp:for_clim_tm_sod} present time-series comparisons of average daily temperature and daily precipitation for Hyytiälä and Sodankyl\"a, respectively. Figures~\ref{fig:supp:for_clim_hm_hyy} and~\ref{fig:supp:for_clim_hm_sod} show the corresponding heatmaps of error metrics. Overall, we observe very little variation in CHELSA performance across resolutions for both temperature and precipitation.

\subsubsection{CARAIB}
CARAIB simulations were performed on a spatially distributed grid covering the Wallonia region in Belgium and the southern Netherlands, with model output provided at monthly resolution. Simulations were run at 30arcsec, 90arcsec, and 1800arcsec (note: no 300arcsec simulations were available).

\paragraph{Observed Data}
Model performance was evaluated at the Vielsalm FLUXNET station by extracting the nearest grid cell from each spatial resolution. Gross primary productivity (GPP) and evapotranspiration (ET) from the flux tower were processed in the same manner as for the coordinated forest sector (Section~\ref{reg_for_1:obs}) to obtain consistent monthly data.

\begin{figure}[ht!]
\centering
\includegraphics[width=\textwidth]{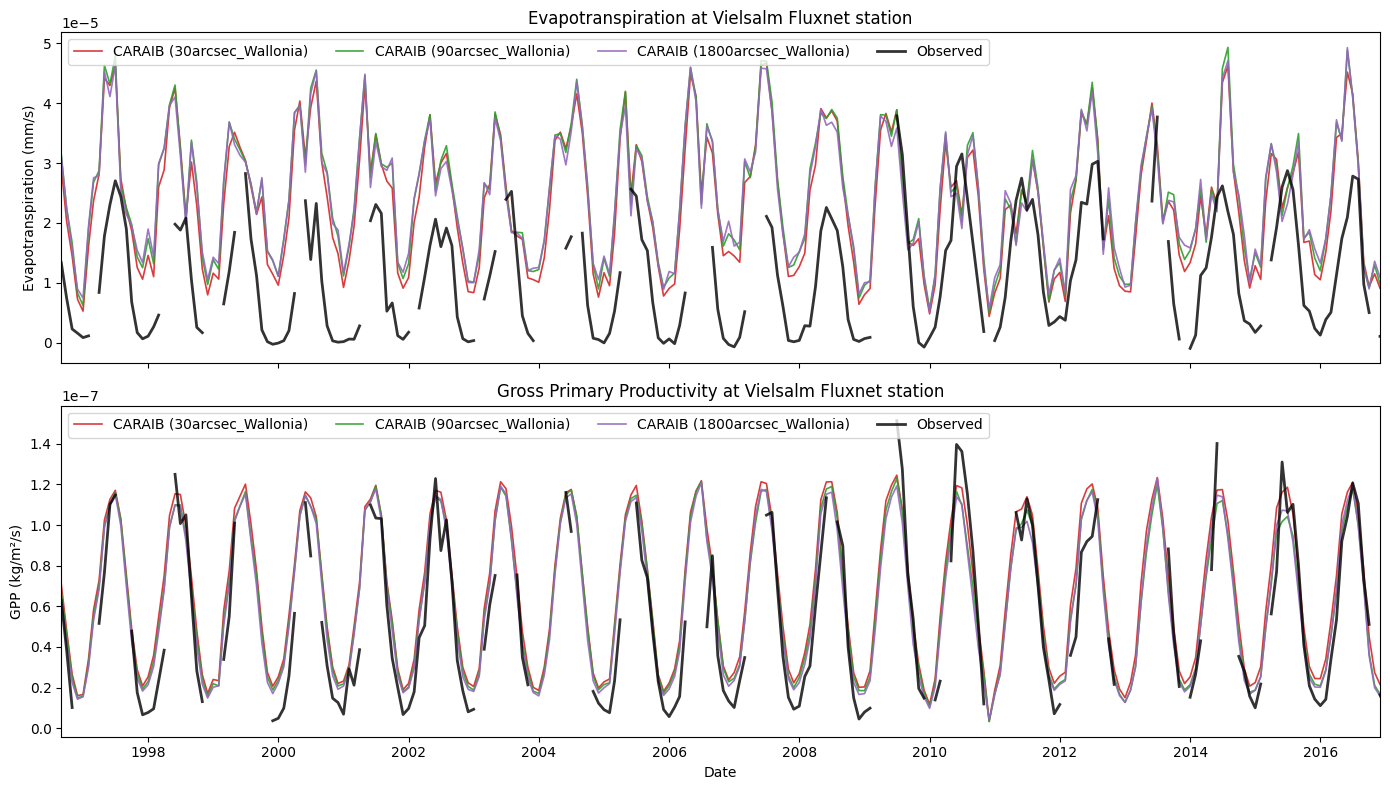}
\caption{Comparison of monthly ET and GPP at three spatial resolutions (30,arcsec, 90,arcsec, 1800,arcsec) as simulated by the forest model CARAIB against observations from the Vielsalm FLUXNET site.}
\label{fig:supp:for_ts_caraib}
\end{figure}

\paragraph{Comparison to Observations}
\paragraph*{Time-Series Comparison}
Figure~\ref{fig:supp:for_ts_caraib} compares monthly simulated and observed ET and GPP. Across resolutions, simulation differences are modest. For ET, the 30arcsec simulation captures the winter dip in observations slightly better than coarser resolutions. For GPP, lower-resolution simulations better match observed minima, whereas the 30arcsec simulation reproduces peak magnitudes more accurately.

\begin{figure}[ht!]
\centering
\includegraphics[width=\textwidth]{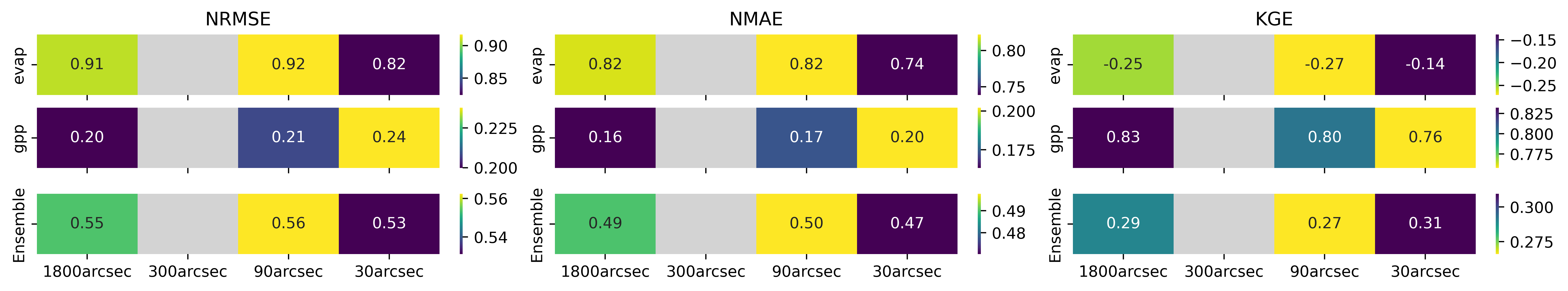}
\caption{Error metrics (NRMSE, NMAE, KGE) for CARAIB simulations at three resolutions (1800'', 90'', 30'') for GPP and ET at the Vielsalm FLUXNET site. The last row shows the ensemble mean, which is used in the main paper.}
\label{fig:supp:for_hm_caraib}
\end{figure}

\paragraph*{Statistical Comparison}
Figure~\ref{fig:supp:for_hm_caraib} shows heatmaps of error metrics (NRMSE, NMAE, KGE) at the Vielsalm site. ET errors decrease slightly at 30, arcsec, while GPP errors increase marginally. KGE values for ET remain low (below zero) across all resolutions.

\begin{figure}[ht!]
\centering
\includegraphics[width=\textwidth]{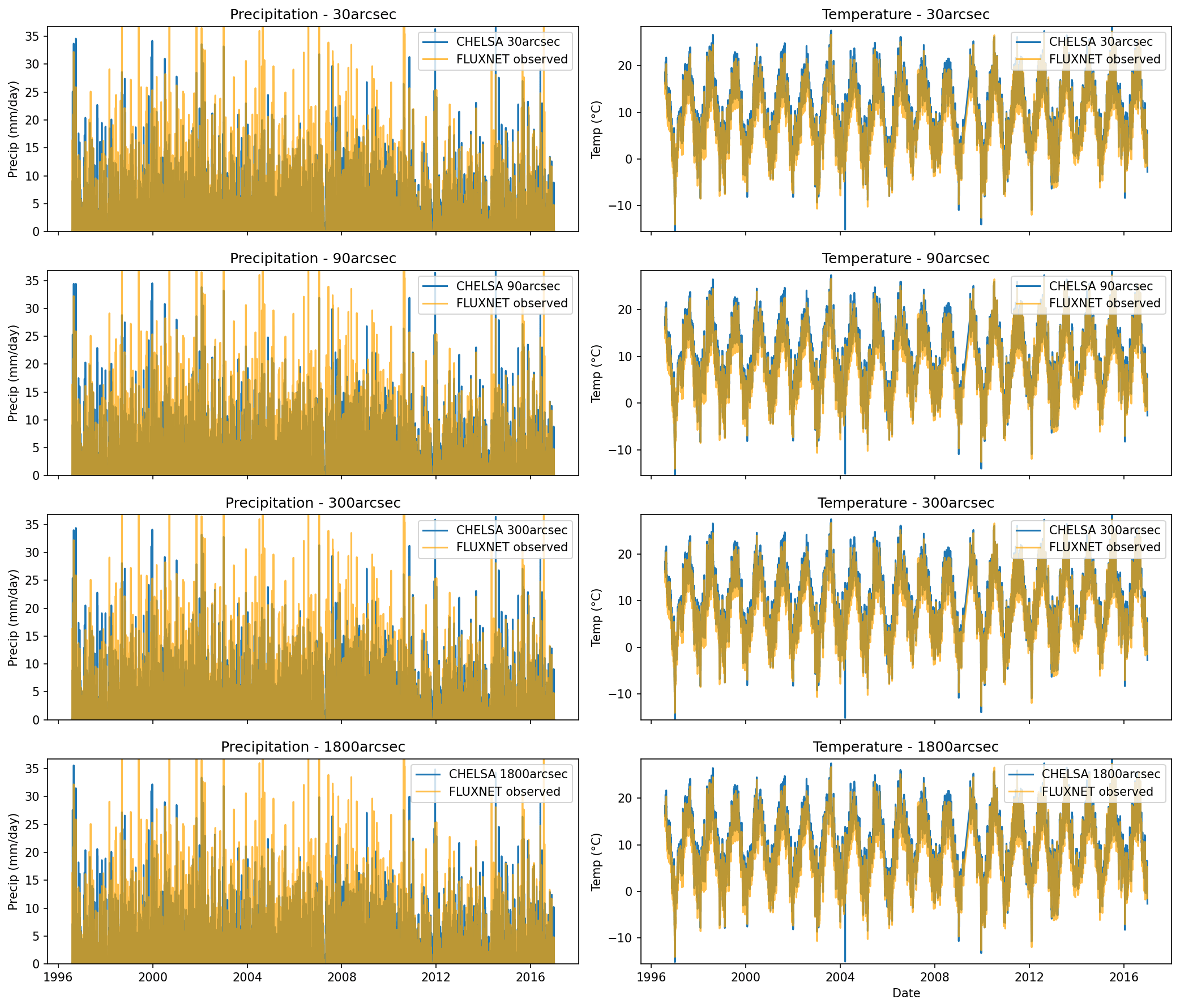}
\caption{Time series of CHELSA daily precipitation and mean daily temperature at four spatial resolutions, compared with FLUXNET observations at Viesalm.}
\label{fig:supp:for_clim_tm_vies}
\end{figure}

\begin{figure}[ht!]
\centering
\includegraphics[width=\textwidth]{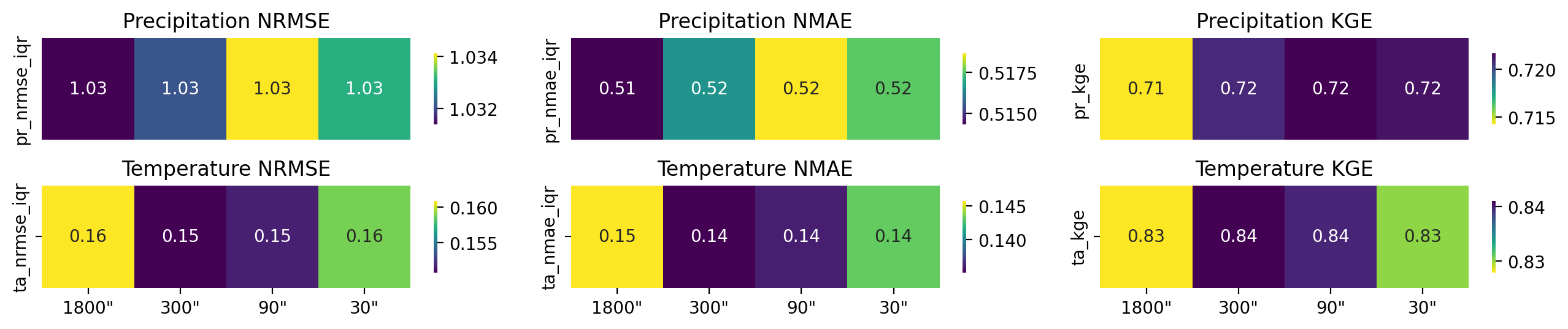}
\caption{Error metrics for precipitation (top row) and temperature (bottom row): first column shows NRMSE, second column MAE, and third column KGE. Location: Viesalm.}
\label{fig:supp:for_clim_hm_vies}
\end{figure}

\paragraph{Local Climate Data}
As for the coordinated forest effort and PREBAS simulations, CHELSA temperature and precipitation data were evaluated at all four resolutions for the Vielsalm site to assess local climate input accuracy. 
Figure~\ref{fig:supp:for_clim_tm_vies} presents time-series comparisons of average daily temperature and daily precipitation for the Viesalm Fluxnet station. Figures~\ref{fig:supp:for_clim_hm_vies} shows the corresponding heatmaps of error metrics. Overall, we observe very little variation in CHELSA performance across resolutions for both temperature and precipitation.

\subsubsection{TreeMig}
TreeMig was applied in a spatially distributed setup across Switzerland, and its performance was evaluated using species-specific basal area data from the Swiss National Forest Inventory (LFI/NFI) of 1983. To enable spatially consistent comparison, LFI measurements were aggregated using a 3km square moving window, retaining only windows with at least three inventory plots. For each LFI plot, absent species were recorded with a basal area of zero.

TreeMig simulations were run without land-use forcing. For comparison with observations, simulation outputs were filtered to match LFI plot locations, and moving-window averages were computed using only cells corresponding to LFI plots.

Species were included in the analysis only if they occurred in at least 100 LFI plots. Twenty-four species were included in this analysis: \textit{Abies alba, Acer campestre, Acer platanoides, Acer pseudoplatanus, Alnus glutinosa, Alnus incana, Betula pendula, Carpinus betulus, Castanea sativa, Fagus sylvatica, Fraxinus excelsior, Larix decidua, Picea abies, Pinus cembra, Pinus montana, Pinus sylvestris, Quercus petraea, Quercus robur, Salix alba, Sorbus aria, Sorbus aucuparia, Tilia cordata, and Tilia platyphyllos}.

\begin{figure}[ht!]
\centering
\includegraphics[width=\textwidth]{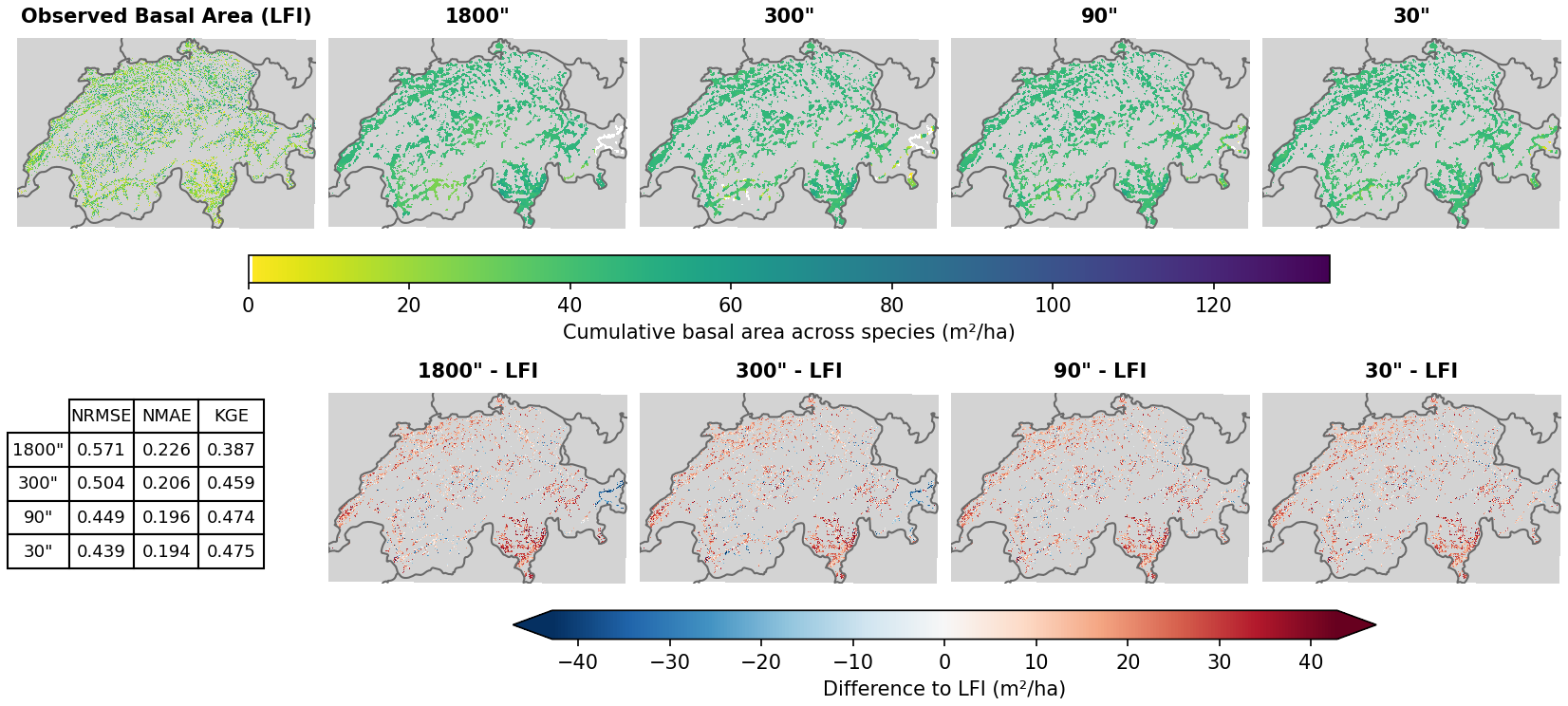}
\caption{Spatial comparison of cumulative basal area across all species. The first row shows observed basal area (LFI) and TreeMig simulations at 1800'', 300'', 90'', and 30''. Zero basal area is indicated in white. The second row shows difference maps between the respective TreeMig simulations and LFI data. The table on the left summarizes error metrics (NRMSE, NMAE, KGE) for each resolution.}
\label{fig:supp:for_treemig_spatial}
\end{figure}

\paragraph{Spatial comparison}
Figure~\ref{fig:supp:for_treemig_spatial} shows a spatial comparison of cumulative basal area across all species: TreeMig simulations at 1800'', 300'', 90'' and 30'' versus observed LFI data. In the eastern part of Switzerland, the coarser simulations (e.g.\ 1800'') occasionally predict zero basal area (i.e.\ no trees), which is not reflected in the observations. Overall, model performance improves steadily with finer spatial resolution, though the resolution effect is more pronounced when considering individual species rather than the cumulative basal area (see also Figure~4 in the main manuscript).

\begin{figure}[ht!]
\centering
\includegraphics[width=\textwidth]{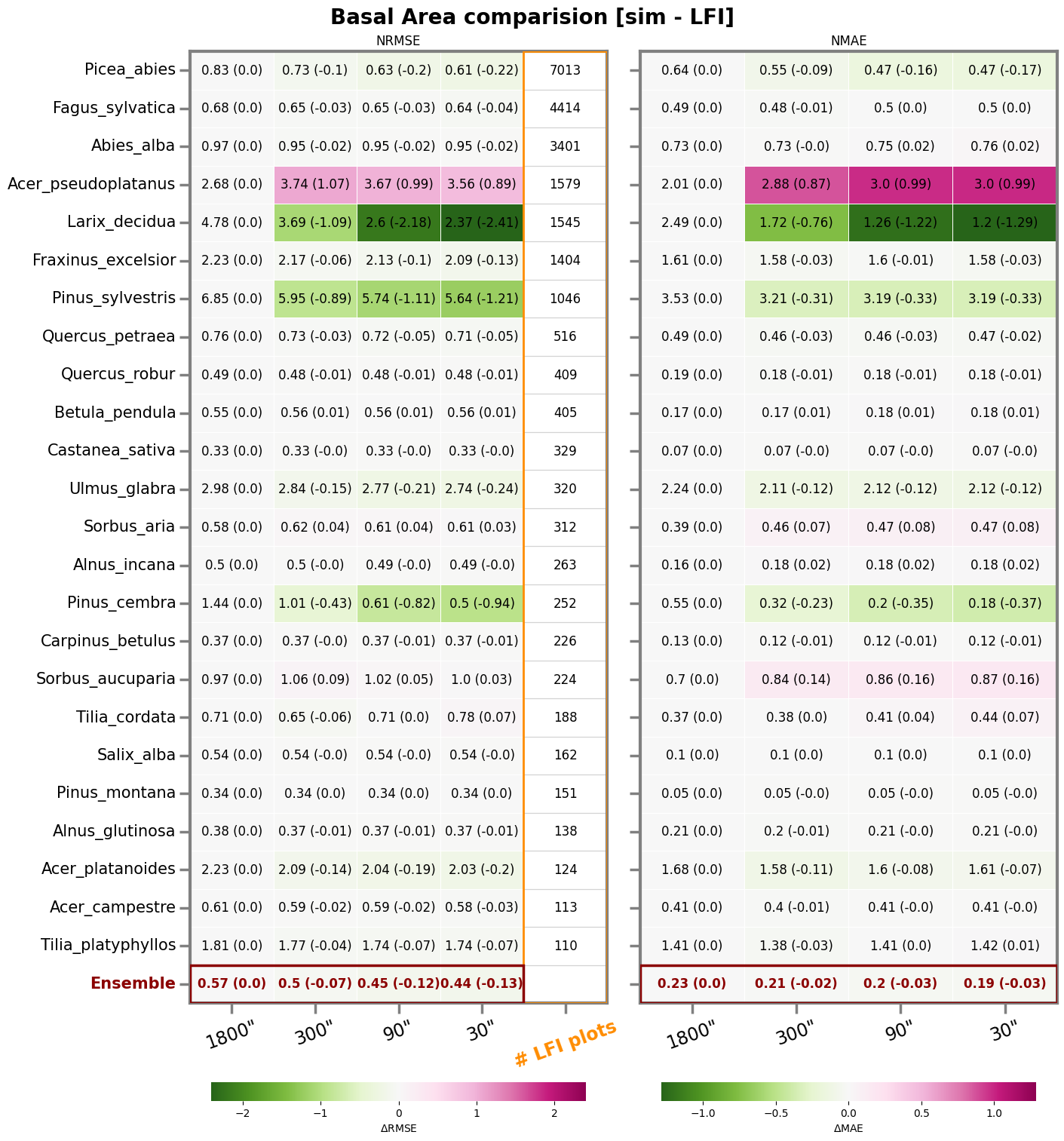}
\caption{Comparison of NRMSE and NMAE per species across the four TreeMig resolutions, and across all species combined (last row, ``Ensemble,'' which is used in the main paper analysis). Species are ordered by their frequency of occurrence in LFI plots.}
\label{fig:supp:for_treemig_stats}
\end{figure}

\paragraph{Statistical comparison and scale effects}
Figure~\ref{fig:supp:for_treemig_stats} shows computed error metrics (NRMSE, NMAE) per species. While errors generally decrease with increasing resolution, one species in particular, \textit{Acer pseudoplatanus}, behaves differently, showing a reduction in performance at finer scales.

This counterintuitive result can be explained by scale effects in regions with strong topographic heterogeneity, such as the Valais. In the observational data, \textit{A.\ pseudoplatanus} occurs only rarely, whereas TreeMig at fine resolutions (e.g.\ 1~km) predicts it frequently. This mismatch is likely related to management effects: \textit{Picea abies} is overrepresented in the data due to forestry practices, which suppresses \textit{A.\ pseudoplatanus}, despite climatic conditions at the lower valley slopes being suitable for the species. 

In contrast, in coarse-resolution simulations (e.g.\ 60~km), the aggregated temperature signal (day-degree sum) falls below the 615\,°C threshold required for \textit{A.\ pseudoplatanus}, causing it not to appear at all. The absence of both observed and simulated occurrences leads to artificially low errors, which is effectively a compensation of two biases.

This scale aggregation issue affects all species to some degree in rugged terrain. Coarse data smooths out valleys, assigning them temperatures corresponding to higher elevations, which favors cold-tolerant species such as \textit{Larix decidua}, \textit{Pinus cembra}, and to some extent \textit{Picea abies}, while suppressing thermophilous species.

\paragraph{Climate Data Accuracy}
We assessed the accuracy of climate data by using all GHCNd station records available within ±0.0125° around each LFI plot location. Daily CHELSA precipitation and average temperature were compared to the measured station values. These comparisons were then aggregated by species, i.e., by considering all LFI plot locations where the respective species occurred.  

\begin{figure}[ht!]
\centering
\includegraphics[width=\textwidth]{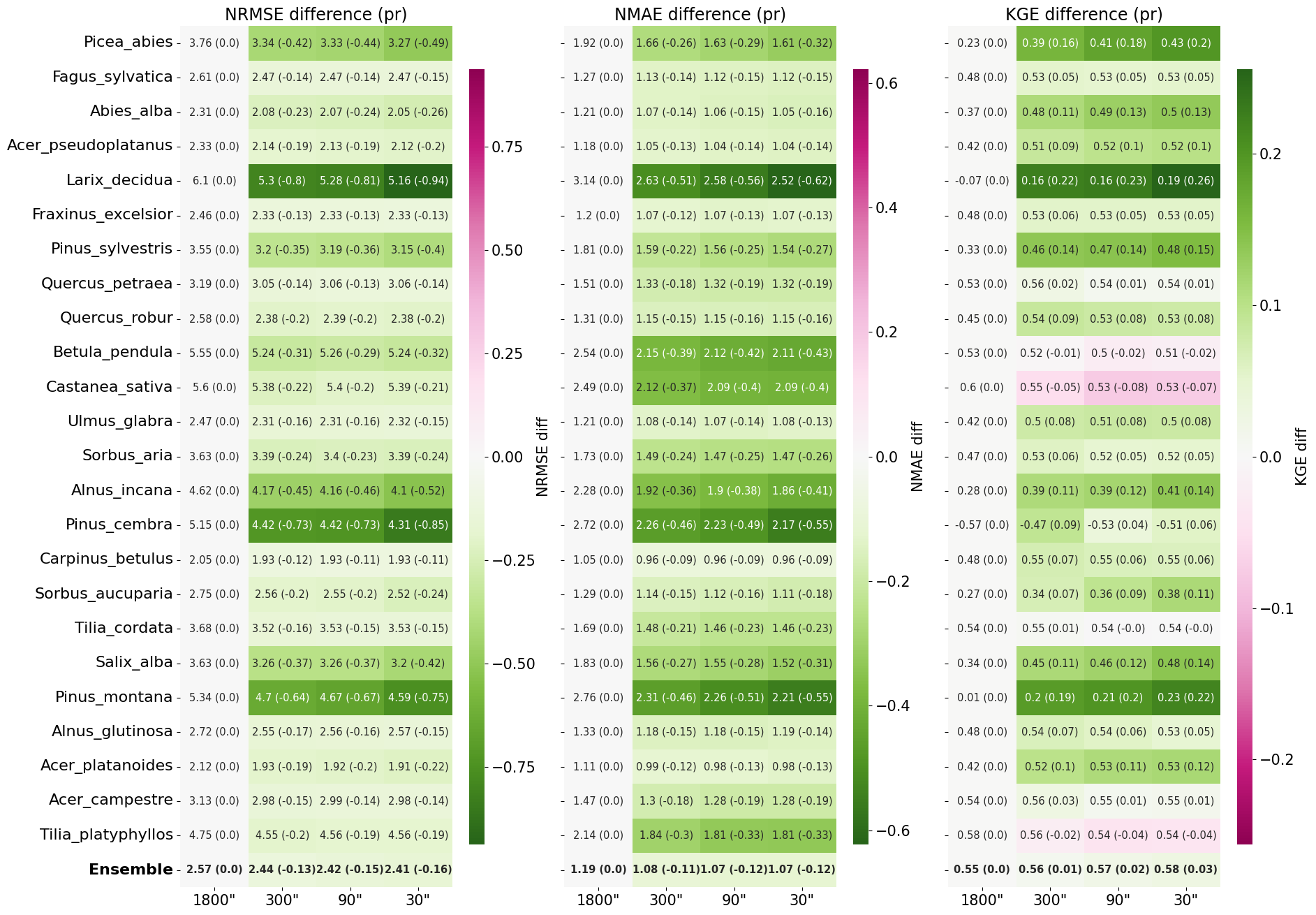}
\caption{Comparison of NRMSE, NMAE, and KGE of CHELSA daily precipitation per species across the four resolutions, and across all species combined (last row, ``Ensemble''). CHELSA daily precipitation is compared to GHCNd station data wherever available within the 0.25° grid cell around each LFI plot, including only plots where the respective species was present. Species are ordered by their frequency of occurrence in LFI plots. Values are shown as absolute errors, with differences to the 1800$^{\prime\prime}$ resolution indicated in brackets. Colors represent how the errors of the 300$^{\prime\prime}$, 90$^{\prime\prime}$, and 30$^{\prime\prime}$ resolution CHELSA daily precipitation product differ relative to the 1800$^{\prime\prime}$ resolution daily precipitation product.}
\label{fig:supp:for_treemig_clim_stats_pr}
\end{figure}

\begin{figure}[ht!]
\centering
\includegraphics[width=\textwidth]{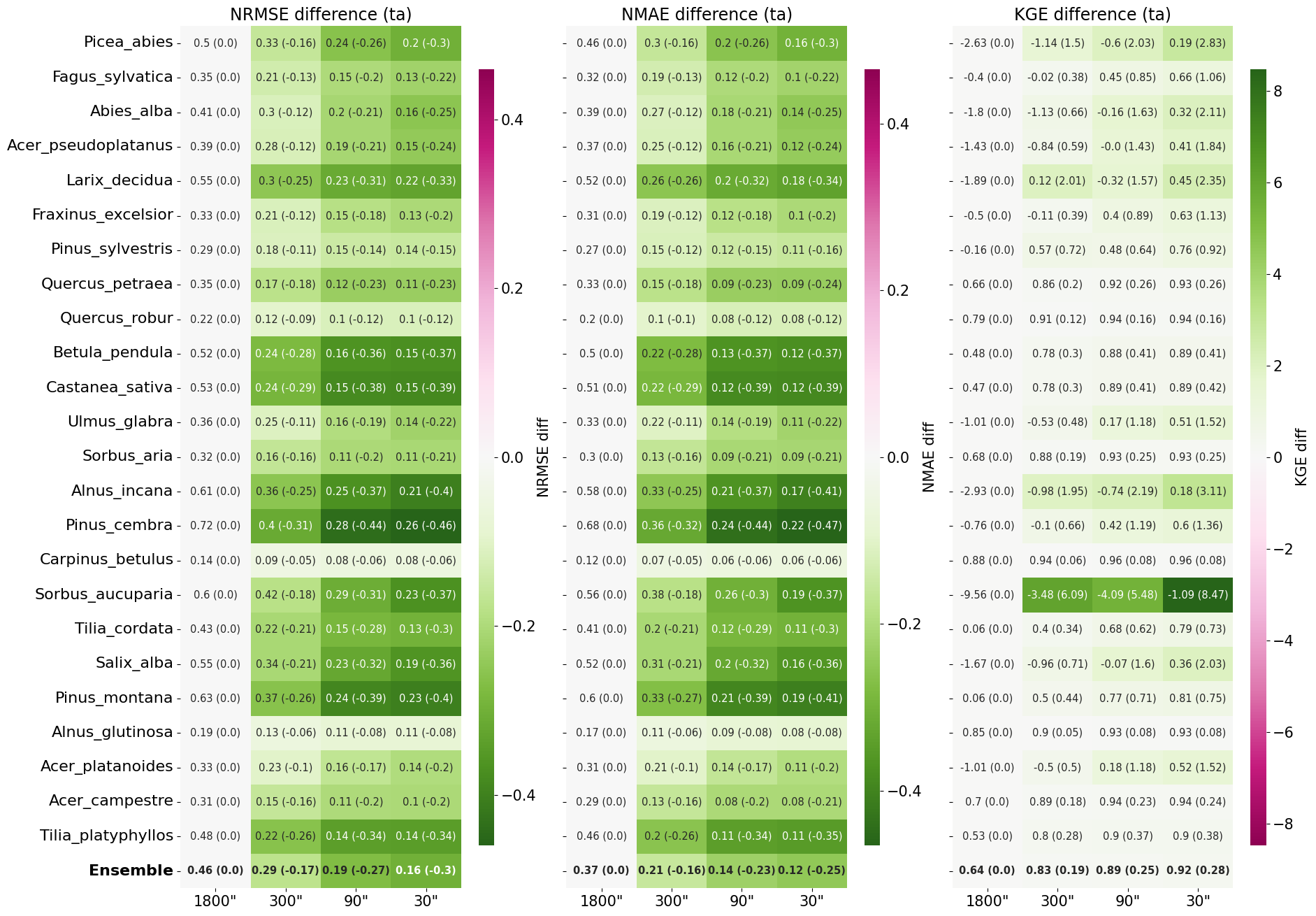}
\caption{Comparison of NRMSE, NMAE, and KGE of CHELSA daily average temperature per species across the four resolutions, and across all species combined (last row, 'Ensemble'). CHELSA daily average temperature is compared to GHCNd station data wherever available within the 0.25° grid cell around each LFI plot, including only plots where the respective species was present. Species are ordered by their frequency of occurrence in LFI plots. Values are shown as absolute errors, with differences to the 1800$^{\prime\prime}$ resolution indicated in brackets. Colors represent how the errors of the 300$^{\prime\prime}$, 90$^{\prime\prime}$, and 30$^{\prime\prime}$ resolution simulations CHELSA daily average temperature product differ relative to the 1800$^{\prime\prime}$ resolution daily average temperature product.}
\label{fig:supp:for_treemig_clim_stats_ta}
\end{figure}

Figures~\ref{fig:supp:for_treemig_clim_stats_pr} and~\ref{fig:supp:for_treemig_clim_stats_ta} show the error metrics per species, as well as the overall results across species. Improvements are generally stronger for temperature than for precipitation. For both variables, accuracy increases continuously with spatial resolution, moving from 1800$^{\prime\prime}$ to 300$^{\prime\prime}$, 90$^{\prime\prime}$, and 30$^{\prime\prime}$.  

\clearpage

\subsection{Biomes}
\subsubsection{ORCHIDEE-MICT}
\paragraph{Observational data vs. modeling results}
Evaluation targeted active layer thickness (ALT), mean annual ground temperature (MAGT), and soil temperatures (TSOI) at various depths across sites on the Qinghai–Xizang (Tibetan) Plateau. Unlike other models, evaluation data were derived from multiple sources, mostly literature, and values were partially extracted via digitization.  

\begin{figure}[ht!]
\centering
\includegraphics[width=\textwidth]{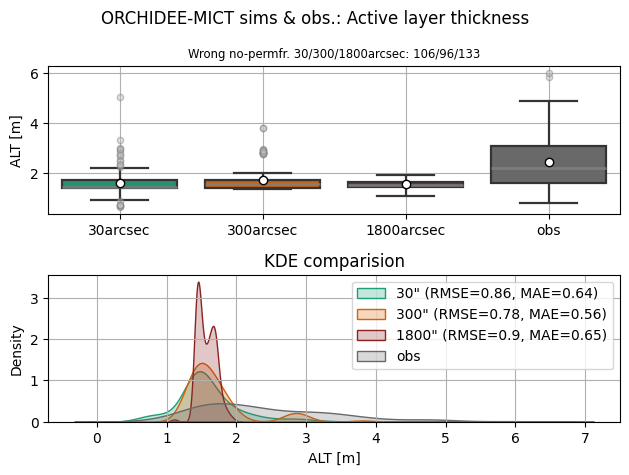}
\caption{Comparison of the distribution of Active Layer Thickness as simulated by the three ORCHIDEE-MICT resolution simulations (30arcsec, 300arcsec, 1800arcsec) versus observed values. The top plot shows boxplots; the bottom plot shows KDEs, including error metrics in the legend.}
\label{fig:supp:biomes_alt}
\end{figure}

\begin{figure}[ht!]
\centering
\includegraphics[width=\textwidth]{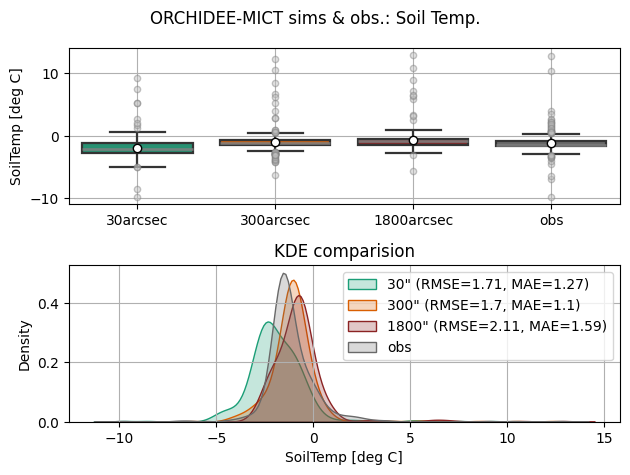}
\caption{Comparison of the distribution of soil temperature at various depths as simulated by the three ORCHIDEE-MICT resolution simulations (30arcsec, 300arcsec, 1800arcsec) versus observed values. The top plot shows boxplots; the bottom plot shows KDEs, including error metrics in the legend.}
\label{fig:supp:biomes_stemp}
\end{figure}

\begin{figure}[ht!]
\centering
\includegraphics[width=\textwidth]{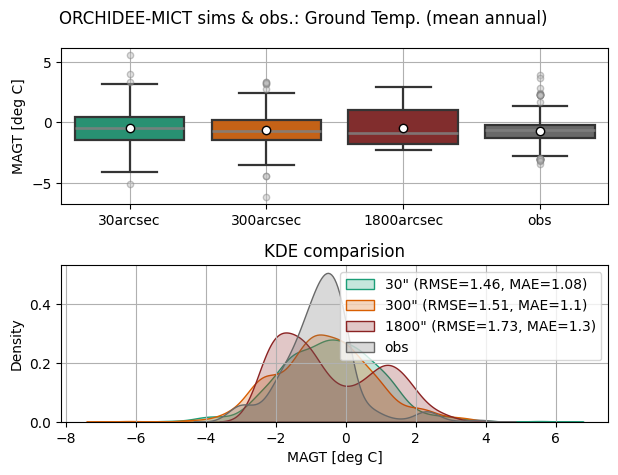}
\caption{Comparison of the distribution of mean annual ground temperature as simulated by the three ORCHIDEE-MICT resolution simulations (30arcsec, 300arcsec, 1800arcsec) versus observed values. The top plot shows boxplots; the bottom plot shows KDEs, including error metrics in the legend.}
\label{fig:supp:biomes_magt}
\end{figure}

\begin{figure}[ht!]
\centering
\includegraphics[width=\textwidth]{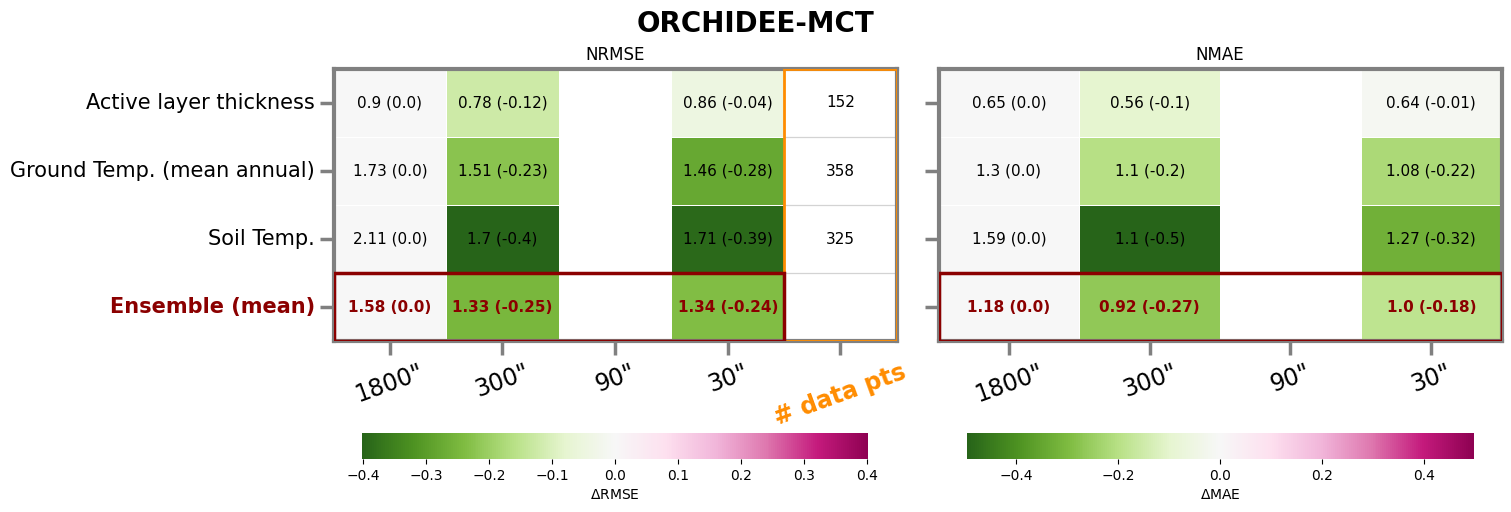}
\caption{Comparison of error metrics across the three available ORCHIDEE-MICT resolutions (30arcsec, 300arcsec, 1800arcsec) for ALT, TSOI, and MAGT. The last row shows the ensemble, as used in the main manuscript.}
\label{fig:supp:biomes_error}
\end{figure}

Figures~\ref{fig:supp:biomes_alt}, ~\ref{fig:supp:biomes_stemp}, and ~\ref{fig:supp:biomes_magt} show distributions of simulated and observed ALT, TSOI, and MAGT, including error metrics. One notable result is that the number of incorrect no-permafrost classifications varied across resolutions: 106 cases for the 30arcsec simulation, 96 cases for the 300arcsec resolution, and 133 cases for the coarsest 1800arcsec resolution.  Figure~\ref{fig:supp:biomes_error} compares error metrics (NRMSE, NMAE) across the resolutions for each of the three variables, visualized as a heatmap. We observe an improvement in accuracy when moving from 1800arcsec to 300arcsec, but only minimal changes when further increasing the resolution to 30arcsec.

\begin{figure}[ht!]
\centering
\includegraphics[width=\textwidth]{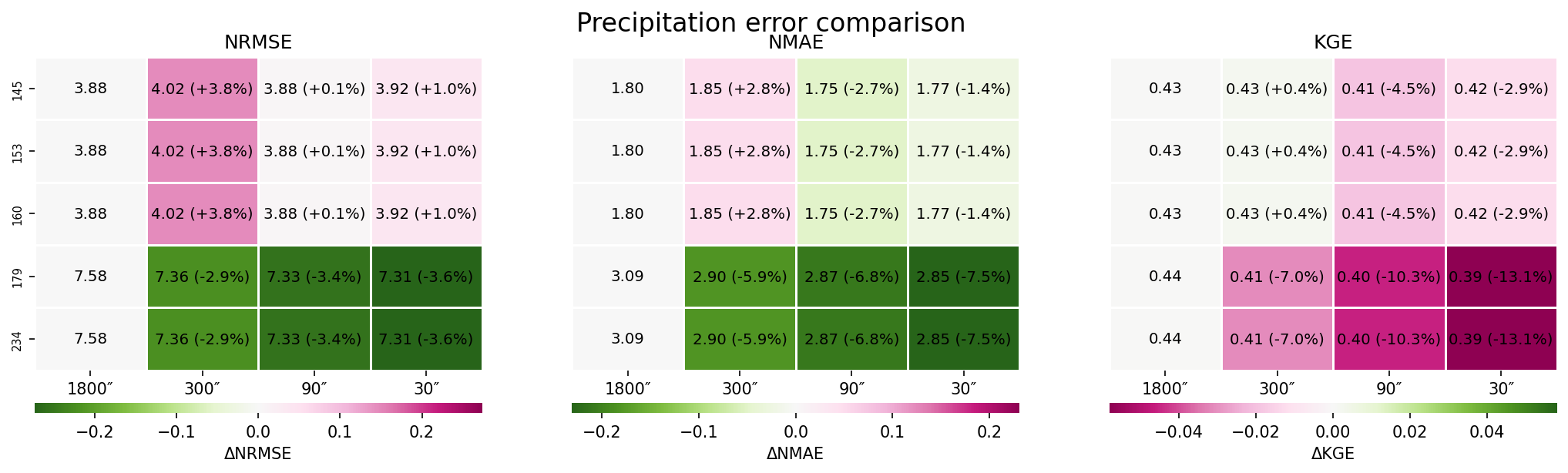}
\caption{Comparison of NRMSE, NMAE, and KGE of CHELSA daily precipitation across the four resolutions at ALT, TSOI, and MAGT observational sites that overlapped with GHCNd station data. In practice, only MAGT observational sites had such overlap, so the results shown correspond exclusively to MAGT sites. Values are shown as absolute errors, with differences to the 1800$^{\prime\prime}$ resolution indicated in brackets. Colors represent how the errors of the 300$^{\prime\prime}$, 90$^{\prime\prime}$, and 30$^{\prime\prime}$ resolution simulations CHELSA daily average temperature product differ relative to the 1800$^{\prime\prime}$ resolution daily precipitation product. }
\label{fig:supp:biomes_clim_pr}
\end{figure}

\begin{figure}[ht!]
\centering
\includegraphics[width=\textwidth]{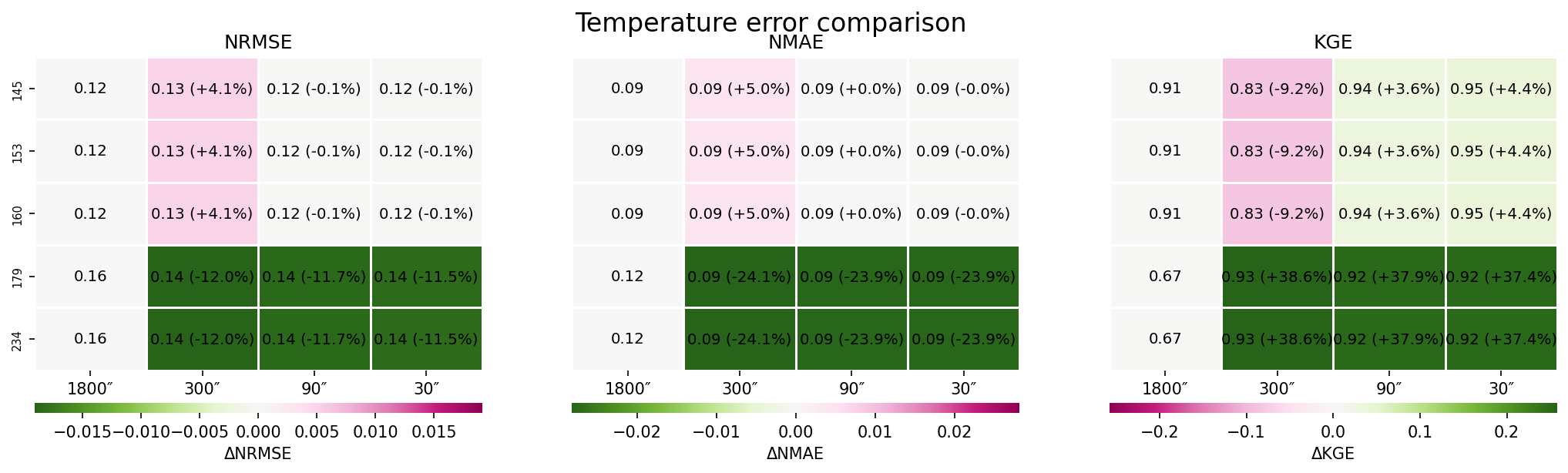}
\caption{Comparison of NRMSE, NMAE, and KGE of CHELSA daily average temperature for all locations of ALT, TSOI and MAGT observational sites across the four resolutions which overlapped with GHCNd station data. In practice, only MAGT observational sites had such overlap, so the results shown correspond exclusively to MAGT sites. Values are shown as absolute errors, with differences to the 1800$^{\prime\prime}$ resolution indicated in brackets. Colors represent how the errors of the 300$^{\prime\prime}$, 90$^{\prime\prime}$, and 30$^{\prime\prime}$ resolution simulations CHELSA daily average temperature product differ relative to the 1800$^{\prime\prime}$ resolution daily average temperature product.}
\label{fig:supp:biomes_clim_tas}
\end{figure}

\paragraph{Climate Data Accuracy}
We evaluated climate data accuracy using all GHCNd station records located within ±0.0125° of each ALT, TSOI, or MAGT observation site. Daily CHELSA precipitation and mean temperature at the different resolutions were compared against the corresponding station measurements by selecting the nearest CHELSA grid value. Of the 265 unique observation sites, only five lay within a 0.25° grid cell containing a precipitation and a temperature station, corresponding to just two distinct GHCNd stations. All of these locations are MAGT locations. Figures~\ref{fig:supp:biomes_clim_pr} and \ref{fig:supp:biomes_clim_tas} illustrate resolution-dependent changes in CHELSA temperature and precipitation accuracy at those two stations / five observational MAGT sites. For temperature, one station shows clear improvement when moving from 1800$^{\prime\prime}$ to 300$^{\prime\prime}$, with little further change at 90$^{\prime\prime}$ and 30$^{\prime\prime}$, while the second station exhibits only minimal resolution effects. Precipitation accuracy shows a similar pattern: at one station, resolution-dependent improvements persist across all metrics (particularly NRMSE and MAE, though KGE decreases slightly), whereas at the other station, resolution effects remain small, with a slight decline at 300$^{\prime\prime}$ followed by minor improvements at finer resolutions.
However, because at each of these five MAGT observational sites only one or two time steps with co-located station data are available, we did not include these locations in the aggregated analysis of impact model performance versus climate accuracy in the main manuscript (Fig.~5).

\clearpage

\subsection{Regional Water}
\subsubsection{CWATm}

\begin{figure}[ht!]
\centering
\includegraphics[width=\textwidth]{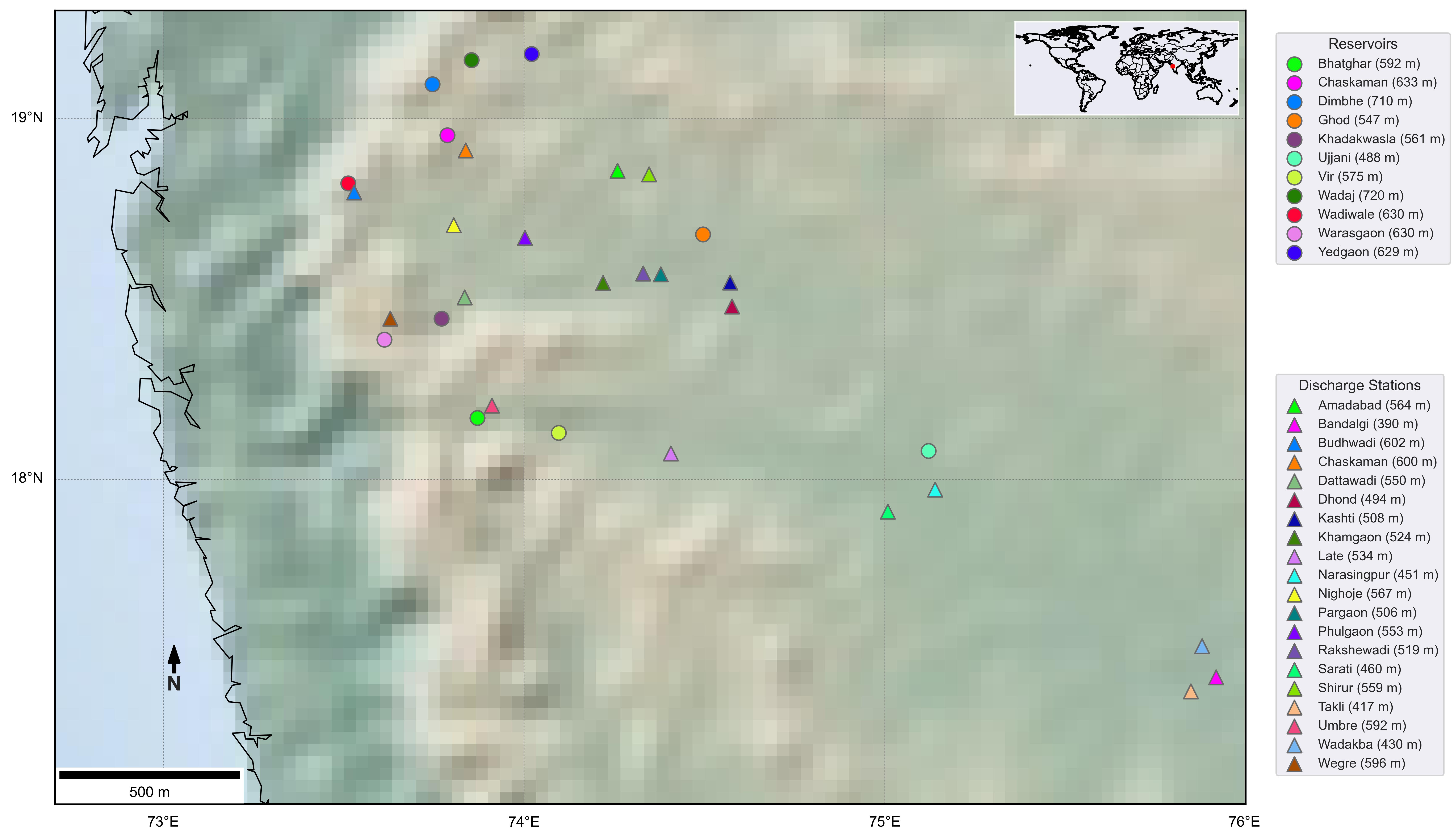}
\caption{Evaluation locations for the CWATm model in the Bhima basin. Circles denote reservoir-level observation sites; triangles denote discharge gauging stations.}
\label{fig:supp:water_cwatm_loc}
\end{figure}

\begin{figure}[ht!]
\centering
\includegraphics[width=\textwidth]{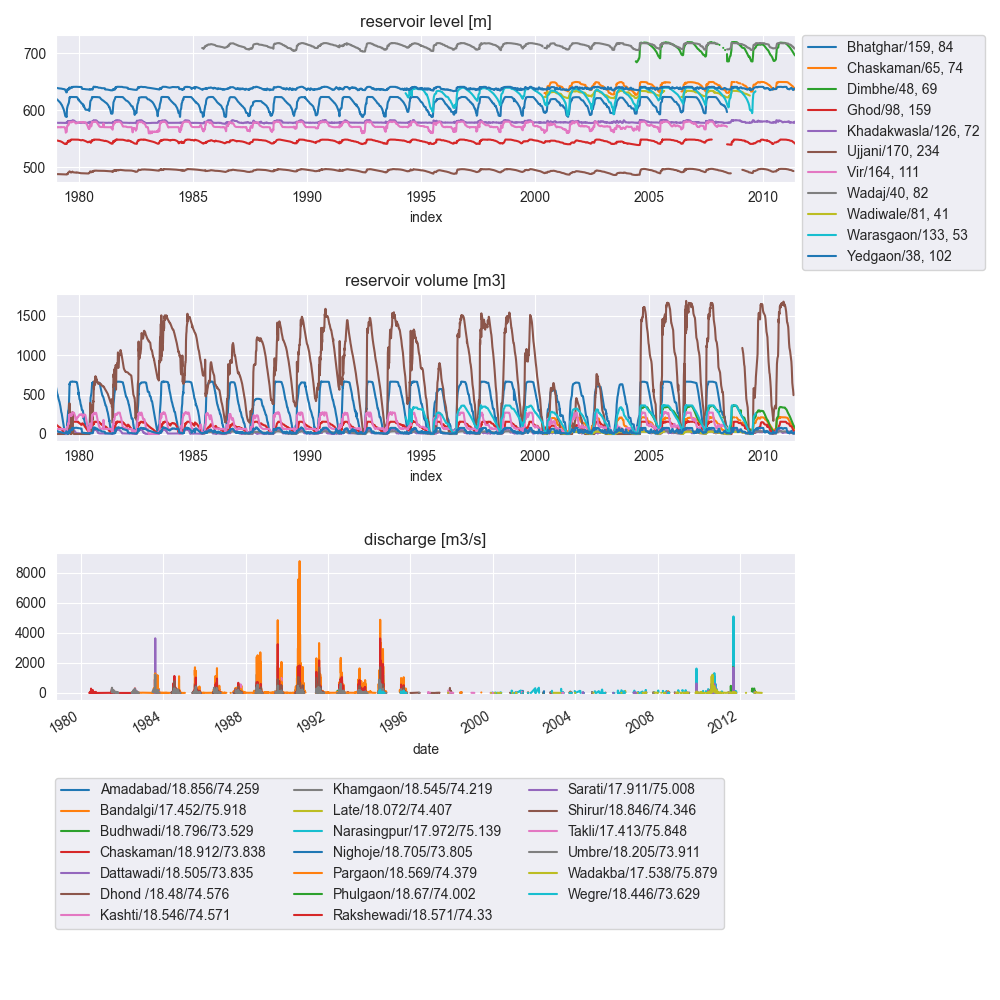}
\caption{Time series of observations at reservoir-level locations (reservoir level and corresponding converted storage volume) and discharge stations. Note that not all sites provided data over the full simulation period.}
\label{fig:supp:water_cwatm_data}
\end{figure}

Evaluation was conducted at 20 discharge gauging stations and 11 reservoir water level observation sites across the Bhima basin, as shown in Figure~\ref{fig:supp:water_cwatm_loc}. Reservoir levels were converted to storage volumes using reservoir-specific polynomial relationships, based on storage–level observations from 2017. Time series of the available observations for reservoir level, converted reservoir volume, and discharge are shown in Figure~\ref{fig:supp:water_cwatm_data}.

\paragraph{Evaluation of accuracy}
Figures~\ref{fig:supp:water_cwatm_met_res_level} and \ref{fig:supp:water_cwatm_met_discharge} show simulation errors (NRMSE, NMAE, KGE) at different model resolutions for all reservoir-level sites and discharge stations, respectively.  

For reservoir levels, accuracy generally improves with increasing resolution at most sites (except Chaskaman), with the largest gains occurring between 1800$^{\prime\prime}$ and 300$^{\prime\prime}$. 

For discharge, 11 of the 20 stations show improved performance at higher resolution. Where accuracy decreased, the effect was relatively small (e.g.\ maximum NRMSE increase of 4.3\%), whereas improvements were larger (e.g.\ maximum NRMSE decrease of 13.3\%).  

When aggregating results into ensembles for reservoir level and discharge, and further averaging across both (Figure~\ref{fig:supp:water_cwatm_met_overall}), higher-resolution simulations consistently outperform the 1800~arcsec baseline. The greatest improvements again occur between 1800$^{\prime\prime}$ and 300$^{\prime\prime}$, while changes between 300$^{\prime\prime}$ and 90$^{\prime\prime}$ are often minor and can occasionally result in slight performance reductions. The bottom row of Figure~\ref{fig:supp:water_cwatm_met_overall}, representing the overall mean, is used for the main analysis in the paper.

\begin{figure}[ht!]
\centering
\includegraphics[width=\textwidth]{figures_supp/reg_water/cwatm/stats_res_level_metrics_relative.png}
\caption{Relative NRMSE, NMAE and KGE of CWATm reservoir-level simulations (1800$^{\prime\prime}$, 300$^{\prime\prime}$, 90$^{\prime\prime}$, 30$^{\prime\prime}$) at all observation sites  relative to the 1800$^{\prime\prime}$ baseline. The printed values are the absolute metric values; percentages in brackets show changes relative to 1800~arcsec. The last row gives the mean across all locations.}
\label{fig:supp:water_cwatm_met_res_level}
\end{figure}

\begin{figure}[ht!]
\centering
\includegraphics[width=\textwidth]{figures_supp/reg_water/cwatm/stats_discharge_metrics_relative.png}
\caption{Relative NRMSE, NMAE and KGE of CWATm discharge simulations (1800$^{\prime\prime}$, 300$^{\prime\prime}$, 90$^{\prime\prime}$, 30$^{\prime\prime}$) at all gauging stations relative to the 1800$^{\prime\prime}$baseline. The printed values are the absolute metric values; percentages in brackets show changes relative to 1800$^{\prime\prime}$. The last row gives the mean across all locations.}
\label{fig:supp:water_cwatm_met_discharge}
\end{figure}

\begin{figure}[ht!]
\centering
\includegraphics[width=\textwidth]{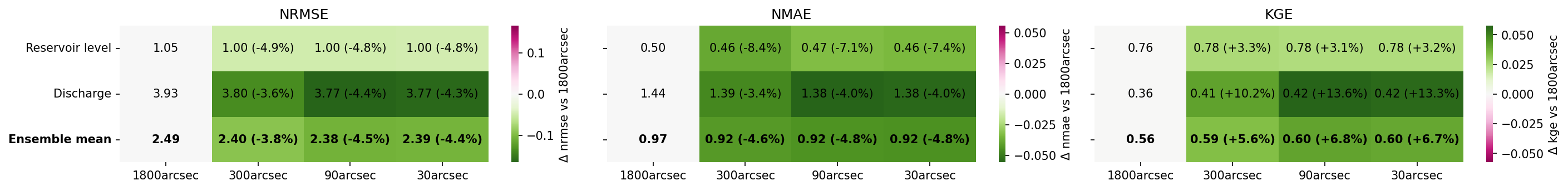}
\caption{Relative NRMSE, NMAE and KGE of CWATm simulations aggregated as follows: mean across reservoir-level sites (top row), mean across discharge stations (middle row), and overall mean across both variables (bottom row). The bottom row is used in the main paper analysis.}
\label{fig:supp:water_cwatm_met_overall}
\end{figure}

\begin{figure}[ht!]
\centering
\includegraphics[width=\textwidth]{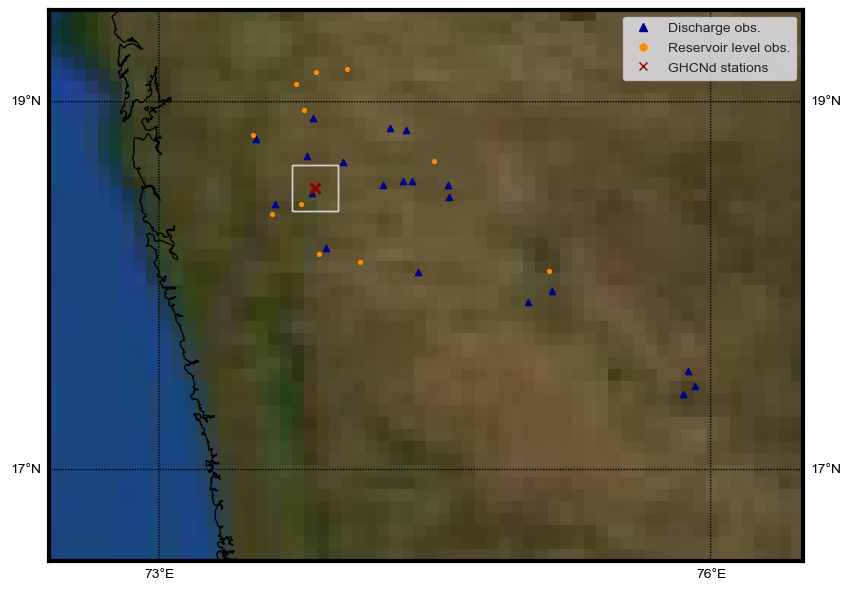}
\caption{Overview of available GHCNd station data in the Bhima basin, relative to reservoir level and discharge station locations. The square around each GHCNd station indicates the defined area (+/- 0.125°) used to link climate data accuracy with model performance.}
\label{fig:supp:water_cwatm_ghcn_locs}
\end{figure}

\paragraph{Climate Data Accuracy}
Figure~\ref{fig:supp:water_cwatm_ghcn_locs} shows the locations of GHCNd stations relative to reservoir level and discharge stations. The square around the GHCNd station represents the defined area (+/- 0.125°) used to link climate data accuracy to model performance. Within this area, only one discharge station (Dattawadi) and one reservoir level station (Khadakwasla) are located. Consequently, we can only assess these two sites for comparisons between model performance across resolutions and climate data accuracy across resolutions.

\begin{figure}[ht!]
\centering
\includegraphics[width=\textwidth]{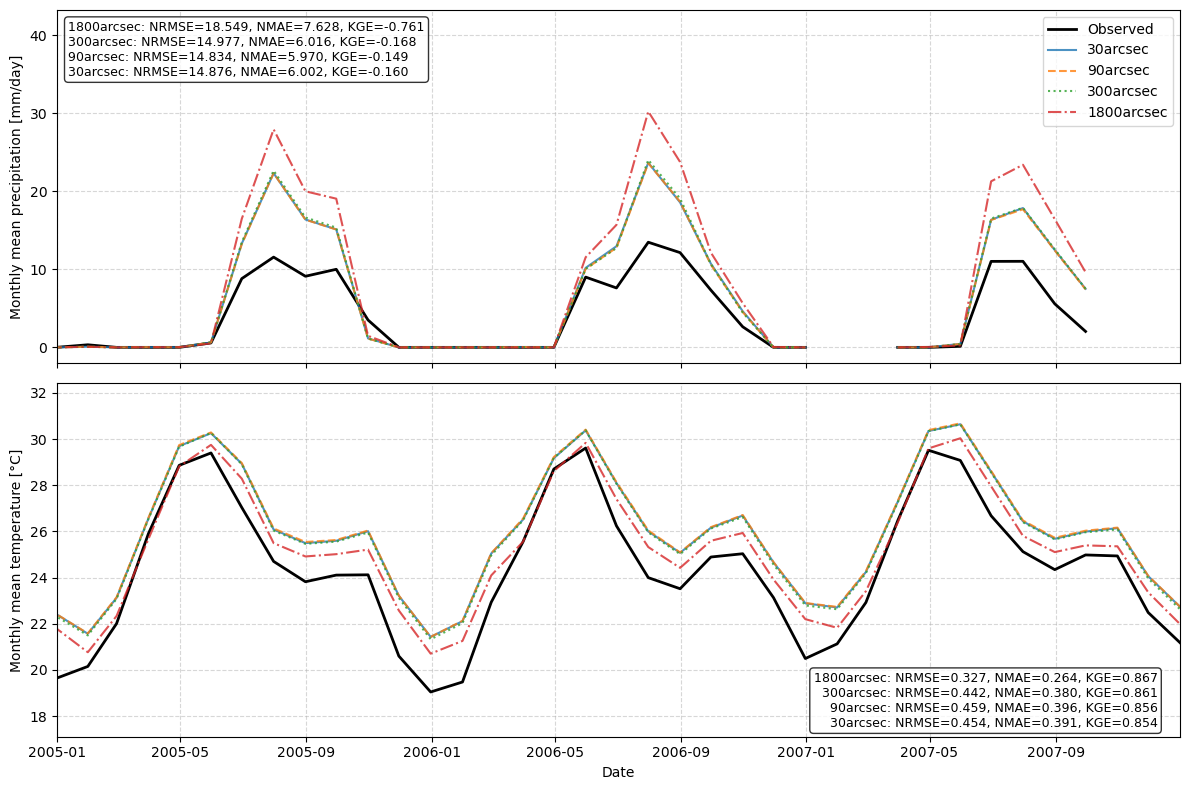}
\caption{Zoomed-in comparison of CHELSA precipitation and average temperature at different spatial resolutions with observed data from the GHCNd station. Error metrics (NRMSE, NMAE, and KGE) for all four resolutions across all available daily timestamps from 1979 to 2016 are shown in the inset box.}
\label{fig:supp:water_cwatm_ghcn_eval}
\end{figure}

Figure~\ref{fig:supp:water_cwatm_ghcn_eval} presents a three-year zoomed-in comparison of CHELSA precipitation and average temperature data at the GHCNd station in the Bhima basin against observed station values. The figure also reports error metrics (NRMSE, NMAE, and KGE) for all four resolutions across all available daily timestamps from 1979 to 2016. 

At this station, only minimal differences are observed between the 30$^{\prime\prime}$, 90$^{\prime\prime}$, and 300$^{\prime\prime}$ CHELSA simulations, both in the time series and in the error metrics (e.g., precipitation NRMSE: 14.977, 14.834, and 14.876 for 300, 90, and 30$^{\prime\prime}$ , respectively). In contrast, the 1800$^{\prime\prime}$ CHELSA data shows larger deviations (NRMSE: 18.549). Notably, for precipitation, the 30/90/300$^{\prime\prime}$ resolutions outperform the 1800$^{\prime\prime}$ resolution, whereas for temperature, the 1800$^{\prime\prime}$ data exhibits slightly better performance.

\clearpage

\subsubsection{GR4J}
\begin{figure}[ht!]
\centering
\includegraphics[width=\textwidth]{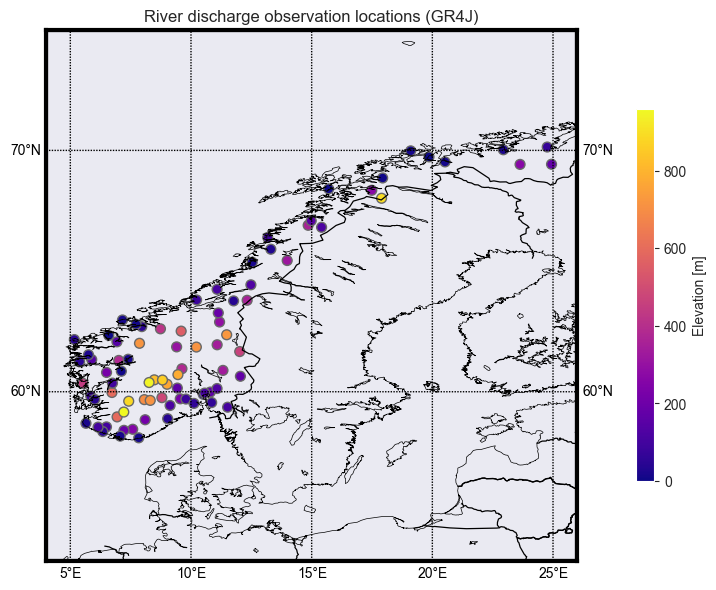}
\caption{Locations of river discharge stations across Norway used to assess GR4J simulations at different spatial resolutions.}
\label{fig:supp:water_gr4j_obs}
\end{figure}

GR4J simulations were evaluated at 86 discharge stations across Norway, as shown in Figure~\ref{fig:supp:water_gr4j_obs}. 

\begin{figure}[ht!]
\centering
\includegraphics[width=\textwidth]{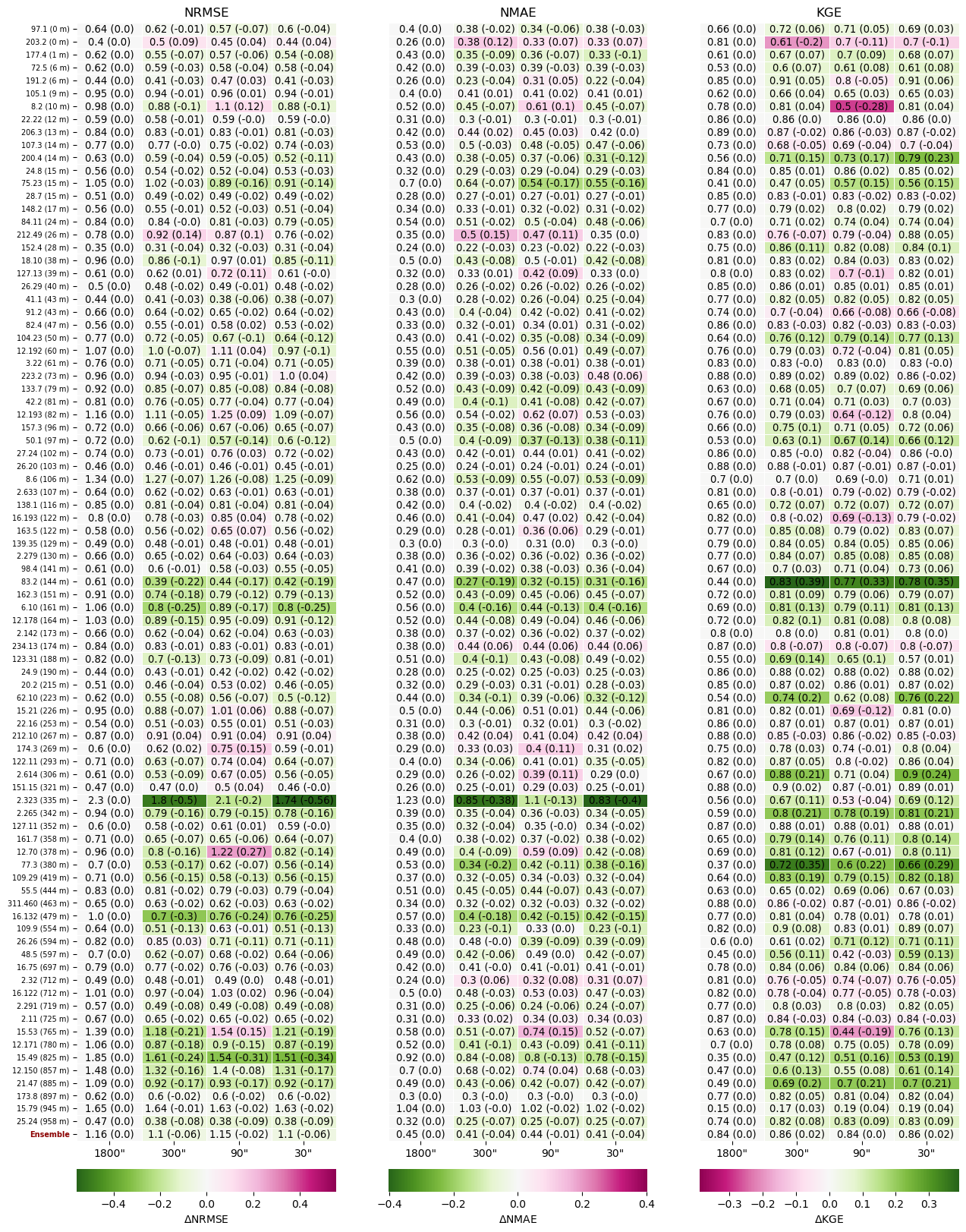}
\caption{Relative NRMSE, NMAE, and KGE of GR4J discharge simulations (1800$^{\prime\prime}$, 300$^{\prime\prime}$, 90$^{\prime\prime}$, 30$^{\prime\prime}$) at all gauging stations, relative to the 1800$^{\prime\prime}$ baseline. Printed values are absolute metrics; percentages in brackets indicate changes relative to 1800$^{\prime\prime}$. The final row gives the mean across all locations. Stations are ordered by increasing elevation.}
\label{fig:supp:water_gr4j_eval}
\end{figure}

\paragraph{Evaluation of accuracy}
Figure~\ref{fig:supp:water_gr4j_eval} shows error metrics (NRMSE, NMAE, KGE) for GR4J simulations at different spatial resolutions, with stations ordered by increasing elevation. Larger differences between resolutions are evident at higher elevations (e.g., compare the top and bottom thirds of Figure~\ref{fig:supp:water_gr4j_eval}). The final row in the figure reports the ensemble mean used in the main article's analysis.

\begin{figure}[ht!]
\centering
\includegraphics[width=\textwidth]{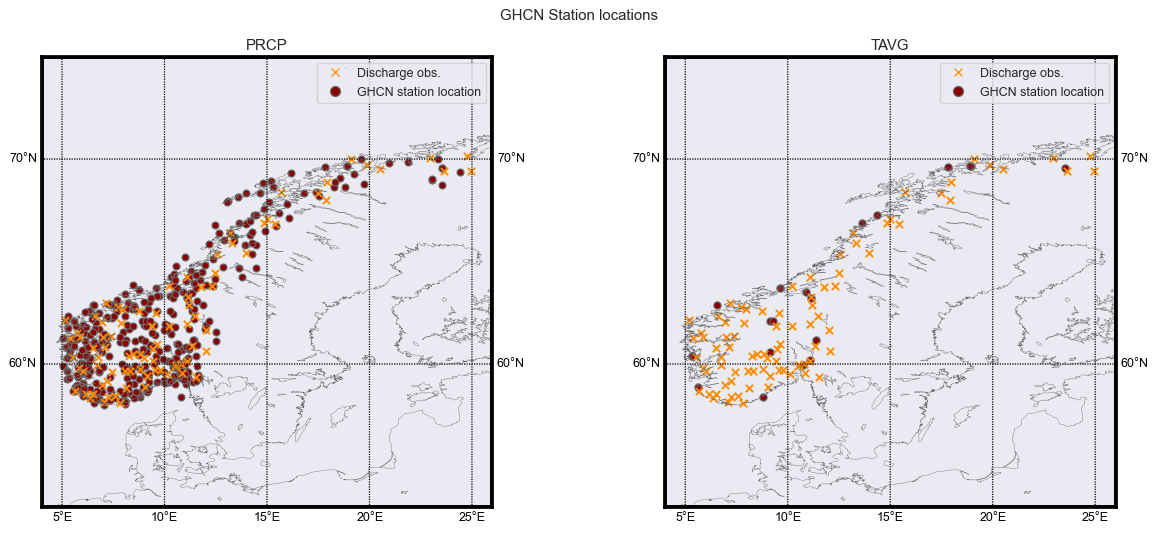}
\caption{Locations of GHCNd precipitation and temperature stations relative to discharge stations used to evaluate GR4J performance.}
\label{fig:supp:water_gr4j_clim}
\end{figure}

\begin{figure}[ht!]
\centering
\includegraphics[width=\textwidth]{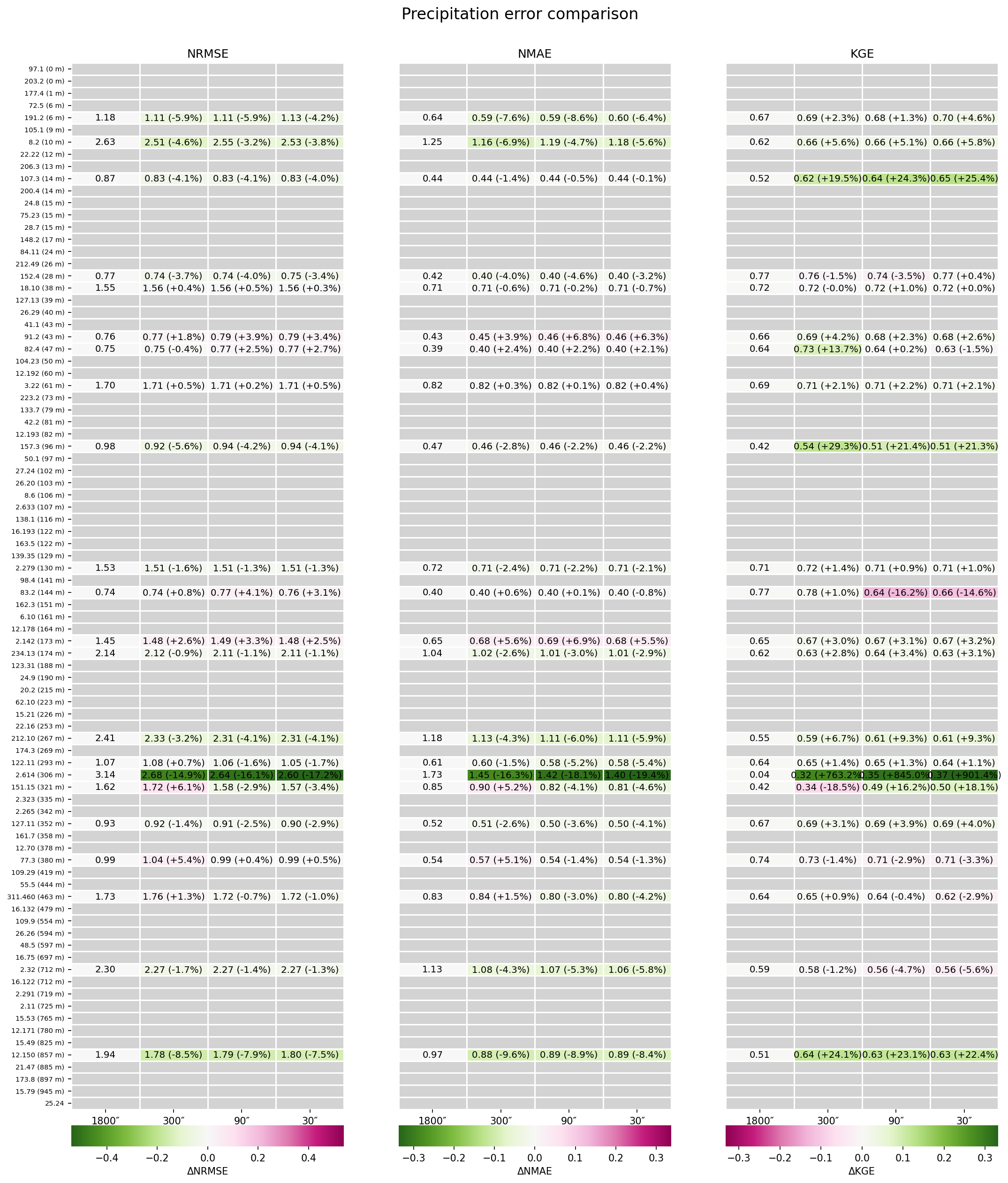}
\caption{Relative NRMSE, NMAE, and KGE of CHELSA daily precipitation at various spatial resolutions, evaluated at discharge stations where GR4J simulations were assessed. The colormap shows values for 300$^{\prime\prime}$, 90$^{\prime\prime}$, and 30$^{\prime\prime}$ relative to 1800$^{\prime\prime}$; annotations show absolute metrics with percentage differences in brackets. Stations are ordered by elevation. Grey indicates no GHCNd precipitation data within the catchment area of the discharge station.}
\label{fig:supp:water_gr4j_clim_precip}
\end{figure}

\begin{figure}[ht!]
\centering
\includegraphics[width=\textwidth]{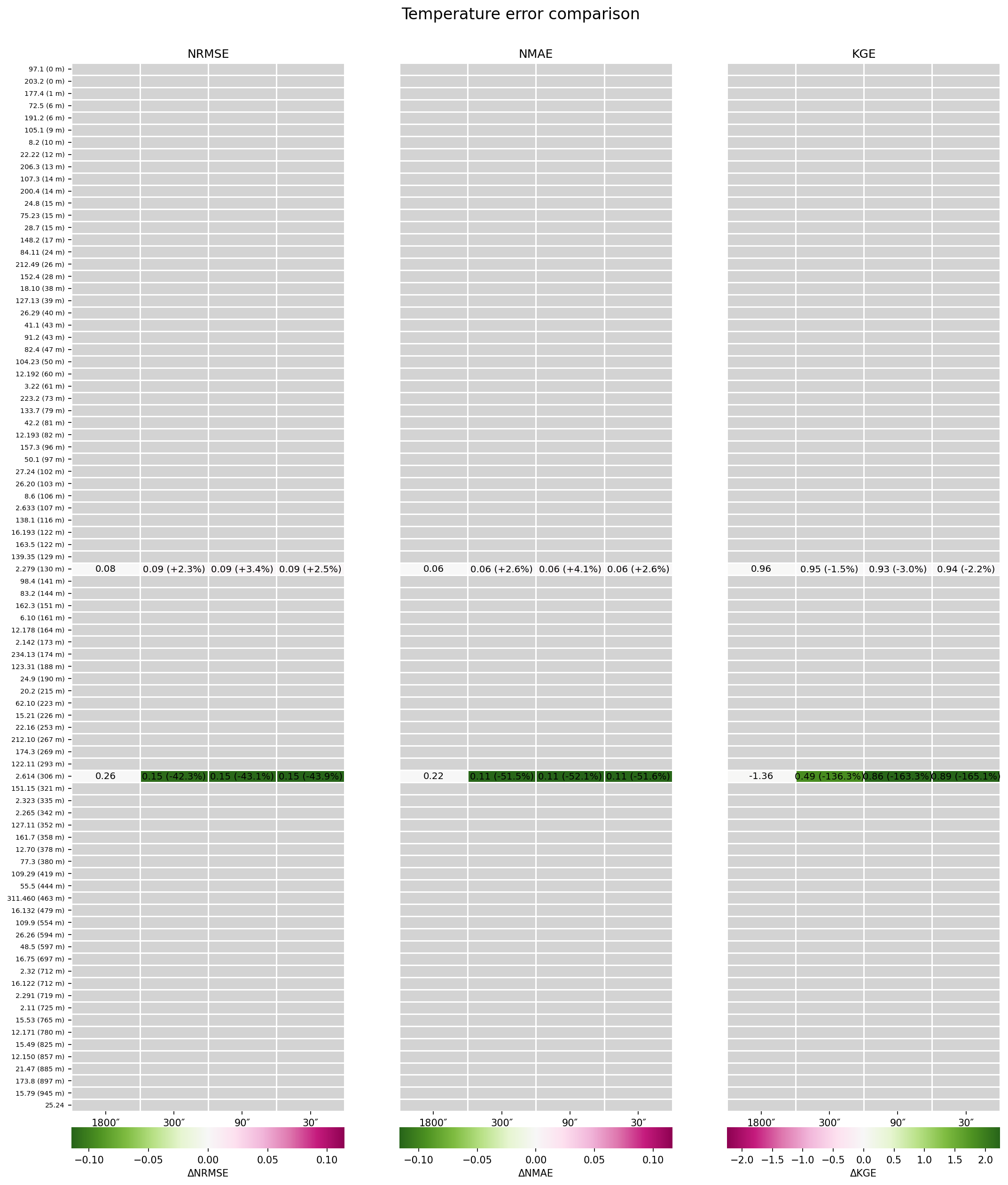}
\caption{Relative NRMSE, NMAE, and KGE of CHELSA daily average temperature at various spatial resolutions, evaluated at discharge stations where GR4J simulations were assessed. The colormap shows values for 300$^{\prime\prime}$, 90$^{\prime\prime}$, and 30$^{\prime\prime}$ relative to 1800$^{\prime\prime}$; annotations show absolute metrics with percentage differences in brackets. Stations are ordered by elevation. Grey indicates no GHCNd temperature data within the catchment area of the discharge station.}
\label{fig:supp:water_gr4j_clim_temp}
\end{figure}

\paragraph{Climate data accuracy}
Figure~\ref{fig:supp:water_gr4j_clim} also shows GHCNd temperature and precipitation stations, while Figures~\ref{fig:supp:water_gr4j_clim_precip} and \ref{fig:supp:water_gr4j_clim_temp} present climate data error of GHCNd stations located within the catchments of each discharge station.

For precipitation, higher-resolution CHELSA data (30/90/300$^{\prime\prime}$) generally perform better than the 1800$^{\prime\prime}$ product. However, apart from a single low-elevation station, there is little difference between the 30, 90, and 300$^{\prime\prime}$ resolutions. 

For temperature, only three discharge stations overlap with the GHCNd perimeter. Among these, one station shows consistent improvement with increasing resolution (from 1800 to 30$^{\prime\prime}$), one shows little change, and one shows degraded performance at higher resolutions (with 300$^{\prime\prime}$ performing worst and both 90 and 30$^{\prime\prime}$ performing worse than 1800$^{\prime\prime}$).

\clearpage

\subsubsection{SWAT}
SWAT simulations were evaluated at four discharge stations on the watershed’s main streams (Hermon, Meshushim, Dishon, and Jordan), as shown in Figure~\ref{fig:supp:water_swat_obs}. 

\begin{figure}[ht!]
\centering
\includegraphics[width=\textwidth]{figures_supp/reg_water/swat/location_obs.png}
\caption{Locations of river discharge stations used to assess GR4J simulations at different spatial resolutions.}
\label{fig:supp:water_swat_obs}
\end{figure}

\paragraph{Evaluation of accuracy}
SWAT primarily simulates surface processes in watersheds, while groundwater contributions (i.e., stream baseflow) are prescribed as inputs. During the dry season (May–October), precipitation contributions are negligible and climate inputs primarily affect evapotranspiration. Therefore, only the rainy season (November–April) was considered, as the model is more sensitive to climate inputs during these months.

Model performance was evaluated using monthly time steps because daily simulations showed inaccuracies in reproducing storm events, which are critical for discharge timing.

It should also be noted that SWAT is a semi-distributed model, divided into sub-units (polygons) of variable size. Each sub-unit is assigned a single climate station (nearest-neighbor approach). Approximately 92\% of these units exceed 1~km$^2$, and 58\% fall between 1 and 3~km$^2$. Consequently, results at 1~km resolution are expected to be very similar to those at 3~km resolution.

Figure~\ref{fig:supp:water_swat_eval} shows error metrics for SWAT simulations at all four discharge stations, as well as the ensemble results used in the main paper’s analysis. At the Hermon and Meshushim stations, model performance improves when moving from 1800$^{\prime\prime}$ to 300$^{\prime\prime}$ resolution, with only minor additional changes at 90$^{\prime\prime}$ and 30$^{\prime\prime}$. In contrast, the Dishon and Jordan stations show only marginal sensitivity to resolution.

\begin{figure}[ht!]
\centering
\includegraphics[width=\textwidth]{figures_supp/reg_water/swat/error_comp_rel_monthly_rainy_1_v2.png}
\caption{Relative NRMSE, NMAE, and KGE of SWAT discharge simulations (1800$^{\prime\prime}$, 300$^{\prime\prime}$, 90$^{\prime\prime}$, 30$^{\prime\prime}$) at all discharge stations, relative to the 1800$^{\prime\prime}$ baseline. Printed values are absolute metrics; percentages in brackets indicate changes relative to 1800$^{\prime\prime}$. The final row gives the mean across all locations.}
\label{fig:supp:water_swat_eval}
\end{figure}

\paragraph{Climate data accuracy}
Only the Dishon discharge station was located within $\pm$0.125$^\circ$ of a GHCNd station, and no additional local station data were available. As shown in Figure~\ref{fig:supp:water_swat_clim_eval}, CHELSA average temperature at the nearby GHCNd station improved with increasing spatial resolution, although the difference between 90$^{\prime\prime}$ and 30$^{\prime\prime}$ was minimal. 

For daily CHELSA precipitation, performance decreased at higher resolutions according to NRMSE and NMAE. KGE improved slightly with higher resolution; however, the absolute KGE values remained low (maximum $\approx$ 0.29).

\begin{figure}[ht!]
\centering
\includegraphics[width=\textwidth]{figures_supp/reg_water/swat/clim_eval_combined.png}
\caption{Relative NRMSE, NMAE, and KGE of CHELSA daily precipitation and average temperature at various spatial resolutions, evaluated at discharge stations where SWAT simulations were assessed. The colormap shows values for 300$^{\prime\prime}$, 90$^{\prime\prime}$, and 30$^{\prime\prime}$ relative to 1800$^{\prime\prime}$; annotations show absolute metrics with percentage differences in brackets. Grey indicates no GHCNd temperature data within $\pm$0.125$^\circ$ of the discharge station.}
\label{fig:supp:water_swat_clim_eval}
\end{figure}

\subsubsection{SWIM}
The SWIM model was set up for the Zeravshan catchment, with the outlet gauge at Dupli serving as the evaluation site (see Figure~\ref{fig:supp:water_swim_obs}). SWIM simulations were available only at 1800$^{\prime\prime}$ and 300$^{\prime\prime}$ resolutions. 

\begin{figure}[ht!]
\centering
\includegraphics[width=\textwidth]{figures_supp/reg_water/swim/location_obs.png}
\caption{Location of the river discharge station used to assess SWIM simulations at 1800$^{\prime\prime}$ and 300$^{\prime\prime}$ spatial resolutions.}
\label{fig:supp:water_swim_obs}
\end{figure}

\paragraph{Evaluation of accuracy}
Figure~\ref{fig:supp:water_swim_eval} shows a time-series comparison between observed discharge measurements and SWIM simulations at 1800$^{\prime\prime}$ and 300$^{\prime\prime}$ resolution. River discharge during low-flow periods was consistently underestimated by both simulations, with little difference between the two resolutions. Peak discharge was also substantially underestimated, particularly after 2000. The 300$^{\prime\prime}$ simulation, however, generally showed less underestimation than the 1800$^{\prime\prime}$ simulation.

As summarized in Figure~\ref{fig:supp:water_swim_eval_metrics}, SWIM simulations at 300$^{\prime\prime}$ resolution were approximately 15--20\% more accurate than those at 1800$^{\prime\prime}$, based on error metrics.

\begin{figure}[ht!]
\centering
\includegraphics[width=\textwidth]{figures_supp/reg_water/swim/overview_temporal.png}
\caption{Time-series comparison of SWIM simulations at 1800$^{\prime\prime}$ and 300$^{\prime\prime}$ resolution against observed discharge at the Dupli station.}
\label{fig:supp:water_swim_eval}
\end{figure}

\begin{figure}[ht!]
\centering
\includegraphics[width=0.8\textwidth]{figures_supp/reg_water/swim/comp_error_relative.png}
\caption{Relative NRMSE, NMAE, and KGE of SWIM discharge simulations (1800$^{\prime\prime}$, 300$^{\prime\prime}$) at the Dupli station, relative to the 1800$^{\prime\prime}$ baseline. Printed values are absolute metrics; percentages in brackets indicate changes relative to 1800$^{\prime\prime}$.}
\label{fig:supp:water_swim_eval_metrics}
\end{figure}

\paragraph{Climate data accuracy}
No GHCNd station data or other local climate data were available within $\pm$0.125$^\circ$ of the Dupli discharge station. Therefore, it was not possible to assess climate data accuracy for the SWIM model.

\clearpage

\subsection{Lakes}
\subsubsection{Simstrat}
Simstrat simulations were performed for 72 regional ISIMIP lakes distributed globally. However, lakes where the calibration did not converge, or where the calibration result was inconsistent across resolutions were not included in this analysis, resulting in 32 lakes (Figure~\ref{fig:supp:lakes_simst_obs}). Table~\ref{tab:lake_overview_sim} provides an overview of the lakes used in the simulations, including their location, elevation, maximum depth, and surface area.

\begin{sidewaystable}
\caption{Overview of lakes used in the simulations and their main morphometric characteristics.}
\label{tab:lake_overview_sim}
\begin{tabular*}{\textheight}{@{\extracolsep\fill}l r r r r r}
\toprule
\parbox[t]{3.2cm}{\raggedright \textbf{Lake} \\ \textbf{(folder name)}} &
\parbox[t]{2.4cm}{\raggedright \textbf{Latitude} \\ ($^\circ$)} &
\parbox[t]{2.4cm}{\raggedright \textbf{Longitude} \\ ($^\circ$)} &
\parbox[t]{2.6cm}{\raggedright \textbf{Elevation} \\ \textbf{(m a.s.l.)}} &
\parbox[t]{2.4cm}{\raggedright \textbf{Max. depth} \\ \textbf{(m)}} &
\parbox[t]{2.6cm}{\raggedright \textbf{Lake area} \\ \textbf{(km$^{2}$)}} \\
\midrule
allequash      &  46.04 &  -89.62 &  494.00 &   8.00 &   1.64 \\
biel           &  47.08 &    7.16 &  429.00 &  74.00 &  39.30 \\
big-muskellunge&  46.02 &  -89.61 &  500.00 &  21.30 &   3.63 \\
black-oak      &  46.16 &  -89.32 &  521.51 &  25.91 &   2.28 \\
burley-griffin & -35.30 &  149.07 &  556.00 &  17.00 &   6.64 \\
crystal-lake   &  46.00 &  -89.61 &  501.00 &  20.40 &   0.38 \\
feeagh         &  53.90 &   -9.50 &   15.00 &  44.00 &   3.90 \\
green          &  43.81 &  -89.00 &  243.00 &  72.00 &  29.48 \\
harp           &  45.38 &  -79.13 &  327.00 &  37.50 &   0.71 \\
kivu           &  -1.73 &   29.24 & 1463.00 & 485.00 & 2700.00 \\
klicava        &  50.07 &   13.93 &  294.60 &  35.00 &   0.55 \\
lower-zurich   &  47.28 &    8.58 &  406.00 & 136.00 &  67.00 \\
mendota        &  43.10 &  -89.41 &  258.50 &  25.30 &  39.61 \\
neuchatel      &  46.54 &    6.52 &  429.00 & 152.00 & 217.00 \\
okauchee       &  43.13 &  -88.43 &  269.00 &  28.65 &   4.90 \\
rotorua        & -38.08 &  176.28 &  280.00 &  52.90 & 425.00 \\
sau            &  41.97 &    2.40 &  425.00 &  65.00 &   5.80 \\
sparkling      &  46.01 &  -89.70 &  495.00 &  20.00 &   0.64 \\
stechlin       &  53.17 &   13.03 &   59.80 &  69.50 &   2.23 \\
sunapee        &  43.23 &  -72.50 &  333.00 &  34.00 &  16.55 \\
tarawera       & -38.21 &  176.43 &  300.00 &  87.50 &  41.30 \\
toolik         &  68.63 & -149.60 &  720.00 &  25.00 &   1.49 \\
trout-lake     &  46.03 &  -89.67 &  491.80 &  35.70 &  15.65 \\
trout-bog      &  46.04 &  -89.69 &  495.00 &   7.90 &   0.00 \\
two-sisters    &  45.77 &  -89.53 &  481.00 &  19.20 &   2.91 \\
washington     &  47.64 & -122.27 &    5.00 &  65.20 &  87.60 \\
wingra         &  43.05 &  -89.43 &  254.00 &   6.70 &   1.36 \\
zlutice        &  50.09 &   13.11 &  507.95 &  24.00 &   1.50 \\
zurich         &  47.28 &    8.60 &  406.00 & 136.00 &  68.20 \\
thun           &  46.70 &    7.70 &  558.00 & 212.00 &  48.30 \\
murten         &  46.93 &    7.08 &  429.00 &  45.00 &  22.80 \\
bryrup         &  56.02 &    9.52 &   66.00 &   9.00 &   0.37 \\
\bottomrule
\end{tabular*}
\end{sidewaystable}

\begin{figure}[ht!]
\centering
\includegraphics[width=\textwidth]{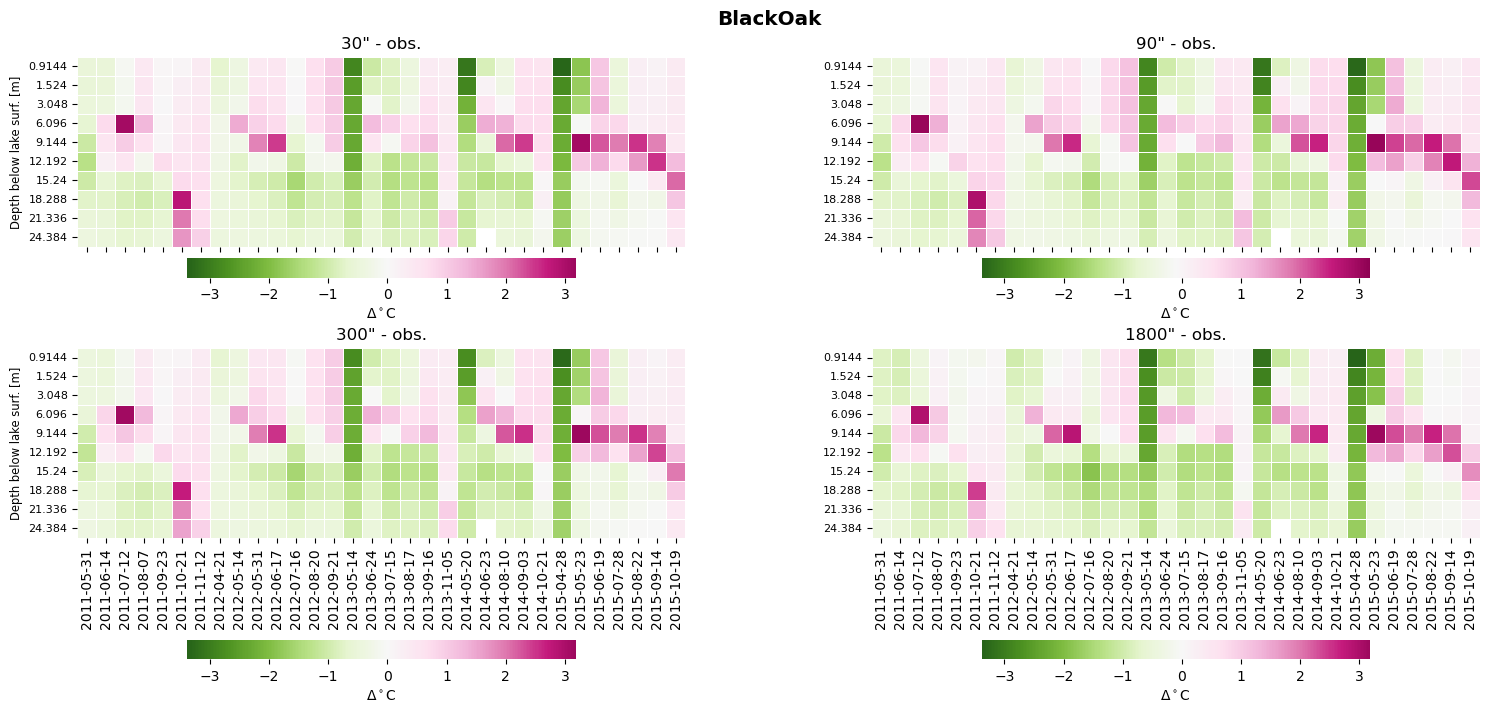}
\caption{Example of depth–time temperature differences between Simstrat simulations and observations for Black Oak Lake, for 1800$^{\prime\prime}$, 300$^{\prime\prime}$, 90$^{\prime\prime}$, and 30$^{\prime\prime}$ spatial resolutions.}
\label{fig:supp:lakes_simst_example_black_oak}
\end{figure}

\begin{figure}[ht!]
\centering
\includegraphics[width=\textwidth]{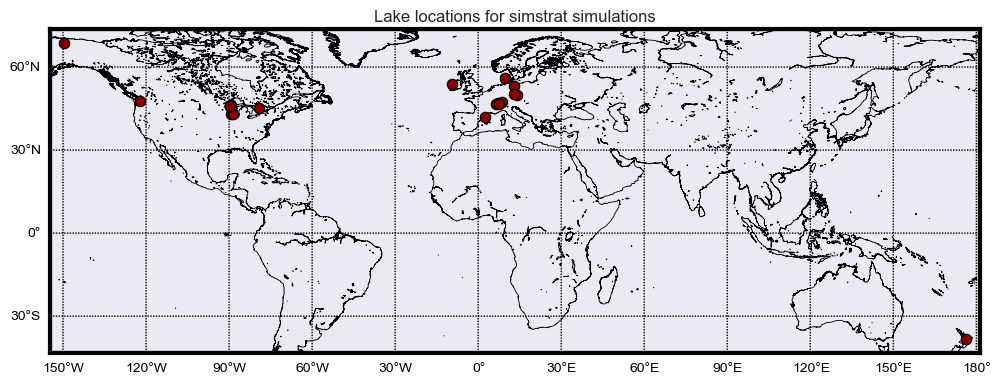}
\caption{Locations of lakes with available temperature profile measurements and Simstrat simulations at 1800$^{\prime\prime}$, 300$^{\prime\prime}$, 90$^{\prime\prime}$, and 30$^{\prime\prime}$ resolutions for the high-resolution sensitivity experiments.}
\label{fig:supp:lakes_simst_obs}
\end{figure}

For each lake, temperature profile observations were available and used to evaluate the performance of Simstrat simulations at multiple spatial resolutions. These observations were not used for the calibration procedure. Rather than aggregating the profiles over time or depth, we compared simulated and observed temperatures at each available time step and depth, and then computed error metrics across all time–depth combinations. Figure~\ref{fig:supp:lakes_simst_example_black_oak} illustrates this comparison for one example lake (Black Oak).

\begin{figure}[ht!]
\centering
\includegraphics[width=\textwidth]{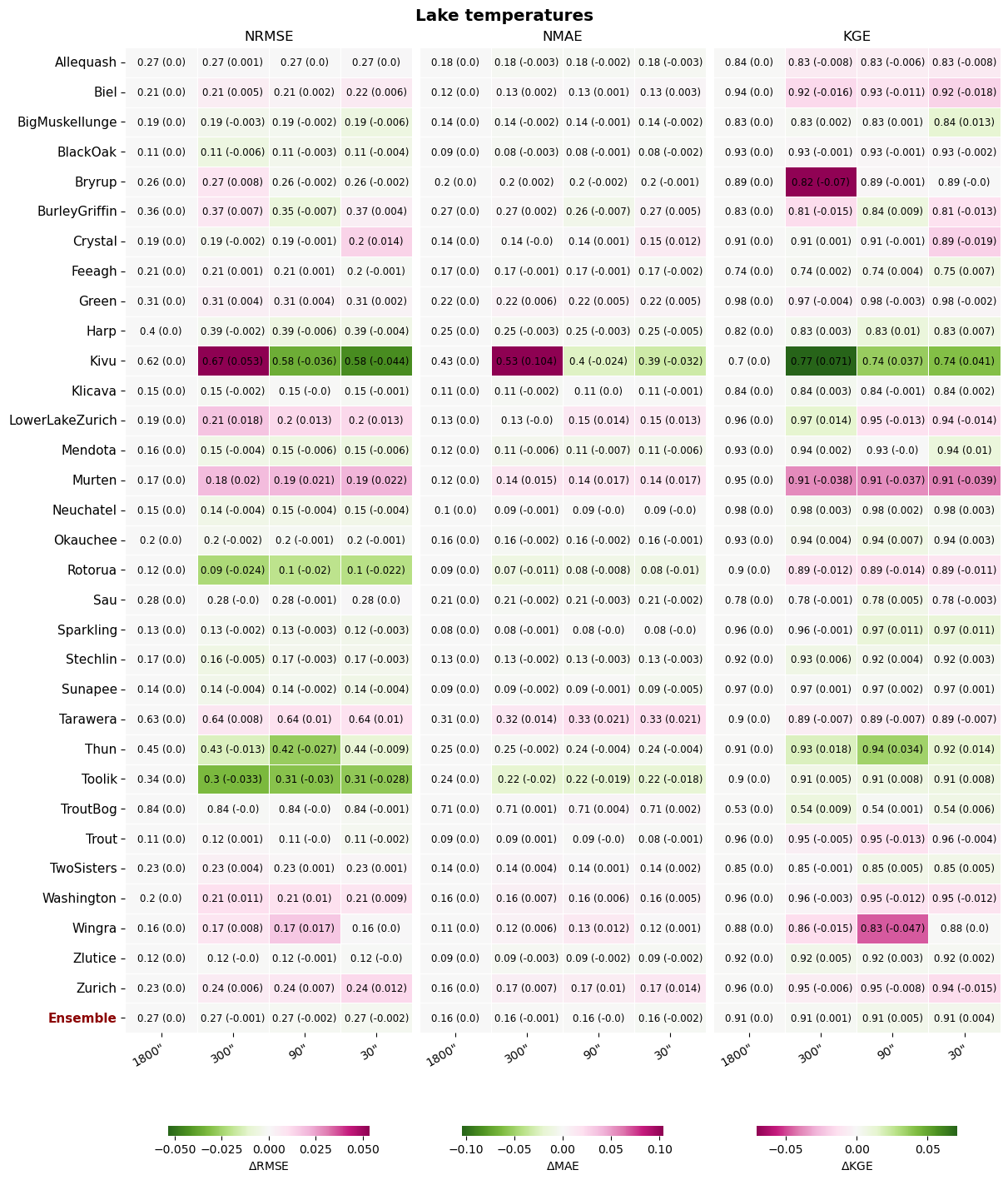}
\caption{Relative NRMSE, NMAE, and KGE of Simstrat lake temperature simulations (1800$^{\prime\prime}$, 300$^{\prime\prime}$, 90$^{\prime\prime}$, 30$^{\prime\prime}$) at all lakes with temperature profile measurements, relative to the 1800$^{\prime\prime}$ baseline. Printed values are absolute metrics; percentages in brackets indicate changes relative to 1800$^{\prime\prime}$. The final row shows the mean across all locations, which is used in the analysis presented in the main paper.}
\label{fig:supp:lakes_simst_eval}
\end{figure}

\paragraph{Evaluation of accuracy}
Figure~\ref{fig:supp:lakes_simst_eval} shows that, overall, the spatial resolution of the input data has little effect on the accuracy of Simstrat lake temperature simulations. While a few lakes (e.g., Toolik, Thun, Rotorua) exhibit clear improvements at higher resolutions, the majority show nearly identical performance across resolutions. This is reflected in the ensemble error metrics, which display virtually no changes between the different resolutions.

\begin{figure}[ht!]
\centering
\includegraphics[width=\textwidth]{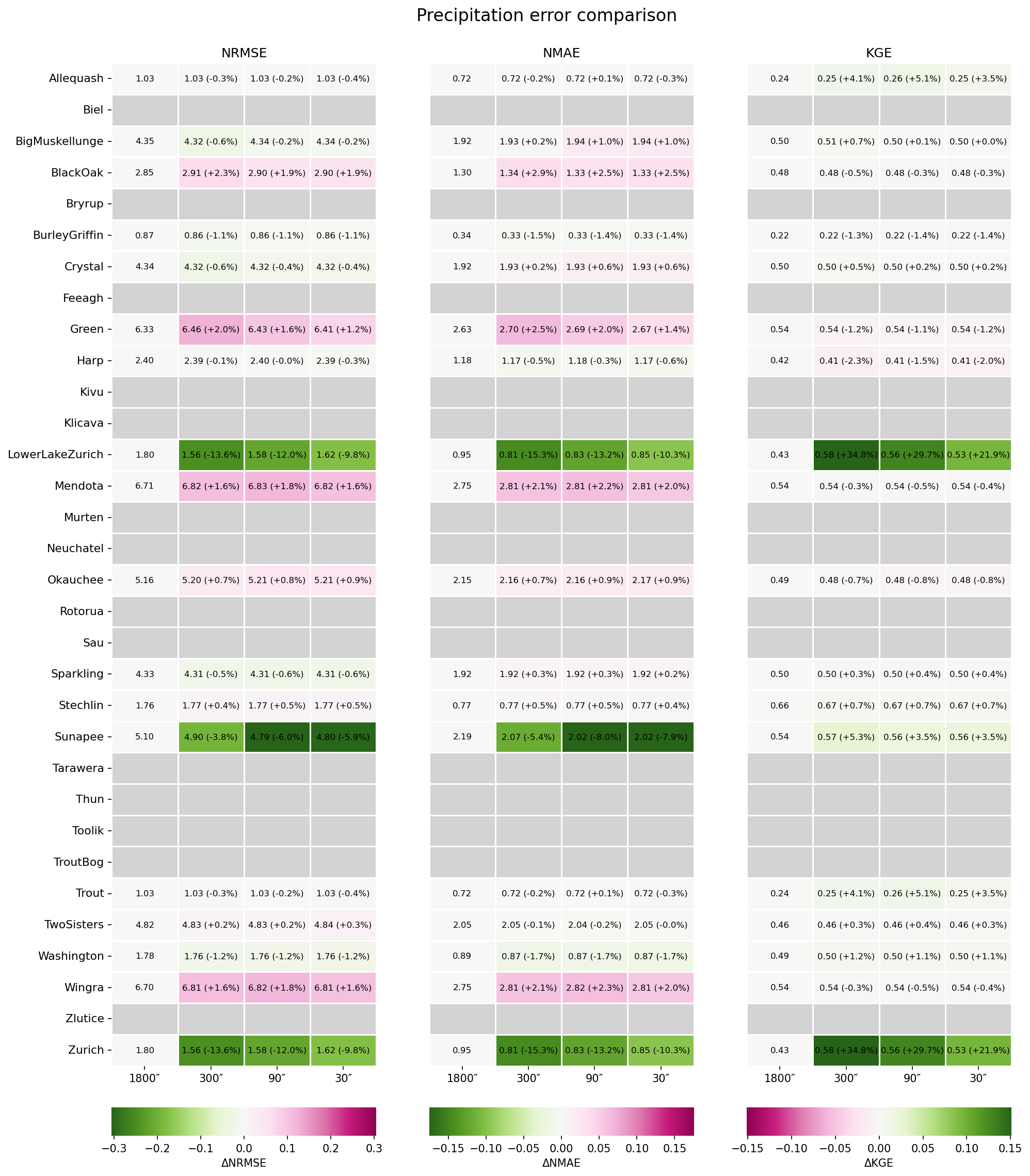}
\caption{Relative NRMSE, NMAE, and KGE of CHELSA daily precipitation at various spatial resolutions, evaluated at lake locations where Simstrat simulations were assessed. The colormap shows values for 300$^{\prime\prime}$, 90$^{\prime\prime}$, and 30$^{\prime\prime}$ relative to 1800$^{\prime\prime}$; annotations show absolute metrics with percentage differences in brackets. Grey indicates no GHCNd precipitation data within $\pm$0.125$^\circ$ of the lake location.}
\label{fig:supp:lake_simst_clim_precip}
\end{figure}

\begin{figure}[ht!]
\centering
\includegraphics[width=\textwidth]{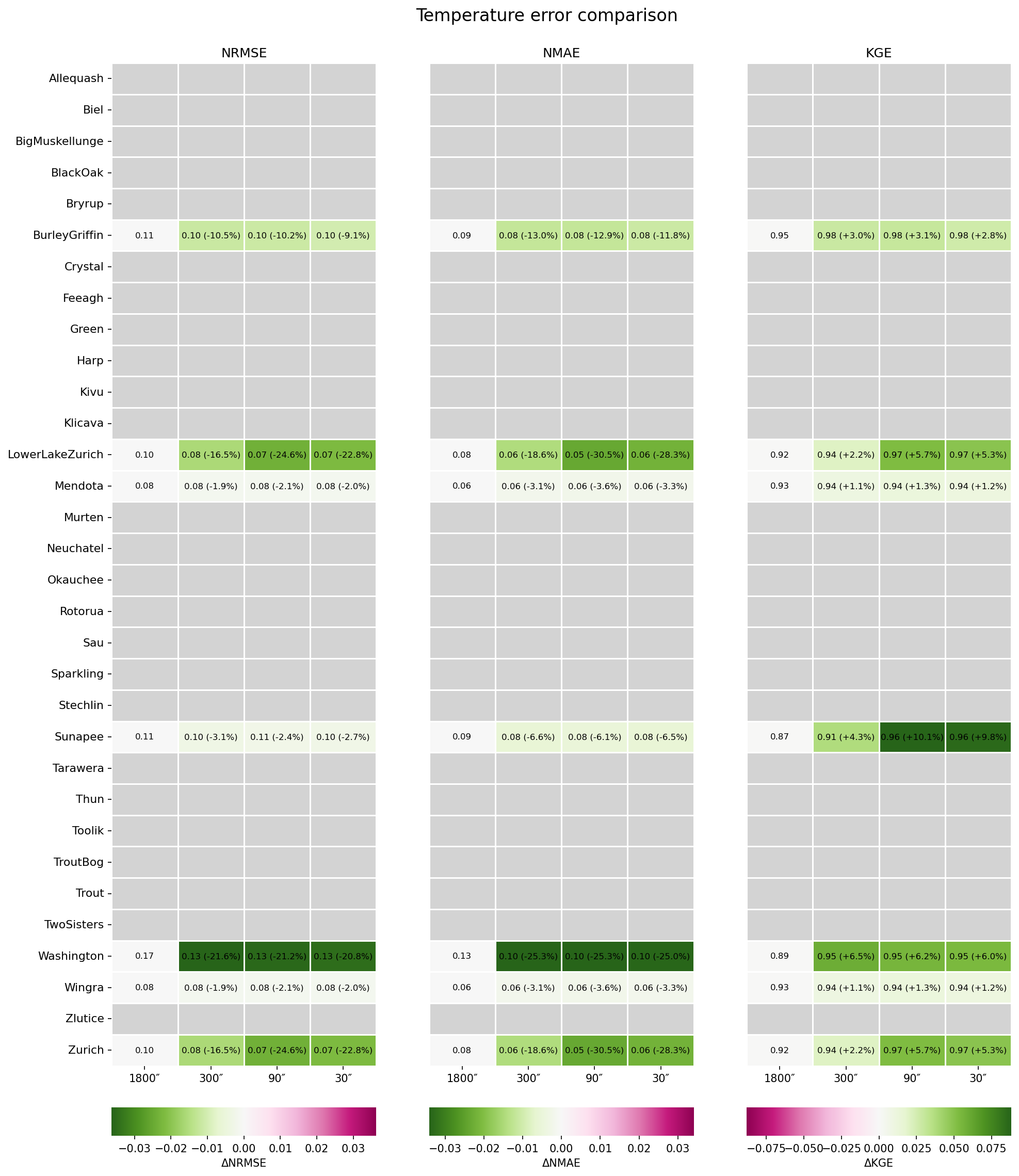}
\caption{Relative NRMSE, NMAE, and KGE of CHELSA daily average temperature at various spatial resolutions, evaluated at lake locations where Simstrat simulations were assessed. The colormap shows values for 300$^{\prime\prime}$, 90$^{\prime\prime}$, and 30$^{\prime\prime}$ relative to 1800$^{\prime\prime}$; annotations show absolute metrics with percentage differences in brackets. Grey indicates no GHCNd temperature data within $\pm$0.125$^\circ$ of the lake location.}
\label{fig:supp:lake_simst_clim_temp}
\end{figure}

\paragraph{Climate data accuracy}
Figures~\ref{fig:supp:lake_simst_clim_precip} and \ref{fig:supp:lake_simst_clim_temp} present the accuracy of various resolution CHELSA precipitation and temperature data at the lake locations used for Simstrat evaluation. For temperature, no location exhibited decreased performance when moving from 1800$^{\prime\prime}$ to higher resolutions. For precipitation, results were more mixed: some locations showed improvements of up to 35\%, while the largest decrease in performance was below 5\%.

\subsubsection{MITgcm}
\begin{figure}[ht!]
\centering
\includegraphics[width=\textwidth]{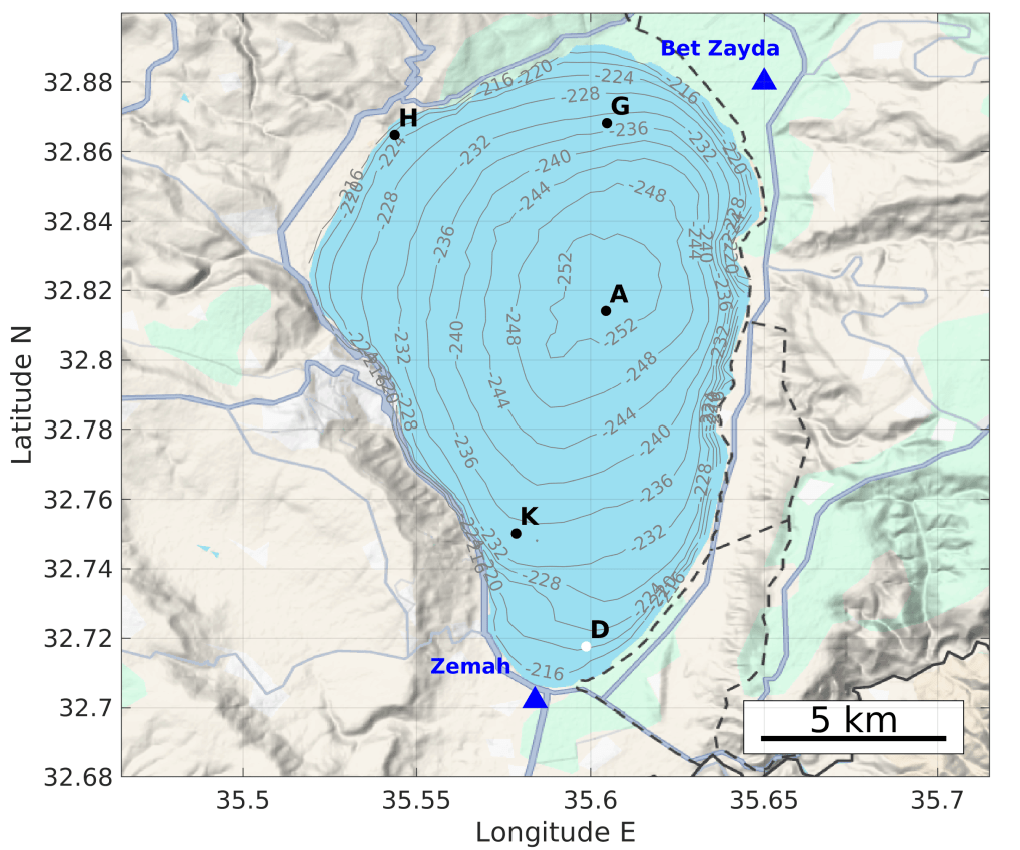}
\caption{Locations of temperature profile measurement stations ("A", "K", "D", "G", "H") within Lake Kinneret used to evaluate MITgcm simulations at 300$^{\prime\prime}$, 90$^{\prime\prime}$, and 30$^{\prime\prime}$ resolutions for the ISIMIP high-resolution sensitivity experiments. Locations of the two meteorological stations (Zemah, Bet Zayda) used for climate data accuracy assessment are also highlighted. Values within the lake show bathymetry.}
\label{fig:supp:lakes_mitgcm_obs}
\end{figure}

MITgcm simulations were performed at 300$^{\prime\prime}$, 90$^{\prime\prime}$, and 30$^{\prime\prime}$ resolutions. Given the size of Lake Kinneret (approximately 21 km long and 13 km wide), 1800$^{\prime\prime}$ simulations were not feasible. Simulations were run on a 400~m grid, with nearest-neighbor interpolation used to obtain input data at the various resolutions. Model performance was evaluated at five measurement locations within the lake ("A", "K", "D", "G", "H") where temperature profile data were available (Figure~\ref{fig:supp:lakes_mitgcm_obs}). Lake Kinneret's maximum depth is 45m and it's exact area is 168 km$^2$.

\begin{figure}[ht!]
\centering
\includegraphics[width=\textwidth]{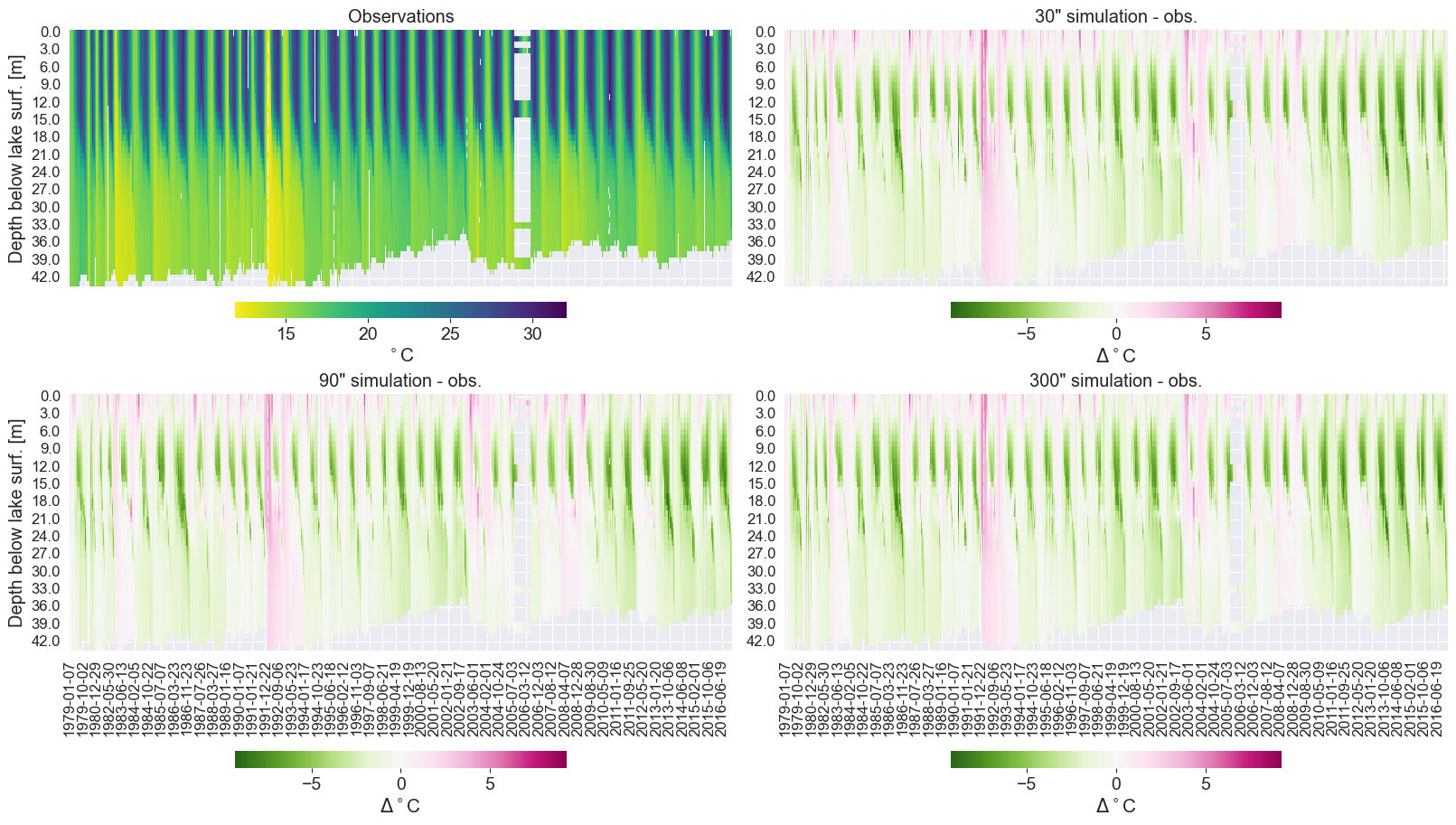}
\caption{Observed temperature per depth and timestep at station A, alongside differences for MITgcm simulations at 30$^{\prime\prime}$, 90$^{\prime\prime}$, and 300$^{\prime\prime}$.}
\label{fig:supp:lakes_mitgcm_eval_A}
\end{figure}

\begin{figure}[ht!]
\centering
\includegraphics[width=\textwidth]{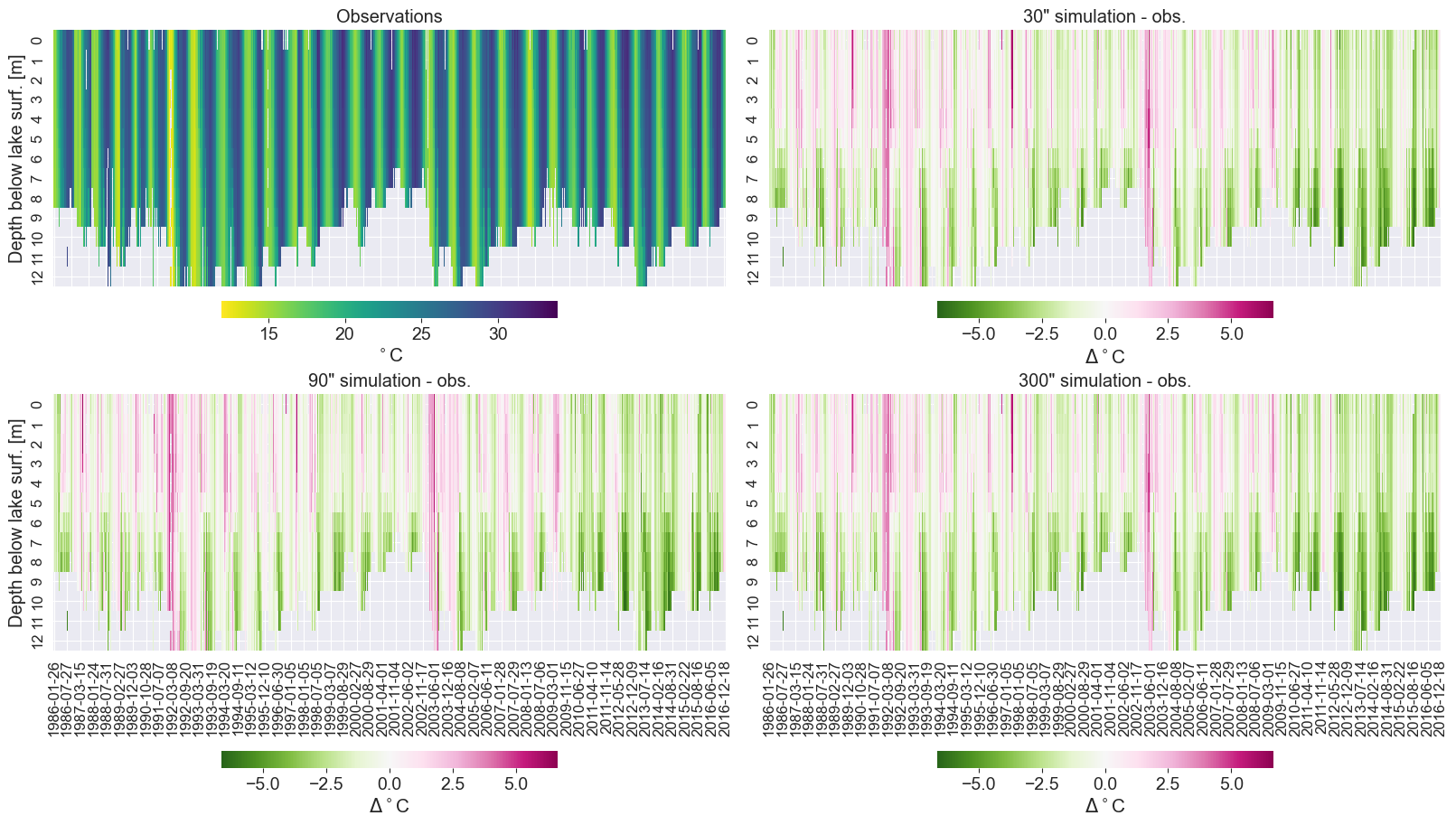}
\caption{Observed temperature per depth and timestep at station D, alongside differences for MITgcm simulations at 30$^{\prime\prime}$, 90$^{\prime\prime}$, and 300$^{\prime\prime}$.}
\label{fig:supp:lakes_mitgcm_eval_D}
\end{figure}

\begin{figure}[ht!]
\centering
\includegraphics[width=\textwidth]{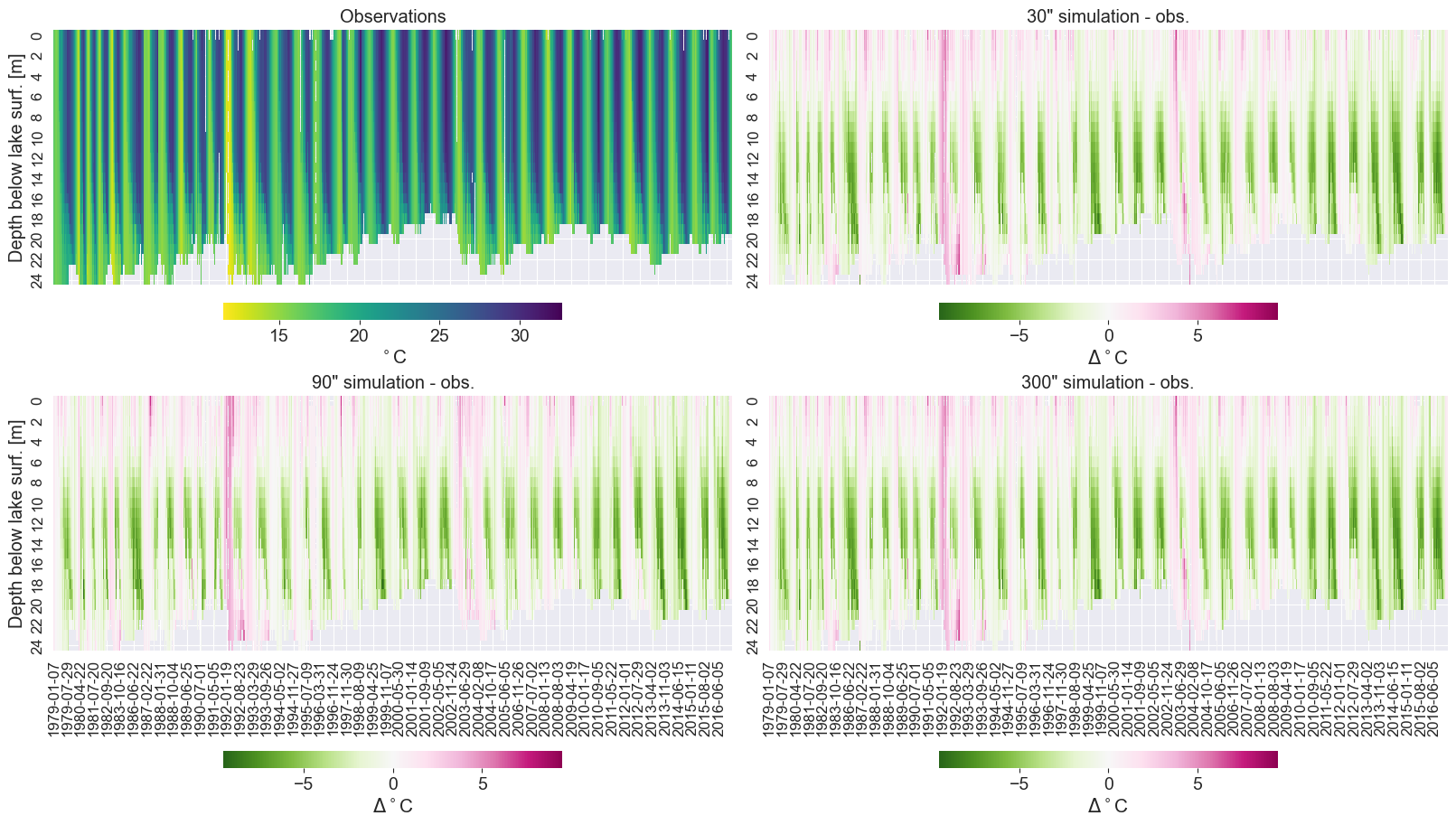}
\caption{Observed temperature per depth and timestep at station G, alongside differences for MITgcm simulations at 30$^{\prime\prime}$, 90$^{\prime\prime}$, and 300$^{\prime\prime}$.}
\label{fig:supp:lakes_mitgcm_eval_G}
\end{figure}

\begin{figure}[ht!]
\centering
\includegraphics[width=\textwidth]{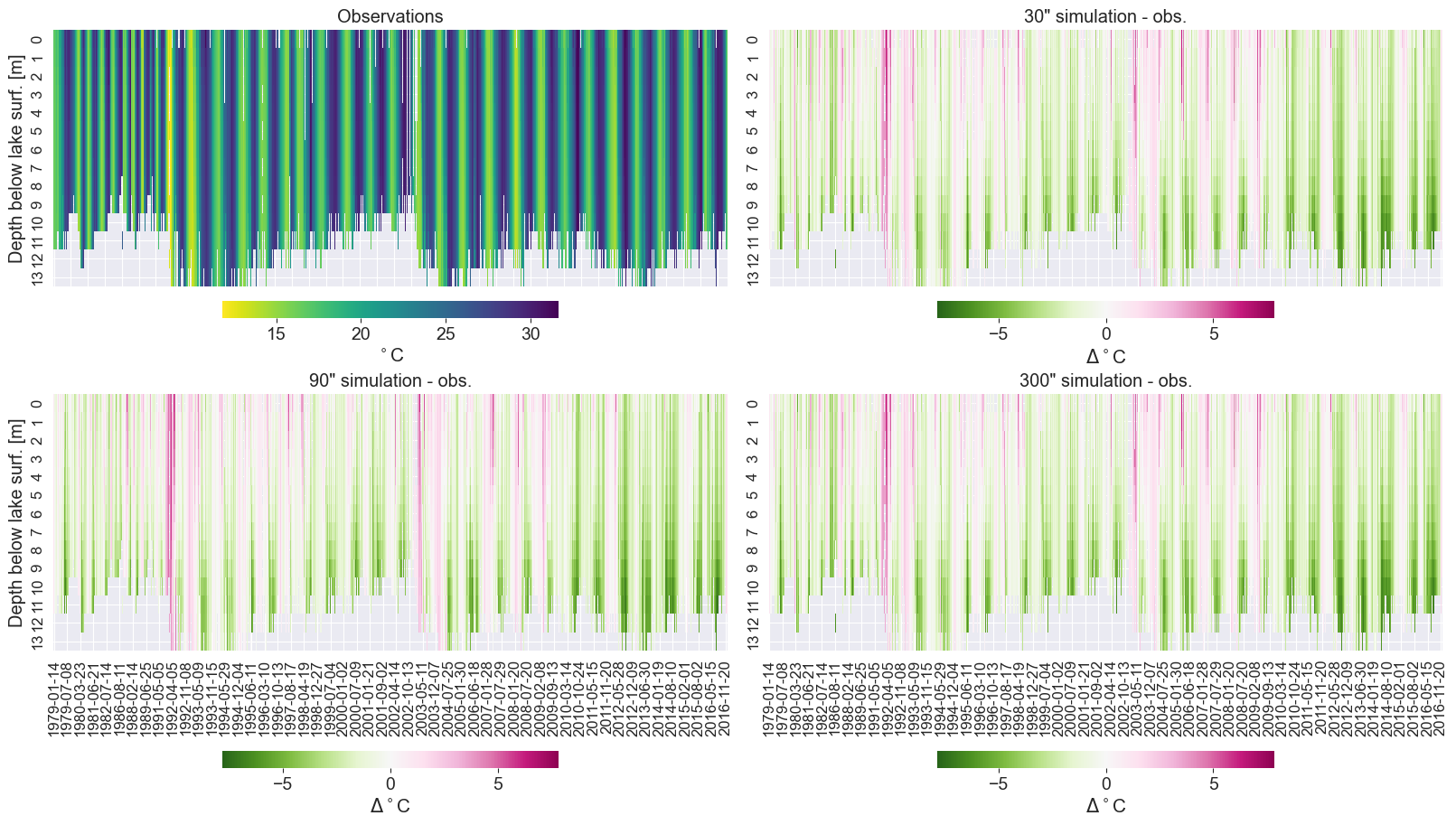}
\caption{Observed temperature per depth and timestep at station H, alongside differences for MITgcm simulations at 30$^{\prime\prime}$, 90$^{\prime\prime}$, and 300$^{\prime\prime}$.}
\label{fig:supp:lakes_mitgcm_eval_H}
\end{figure}

\begin{figure}[ht!]
\centering
\includegraphics[width=\textwidth]{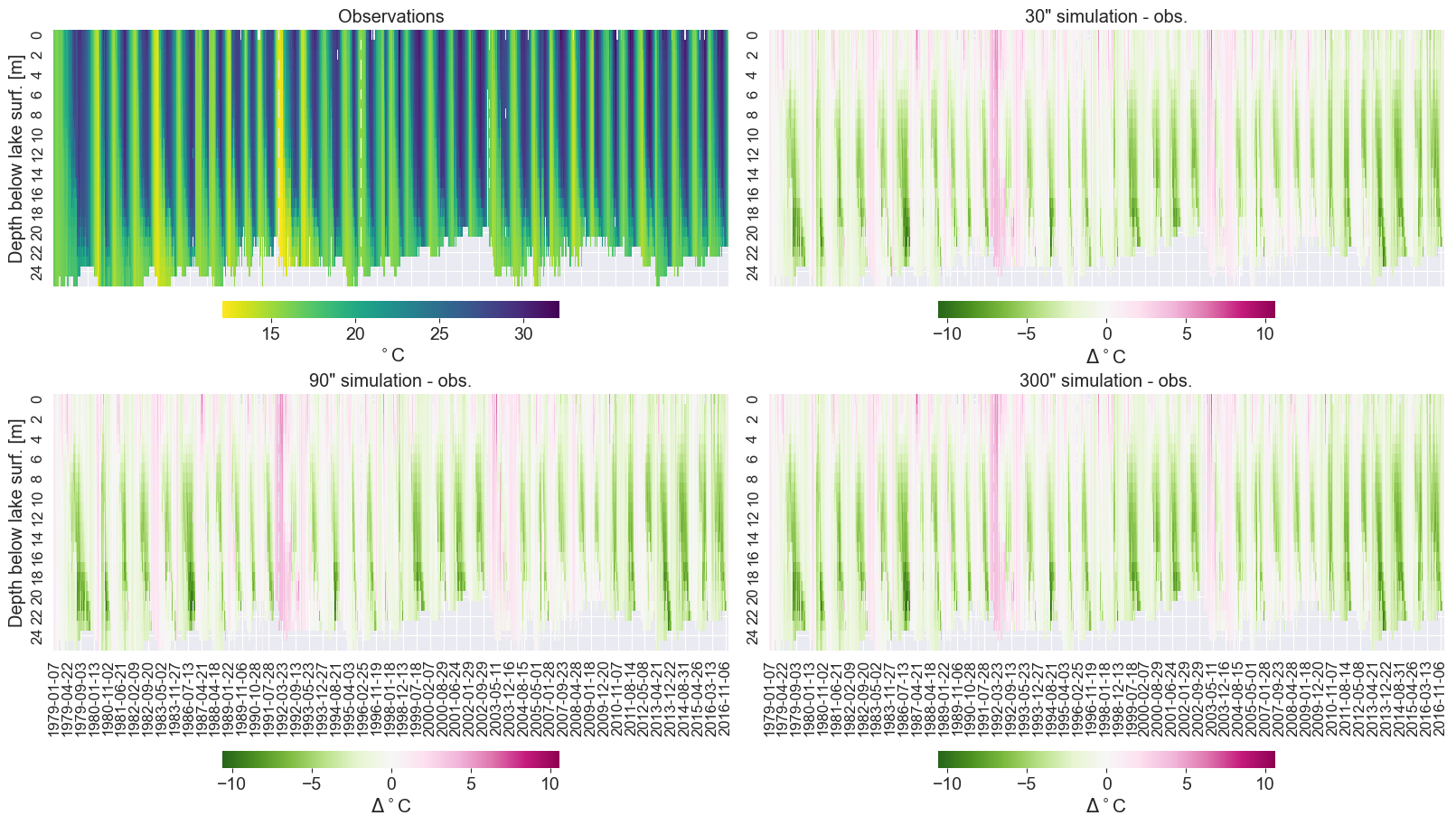}
\caption{Observed temperature per depth and timestep at station K, alongside differences for MITgcm simulations at 30$^{\prime\prime}$, 90$^{\prime\prime}$, and 300$^{\prime\prime}$.}
\label{fig:supp:lakes_mitgcm_eval_K}
\end{figure}

\paragraph{Evaluation of accuracy}
Figures~\ref{fig:supp:lakes_mitgcm_eval_A}--\ref{fig:supp:lakes_mitgcm_eval_K} show observed temperature data per depth and timestep alongside the difference between MITgcm simulations at 30$^{\prime\prime}$, 90$^{\prime\prime}$, and 300$^{\prime\prime}$ and the observations for stations A, D, G, H, and K. Visual inspection indicates that variability between the resolutions is small. 

\begin{figure}[ht!]
\centering
\includegraphics[width=\textwidth]{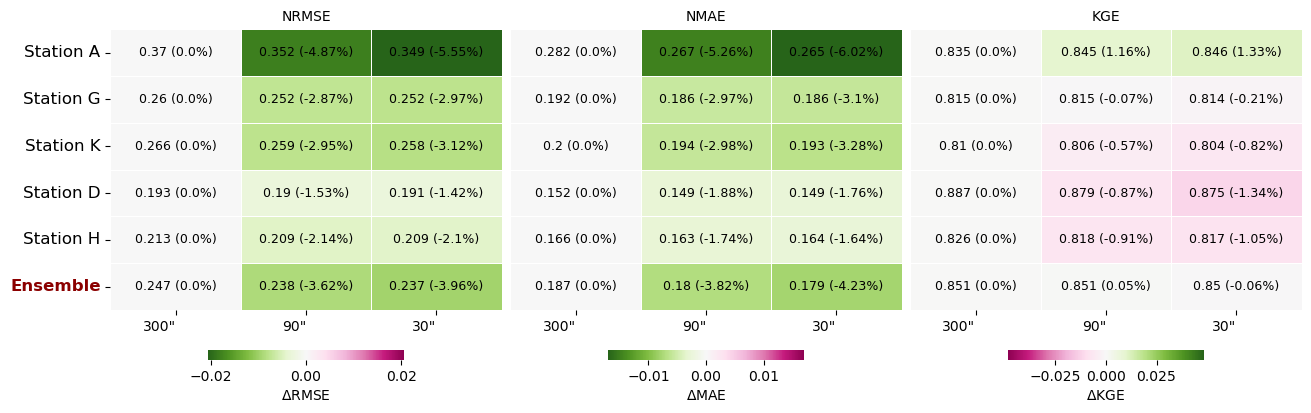}
\caption{Relative NRMSE, NMAE, and KGE of MITgcm lake temperature simulations (300$^{\prime\prime}$, 90$^{\prime\prime}$, 30$^{\prime\prime}$) at all measurement stations, relative to the 300$^{\prime\prime}$ baseline (1800$^{\prime\prime}$ simulations were not available). Printed values are absolute metrics; percentages in brackets indicate changes relative to 300$^{\prime\prime}$. The final row shows the mean across all locations, which is used in the main paper analysis.}
\label{fig:supp:lakes_mitgcm_eval_stats}
\end{figure}

The statistical comparison (Figure~\ref{fig:supp:lakes_mitgcm_eval_stats}) shows a general improvement in NRMSE and NMAE as resolution increases from 300$^{\prime\prime}$ to 90$^{\prime\prime}$ and 30$^{\prime\prime}$, with maximum improvement around 6\%. KGE shows slight decreases at 4 of the 5 stations, with changes ranging from 0 to 1.3\%.

\begin{figure}[ht!]
\centering
\includegraphics[width=\textwidth]{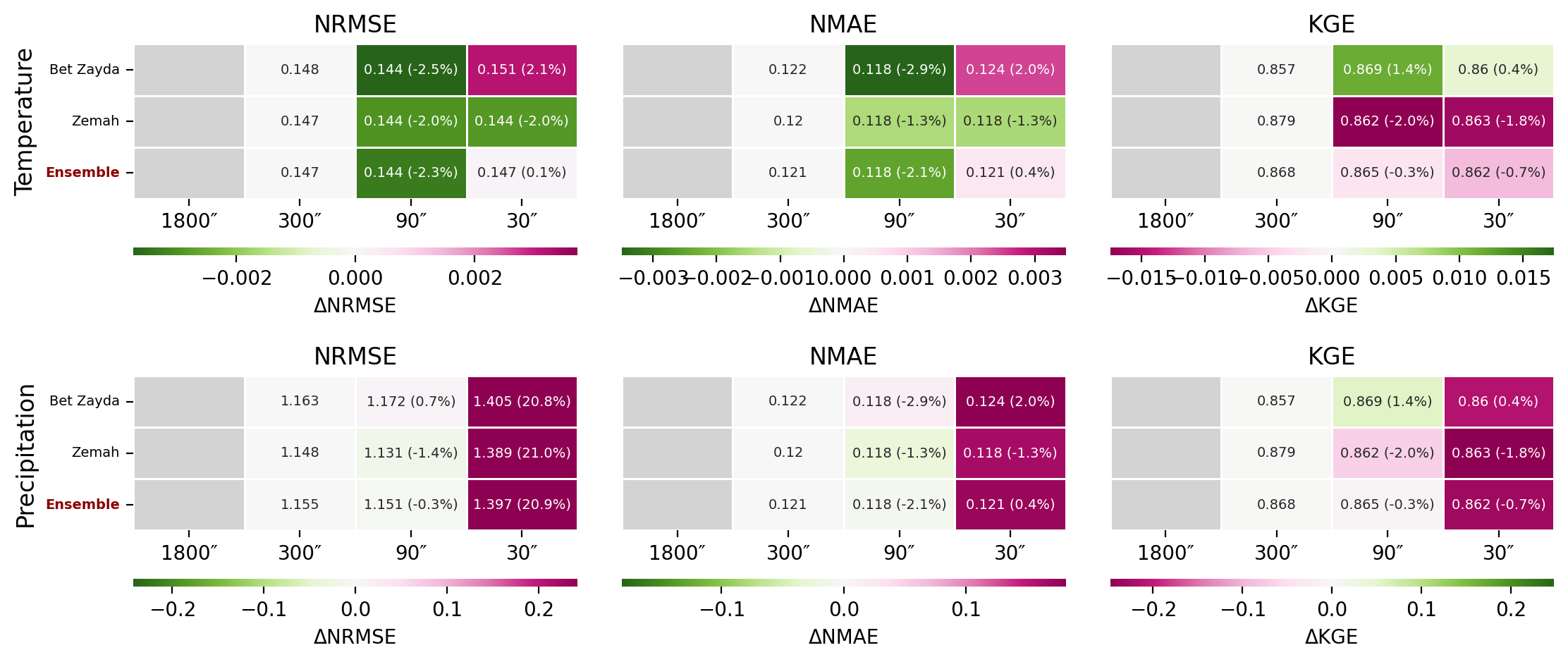}
\caption{Relative NRMSE, NMAE, and KGE of CHELSA daily average temperature at various spatial resolutions, evaluated at two local meteorological stations near Lake Kinneret. The colormap shows values for 90$^{\prime\prime}$ and 30$^{\prime\prime}$ relative to 300$^{\prime\prime}$ (no 1800$^{\prime\prime}$ MITgcm simulations were available); annotations show absolute metrics with percentage differences in brackets.}
\label{fig:supp:lake_mitgcm_clim_eval}
\end{figure}

\paragraph{Climate data accuracy}
Local climate data were available from two meteorological stations near Lake Kinneret: Zemah (Lat: 32°42'7.20"N, Lon: 35°35'2.40"E) and Bet Zayda (Lat: 32°52'48.00"N, Lon: 35°39'0.00"E), shown in Figure~\ref{fig:supp:lakes_mitgcm_obs}. Figure~\ref{fig:supp:lake_mitgcm_clim_eval} shows that differences in temperature accuracy across resolutions were small, with a maximum change of 2.9\%. At Bet Zayda, NRMSE and NMAE slightly decreased when moving from 90$^{\prime\prime}$ to 30$^{\prime\prime}$, while at Zemah, values remained effectively unchanged. For precipitation, performance decreased from 90$^{\prime\prime}$ to 30$^{\prime\prime}$, particularly for NRMSE.

\clearpage

\subsection{Labour}
Labour simulations were conducted at the global scale. However, due to the lack of observational labour data at high-spatial resolution, direct validation is not feasible. This is the case with most if not all socioeconomic data. As such, we used temperature data from the Global Historical Climatology Network daily (GHCNd) as a proxy for evaluation. GHCNd provides climatic data from more than 100,000 stations in 180 countries and territories. Simulations were re-run using GHCNd data, which then served as the reference for model evaluation, since observed labour force data at a 10 km resolution are not available. We spatially merged the station-based data to the geo-references of the labour force dataset to the second administrative level. We use the GHCNd data to compute the change in labour force outcomes at the sub-national level. 

\begin{figure}[ht!]
\centering
\includegraphics[width=0.7\textwidth]{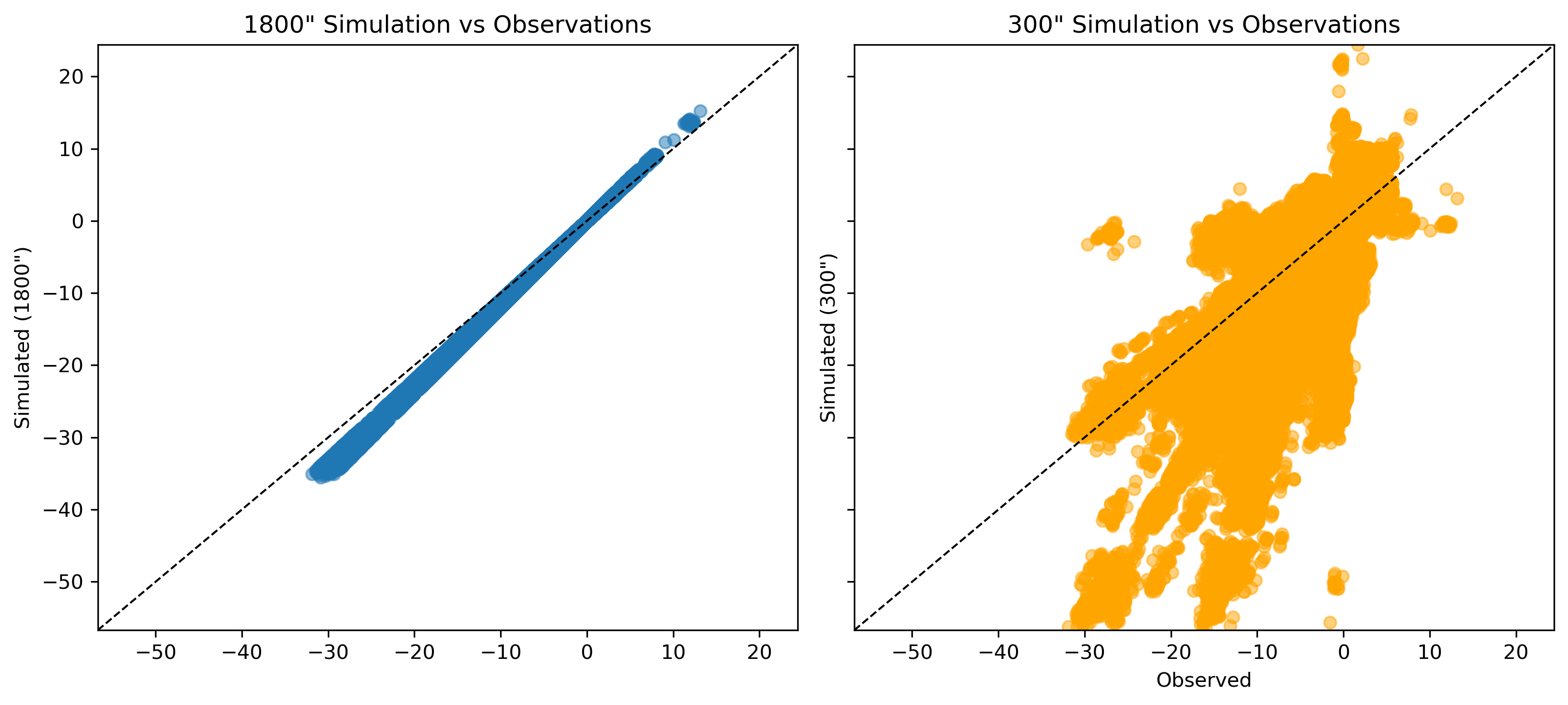}
\caption{Direct comparison of a) 1800$^{\prime\prime}$ labour simulation vs observations and b) 300$^{\prime\prime}$ vs observations.}
\label{fig:supp:labour_comp}
\end{figure}

\paragraph{Evaluation of accuracy}
As shown in Figure~\ref{fig:supp:labour_comp}, simulations at 1800$^{\prime\prime}$ resolution were closer to the reference dataset compared with the 300$^{\prime\prime}$ simulations, indicating that coarser-resolution runs captured effective labour effects more accurately under the current evaluation framework. Nevertheless, for the labour force, high-resolution data offer the advantage of estimating more robust impacts compared with coarser resolutions, and in this context, the increased uncertainty associated with finer resolution compared to the other sectors is not considered problematic.

\section{Part C}

\begin{figure}[htbp]
\centering
\includegraphics[width=\textwidth]{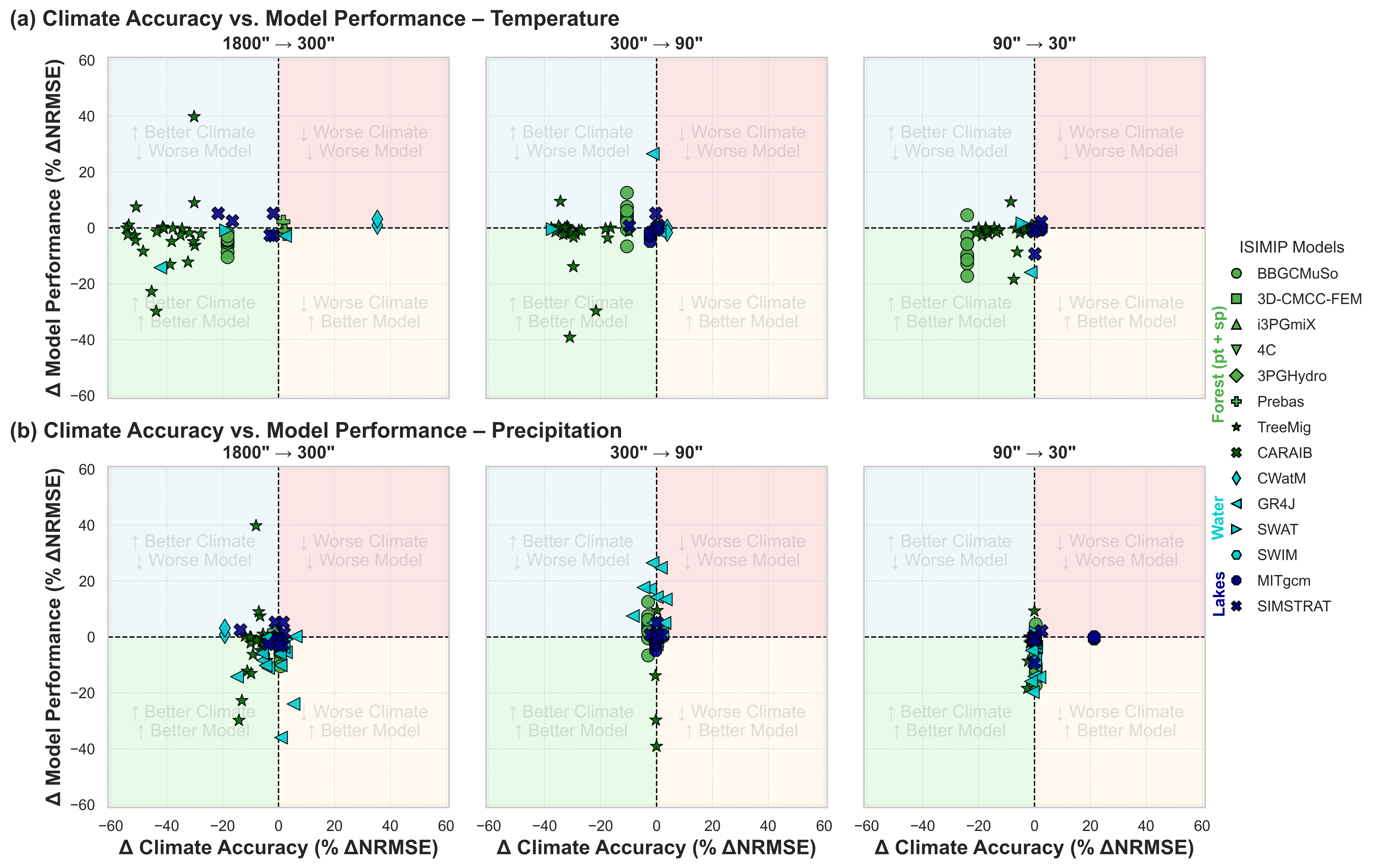} 
\caption{
Per-model comparison of percentage changes in NRMSE for model performance and climate accuracy across spatial resolutions at all model evaluation locations where climate accuracy reference data is available. Panel (a) uses GHCNd temperature station data, and panel (b) uses GHCNd precipitation station data for climate accuracy assessment. In both panels, the subplots from left to right correspond to resolution changes from 1800$^{\prime\prime}$ (approximately 60 km) to 300$^{\prime\prime}$ (approximately 10 km), from 300$^{\prime\prime}$ to 90$^{\prime\prime}$ (approximately 3 km), and from 90$^{\prime\prime}$ to 30$^{\prime\prime}$ (approximately 1 km). Each data point represents a model-variable evaluation pair with co-located GHCNd or local station data. The plots are divided into four quadrants: the green lower-left quadrant represents simultaneous improvement in both climate accuracy and model performance; the blue upper-left quadrant indicates improved climate accuracy but degraded model performance; the yellow lower-right quadrant indicates worse climate accuracy but better model performance; and the red upper-right quadrant indicates a decline in both climate accuracy and model performance.
}
\label{fig:app_all_comp}
\end{figure}

\begin{figure}[htbp]
\centering
\includegraphics[width=\textwidth]{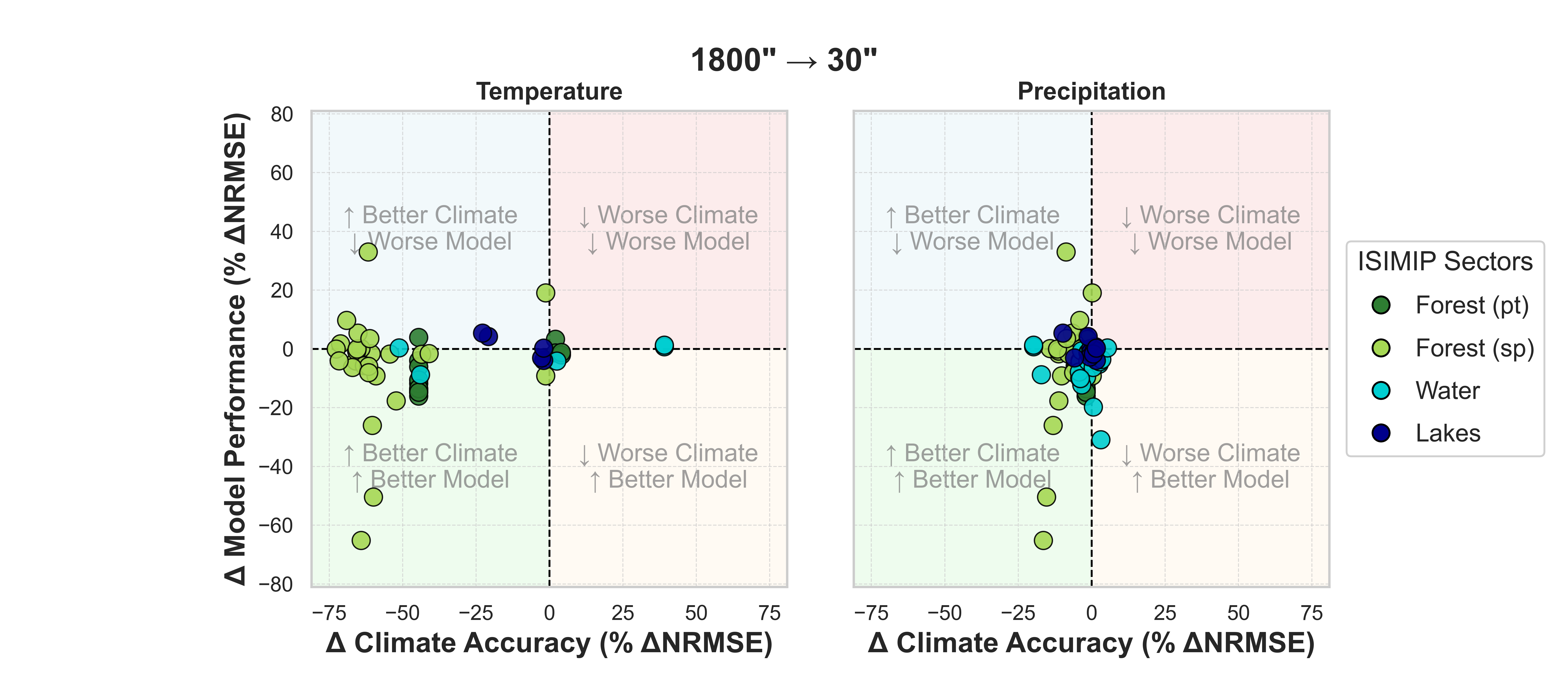} 
\caption{
Per-sector comparison of percentage changes in NRMSE for model performance and climate accuracy across spatial resolutions at all model evaluation locations where climate accuracy reference data is available. Panel (a) uses GHCNd temperature station data, and panel (b) uses GHCNd precipitation station data for climate accuracy assessment. In both panels, we show resolution changes from 1800$^{\prime\prime}$ (approximately 60 km) to 30$^{\prime\prime}$ (approximately 1 km). Each data point represents a model-variable evaluation pair with co-located GHCNd or local station data. The plots are divided into four quadrants: the green lower-left quadrant represents simultaneous improvement in both climate accuracy and model performance; the blue upper-left quadrant indicates improved climate accuracy but degraded model performance; the yellow lower-right quadrant indicates worse climate accuracy but better model performance; and the red upper-right quadrant indicates a decline in both climate accuracy and model performance.
}
\label{fig:app_clim_60_1}
\end{figure}

Part C presents additional diagnostic figures that complement the main manuscript (specifically Fig. 5) by illustrating how changes in climate forcing accuracy with spatial resolution relate to changes in sectoral model performance, now at the per-model rather than per-sector level. First, we show per-model patterns at all locations with co-located climate reference data (Fig.~\ref{fig:app_all_comp}), and then we show a direct comparison between 1800$^{\prime\prime}$ (60 km) and 30$^{\prime\prime}$ (1 km) resolution (Fig.~\ref{fig:app_clim_60_1}).

\clearpage
\bibliography{isimip_high_res_exp}